\documentclass[11pt]{article}

\usepackage{geometry}
\usepackage{fullpage}
\usepackage{graphicx}
\usepackage{pdflscape}
\usepackage{multirow}
\usepackage{ragged2e}

\usepackage{dcolumn}
\usepackage{booktabs,calc}
\usepackage{longtable}
\usepackage{array}
\usepackage{paralist}
\usepackage{verbatim}
\usepackage{subfig}
\usepackage{setspace}
\usepackage{amsmath}
\usepackage{amssymb}
\usepackage{amsthm}
\usepackage{natbib}
\usepackage[linktocpage=true, colorlinks=true, linkcolor=blue, citecolor=black]{hyperref}
\usepackage{enumerate}
\usepackage{authblk}
\usepackage{placeins}
\usepackage{multicol}
\usepackage{array}
\usepackage[usenames, dvipsnames]{xcolor}
\usepackage{url}
\usepackage{listings}

\lstdefinestyle{myListingStyle} 
    {
        basicstyle = \small\ttfamily,
        breaklines = true,
    }

\usepackage{threeparttable}

\newcommand{\Figtext}[1]{%
 \begin{tablenotes}[para,flushleft]
 \hspace{6pt}
 \hangindent=1.75em
 #1
 \end{tablenotes}
}

\newcommand{\Fignote}[1]{\Figtext{\emph{Note:~}~#1}}

\let\estinput=\input

\newcommand{\estwide}[3]{
		\vspace{.75ex}{
			\begin{tabular*}
			{\linewidth}{@{\hskip\tabcolsep\extracolsep\fill}l*{#2}{#3}}
			\\[-1.8ex]\hline 
			\hline \\[-1.8ex]
			\estinput{#1}
			\\[-1.8ex]\hline 
			\hline \\[-1.8ex]
			\addlinespace[.75ex]
			\end{tabular*}
			}
		}

\usepackage{siunitx} 
    \defcitealias{ETH1}{CSA, 2014}
    \defcitealias{ETH2}{CSA, 2015}
    \defcitealias{ETH3}{CSA, 2017}
    
    \defcitealias{MWI1}{NSO, 2012}
    \defcitealias{MWI2}{NSO, 2015}
    \defcitealias{MWI3}{NSO, 2017}
    
    \defcitealias{NGR1}{NIS, 2014}
    \defcitealias{NGR2}{NIS, 2016}
    
    \defcitealias{NGA1}{NBS, 2012}
    \defcitealias{NGA2}{NBS, 2014}
    \defcitealias{NGA3}{NBS, 2019}
    
    \defcitealias{TZA1}{TNBS, 2011}
    \defcitealias{TZA2}{TNBS, 2012}
    \defcitealias{TZA3}{TNBS, 2015}
    
    \defcitealias{UGA1}{UBOS, 2014a}
    \defcitealias{UGA2}{UBOS, 2014b}
    \defcitealias{UGA3}{UBOS, 2016}

\begin{document}
	
\title{Estimating the Impact of Weather on Agriculture\thanks{Correspondence to \href{mailto:tkilic@worldbank.org}{tkilic@worldbank.org} and \href{mailto:smurray@worldbank.org}{smurray@worldbank.org}. A pre-analysis plan for this research has been filed with Open Science Framework (OSF): \href{https://osf.io/8hnz5/}{https://osf.io/8hnz5/}. We gratefully acknowledge funding from the World Bank Knowledge for Change Program (KCP). This paper has been shaped by conversations with Leah Bevis as well as seminar participants at the AAEA annual meetings in Chicago and Atlanta, and participants in presentations at Arizona State University in September 2019, the University of Minnesota in November 2019, the World Bank in January 2020, the $31^{st}$ triennial ICAE conference in August 2021, and Virginia Tech in September 2021. We are especially grateful Alison Conley, Emil Kee-Tui, and Brian McGreal for their diligent work as research assistants and to Oscar Barriga Cabanillas and Aleskandr Michuda for the early help in developing the Stata weather package. We are solely responsible for any errors or misunderstandings.}}

	\author[1]{Jeffrey D. Michler}
	\author[1]{Anna Josephson}
	\author[2]{Talip Kilic}
	\author[2]{Siobhan Murray}
	\affil[1]{\small \emph{Department of Agricultural and Resource Economics, University of Arizona}}
	\affil[2]{\small \emph{Development Data Group (DECDG), World Bank}}

\date{}
\maketitle

\thispagestyle{empty}

\begin{center}\begin{abstract}
		\noindent This paper quantifies the significance and magnitude of the effect of measurement error in remote sensing weather data in the analysis of smallholder agricultural productivity. The analysis leverages 17 rounds of nationally-representative, panel household survey data from six countries in Sub-Saharan Africa. These data are spatially-linked with a range of geospatial weather data sources and related metrics. We provide systematic evidence on measurement error introduced by 1) different methods used to obfuscate the exact GPS coordinates of households, 2) different metrics used to quantify precipitation and temperature, and 3) different remote sensing measurement technologies. First, we find no discernible effect of measurement error introduced by different obfuscation methods. Second, we find that simple weather metrics, such as total seasonal rainfall and mean daily temperature, outperform more complex metrics, such as deviations in rainfall from the long-run average or growing degree days, in a broad range of settings. Finally, we find substantial amounts of measurement error based on remote sensing product. In extreme cases, data drawn from different remote sensing products result in opposite signs for coefficients on weather metrics, meaning that precipitation or temperature draw from one product purportedly increases crop output while the same metrics drawn from a different product purportedly reduces crop output. We conclude with a set of six best practices for researchers looking to combine remote sensing weather data with socioeconomic survey data.
	\end{abstract}\end{center}

	{\small \noindent\emph{JEL Classification}: C83, O13, Q12 \\
	\emph{Keywords}: Remote Sensing, Agricultural Production, Crop Yield Estimation, Sub-Saharan Africa}

\newpage
\onehalfspacing


\section{Introduction}

Accurate measurement in agricultural survey data is key to official agricultural statistics and central to tracking progress towards national and international development goals. Recent work has shown that there is systematic measurement error in agricultural survey data on a range of topics, including cultivated area, crop production, yields, and crop variety, among others \citep{Carlettoetal17, AbayEtAl19, KosmowskiEtAl19, Lobelletal19, GollinUdry21, KilicEtAl21}. Such mismeasurement creates challenges for generating unbiased point estimates, making valid inferences, and, ultimately, for providing sound policy recommendations.

Lacking from this strand of empirical research is an exploration of the consequences of measurement error in remote sensing weather data. The goal of a remote sensing weather product is to document an objective fact: that is, the volume of precipitation or the temperature in a given location at a given time. Inaccuracies introduced by either the sensor (e.g. infrared, microwave, optical), the algorithm used to convert sensor data into rainfall or temperature (e.g. reanalysis, interpolation), or the resolution of the data (e.g. spatial, temporal) means remote sensing products may mismeasure the objective fact. Simply with respect to the ``raw'' weather data, there can be substantial variation in what a remote sensing product reports as the actual rainfall or temperature in a given location. Figures~\ref{fig:rain_res} and \ref{fig:temp_res} show this variation across six remote sensing precipitation products and three temperature products. One precipitation product reports rainfall of 0-5mm in the southeast corner of the grid cell while a different product reports 47-64mm for the same location on the same day. Temperature also varies by remote sensing product, with one product reporting a maximum temperature of 23 degrees Celsius while another reports the maximum temperature that day as 27 degrees Celsius.

In this paper, we quantify the significance and magnitude of the effect of measurement error in remote sensing data from each of the above sources. We test this by modeling the relationship between weather and smallholder agricultural productivity, as measured through nationally-representative, panel household surveys. Besides being a topic of research itself, agricultural production is often used to proxy for a variety of economic outcomes, including economic growth \citep{DescheneGreenstone07}, intra-household bargaining power \citep{CornoEtAl20}, and migration \citep{Jayachandran06}. We combine nine geospatial weather data sets (six precipitation, three temperature) with the geo-referenced household survey data from six Sub-Saharan African countries that are being supported by the World Bank Living Standards Measurement Study – Integrated Surveys on Agriculture (LSMS-ISA) initiative. The objective is to provide systematic evidence on mismeasurement in remote sensing data due to methods used to obfuscate exact household coordinates, metrics used to quantify the weather, and remote sensing data source.

Our goal is to provide guidance to researchers looking to use remote sensing weather data in economic applications regarding what data sources and weather metrics have strong predictive power over a large set of contexts along with which are only useful in highly specific settings. First, we find no clear evidence that different obfuscation methods have an impact on estimates of agricultural production. At this time, publicly available remote sensing weather products are too coarse a resolution for any of the ten obfuscation methods tested to make a substantial difference in which pixel a household ends up in. Second, we find mixed evidence regarding how different metrics used to quantify precipitation and temperature impact estimates of production. Of the 22 metrics (14 rainfall and eight temperature) that we test, only six performed consistently well across different models and countries. These tended to be simpler metrics, such as total seasonal rainfall or mean daily temperature instead of more complicated metrics, such as deviations in rainfall from the long run average or growing degree days (GDD). That said, some metrics performed particularly well in specific circumstances, such as longest dry spell in Niger and the variance of daily temperature in Tanzania. Lastly, we find substantial evidence of variation in how data draw from different remote sensing products correlates with agricultural production. Remote sensing precipitation products that merge gauge and satellite data, such as ARC2, TAMSAT, and CHIRPS, all perform consistently well across a wide variety of settings and tend to produce results similar to each other.\footnote{See Section~\ref{sec:weatherdata} for a full description of each of these products.} Precipitation products that rely on assimilation models, such as ERA5 and MERRA-2, tend to report much higher volumes of precipitation compared to the other remote sensing products. In some cases, data drawn from these different products result in opposite signs for coefficients on rainfall metrics, meaning that precipitation as measured by ARC2, TAMSAT, and CHIRPS purportedly increases crop output while the same metrics drawn from ERA5 and MERRA-2 purportedly reduces crop output. The starkly different results, between precipitation products that merge gauge and satellite data and those that rely on assimilation models, are not present in interpolated gauge data products or in any of the temperature products.

That measurement error exists in remote sensing data is important. There is a large body of literature that relies on remote sensing weather data for identification of causal effects \citep{DellEtAl14, DonaldsonStoreygard16}. This includes important contributions to our understanding of human capital formation \citep{MacciniYang09, ShahSteinberg17, GargEtAl20}, labor markets \citep{Jayachandran06, ChenEtAl17, Kaur19, Morten19}, conflict and institutions \citep{BruckerCiccone11, Sarsons15, KonigEtAl18}, agricultural production and economic growth \citep{MiguelEtAl04, DescheneGreenstone07, BarriosEtAl10, DellEtAl12, YehEtAl20}, intra-household bargaining \citep{CornoEtAl20}, technology adoption \citep{Suri11, Taraz18, JagnaniEtAl21, AragonEtAl21, TesfayeEtAl21}, and extreme weather impacts \citep{WinemanEtAl17, MichlerEtAl19, McCarthyEtAl21a}. Although these studies take care to establish the robustness of their results to different modelling assumptions and potential measurement errors in administrative or survey data, none of them examine the robustness of their results to potential mismeasurement in their choice of remote sensing data source.\footnote{The question of whether weather data is mismeasured is distinct from the question of whether weather is exogenous and thus a valid instrument. This latter question has been addressed by \cite{Deaton10} and \cite{Sarsons15}.}

Complicating these matters is the degrees of freedom a researcher has regarding which of many possible weather metrics to use in their analysis. The choice of how to quantify precipitation or temperature can result in the increased likelihood of Type I errors. The opportunity for $p$-hacking is especially pernicious in economic research, where there is no clear theory on just how rainfall or temperature may increase migration or spur conflict or discourage adoption of a new technology. Should the percentage change in rainfall from one year to the next be used to predict conflict, as in \cite{MiguelEtAl04}, or should total rainfall in a year be used, as in \cite{BruckerCiccone11}? Should deviations in rainfall in the year of one's birth be used to predict human capital formation, as in \cite{MacciniYang09}, or should days when the temperature exceeded 29 Celsius be used, as in \cite{GargEtAl20}? Even in estimating agricultural production, where one might assume a clear and well defined agronomic response, there is little consensus on what matters. Anything from seasonal rainfall to growing degree days (GDD) to deviations in these from the long-run average to a combination of these and their higher moments can and have been used to predict yields \citep{Stallings61, Shaw64, Oury65, DescheneGreenstone07, OrtizBobeaJust12, TackEtAl12, BurkeEtAl15, MichlerEtAl19, McCarthyEtAl21a}. The choice of how rainfall, temperature, or their combination enters the production function appears \emph{ad hoc} and is almost never justified by the researcher. This raises the concern that, absent pre-analysis plans, researchers may engage in data dredging or $p$-hacking in order to find the weather metric that generates the results they want. This issue is particularly problematic in instrumental variable (IV) estimation \citep{BrodeurEtAl16, BrodeurEtAl20}, which is a common use of weather data in economics research.

A final potential source of concern in terms of mismeasurement is that the process of spatially linking remote sensing data with public use data on plots, households, or communities creates a another source of measurement error. This source is introduced by the obfuscation of GPS coordinates of sampled households in order to protect the privacy of survey respondents. Figure~\ref{fig:features} visualizes the mismeasurement introduced by a number of different obfuscation methods, which may shift a unit record's true GPS coordinates from a location experiencing high rainfall to one experiencing drought conditions. While this is not an issue when the unit record is the country, or when the researcher conducted the data collection herself, the economics literature increasingly relies on census, administrative, or third-party data in which the true location of the unit records are obscured. Obfuscating coordinates for the sake of privacy sacrifices accuracy and precision, with respect to estimates, and may result in Type II errors.

In relation to the above, we investigate three specific hypotheses:

\begin{enumerate}
    \item $H_0^1$ - different obfuscation procedures implemented to preserve privacy of farms or households have no impact on estimates of agricultural productivity.
	\item $H_0^2$ - different weather metrics have the same impact on estimates of agricultural productivity.
	\item $H_0^3$ - different measurement technologies for precipitation and temperature have the same impact on estimates of agricultural productivity.
\end{enumerate}

\noindent To test the first hypothesis, we extract remote sensing weather data for LSMS-ISA survey locations using ten different combinations of spatial feature and extraction method. The spatial features represent different obfuscation techniques (aggregation and displacement) and extraction methods (simple, bilinear, and zonal statistics) represent different approaches to dealing with both coarse spatial resolution and uncertainty in location. These data are then combined with household-level data on agricultural production. Following the model specification of \cite{DescheneGreenstone07}, we then estimate agricultural yield functions. This allows us to test if there are differences in the predicted effect of a weather metric on yield across the different obfuscation procedures. Similarly, to test the second hypothesis, we calculate 22 different weather metrics that are commonly used in the economics literature. For rainfall, these include, but are not limited to, mean daily rainfall, total seasonal rainfall, deviations from the long-run average of seasonal rainfall, and the longest intra-season dry spell. For temperature, these include, but are not limited to, measurements such as mean daily temperature, GDD, and long-run deviations in GDD. By testing each of these weather metrics individually and in combination with each other, we can determine which measure of rainfall or temperature is a consistent predictor of yields. And, by extension, which have poor predictive power or are predictive for only a certain crop or in a certain country. Finally, to test the third hypothesis, we conduct all of the above analysis on data from six different remote sensing precipitation data sets and three different remote sensing temperature data sets. This allows us to see if all data sources provide essentially the same results, or if results vary by data source. All told, we run more than 129,600 regressions to identify which weather metrics have strong predictive power over a large set of crops, countries, obfuscation methods, and data sources.

While our approach allows us to compare various combinations of obfuscation or metric or source to another, it does not allow us to compare any combination to the objective fact that a data source is trying to measure. This is because there is no data source that records the objective fact for the households in our data set - or in most data sets used by economists. Unlike with studies that seek to define mismeasurement in land area, harvest, or seed variety planted, we are unable to measure the objective fact of precipitation and temperature. The equivalent to GPS traces of a plot, crop cutting of harvest, or DNA fingerprinting of seed would, in our case, be a rain gauge and thermometer at every household with data recorded daily. While it is clearly technically feasible to collect this sort of data, it is highly atypical data for collection as part of economic research. Hence the reason why economists have come to rely on remote sensing weather data.

As we lack a record of the objective facts of precipitation and temperature for each household, all of our comparisons of mismeasurement are comparisons of one obfuscation/metric/source combination to another. In order to make these comparisons meaningful, we adopt the prior that both rainfall and temperature have a significant impact on crop production. Some metrics, such as the number of days without rain or the variance in temperature, may have a negative impact on yields but we maintain the assumption that this impact will be statistically significant. Our empirical results then allow us to update our prior based on where the weight of the evidence lays. If the vast majority of a set of obfuscation/source combinations show that the number of rainy days significantly increases yield, then we retain our prior belief. We can then claim that an obfuscation/source combination that does not result in a significant coefficient, or results in a significant negative coefficient, on the number of rainy days suffers from mismeasurement. Conversely, if the vast majority of a set of obfuscation/source combinations show that average maximum daily temperature in the growing season has no significant impact on yield, then we reject our prior belief. We can then claim that an obfuscation/source combination that shows a significant impact on yields suffers from mismeasurement.

Our lack of data on the objective facts of precipitation and temperature informed how we implemented our research design. First, we developed a pre-analysis plan and registered it at Open Science Framework \citep{PAP}. While pre-analysis plans have become common in experimental economics, they are still relatively uncommon for binding a researcher's hands when using observational data \citep{JanzenMichler21}. The use of a pre-analysis plan allowed us to pre-define the sources of data for inclusion in the study, what metrics would be tested using what functional forms, and how we would compare results across models in the absence of formal statistical tests. Second, we adopted a blinding strategy to help ensure objectivity in the analysis. In this blinding strategy the authors were divided into two groups: the Data Generating Group  and the Data Analysis Group. Kilic and Murray were in the Data Generating Group  and had full responsibility for extracting the remote sensing data and matching it to the household records in the LSMS-ISA data to create a number of different paired weather-LSMS-ISA data sets.\footnote{For example, in one data set the remote sensing weather data product may be matched with the exact household GPS coordinates in the LSMS-ISA, while in another data set the remote sensing weather data may be matched with low-level administrative centroid.} In these data sets, the source of the weather data and the obfuscation methods was anonymized prior to sharing with the Data Analysis Group. Josephson and Michler made up the Data Analysis Group  and had full responsibility for cleaning the LSMS-ISA production data, running the regressions, and conducting and writing the analysis. The pre-specified analysis was carried out on the anonymized data sets and these results were posted to arXiv.org prior to unblinding \citep{MichlerEtAl20}.\footnote{An initial incomplete draft of the paper was posted to arXiv.org on 22 December 2020 (\href{https://arxiv.org/abs/2012.11768v1}{arXiv:2012.11768v1}). This draft was a placeholder and posted to satisfy project reporting requirements at the World Bank. On 19 August 2021 the complete anonymized draft was posted (\href{https://arxiv.org/abs/2012.11768v2}{arXiv:2012.11768v2}). On 23 August 2021 the Data Generating Group  shared the key to de-anonymize the data with the Data Analysis Group. The version of the paper in hand is the updated paper (v3) containing the same analysis as the anonymized v2 of the paper, just replacing the anonymized placeholders with the actual de-anonymized names.} The generation of data sets in this manner preserves the objectivity of any findings regarding differences in outcomes between different remote sensing products and types of obfuscation. 

Our results have implications for two distinct streams of literature. The first is for the literature on measurement error in economic data. Most research in this stream has focused on  mismeasurement in the context of agricultural production \citep{Carlettoetal17, AbayEtAl19, KosmowskiEtAl19, Lobelletal19, Abay20, AbayEtAl21, GollinUdry21, KilicEtAl21} but a related literature focuses on mismeasurement in everything from time preferences \citep{AndersonEtAl08} to contingent valuation \citep{FlachaireHollard06}. Unlike most of the work on mismeasurement in agricultural production, we lack data on the objective fact that may be measured. As such, we develop new methods to assess the degree of measurement error relying on the richness of the LSMS-ISA data and a combination of a number of different remote sensing data sets. By quantifying the magnitude and significance of measurement error, we can better understand the potential effects of mismeasurement in the remote sensing products commonly relied upon by economists.

Our results also have implications for the large and varied stream of literature that relies on remote sensing weather data for causal identification. This includes seminal and recent work on human capital formation \citep{MacciniYang09, ShahSteinberg17}, labor markets \citep{Jayachandran06, ChenEtAl17, Kaur19, Morten19}, conflict and institutions \citep{BruckerCiccone11, Sarsons15, KonigEtAl18}, agricultural production and economic growth \citep{MiguelEtAl04, DescheneGreenstone07, BarriosEtAl10, DellEtAl12}, intra-household bargaining \citep{CornoEtAl20}, technology adoption \citep{Suri11, JagnaniEtAl21, AragonEtAl21}. Our finding suggests that economists need to be much more careful about the remote sensing data source and the metric they use to measure weather. While our analysis provides practical guidance about what sources and metrics are generally reliable, researchers may need to demonstrate the robustness of their results to different sources and metrics when these are key to their identification strategy.

The paper is organized as follows: in Section 2 we discuss the sources and characteristics of the weather data and the household data. We also provide details on how data was integrated, including specifics on how the blinded data was combined. The section concludes by presenting some descriptive evidence of mismeasurement in the weather data. Section 3 provides details of the pre-analysis plan, specifically our estimation strategy and approach to inference. Section 4 discusses results, first covering differences by obfuscation method, then by weather metric, and finally by remote sensing product. Section 5 provides a summary of the results and recommends six best practices for researchers looking to use remote sensing data in combination with socioeconomic data. We also outline future work and then, in Section 6, conclude.


\section{Data} \label{sec:data}

We use existing, publicly available satellite-based weather data products combined with publicly available unit-record survey data that have been generated as part of the World Bank LSMS-ISA initiative and that are made available through the World Bank Microdata Library. In this section, we first briefly describe the weather data and household data. We then discuss the blinding of the research team and the data integration process. We conclude with a discussion of some descriptive statistics for the combined weather-household data sets.


\subsection{Weather Data} \label{sec:weatherdata}

We use a variety of public domain sources of weather data sets representing different modeling types, input sources and spatial resolutions. Table~\ref{tab:weather} briefly describes each data sources, including the length of record, spatial and temporal resolution, and the type of data recorded. Although there are many possible weather products to consider, we sought to include the remote sensing data products most commonly used by economists. However, to ensure consistency and enable the production of common metrics across the analysis, we imposed two inclusion criteria. The source had to have (1) high temporal resolution (daily) and (2) a minimum 30-year length of record (1987-2017). Daily resolution was necessary for us to calculate some of the most common weather metrics used in the literature today, such as growing degree days (GDD). A consistent length of record was necessary to cover all years of the LSMS-ISA, provide sufficient historical data to measure seasonal deviations in long-term trends, and provide weather data sets all of the same length. This last point was necessary in order to maintain the blinding of the Data Analysis Group  from the data sources. Unfortunately, this criteria meant that some data sources used by economists, such as the various versions of the monthly \emph{Terrestrial Air Temperature and Precipitation} from the Center for Climatic Research at the University of Delaware was excluded.\footnote{Despite its lack of daily data, the product from the Center for Climatic Research at the University of Delaware remains popular among economists. Papers which rely on these data include \cite{CornoEtAl20}, \cite{DellEtAl12}, \cite{ItoEtAl09}, \cite{Jayachandran06}, \cite{Kaur19}, \cite{Sarsons15}, and \cite{ShahSteinberg17}.} While all weather product data used in this analysis are publicly available, accessing the data for analysis remains challenging due to differences in formats and platforms. See Appendix~\ref{sec:appRS} for more details on each remote sensing product and guidance for economists on merging these data with survey data.


\subsubsection{Merged Gauge and Satellite Data}

In the past two decades, improvements in the accuracy of rainfall estimation in data scarce environments have been achieved by merging rain gauge data, which provide site-level observations, with data from meteorological satellites, which provide valuable indirect information at full coverage. We make use of three data products in this category. 

First, the African Rainfall Climatology version 2 (ARC2) was developed to provide long-term daily rainfall monitoring for improved famine early warning systems over Africa \citep{ARC2}. Input sources are daily Global Telecommunications System (GTS) rain gauge data and the Geostationary Operational Environmental Satellite (GOES) precipitation index (GPI) calculated from cloud-top infrared (IR) temperatures. Daily data are produced at $0.1^{\circ}$ resolution. A number of papers use ARC2, including \cite{ArslanEtAl15}, \cite{AmareEtAl18}, \cite{AsfawMaggio18}, \cite{AsfawEtAl19}, \cite{Coromaldi20}, \cite{AlfaniEtAl21}, and \cite{AragonEtAl21}.

A second data product, Tropical Applications of Meteorology using SATellite data and ground-based observations (TAMSAT), similarly makes use of rain gauge information from the GTS and Meteosat thermal infrared (TIR) data \citep{TAMSAT}. Input gauge data are supplemented by collection from local sources and the relationship of rainfall to cloud cover duration (CCD) derived from TIR is optimized for spatio-temporal calibration zones. Data are produced at $0.375^{\circ}$ resolution. \cite{Eberhard-RuizMoradi19}, \cite{MamoEtAl19}, and \cite{MorganEtAl19} are some of the papers which rely on TAMSAT data.  

Finally, the Climate Hazards group InfraRed Precipitation with Station Data (CHIRPS) data set is a global product that builds on the same blending approach of ARC2 and TAMSAT \citep{CHIRPS}. Enhancements include use of a phased approach, incorporating additional climatological products, and ingestion of gauge data from national sources. Daily data are produced at $0.05^{\circ}$ spatial resolution. A large number of papers utilize CHIRPS, including \cite{MichlerEtAl19}, \cite{SheeEtAl19}, \cite{JagnaniEtAl21}, and \cite{MarteyKuwornu21}.


\subsubsection{Assimilation Models}

Assimilation models combine a large number of observations from different sources (e.g. satellites, weather stations, ships, aircraft) to produce a model of the global climate system or a particular atmospheric phenomenon. Reanalyses are assimilation models applied retrospectively over a given time period, to incorporate improvements in algorithms, data processing, and new data sets to support analysis of climate over the long-term. Outputs are inferred or predicted based on the system state and understanding of interactions between model variables. Outputs are differentiated at higher temporal resolution (sub-daily), as well as vertical/pressure levels, but typically lower spatial resolution than other weather data sets. We use two reanalysis data sets for both rainfall and temperature in this analysis: The European Centre for Medium-Range Weather Forecasts ERA5, gridded at $0.28^{\circ}$, and the NASA Modern-Era Retrospective analysis for Research and Applications (MERRA-2), on an irregular grid of $0.625^{\circ}\times0.5^{\circ}$ \citep{ERA5, MERRA2}. The two sources have many inputs in common, although differ in processing routines and models. Papers which use ERA5 include \cite{ArslanEtAl15}, \cite{AsfawMaggio18}, \cite{KalkuhlWenz20}, \cite{SedovaKalkuhl20}, and \cite{JagnaniEtAl21}. Those which use MERRA-2 include \cite{ChenEtAl17}, \cite{ChenEtAl18}, and \cite{LettaEtAl18}.


\subsubsection{Interpolated Gauge Data}

Last, we consider data products produced primarily from gauge data, using only spatial interpolation techniques to produce a continuous surface from observed measurements. The NOAA Climate Prediction Center (CPC) Unified Gauge-Based Analysis of Daily Precipitation and Temperature data sets were created using all information sources available at CPC, GTS daily reports and CPC special collections \citep{CPC}. Extensive pre-processing and cleaning of gauge data includes comparison with contemporaneous data from satellite and other sources. Precipitation data are gridded at $0.5^{\circ}$ resolution using an Optimal Interpolation (OI) technique, which accounts for orographic effects. The global temperature data set is gridded at the same resolution using the Shepard Algorithm. A series of papers use CPC data, including \cite{TennantGilmore20} and \cite{WilliamsTravis19}. 


\subsection{Household Survey Data} \label{sec:householddata}

The World Bank Living Standards Measurement Study - Integrated Surveys on Agriculture (LSMS-ISA) is a household survey program that provides financial and technical assistance to national statistical offices in Sub-Saharan Africa for the design and implementation of national, multi-topic longitudinal household surveys with a focus on agriculture. As detailed below, the analysis leverages data from several rounds of panel household surveys conducted over the last decade by the respective national statistical office in Ethiopia, Malawi, Niger, Nigeria, Uganda and Tanzania, with support from the LSMS-ISA initiative.\footnote{LSMS-ISA has supported nationally-representative cross-sectional surveys in Mali in 2014 and 2017. The data from Mali will be incorporated as part of future work, per the analysis plan \citep{MichlerEtAl19}. While Burkina Faso has also been supported by the LSMS-ISA, the resulting survey data cannot be used in our analysis since the sampled households were not georeferenced.} Table~\ref{tab:lsms} provides a summary of the countries, years, and observations used in the analysis. Appendix~\ref{sec:appHH} provides greater details on each country's sampling frame and data collection process.

In Ethiopia, we use the data from the 2011/12, 2013/14 and 2015/16 rounds of the Ethiopia Socioeconomic Survey (ESS), which has been conducted by the Central Statistical Agency of Ethiopia \citepalias{ETH1, ETH2, ETH3}. The Wave 1 data is representative at the regional level for the most populous regions in the country while Wave 2 and 3 expanded to include 1,500 households in urban areas. After data cleaning to remove urban and non-agricultural rural households, we are left with 7,272 household observations across three survey waves.

In Malawi, the LSMS-ISA data includes two separate surveys: the cross-sectional Integrated Household Survey (IHS), and the longitudinal Integrated Household Panel Survey (IHPS) \citepalias{MWI1, MWI2, MWI3}. This analysis relies on the data from the IHPS, which is representative at the national-, urban/rural-, and regional-level. Data comes from 2010/11, 2013, and 2016/17. A key IHPS design feature is that starting in 2013, the survey attempted to track all individuals that changed locations between the survey waves, and brought into the sample the new households that the movers formed/joined. Our analysis relies on households that did not move vis-\`{a}-vis the baseline interview location, since it is not obvious what the appropriate reference location historical weather data should be drawn from for calculating seasonal deviations from long-term trends. Should the long-term trend be the weather in the location the households lived in the past or the long-term trend in the location the households now resides? After data cleaning to remove tracked and non-agricultural households, we are left with 3,250 household observations across three survey waves.

The LSMS-ISA data from Niger includes two waves, the first from 2011 and the second from 2014 \citepalias{NGR1, NGR2}. The sample is representative at the national and urban/rural-level. Data cleaning and removal of non-agricultural households gives us 3,913 household observations across two survey waves.

In Nigeria, we use the data from the 2010/11, 2012/13, and 2015/16 rounds of the General Household Survey - Panel, which is representative at the national and urban/rural-level \citepalias{NGA1, NGA2, NGA3}. Data cleaning and removal of non-agricultural households yields 8,384 household observations across three survey waves.

In Tanzania, the data stem from the 2008/09, 2010/11, and 2012/13 rounds of the Tanzania National Panel Survey (TZNPS) \citepalias{TZA1, TZA2, TZA3}. Similar to Malawi, the LSMS-ISA from Tanzania sought to track households the split and moved locations. As a result of this, the sample size expands with each wave. Focusing on rural, crop producing households that do no move, we have 5,669 household observations across three survey waves.

In Uganda, we use the data from the 2009/10, 2010/11, and 2011/12 rounds of the Uganda National Panel Survey (UNPS) \citepalias{UGA1, UGA2, UGA3}. As with the other LSMS-ISA data, the Uganda sample was designed to be representative at the national-, urban/rural- and regional-level. We include 5,250 household observations after cleaning and removing non-agricultural households.

For the analysis, we combine data from the six countries and all waves to generate a single cross-country panel data set containing 33,738 household observations. For estimation, we include two measures of agricultural production: yield (kg/ha) of the primary cereal crop and the value (2010 USD/ha) of all seasonal crop production on the farm. As coviariates in our regressions we include a set of inputs: labor, fertilizer, seed, pesticide, herbicide, and irrigation. Exact definitions of output and input variables are in Table~\ref{tab:HHvar}. All cleaning code for each of the 17 data sets is available on \href{https://github.com/AIDELabAZ}{Github}.


\subsection{Data Integration} \label{sec:integration}

Methods of data integration are often overlooked in the process of merging spatial data, in particular weather data, with household surveys. Publicly available data sets obfuscate the exact GPS coordinates of unit-records to ensure privacy. If underlying data sets are fairly smooth and areas of interest are small relative to the resolution of spatial data, then the effect of integration method could be negligible. However, this is not known. As the spatial resolution of remote sensing data improves, these obfuscation methods are likely to matter more. Therefore, we include this as a parameter in our analysis, as well as the choice of spatial feature representation or abstraction.


\subsubsection{Spatial Feature Abstraction}

The minimum spatial resolution of weather data sets in this analysis is 0.0375 decimal degrees, or approximately four kilometers near the equator. At the same time, we find typically low dispersion of households and plots within enumeration areas (less than two kilometers from an EA centerpoint) across surveys used in this analysis. With this in mind, we expect the weather data sets to provide landscape-level contextual information, but not to capture field-level variation.

The most accurate spatial representation used in this analysis is household location. We assess five alternative representations, which will shed light on the effect of choices made in public dissemination of microdata and the usefulness of such data for research. In addition to household location, we use (1) the average of household locations within EA, (2) an anonymized (offset) EA location, (3) the full extent of anonymizing region (buffer of anonymized EA location), (4) the administrative unit associated with lowest-level locality variable in the public microdata, and (5) the administrative centerpoint. 


\subsubsection{Extraction Method}

The spatial features discussed above are a mix of point and polygon, or area, representations (see  Figure~\ref{fig:features}). We evaluate two commonly employed techniques for merging values from raster data to household roster records using these feature types. For point features we extract weather time series using both simple and bilinear methods. The simple method extracts raster cell values by spatial intersection alone, not accounting for the point location within often arbitrary cell boundaries. The bilinear method computes the distance weighted average of values at four nearest cell centers. It is important to note that the bilinear method would be preferred for integration of continuous data like precipitation and temperature. However, as we are aiming to assess the added value of the more complex calculations in this context, both are evaluated. For polygon features we extract values using a zonal mean, or average of all cells overlapped by the polygon. 


\subsubsection{Combining Blinded Data}

As mentioned in the introduction and in the pre-analysis plan, the authors divided themselves into two groups in order to blind the Data Analysis Group from the identity of the remote sensing data \citep{PAP}. The entire team participated in the development and registration of the pre-analysis plan, which included defining the remote sensing products to be used and the extraction methods to be employed. At that point, the Data Generating Group accessed the publicly available remote sensing data for use in the study. They also used the privately available household coordinate data to generate the ten different sets of extraction methods to be used. The actual GPS household location is not part of the publicly available LSMS-ISA data and is known only to a limited number of individuals at the World Bank. 

After pre-processing, the Data Generating Group extracted the relevant remote sensing data for the LSMS-ISA households based on the ten data extraction methods for all nine remote sensing sources. This generated time series data sets of daily precipitation or temperature from January 1, 1983 until December 31, 2017. For each country in each of these years, a growing season was defined based on FAO recommendations.\footnote{For more details on the definitions of growing seasons in each country, see Appendix~\ref{sec:appRS_gs} and Table~\ref{tab:growseason}.} And so, for each of the 17 LSMS-ISA country-wave household data sets, this generated 90 remote sensing weather data sets (six precipitation sources $+$ three temperature sources $\times$ ten extraction methods). The time series weather data sets include daily observations and the unique household identifiers made part of the publicly available LSMS-ISA data. Data sets were named and labelled \texttt{x0, ..., x9} for each extraction method, \texttt{rf1, ..., rf6} for each precipitation data source, and \texttt{tp1, ..., tp3} for each temperature data source. These 1,530 blinded time series data sets were then shared, via a secure server, with the Data Analysis Group. 

The Data Analysis Group then processed each of the time series weather data sets using a user-written Stata package \texttt{wxsum} which is available via \href{https://github.com/AIDELabAZ}{Github}. This package processes daily precipitation or temperature data and outputs up to 22 different weather metrics. See Table~\ref{tab:Wvar} for a complete list of weather metrics used in the analysis. These weather metrics from each of the 1,530 weather data sets were then merged to the relevant country-wave LSMS-ISA data set using the unique household identifier (90 weather data sets per country-wave data set). All country-wave data sets containing the production data and the weather metrics from each remote sensing source and extraction method were then appended to create a single panel data set covering all countries, waves, remote sensing sources, and extraction methods. Table~\ref{tab:sourcesetc} summarizes the scope of the resulting data.

The Data Analysis Group then conducted all of the analysis on the blinded data set, posting the results to arXiv.org on 19 August 2021. On 23 August 2021, the Data Generating Group shared the key so that the Data Analysis Group could de-anonymize the data. Version 2 of this paper (\href{https://arxiv.org/abs/2012.11768v2}{arXiv:2012.11768v2}) refers to all results based on their randomly assigned identifier (\texttt{x0, ..., x9}; \texttt{rf1, ..., rf6}; and \texttt{tp1, ..., tp3}). The current version of this paper (version 3) presents the same analysis as version 2 but replaces the randomly assigned identifiers with the actual extraction methods and names of remote sensing sources.


\subsection{Descriptive Statistics} \label{sec:summarystats}

Based on the combined household and weather data, we provide some general descriptive statistics on each of the two data sources, broadly defined.


\subsubsection{Weather Descriptives}

In addition to estimating mismeasurement from the choice of remote sensing products and extraction method, we examine variation arising from the choice of different weather metrics. In total, we test 22 different ways to measure precipitation and temperature. A complete list of these variables with their exact definitions are in Table~\ref{tab:Wvar}. 

To provide a sense of the variation in measurement induced by the remote sensing product, we present summary figures for a subset of the 22 weather metrics we examine. While our focus in the econometric analysis is country, because that is the relevant unit of analysis for most economists, we present descriptive evidence by global agro-ecological zone as these agro-climatic divisions are more relevant when seeking to assess difference in weather patterns (absent the connection to household data). The Food and Agriculture Organization of the United Nations (FAO) and the International Institute for Applied Systems Analysis (IIASA) develop agro-ecological zones to assess agricultural resources and potential. The measurement integrates available land and water resources, agro-climatic resources, and general conditions about agricultural suitability to determine a set of zones, which differ across these metrics. We use a revised AEZ surface for Africa \citep{Sebastian09} that incorporates high resolution climatology (WorldClim) and elevation (SRTMv4) surfaces to produce a downscaled version of the agro-ecological zones. In our data, six agro-ecological zones are covered, including (1) tropic-warm/semi-arid, (2) tropic-warm/sub-humid, (3) tropic-warm/humid, (4) tropic-cool/semi-arid, (5) tropic-cool/sub-humid, and (6) tropic-cool/humid.

Figure~\ref{fig:density_aez_rf} presents the distribution of total season rainfall (measured in millimeters), by remote sensing product and agro-ecological zone. Within any given agro-ecological zone there are substantial differences in the distribution of rainfall as reported by each remote sensing product. In tropic-warm/semi-arid regions all products report about the same mean (between 550 and 650mm) but some report a bimodal distribution while others report a unimodal distribution. At the upper end of the distribution, these differences grow, with MERRA-2 and ERA5 reporting maximum values of greater than 1700mm while all other sources report maximums of less than 1400mm.

Among the six agro-ecological zones, there tends to be the greatest degree of agreement between remote sensing sources for tropic-warm regions. In the three tropic-warm sub-regions, means from each remote sensing source tend to be close to each other and the largest maximum value is usually less than double the smallest maximum value. By comparison, there is substantial variation in reported rainfall for the three tropic-cool sub-regions. In these agro-ecological zones, the largest mean is nearly double the smallest mean. In tropic-cool/semi-arid locations, ERA5 reports a mean seasonal rainfall of 859mm (maximum 2,579mm) while CPC reports a mean of 454mm (maximum 1,006mm). In tropic-cool/sub-humid locations, ERA5 reports a mean of 1446mm (maximum 6968mm) while CPC reports a mean of 706mm (maximum 1,714mm). In tropic-cool/humid locations, ERA5 reports a mean of 1435mm (maximum 4,363mm) while CPC reports a mean of 766mm (maximum 1,684mm). These are differences in mean value of several feet and differences in maximum value of up to 18 feet.

Figure~\ref{fig:norain_aez_rf} further explores these differences by estimating the mean number of days without rain reported by each remote sensing product in each season. Mean estimates are generated using a fractional-polynomial and graphs include $95\%$ confidence intervals on the mean estimates. CHIRPS, CPC, and ARC2 frequently report a similar number of days without rain, as do MERRA-2 and ERA5. TAMSAT is often similar to CPC and ARC2, though sometimes deviates from the the other remote sensing products. Typically, the measurements from CHIRPS, CPC, ARC2, and TAMSAT suggest that there are more days without rain, relative to the measurements from MERRA-2 and ERA5. Unlike total seasonal rainfall, where we saw differences in accuracy based on warm or cool regions, for days without rain we see the largest differences in accuracy between semi-arid sub-regions and the sub-humid and humid sub-regions. For tropic-cool/semi-arid and tropic-warm/semi-arid, MERRA-2 and ERA5 report about $50\%$ fewer days without rain relative to the other three products (about 120 versus 60 days in tropic-cool/semi-arid and about 200 versus 100 days in tropic-warm/semi-arid). For the four sub-humid and humid zones, MERRA-2 and ERA5 report about $70\%$ fewer days without rain, meaning the other products are reporting three times as many days without rain. As an example, in tropic-cool/sub-humid regions, MERRA-2 reports an average of only 33 days without rain while ARC2 reports an average of 107 days without rain.

In Figure~\ref{fig:density_aez_tp} we present the distribution of mean season temperature (measured in degrees Celsius), by remote sensing product and agro-ecological zone. Compared to the distribution of total seasonal rainfall, the figures show much tighter distributions around average temperature. This is especially true in tropic-warm/semi-arid and tropic-cool/sub-humid agro-ecological zones. While there is variation in the distributions, the three temperature products report very similar means, minimums, and maximums. Values are always within two degrees of each other and in the vast majority of cases they vary by less than a degree.

In Figure~\ref{fig:gdd_aez_tp} one can see that there is more variation in reported temperatures by remote sensing product when we calculate growing degree days (GDDs). But even here, the differences are not as substantial as was seen in terms of rainfall. In the three tropic-cool sub-regions, all three temperature products report essentially the same number of GDDs. In the tropic-warm sub-regions predicted mean GDDs do differ significantly, with CPC consistently reporting fewer GDDs than the other two products. However, these differences may not be agronomically meaningful. Mean GDDs differ by about 10 days out of around 180 days in tropic-warm zones.

Summarizing the descriptive evidence, there is clearly more mismeasurement of rainfall than of temperature. Given that the amount of precipitation and the temperature in a specific location on a specific day is an objective fact, any difference in reported values is evidence of mismeasurement. We lack documentation of what the true rainfall and temperature values are for any of the households in our data, so we cannot determine which remote sensing product is wrong or which product is less accurate or which product contains greater measurement error. But what we can describe is how remote sensing products compare to each other. MERRA-2 and ERA5 tend to report more rainfall than the other three products. Differences between products can be more pronounced when comparing across warm/cool zones or when comparing across semi-arid/sub-humid/humid sub-regions, depending on which weather metric one adopts. The takeaway is that one may end up with very different outcomes depending on what precipitation product one selects, what metric is used to measure precipitation, and what region of Africa is being studied.

The mismeasurement we see in rainfall is not as pronounced when we look at temperature. Here the three temperature products tend to agree with each other on what the temperature was in a given location on a given day. That is not to say that there is no mismeasurement in remotely sensed temperature, since all three products could be off in the same way. Short of knowing the actual temperature, what we can say is that one would expect results to be similar when using temperature, regardless of the temperature products or region of Africa is being studied.


\subsubsection{Household Descriptives}

We seek to provide systematic evidence on mismeasurement in remote sensing data  by modeling the relationship between weather and smallholder agricultural productivity. Because of this, our focus within the LSMS-ISA household data is on variables relevant to crop production. We examine both total farm production and production of the primary cereal crop within a country. The primary crop is maize in all countries except for Niger, where millet is the primary crop.

We present summary statistics of our dependent variables and controls by country, aggregated across all households for all waves in Table~\ref{tab:sumstattab}. Panel A presents summary statistics on output and inputs for the total farm (aggregating across all seasonal crops). The average value of total farm production varies from a low of $\$168$ (Uganda) to a high of $\$664$ (Nigeria), with most countries falling in the $\$100$ to $\$300$ range.\footnote{All monetary values are converted to U.S. dollars at current exchange rates and then appreciated/depreciated to 2010 dollars. We use data from the World Bank Development Indicators, last updated on 9 April 2020, to make the exchange rate and inflation conversions.} The value of total farm production is not particularly informative given the large differences in farm size. On average, farms in Ethiopia, Malawi, and Tanzania tend to be small - less than two hectares (ha). By comparison, the average farm in Uganda is over five ha and the average farm in Niger is 11 ha. Because of these variations, farms in Niger are, on average, the least productive among the LSMS-ISA countries, with a per ha value of only $\$60$. By comparison, farms in Nigeria, on average, produce $\$680$ in value per ha. Labor also varies substantially across countries, with farms in Ethiopia using an average of 434 days per ha, while farms in Niger use less than 100 days of labor per ha. Similarly, fertilizer use rates vary across countries, with less than one kg per ha used, on average, in Uganda compared to nearly 100 kg per ha used, on average, in Nigeria. In terms of other purchased inputs, Nigerian farms tend to apply more than the other countries. About $20\%$ of farmers in Nigeria apply pesticide and herbicide, while less than $10\%$ of farmers in most of the other countries apply these chemicals. Across all countries, irrigation levels are very low, typically less than five percent.

Panel B presents summary statistics on the primary crop production (millet in Niger, maize in all other countries). The patterns present in our cross country comparisons of total farm production are mostly borne out in our comparison of primary crop production. Nigeria has, on average, the largest harvests, highest yields, and the highest use of purchased inputs like fertilizer, pesticide, and herbicide. On average, Niger has the largest land area dedicated to the primary crop and Ethiopia, on average, expends the most labor per ha. Uganda has, on average, the smallest harvests and yields and applies the least amount of fertilizer.

Since total farm values are given in constant USD and primary crop values are in kg, we cannot directly compare most values. However, we can compare land area for the farm with land area dedicated to the primary crop. In Ethiopia, Nigeria, and Uganda, land area to the primary crop makes up around $30\%$ of the total farm area, implying fairly diversified production practices. By contrast, the primary crop in Malawi, maize, commands $92\%$ of total farm land, suggesting Malawian agriculture is highly specialized. Niger $(44\%)$ and Tanzania $(58\%)$ are in between these two extremes. The relative degree of specialization (mono-cropping) compared to diversified agricultural production has implications for the use of remote sensing weather data to predict agricultural production as well as serve as an instrument for other economic indicators. In Malawi, where $92\%$ of the farm is dedicated to producing a single crop, it may be fairly easy to settle on a single rainfall and temperature variable that do a good job at predicting yields. However, in highly diversified agricultural settings, like Ethiopia, Nigeria, and Uganda, it may be difficult to find a single set of variables that predict outcomes, since the rainfall and temperature metrics that are highly correlated with maize production may not be correlated with teff, cassava, or bean production. The need to account for numerous crop-weather relationships when choosing a weather metric to proxy for economic outcomes is an issue not adequately addressed in the literature.


\section{Analysis Plan}  \label{sec:analysisplan}

The following analysis, and the associated results, was pre-specified in our pre-analysis plan \citep{MichlerEtAl19} and was registered with Open Science Framework (OSF). If methods, approaches, or inference criteria differ from our plan, we highlight these differences. Results arising from these deviations in our plan should be interpreted as exploratory.


\subsection{Estimation}\label{sec:eststart}

Our basic model specification follows \cite{DescheneGreenstone07}:

\begin{equation}
Y_{ht} = \alpha_{h} + \gamma_{t} + X_{ht} \pi + \sum_{j}^{J} \beta_{j} f_{j} \left( W_{jht} \right) + u_{ht}
\end{equation}

\noindent where $Y_{ht}$ is our outcome variables from the LSMS-ISA, described above, for household $h$ in year $t$ and $X_{ht}$ is a matrix of input variables from the LSMS-ISA.\footnote{Where relevant, all continuous variables from the LSMS-ISA are log transformed using the inverse hyperbolic sine.} We control for year fixed-effects $(\gamma_{t})$ and include household fixed-effects $(\alpha_{h})$ in some specifications. The function $f_{j} \left( W_{jht} \right)$ represents our weather variables of interest where $j$ represents a particular measurement of weather. Finally, $u_{ht}$ is an idiosyncratic error term clustered at the household-level.

From this general set-up, we estimate six versions of the model: three linear and three quadratic. For each model, a single weather variable is considered. For the linear specification:

\begin{subequations}
\begin{align}
Y_{ht} &=  \beta_{1} W_{ht} + u_{ht} \label{eq:linear} \\
Y_{ht} &= \alpha_{h} + \gamma_{t} +  \beta_{1} W_{ht} + u_{ht} \label{eq:linearFE} \\
Y_{ht} &= \alpha_{h} + \gamma_{t} +  X_{ht} \pi + \beta_{1} W_{ht} + u_{ht}  \label{eq:linearCOV}
\end{align}
\end{subequations}

\noindent For the quadratic specification:

\begin{subequations}
\begin{align}
Y_{ht} &=  \beta_{1} W_{ht} + \beta_{2} W^{2}_{ht} + u_{ht} \label{eq:quad} \\
Y_{ht} &= \alpha_{h} + \gamma_{t} +  \beta_{1} W_{ht} + \beta_{2} W^{2}_{ht} + u_{ht} \label{eq:quadFE} \\
Y_{ht} &= \alpha_{h} + \gamma_{t} +  X_{ht} \pi + \beta_{1} W_{ht} + \beta_{2} W^{2}_{ht} + u_{ht} \label{eq:quadCOV}
\end{align}
\end{subequations}

\noindent All of the regression models are estimated for each permutation of the data (see Table~\ref{tab:sourcesetc}). This is a substantial number of regressions, given the number of variables defined (14 rainfall, eight temperature variables), the number of countries (six), the number of remote sensing products (six rainfall, three temperature), the number of extraction methods (ten), and the number of outcomes (two). This gives us a total of 77,760 different regressions: each of our six models on the 12,960 different data sets.\footnote{We also test a series of 51,840 linear combinations of temperature and rainfall data for each of the possible combinations of rainfall and temperature weather products (six data products for rainfall with three data products for temperature). We only test combinations of rainfall and temperature for the same extraction method. For the linear combinations, we estimate each with the linear specification, but estimate two linear combinations with the quadratic specification. These are presented in Appendix~\ref{sec:aplinearcomboresults}. The 77,760 standard regressions plus the 51,840 linear combinations give us a total of 129,600 regressions results presented in this paper.} By varying both specifications and data, we seek to define a robust set of outcomes by combining the multiple analysis approach of \cite{SimonsohnEtAl20} with the multiverse approach of \cite{SteegenEtAl16}.


\subsection{Inference}

In a typical economics paper, empirical results would be presented in a table, which would include coefficient estimates and some statistic for inference, such as standard errors, $p$-values, $t$-statistics, or confidence intervals. In our case, because of the large number of regressions that we estimate, standard modes of inference and traditional presentations of results are not appropriate. Instead, per our pre-analysis plan, we rely on a series of methods and criteria to make inference, evaluate the results, and present our findings.\footnote{Per our pre-analysis plan, we intended to examine the CDFs of coefficient estimates, following \cite{SaliMartin971, SaliMartin972}. However, using this approach in our context did not yield informative results. As such, we instead graph coefficients and confidence intervals ordered by the size of the coefficient estimate in specification charts. While not the same as the CDFs of coefficients in \cite{SaliMartin972, SaliMartin971}, the graphs communicate roughly the same information and are more appropriate for the variation in metrics, data products, extraction methods, etc. which are relevant for this analysis.}

Since no formal statistical test exists to compare results across models, we develop three heuristics that allow us to describe similarities and differences in our results. Before describing these heuristics, it is useful to reflect on what sort of characteristics a heuristic would need to be useful for our purposes (i.e., comparing across tens of thousands of models). First, some weather metrics that we test are likely to be positively correlated with outcomes (mean rainfall) while others are likely to be negatively correlated (longest dry spell). So, a heuristic should be agnostic about the sign of the coefficient. Second, our prior is that weather is significantly correlated with outcomes, regardless of direction. This maintained assumption is based on the frequency with which weather is used in the economics literature to predict all sorts of outcomes, from crop production to migration to economic growth. So, one would want a heuristic that is able to determine when a weather metric is significantly correlated with outcomes and when it is not. Finally, and in line with our prior, we expect weather to reduce the amount of unexplained variance in a model, all else being equal. So, one would want a heuristic that can measure the amount of unexplained variance in the model after controlling for weather.

With these three characteristics in mind, we adopt three general metrics to evaluate our results and two methods to test differences between these metrics. The three metrics are (1) mean adjusted R$^2$ values, (2) share of coefficient $p$-values significant at standard levels (0.01, 0.05, and 0.10), and 3) coefficients with 95\% confidence intervals. To compare our metrics across regressions, we apply two tests:

\begin{enumerate}
    \item Weak difference test: the value of a result (either mean adjusted R$^2$, share of significant $p$-values, or coefficients) from one model lies outside the 95\% confidence interval on the value of a result from a competing model. The confidence intervals \emph{can} overlap.
    \item Strong difference test: the 95\% confidence interval on the value of a result (either mean adjusted R$^2$, share of significant $p$-values, or coefficients) from one model lies outside the 95\% confidence interval on the value of a result from a competing model. The confidence intervals \emph{cannot} overlap.
\end{enumerate}

\noindent Our approach builds on the extreme bounds approach to assessing difference in estimates from \cite{LevineRenelt92} and the graphical methods to visualize these differences in \cite{SaliMartin971, SaliMartin972}.

While the three metrics are formal statistics, our weak and strong tests are not and we do not treat them that way. Rather, we use the combination of metrics and informal tests as heuristics in evaluating what weather metrics matter, for what countries, from what remote sensing products, and from what extraction method. Again, since we lack data on the objective fact which each data source is trying to measure, all comparisons of one obfuscation/metric/source combination are made relative to a different obfuscation/metric/source combination. Our data and heuristics do not allow us to make claims regarding the accuracy of a remote sensing source. Rather, we quantify the significance and magnitude of measurement error in remote sensing data by comparing results from each product with results from the other sources always bearing in mind that, for a given metric and country, if there was no measurement error the results from our tens of thousands of regressions would be exactly the same regardless of the obfuscation/source combination.

An important caveat to bear in mind with respect to our results, and all results focused on $p$-values, is that the significance of a point estimate does not imply that the model is correctly specified, that the point estimate is agronomically meaningful, or that the point estimate has the correct sign. It may very well be the case that the skew of temperature does not matter for agricultural production and that remote sensing products which result in a significant coefficient are therefore products with measurement error, while products which do not result in a significant coefficient are the more accurate products. These results and the associated figures simply allow us to visualize the variability in the number of significant coefficients across these specifications of interest. And any variability in results is a sign that obfuscation/source combinations provide different measures of weather and therefore measurement error exists.


\section{Results}

Due to the large number of regressions and estimated values produced from our analysis, we do not present our results as one would for a traditional economics paper, with coefficients, standard errors, and stars signifying $p$-values, all reported in tables corresponding to each regression. Instead, we present a series of figures, which allow us to evaluate the significance and magnitude, per our heuristics, of the effect of various sources of measurement error in remote sensing weather data. We start by examining measurement error due to extraction methods used to preserve the privacy of households. We then examine measurement error in relation to the choice of weather metrics to measure precipitation and temperature. Finally, we investigate the degree to which specific remote sensing products mismeasure precipitation and temperature.


\subsection{Extraction Method}\label{sec:ext}

We begin by examining extraction method, following Hypothesis $1$ ($H_0^1$ - different obfuscation procedures implemented to preserve privacy of farms or households have no impact on estimates of agricultural productivity). Our null hypothesis rests on two assumptions. First, existing publicly available remote sensing weather products are too coarse a resolution for any of the extraction methods to make a substantial difference in which pixel a household ends up in. Second, even if the obfuscation technique moves a household location from one pixel to another, weather is sufficiently spatially correlated that the shift will not matter. The alternative hypothesis is that obfuscating a household's true GPS coordinates introduces substantial mismeasurement resulting in researchers matching the obfuscated coordinates with gridded remote sensing data that does not accurately reflect the true weather experienced by the household.

To test Hypothesis 1, we pool the results from the 77,760 regressions (those that do not include linear combinations) and then divide the pool into ten bins, one for each extraction method. We then calculate descriptive statistics for each bin of results. These include the mean adjusted R$^2$ value and the share of coefficients $(\beta_{1})$ with $p$-values of $p>0.90, p>0.95$ or $p>0.99$. For each of these values, we calculate the $95\%$ confidence interval on the mean.\footnote{We also create graphs which examine the differences in coefficients $(\beta_{1})$ and their relative significance by extraction method. For parsimony, as these do not reveal any further information, they are presented in the Appendix, Figure~\ref{fig:ext_moment_rf} through Figure~\ref{fig:ext_total_tp}.} We then compare mean adjusted R$^2$ values or the share of $p>0.95$s across all ten extraction methods and use the $95\%$ confidence interval on the mean to evaluate differences using our weak and strong test criteria. This allows us in turn to make a determination about our Hypothesis: if the preponderance of evidence is such that we see no differences in our heuristics, then we fail to reject the null and conclude that obfuscation of household GPS coordinates does not introduce substantial mismeasurement into the analysis. Alternatively, if the preponderance of evidence is such that our heuristics are either weakly or strongly different, then we reject the null and conclude that obfuscation does introduce measurement error into the analysis.


\subsubsection{Adjusted R\texorpdfstring{$^2$}{2}}\label{sec:extR2}

We use a specification chart to examine adjusted R$^2$ values across the ten types of extraction method. Figure~\ref{fig:r2_ext} shows the mean adjusted R$^2$ and the $95\%$ confidence interval on the mean by extraction method. We further disaggregate results by model specification, since a model with covariates or fixed effects will have a different adjusted R$^2$ value than a model without covariates or fixed effects. The northwest panel of Figure~\ref{fig:r2_ext} displays results from model specifications~\eqref{eq:linear} and~\eqref{eq:quad}, which are the linear and quadratic models without coviariates and without household fixed effects. The northeast panel displays results from model specifications~\eqref{eq:linearFE} and~\eqref{eq:quadFE}, which include household fixed effects, but not covariates. The southwest panel displays results from model specifications~\eqref{eq:linearCOV} and~\eqref{eq:quadCOV}, which include fixed effects and covariates. At the top of each ``column'' in the specification chart is the mean adjusted R$^2$ and the $95\%$ confidence interval on the mean for the set of 1,296 regressions run. Below, the chart indicates the extraction method and model specification associated with the statistics.

Considering first the northwest panel: mean adjusted R$^2$ values are not substantially different across extraction method within model specifications~\eqref{eq:linear}. The mean adjusted R$^2$ value for any one extraction method passes neither our weak nor strong difference test when compared to any of the other extraction methods. Similarly, when comparing across extraction method within model specification~\eqref{eq:quad}, no mean adjusted R$^2$ is weakly or strongly different from any other. Though not a formal statistical test, our heuristic fails to reject the null for these two model specifications.

We conduct the same exercise for results from the extraction methods and model specifications presented in the northeast and southwest panels. As with the first panel, the mean adjusted R$^2$ value for any one extraction method is neither weakly nor strongly different than any other method. Our heuristic fails to reject the null for any model specification. Based on this, we conclude that remote sensing weather data from any one extraction method does not explain a substantially larger amount of the variance in our outcome variables relative to any other extraction method.

Despite the failure to reject the null based on either the strong or weak criteria, the pattern of which extraction methods result in the largest adjusted R$^2$ values is remarkably consistent. Bilinear extraction methods for household coordinates, EA centerpoint, and modified EA centerpoint always make up three of the top four models. Recall that the bilinear method computes the distance weighted average of values at the four nearest cell centers. Thus, unlike the simple extraction method, the bilinear method accounts for the point location within the arbitrary cell boundaries of the gridded data product. This approach seems to produce better results than the simple extraction method or the zonal means for small areas (households or EAs). Administrative area appears to be too large of an area to produce strong results, as using Administrative area, regardless of extraction method (simple, bilinear, or zonal mean), tends to produce the smallest adjusted R$^2$ values. While the pattern is consistent, it should be remembered that differences between each extraction method is not substantial enough to pass even our weak test, and we fail to reject the null.


\subsubsection{\texorpdfstring{$p$}{p}-values}\label{sec:extpval}

Next, we consider if different extraction methods produce substantially different numbers of significant coefficients. Recall that our maintained assumption is that the weather metrics we examine are significantly correlated with the outcome variables as these weather metrics are commonly used in the literature for just such a purpose. While when examining adjusted R$^2$ values we disaggregated each bin of regression results by model specification, when examining $p$-values we disaggregate by whether the remote sensing data is rainfall or temperature. Figure~\ref{fig:pval_ext} presents the share of significant coefficient estimates for three standard $p$-values: for $p>0.90, p>0.95$ or $p>0.99$. To these bars we add the $95\%$ confidence interval on the mean number of significant coefficients. The north panel presents results from precipitation products while the south panel presents results from temperature products. Each bar and confidence interval in the rainfall panel is based on 6,048 regressions while each bar and confidence interval in the temperature panel is based on 1,728 regressions. To facilitate comparison, we draw red lines to designate the top and bottom of the confidence interval on the mean for Household bilinear, the preferred method.

A quick visual inspection of the results in the top panel of Figure~\ref{fig:pval_ext} does not reveal many, if any, differences across extraction method. Comparing numerical values for the share of significant coefficients from one extraction to the $95\%$ confidence interval on the mean of any other extraction does reveal some small differences. The share of significant coefficients produced by regressions using data from modified EA simple and Administrative area bilinear are weakly different from EA bilinear. The only other pairwise comparison that is weakly difference is between Household bilinear and Household simple. There are no comparisons that are strongly different from each other. The results in the lower panel on temperature look similar to those with rainfall. No obvious differences exist, though a numerical comparison reveals that Household bilinear is weakly different from Administrative area bilinear. No other pairwise comparisons are weakly different and no comparisons are strongly different. As with our examination of adjusted R$^2$ values, the preponderance of evidence here implies that different extraction methods do not introduce substantial mismeasurement into the analysis. We fail to reject Hypothesis 1.

However, there is the possibility of heterogeneity across or within countries. As such, we next consider this same metric, disaggregated by country. Figures~\ref{fig:pval_ext_rf} and~\ref{fig:pval_ext_tp} present different extraction methods across all rainfall and temperature metrics, for each of the six countries. Now that we have divided the results by extraction method, rainfall/temperature, and country, each bar represents the share of significant coefficients from 1,008 regressions for rainfall and 288 for temperature. We simplify the graph by only presenting the share of coefficients with $p > 0.95$.

We see more variation within a country based on extraction method than in our previous graphs. In Ethiopia, the two extraction methods based on Administrative area centerpoints are strongly different from Household bilinear. While there are no other pairwise comparisons in Ethiopia, or any of the other countries, that are strongly different, a number of pairwise comparisons are weakly different. However, there seems to be no pattern to which extraction methods are different.\footnote{In Malawi, Household simple and the two zonal means are weakly different from EA bilinear and Administrative area bilinear. In Niger, Administrative area simple is weakly different from modified EA bilinear. In Nigeria, Administrative zonal mean is weakly different from bilinear Household and modified EA. In Tanzania, EA simple and EA zonal mean are weakly different from bilinear Household and EA, as well as Administrative simple and zonal mean. In Uganda, no pairwise comparison is weakly different.} The evidence for differences in extraction method using temperature data is equally noisy and without any apparent pattern of one extraction method over/under performing the others.

There is, however, some pattern to the variation across countries, with respect to the share of significant values. Tanzania produces fewer significant estimates on rainfall, relative to the other five countries. These differences are strongly significant. For temperature, Ethiopia, Tanzania, and Uganda all produce strongly different results compared to Malawi, Niger, and Nigeria. While this does not speak to mismeasurement due to differences in extraction method, the pattern is interesting to note for the discussion of cross-country differences in weather metrics and remote sensing data source in the subsequent sections.

Taken together, the preponderance of evidence regarding our $p$-value and adjusted R$^2$ heuristics lead us to conclude that we cannot reject the null for Hypothesis 1. We find no clear evidence that different obfuscation procedures implemented to preserve privacy of farms or households have substantially different impacts on estimates of agricultural production. This suggests that any measurement error which may arise from the use of different extraction methods does not substantially affect estimates. While Household, EA, and Modified EA bilinear appear to provide slightly better results than the other extraction methods, when researchers use publicly available data with obfuscated GPS information, they should feel confident that matching those coordinates with remote sensing data will not introduce substantial measurement error into the analysis.


\subsubsection{Selecting a Single Extraction Method}\label{sec:randnum}

Concluding that different extraction methods do not produce substantially different results, we narrow the set of results used from this point forward. For the blinded analysis, the Data Analysis Group could not simply select the ``true'' or most preferred extraction method. To deal with this, the Data Analysis Group used a random number generator in Stata to select one extraction method to continue the analysis with. This resulted in the selection of extraction method 3 (EA bilinear). In version 2 of this paper on arXiv.org, all the analysis was conducted using this extraction method. See Appendix~\ref{sec:apextraction} for more details on this process, including code used for the random number generation.

For the current version of the paper (version 3), the Data Analysis Group is no longer blinded and we can update the previous analysis using Household bilinear, which is the most accurate choice for extraction method given our setting. If our conclusion that extraction method does not introduce substantial mismeasurement, then the results and analysis using a randomly selected extraction method in version 2 should be qualitatively the same as results and analysis using Household bilinear in this version of the paper. As one can see from comparing drafts of the paper, results are in fact qualitatively similar. We point out differences, where they exist.


\subsection{Weather Metrics}\label{sec:metrics}

Having demonstrated that extraction method does not appear to introduce substantial measurement error into analysis combining remote sensing weather data with georeferenced household data, we next consider the particular weather metric used in estimation, following Hypothesis $2$ ($H_0^2$ - different weather metrics have the same impact on estimates of agricultural productivity). Here, our null hypothesis is based on the assumption that most measures of precipitation or temperature are nominally similar. This assumption would likely be demonstrably false to an agronomist or plant scientist familiar with the specific and particular water and temperature requirements of plants at each growth stage.\footnote{In fact, the authors have received reviewer comments on papers that highlight just this point. One example, from an anonymous reviewer:

\begin{quote}
Strictly speaking, what matters [for maize production] is agricultural drought - the sufficiency of water availability for plant growth - not meteorological drought - commonly based on average annual rainfall. Unless you can measure rainfall during the most critical period of crop flowering, for maize this occurs at roughly 1/2 the $\#$ of maturity days, and the period to maturity is 110 days, then you cannot measure the relevant drought.
\end{quote}

\noindent While we do not contend with the veracity of reviewer's point in measuring the biological response of plants to water, economists tend to ignore or remain ignorant of these arguments.} However, economists tend to ignore, omit, or simply remain unaware of these conditions, requirements, and related arguments. 

The fact that economists use so many different weather metrics provides \emph{prima facie} evidence that those in the profession adhere to the assumption that all rainfall or all temperature metrics perform about the same. \cite{MiguelEtAl04} find the percentage change in rainfall a good predictor of conflict while \cite{BruckerCiccone11} find total rainfall in the previous year a good predictor of conflict. \cite{MacciniYang09} find deviations in rainfall in the year of one's birth a good predictor of human capital formation while \cite{GargEtAl20} find the number of days with temperature great than $29^{\circ}$ Celsius to be a good predictor of human capital formation. A review of the literature, with the myriad weather metrics used to predict the same or similar outcomes, gives the impression that economists hold to the assumption that many weather metrics are equally adequate. The alternative hypothesis is that the choice of weather metric does matter. If the alternative hypothesis is true, it suggests that economists should justify their choice of one metric over the other on theoretical or agronomic grounds - or absent that - provide robustness checks to demonstrate that results are not contingent on the specific choice of weather metric, especially if that choice was not pre-specified prior to the analysis.


\subsubsection{Adjusted R\texorpdfstring{$^2$}{2}} \label{sec:metricR2}

As with extraction method, we use a specification chart to examine adjusted R$^2$ values across weather metrics. For ease of comparison, we present rainfall and temperature in separate graphs, Figures~\ref{fig:r2_rf} and~\ref{fig:r2_tp}, respectively. These figures presents the mean adjusted R$^2$ and 95\% confidence interval on the mean by weather metric for each model specification. As before, at the top of each ``column'' is the mean adjusted R$^2$ and the $95\%$ confidence interval on the mean for the set of 72 rainfall regressions and 36 temperature regressions. Below, the chart indicates the extraction method and model specification associated with the statistics.

Starting in the northwest panel, mean adjusted R$^2$ values can differ substantially across weather metrics within model specification. For the linear model~\eqref{eq:linear}, median daily rainfall is strongly different from most other metrics, explaining substantially less variance in the outcome variables than other rainfall metrics. Longest dry spell and the skew of daily rainfall also perform poorly, being weakly different in the amount of variance they explain relative to most other rainfall metrics. The remaining metrics, while having higher adjusted R$^2$ values than those already mentioned, are not weakly different from each other. The results for the quadratic model~\eqref{eq:quad} are qualitatively similar but the differences are more dramatic. The same five rainfall metrics explain the largest amount of variance in the outcome variables but this difference is now strongly greater than the three worst performing metrics and weakly greater than the remaining six metrics. Based on the evidence in this panel, it appears that while many metrics explain a similar amount of variance in the outcome variables, some metrics explain substantially less and some explain substantially more. Based on evidence in this panel alone, mean daily rainfall, total seasonal rainfall, the number of days without rain, the number of rainy days, and the percentage of rainy days all explain substantially more variance in the outcome variables relative to median daily rainfall, skew of daily rainfall, and the longest dry spell.

This interpretation of the results becomes muddled when we look at model specifications that include household fixed effects and fixed effect as well as inputs. There are no longer any weak or strong differences across rainfall metrics once we control for time-invariant household unobservables. The results do not change qualitatively when we control for time-varying inputs. Examining our R$^2$ heuristic, the preponderance of evidence suggests that we cannot reject the null, at least in panel data. In cross-sectional data, where one cannot control for household unobservables, the argument of the anonymous reviewer appears to be correct: one must take care in selecting the relevant rainfall metrics. However, if one is comparing the outcome of a household to itself (as one does in a household fixed effect panel setting) the choice of rainfall metric appears to no longer matter. Skew of daily rainfall explains just as much variance in outcome as total seasonal rainfall. This suggests that, once time-invariant household characteristics and time-variant input choices are controlled for, rainfall matters little. This is a point made in \cite{MichlerEtAl21}, who demonstrate that for households in India, time-invariant household ability and seasonal input choices allow farmers to adapt to weather, mitigating the impact of year-to-year variations in weather on crop output.

Results for temperature in Figure~\ref{fig:r2_tp} tell a similar story. Starting in the northwest panel, where specifications lack fixed effects or inputs, there are substantial differences in R$^2$ values by temperature metric. R$^2$ values for regressions that include skew of daily temperature, variance of daily temperature, or the z-score of GDD are strongly different from those that include maximum daily temperature, mean daily temperature, and median daily temperature. All of these differences go away once we control for household fixed effects and fixed effects plus inputs.

Synthesizing the results based on our R$^2$ heuristic, the preponderance of evidence indicates that we cannot reject our null hypothesis: different weather metrics have the same impacts on our outcome variables. But, this result applies only when we include household fixed effects to control for time-invariant unobservables. The implication of this is that in cross-sectional data, one must take care in selecting the appropriate weather metric but that this choice may become irrelevant in a panel setting where a household is compared itself.


\subsubsection{\texorpdfstring{$p$}{p}-values} \label{sec:metricpval}

To provide additional guidance to economists on the selection of weather metrics, we consider a series of figures which present $p$-values by weather metric, disaggregated into rainfall and temperature. Figure~\ref{fig:pval_varname} presents the share of significant coefficient estimates and the $95\%$ confidence interval on the mean number of significant coefficients for each weather metrics. The north panel displays results from precipitation products while the south panel displays results from temperature products. Each bar and confidence interval pair in the rainfall panel are based on 432 regressions while each bar and confidence interval pair in the temperature panel are based on 216 regressions. Note that, unlike the R$^2$ specification charts, the $p$-value graphs aggregate all six specifications. This is because we are looking at the significance of a single coefficient, not the overall explanatory power of a model and its weather metric.

Mean daily rainfall, total seasonal rainfall, the number of rainy days, and the percentage of rainy days have the largest number of coefficients significant at $p>0.95$. The share of significant coefficients for these four metrics is strongly different than the share for median and skew of daily rainfall, the z-score of total seasonal rainfall, and the longest dry spell. A clear ranking also emerges when we consider the temperature metrics. Mean and median daily temperature are no different from the variance of daily temperature and the maximum of daily temperature. Mean and median are weakly different from GDD and strongly different from skew of daily temperature, deviation in GDD, and z-score of GDD.\footnote{This result using Household bilinear differs from the result using extraction method 3 (EA bilinear). In the anonymized version 2, mean and median daily rainfall were weakly different from variance and max and strongly different from GDD. In switching from EA to Household bilinear, this weak difference disappears and the four metrics are no different from each other. Additionally, the strong difference with GDD is now only a weak difference.} Here it bears repeating that we do not know the objective fact of how much precipitation occurred or what was the temperature: we do not know which weather metrics truly should matter. All we can do is maintain the assumption that, as all 22 of our metrics have been used in the economics literature previously, each one should be significantly correlated with our outcome variables.

As with extraction method, we next investigate if there is heterogeneity across or within countries with respect to which weather metrics matter. Figures~\ref{fig:pval_varname_rf} and~\ref{fig:pval_varname_tp} present the share of coefficients significant at $p>0.95$ for each of the six countries. Results are based on 1,008 rainfall regressions for each country, or 72 regressions for each metric in each country. For temperature, there are 288 regressions for each country and 36 regressions for each metric in each country. To facilitate comparison, we draw red lines to demarcate the top and bottom of the confidence interval for mean daily rainfall and mean daily temperature, as these two metrics most frequently produce the greatest share of significant estimates.

Focusing on rainfall metrics in Figure~\ref{fig:pval_varname_rf}, we again see that weather seems to matter less in Tanzania than in the other five countries. While Ethiopia, Malawi, Niger, Nigeria, and Uganda have several rainfall metrics that are significant predictors of outcome over $60\%$ of the time, Tanzania has none. In Ethiopia, mean daily rainfall is not different from variance, total, rainy days, days without rain and percent rainy days. Mean rainfall is weakly different from the eight other metrics and is not strongly different from any. In Malawi, mean daily rainfall is not different from skew, total, or longest dry spell. Unlike in Ethiopia, in Malawi mean is strongly different from several metrics, including variance, deviation in total seasonal rainfall, and the z-score of total seasonal rain. Mean daily rainfall is weakly different from the remaining seven metrics. In Niger, mean rain is weakly different from only four other metrics and strongly different only from median.\footnote{This result using Household bilinear differs from the result using extraction method 3 (EA bilinear). In the anonymized version 2, mean daily rainfall was strongly different in Niger from skew of daily rainfall. In switching from EA to Household bilinear, this strong difference is now only a weak difference.} Here deviation in total rainfall and longest dry spell perform extremely well, being no different than mean daily rainfall. This is a substantial difference compared to their performance in Ethiopia and Malawi. In Nigeria, mean rainfall is not strongly different from any other metric and is weakly different from only skew and longest dry spell. In fact, in Nigeria, all of the metrics, except for skew and longest dry spell, all perform similarly well. In Tanzania, mean daily rainfall weakly out performs only median daily rainfall.\footnote{This again is a difference in results based on extraction method. In the anonymized version, mean was weakly different from z-score of total rain and not different from median. Using Household bilinear, mean is now weakly different from median and not different from z-score of total rain.} Unlike in the other five countries, mean daily rainfall performs poorly in Tanzania, as does total seasonal rainfall. Instead, the measures of rainy and non-rainy days perform extremely well. Rainy days, number of days without rain, share of days with rain, and all of their deviations are the strongest performing metrics in Tanzania. Again, the evidence suggests that weather in Tanzania impacts crops in ways that are substantially different than in the other LSMS-ISA countries. Finally, in Uganda, mean daily rain is weakly different from median and skew but not weakly different from the other metrics. Like Nigeria, in Uganda there are very few differences in how rainfall metrics perform.

Similar to rainfall, the temperature metrics in Figure~\ref{fig:pval_varname_tp} show cross-country heterogeneity, though stronger patterns are detectable. In Niger, Nigeria (with the exception of skew), and Uganda, none of the temperature metrics are even weakly different from each other. This suggests that the common measures of temperature preform relatively similar to each other and the choice of how to measure temperature in these countries is not a critical one. By contrast, we see several strong differences between temperature metrics in Ethiopia, Malawi, and Tanzania. Differences are most pronounced in Ethiopia, where mean, median, and skew are all weakly different from the other five and are strongly different from GDD, deviation in GDD, and z-score of GDD. In Malawi, mean and median temperature are at least weakly different from the other six metrics. In Tanzania, variance of temperature performs particularly well, being weakly different from mean, median, skew, and max temperature, and strongly different from the remaining three metrics.\footnote{This is a difference in results based on extraction method. In the anonymized version, variance was strongly different from skew. But, using Household bilinear, variance is now weakly different from skew.} Across five of the six countries, mean and median rank among the top three metrics with the largest share of significant coefficients. The lone exception is Nigeria, where mean and median are no less likely to produce significant coefficients than the other metrics.

Summarizing the results pooled by country, there are some weather metrics that are consistently significant predictors of our outcomes. For rainfall, these include mean daily rainfall, total seasonal rainfall, the number of rainy days in a season, and the share of days in a season in which it rains. For temperature, this set is mean and median daily temperature. This suggests that in broad or general geographic settings, or if there is no existing body of evidence about what metric matters for a given location, these variables could be used with confidence. More likely than not, these metrics will be significantly correlated with outcomes of interest in an agricultural setting. When we disaggregate results by country, we see that some weather metrics perform well in one country but not in others. Either mean, total, rainy days, or percent rainy days is almost always the top performer, but in some contexts (Niger) longest dry spell or variance in rain (Nigeria and Uganda) can be equal or better predictors. For temperature, either mean or median is almost always the top performer, with single exceptions for GDD (Nigeria) and variance (Tanzania). In places like Nigeria and Uganda, there are no meaningful differences between rainfall metrics. Similarly, in Niger and Uganda, there are no meaningful difference between temperature metrics.


\subsubsection{Coefficients}\label{sec:metriccoef}

Although the share of explained variance in a regression or the significance of a coefficient to predict an outcome are both valuable heuristics by which to judge the quality of a weather metric, the sign of the coefficient is also of interest to economists. With our maintained assumption that rainfall and temperature are significantly correlated with outcome, we also assume that a given metric should have the same direction of effect across a variety of situations. Said another way, we expect significant coefficients on a weather metric to share the same sign. The expected sign on some metrics are intuitive or supported by existing evidence.\footnote{There are diminishing returns to rainfall, so that at the upper end of the rainfall distribution we should expect a marginal increase in rainfall to be negatively correlated with outcomes. Our use of the quadratic specification helps control for these situations. But overall, we believe it is reasonable to expect the trend to hold in general given the number of households, years, countries, specifications, etc. that we examine.} As an example, it is reasonable to expect mean daily rainfall, total seasonal rainfall, or GDD to be positively correlated with agricultural outcomes. Similarly, it is reasonable to expect the variance in daily rainfall, the z-score in total seasonal rainfall, or the z-score in GDD to be negatively correlated with the same outcomes, since they measure variability or deviations from the average. The expected sign for other metrics, such as skew, are less obvious.\footnote{Appendix Table~\ref{tab:Wvarsign} defines our assumption for the expected sign for each of the 22 metrics.}

In this section, we present plots of coefficients and confidence intervals for individual regressions by weather metrics and country. The goal is to determine which weather metric conforms to our priors about consistency in sign. In attempting to determine which weather metrics perform better or worse than others, we classify a weather metric as more accurate if a larger share of its significant coefficients share the same sign. A weather metric is classified as suffering from mismeasurement if a nearly equal proportion of its significant coefficients are negative and positive. In this exercise, we interpret a non-significant coefficient as a zero, neither positive nor negative. As stated previously, as we lack data on the objective facts regarding rainfall and temperature, our goal is to draw generalized principles about which weather metrics adhere to the reasonable assumption regarding consistency in sign. For brevity, we discuss only a select number of the 22 metrics in the body of the paper, with the remaining presented in Appendix~\ref{sec:apmetrics}.

Figure~\ref{fig:v01_cty} presents coefficients and $95\%$ confidence intervals for single regressions of our outcome variable on mean daily rainfall, by country. Overall, there is general consistency in the coefficient sign for mean rainfall. In Ethiopia, mean rainfall is negatively correlated with outcome in eight regressions and positively correlated with outcome in 37 regressions (the remaining 27 are not significant). In Malawi, the coefficient on mean rainfall is negative in only two regressions while it is positive in 44 regressions (and not significant in 26 regressions). In Niger, coefficients are negative in eight regressions and positive in 35. In Nigeria, coefficients are negative in 12 and positive in 29. In Tanzania, coefficients are negative in 14 and positive in 14, indicating that mean daily rainfall is not significant in 44 regressions. In Uganda, there is only one regression with a negative coefficient while there is a positive coefficient in 38 regressions. The overall impression from these results is that mean daily rainfall is a consistent and positive predictor of outcome for all of the LSMS-ISA countries except Tanzania.

In Tanzania, the majority of coefficient estimates are not significant, while those that are significant are balanced nearly equally between positive and negative values. This leads us to conclude that in Tanzania, mean daily rainfall is not substantially correlated with outcomes. We cannot definitively conclude whether the non-results in Tanzania are a reflection of the true agronomic non-relationship between mean daily rainfall and outcomes or a reflection of remote sensing measurement error that occurs in Tanzania. But, the fact that we see consistent positive and significant relationships in the other five countries suggests it may be the latter. We hence can conclude that remote sensing products produce precipitation data with substantial measurement error in Tanzania that obscures a positive relationship between mean daily rainfall and outcomes.

Turning next to median rainfall, we conduct a similar examination of Figure~\ref{fig:v02_cty}. On inspection, it is immediately obvious that a substantial number of coefficients are not significant. While for mean daily rainfall, $56\%$ of the 432 coefficients were significant, for median, however, only $40\%$ are significant. Among this smaller number of significant coefficients for median rainfall, $65\%$ are positive and $35\%$ are negative. Compare this to mean, where $81\%$ of the larger share of significant coefficients are positive and only $19\%$ are negative. The overall impression for median rainfall is that it is not correlated with outcomes, and that the relative balance between positive and negative signs is a reflection of noise around an expected outcome of zero. As with mean rainfall in Tanzania, we cannot definitively state the reason for this. It could be that median rainfall does in fact have no relationship with our measures of agricultural production. Or it could be that there is a significant correlation, but measurement error introduces noise into the relationship so that we cannot observe this relationship in a statistically meaningful way. To the extent that an economist might want to use median daily rainfall in their analysis, the actual reason for this lack of significance does not matter. Whatever the cause, median daily rainfall is a poor weather metric.

Among the considered rainfall metrics, total seasonal rainfall (Figure~\ref{fig:v05_cty}), rainy days (Figure~\ref{fig:v08_cty}), and percent rainy days (Figure~\ref{fig:v12_cty}) all present consistent results, similar to mean daily rainfall.\footnote{As we have seen previously, there is some variation in results by country. For example, in Tanzania, total rainfall tends to be more non-significant $(61\%)$ than significant $(39\%)$. For the rainy days and share of rainy days, coefficients are as likely to be significant in Tanzania as in any other country (around $55\%$ to $60\%$.} For total rainfall, $57\%$ of regression coefficients are significant, and of those $80\%$ are positive. For rainy days, $58\%$ of regression coefficients are also significant, and of those $82\%$ are positive. Finally, for share of rainy days, $60\%$ of regression coefficients are significant, and of those $83\%$ are positive. As a counterpoint, for longest dry spell (Figure~\ref{fig:v14_cty}) only $47\%$ of regression coefficients are significant and among those they are relatively evenly split between positive $(40\%)$ and negative $(60\%)$. 

For temperature metrics, mean daily temperature (Figure~\ref{fig:v15_cty}) and median daily temperature (Figure~\ref{fig:v16_cty}) present results similar to those from our $p$-value graphs. For mean temperature, $60\%$ of the 216 regression coefficients are significant while $63\%$ of regression coefficients on median temperature are significant. However, as one can see from the figures, there is a fairly equal balance between significant positive and negative coefficients. So, while these two weather metrics matter more than the others, their correlation with outcomes is not consistent. For mean temperature, significant coefficients are positive $46\%$ of the time and negative $54\%$ of the time. For median temperature, significant coefficients are positive $49\%$ of the time and negative $51\%$ of the time. Obviously, these fairly even splits may reflect the true agronomic-temperature interactions but they contradict our assumption that a single weather metric will have a consistent sign. It is interesting to note that in many countries positive and negative coefficients are of a markedly different magnitude. For both temperature variables in Malawi, Nigeria, and Uganda, negative coefficients on mean tend to be extremely small while positive coefficients are much larger. By contrast, in Niger, negative coefficients for mean are substantially larger than positive coefficients. This, however, is not true for median temperature in Niger, when coefficient size is balanced across sign. This balance in coefficient size is seen in Ethiopia for both mean and median daily temperature. Finally, for Tanzania, the relationship seems to be the opposite, with larger negative coefficients and smaller positive coefficients.

Though mean and median temperature are the temperature metrics that were significant across the most regressions, the fact that they lack consistency in sign suggests they may not be ideal metrics. Here we explore two alternatives, variance and GDD, while relegating the discussion of the other metrics to Appendix~\ref{sec:apmetrics}. Coefficients and $95\%$ confidence intervals for variance of daily temperature are presented in Figure~\ref{fig:v17_cty}. While only $55\%$ of regression coefficients are significant, there is a strong consistency in the sign of those significant coefficients. For variance of daily temperature, $64\%$ of significant coefficients are negative and only $36\%$ are positive. By contrast, for GDD (Figure~\ref{fig:v19_cty}), which is commonly used in the economics literature, less than half $(46\%)$ of coefficients are significant. But, among that minority of coefficients, $68\%$ are positive. Adhering to our heuristics, temperature appears to be more problematic in generating accurate metrics. While mean and median daily temperature are frequently significant, coefficients are as likely to be positive as negative. GDD is more likely to be not significant than it is to be a significant predictor of outcome, yet among the minority set of significant coefficients it is strongly positive. Finally, variance of temperature, while not as frequently significant as mean or median, does show consistency in the sign of its significant coefficients.


\subsubsection{Selecting a Subset of Weather Metrics}\label{sec:metricselect}

Summarizing our findings about the weather metrics, there is strong evidence for rejecting Hypothesis 2, that different weather metrics will have the same impacts on estimates of agricultural productivity. Based on our heuristics, some weather metrics perform better than others. But there are also a subset of weather metrics that appear to perform equally well. This suggests that specificity of certain metrics promoted by agronomists and others, which demand a single, particular metric are likely unnecessary in economic applications. Although there are clearly metrics that should not be used, one could substitute several metrics for each other without appreciable differences in results.

For rainfall, the metrics that perform best, based on our heuristics, are (1) mean daily rainfall, (2) total seasonal rainfall, (3) the number of rainy days, and (4) the share of rainy days in a season. All of these perform well in explaining variance in outcomes (R$^2$), being strongly correlated with outcomes ($p$-values), and have consistent coefficient signs. For each of these criteria, the four rainfall metrics are frequently weakly or strongly different from the other rainfall metrics. While there is some heterogeneity across countries, these four are strong performers in all six countries. Median daily rainfall and skew of daily rainfall are particularly poor performers, being significantly correlated with outcomes in only $40\%$ of regressions.

For temperature, the evidence for rejecting Hypothesis 2 is equally strong, with substantial differences between temperature metrics based on our heuristics. However, it is not as clear which temperature metrics perform the best. Based on R$^2$ and $p$-values, mean and median daily temperature appear to be the best predictors of outcomes across many different cases. However, neither do particularly well in being consistently positively or negatively correlated with outcomes. Coefficients on both variables are as likely to be associated with better outcomes as worse outcomes. The different signs do not appear to be more or less likely in one country over another, suggesting the reason for the inconsistency is measurement error and not aggregation over regions, countries, or crops. While not as strong as mean or median in explaining variance in outcomes or correlating with outcomes, coefficients on variance of daily temperature are consistently negative, making it among our preferred measure for temperature. Somewhat surprisingly, given its frequency of use in the economics literature, GDD and deviations from GDD are poor metrics for temperature as measured by our heuristics. While GDD may be appropriate for predicting crop yields in temperate climates such as the United States, it appears to be a poor choice for use in the tropical regions of Sub-Saharan Africa.

Based on this evidence, we proceed to analyze measurement error in remote sensing products focusing on four rainfall metrics and three temperature metrics. These are mean rainfall, total rainfall, number of rainy days, share of rainy days, mean temperature, median temperature, and variance of temperature. Without additional evidence, these appear to be the best metrics for measuring precipitation and temperature. When researchers are considering how to include precipitation or temperature in they economic analysis (at least in Sub-Saharan Africa), we recommend they use one or more of the above seven metrics.\footnote{See Appendix~\ref{sec:aplinearcomboresults} for an exploration of linear combinations of rainfall and temperature metrics.} The choice of a different weather metric should be accompanied by a theoretical or agronomic justification and likely warrant robustness checks to ensure results are not driven by the choice of any one specific weather metric.


\subsection{Remote Sensing Products}\label{sec:satellite}

By narrowing in on a single extraction method and a subset of weather metrics that are correlated with outcomes, we can better investigate which remote sensing products mismeasure precipitation and temperature. Specifically, we want to examine the evidence around Hypothesis 3 ($H_0^3$ - different measurement technologies for precipitation and temperature have the same impact on estimates of agricultural productivity.) As noted previously, mismeasurement in the objective fact of precipitation and temperature may be introduced by inaccuracies in the sensor (e.g. infrared, microwave, optical), the algorithm used to convert sensor data into rainfall or temperature (e.g. reanalysis, interpolation), or the resolution of the data (e.g. spatial, temporal). Our null hypothesis rests on the assumption that these inaccuracies either do not exist (that all remote sensing products accurately report the objective facts) or that any mismeasurement is insubstantial and will not impact the analysis. The alternative hypothesis is that there is substantial mismeasurement in remote sensing weather data, as Figures~\ref{fig:rain_res} and~\ref{fig:temp_res} seem to imply. If the alternative is true, then care must be taken in selecting a remote sensing product and the choice of a product needs to be justified or robustness checks to ensure variance across products does not cause significant or substantial differences in results.

To test Hypothesis 3, we pool results from 2,376 regressions (those that result from our single extraction method and seven weather metrics) and then divide them by remote sensing source. As with our previous hypotheses, we seek to determine if a remote sensing product is weakly or strongly different from other products. To do this, we use our heuristics for adjusted R$^2$ and $p$-values to ascertain if the weather metrics from one remote sensing product explain more of the variance in outcomes or tend to be more correlated with outcomes. We also examine coefficient significance and sign by weather metric and country to determine if remote sensing products are sensitive to which region or country from where they are recording data.

If we fail to reject the null, then we can conclude that measurement error in remote sensing products is not substantial and that all products either accurately report the objective facts or misreport in the same way. If, we instead reject the null, and determine that the alternative hypothesis is true, this means that remote sensing products disagree on what the actual precipitation and temperature was in a given location on a given day. As the precipitation and temperature at a location on a single day are objective facts, then any disagreement must be the result of measurement error in the remote sensing product. Again, as we lack data on the object fact, we cannot definitively determine which product or products contain measurement error. Rather, we can categorize a product as likely containing measurement error if it produces substantially different outcomes from the other products, or if a coefficients on weather metrics derived from a product lack consistency across regressions.

We fully appreciate and understand the weaknesses in this approach. Without data on the objective facts we cannot measure the accuracy of a product against those objective facts. Yet, such data does not exist and therefore our approach should be viewed as a second best approach, given the missing data problem. Knowing that we cannot compare the measurement error in remote sensing products against the object facts is one reason for the blinding of the Data Analysis Group to the identity of the remote sensing products. It helps ensure the objectivity of our findings in the absence of data on the objective facts regarding precipitation and temperature.


\subsubsection{Adjusted R\texorpdfstring{$^2$}{2}} \label{sec:satelliteR2}

We start by looking at R$^2$ values for remotely sensed precipitation aggregating over our four preferred rainfall metrics. This gives us 1,728 regressions from six precipitation products. As previously, we average the adjusted R$^2$ value for regressions from a single product for a single model specification and then calculate the $95\%$ confidence interval on the mean. This results in 48 regressions per precipitation product, per model specification.

Figure~\ref{fig:r2_sat_rf} presents a specification chart for mean adjusted R$^2$ for precipitation products. The northwest panel displays results from model specifications~\eqref{eq:linear} and~\eqref{eq:quad}, the northeast panel displays results from model specifications~\eqref{eq:linearFE} and~\eqref{eq:quadFE}, and the southwest panel displays results from model specifications~\eqref{eq:linearCOV} and~\eqref{eq:quadCOV}. In the northwest panel, for specifications that include only weather, the adjusted R$^2$ values for CPC and MERRA-2 are strongly different from ERA5 and weakly different from CHIRPS and TAMSAT. No other pairwise comparisons are substantially different. For the quadratic specifications, CPC is weakly different from CHIRPS, TAMSAT, and ERA5 but MERRA-2 is no longer substantially different from any other product. This suggests that CPC and, to a lesser extent, MERRA-2, mismeasure precipitation in rainfall metrics calculated using data from these two products. Metrics from these two products explain substantially less of the variation in outcomes relative to the other four products. As with previous specification charts of adjusted R$^2$, the only substantial differences exist in model specifications that do not include household fixed effects. When we examine the other two panels of Figure~\ref{fig:r2_sat_rf} we see no substantial differences across precipitation products.

Results for temperature in Figure~\ref{fig:r2_sat_tp} tell a different story. Here there are no substantial differences between remote sensing products in any of the panels. While CPC has the smallest adjusted R$^2$ value in all six model specifications this difference is not substantial in terms of both the strong and weak test.

This quantitative evidence supports the qualitative evidence in Figures~\ref{fig:rain_res} and~\ref{fig:temp_res} and the descriptive evidence in Figures~\ref{fig:density_aez_rf} through~\ref{fig:gdd_aez_tp}. Differences in reported rainfall across precipitation products can be substantial while differences in temperature across temperature products appear to be relatively similar. The evidence thus far suggests that CPC and MERRA-2 may mismeasure precipitation relative to values reported in the other four precipitation products. For temperature, all three products report similar values, suggesting none of them suffer from measurement error.


\subsubsection{\texorpdfstring{$p$}{p}-values} \label{sec:satellitepval}

Turning to our second heuristic, Figure~\ref{fig:pval_sat} presents the share of significant $p$-values by remote sensing product.\footnote{Tables~\ref{tab:sig_var_eth_rf} through~\ref{tab:sig_var_uga_tp} in Appendix~\ref{sec:apsatellite} present share of significant $p$-values for all 22 weather metrics disaggregated by country and remote sensing product.} As with adjusted R$^2$, we aggregate over weather metric and focus just on the overall share of coefficients with significant $p$-values by remote sensing product.

Here, ERA5 performs best, with coefficients significant at $p>0.95$ in $64\%$ of regressions. MERRA-2 performs the worst, with significant coefficients in $53\%$ of regressions. ERA5 is weakly different from all other products, except ARC2, at $p>0.95$.\footnote{This is a difference in results based on extraction method. In the anonymized version, ERA5 was weakly different from ARC2 but not weakly different from CPC. But, using Household bilinear, ERA5 is now weakly different from CPC and is no longer weakly different from ARC2.} ERA5 is strongly different than MERRA-2 for $p>0.99$ (but not at other significance levels).

For temperature, unlike in previous figures, we see a substantial difference between remote sensing products. ERA5 is weakly different than both MERRA-2 and CPC at all significance levels, with significant coefficients at $p>0.95$ in $65\%$ of regressions compared to $56\%$ for MERRA-2 and CPC. We saw in Figure~\ref{fig:r2_sat_tp}, ERA5 had slightly higher adjusted R$^2$ values compared to the other two, though these differences did not satisfy our weak criteria.

These results raise the question of whether CHIRPS, CPC, MERRA-2, ARC2, and TAMSAT contain measurement error in precipitation and ERA5 is accurate - or whether the reverse is true. Similarly, the question exists if MERRA-2 and CPC contain measurement error in temperature and ERA5 is accurate - or, again, if the reverse is true. Based on our maintained assumption that weather metrics are significantly correlated with outcomes, we should conclude that ERA5 is more accurate than the other products in terms of both rainfall and temperature, while all the other remote sensing sources produce weather data containing measurement error. However, as CHIRPS, CPC, MERRA-2, ARC2, and TAMSAT, report roughly the same results, it could be that they are the more accurate products and ERA5 is the outlier.

This latter interpretation then begs the question of whether, if ERA5 suffers from measurement error in both rainfall and temperature, is the error classical or non-classical. Based on the descriptive evidence, it is possible that ERA5 suffers from non-classical measurement error, in that it consistently reports higher rainfall levels than most of the other precipitation products, suggesting the measurement error is not random but biased. That ERA5 produces the largest share of significant $p$-values on rainfall metrics also suggests that if there is measurement error in the product it may be non-classical in nature. This is because the general understanding of the effects of classical measurement error is to introduce noise, not bias, meaning we would expect a product suffering from classical measurement error to produce fewer, not more, significant $p$-values. For ERA5 temperature measurements, the picture is somewhat more muddled. The descriptive evidence for temperature from ERA5 does not reveal a larger variance, which a common indicator of classical measurement error. As with rainfall from ERA5, that the large number of significant $p$-values produced by temperature metrics from ERA5 is the opposite of what one expects from classical measurement error, where noise introduces imprecision in coefficient estimates. Yet, neither of the descriptive statistics for temperature from ERA5 demonstrate a tendency to under- or over-report temperatures, relative to the other products, as ERA5 does for rainfall.  

This leaves us with one of two options: either ERA5 is accurate and the other five precipitation products mismeasure precipitation or, alternatively, ERA5 suffers from non-classical measurement error, overestimating precipitation and thereby biasing regressions results so that rainfall appears more correlated with outcomes than it actually is in reality. Similarly, either ERA5 is accurate and MERRA-2 and CPC suffer from measurement error or MERRA-2 and CPC accurately report the objective fact of temperature, while ERA5 suffers from non-classical measurement error that results in biased coefficients that are more strongly correlated with outcomes than they should be. If the latter is true for either rainfall or temperature, then we cannot determine if those products suffer from classical or non-classical measurement error. Because we do not observe the actual precipitation and temperature, we cannot determine which of the two types of measurement error CHIRPS, CPC, MERRA-2, ARC2 and TAMSAT suffer from, if they in fact suffer from measurement error.

To dig into these issues further, we disaggregate the $p$-value results by weather metric. Figure~\ref{fig:pval_rf} presents the share of significant $p$-values by mean, total, rainy days, and percent rainy days. For mean daily rainfall and total seasonal rainfall, ERA5 reports weakly different results from all the other products except CPC.\footnote{This is a difference in results based on extraction method. In the anonymized version, ERA5 was weakly different from CPC. But, using Household bilinear, ERA5 is no longer weakly different from CPC.} This is inline with what we saw previously and the long tails of the distribution of rainfall in Figure~\ref{fig:density_aez_rf}. But, when we consider the number of rainy days or the percentage of rainy days, we see that the share of significant $p$-values produced by ERA5 is no longer different from any of the other products. CHIRPS through ERA5 are all by and large the same for both metrics, with TAMSAT being weakly different from MERRA-2. This suggests that ERA5 does indeed suffer from non-classical measurement error, in that it consistently over reports the volume of rainfall on a given day, but not the number of days that it rains.

Figure~\ref{fig:pval_tp} presents the share of significant $p$-values by mean, median, and variance of daily temperature. Here the evidence is less clear. ERA5 reports weakly greater significant $p$-values for mean and variance of temperature at most significant levels relative to CPC. ERA5 is weakly different from MERRA-2 for variance of temperature but not for mean. All three temperature products report substantially similar results for median daily temperature. What this all means regarding which temperature products might be mismeasured is unclear. Given the similarities across temperature products in terms of descriptives and adjusted R$^2$, and as the differences in share of significant $p$-values fails to follow a clear pattern, it may be the case that the differences we observe are simply due to random chance. We explore these points below by examining the sign and significance of specific regression coefficients.


\subsubsection{Coefficients} \label{sec:satellitecoef}

By narrowing our focus to a single extraction method and only four rainfall metrics, it now becomes feasible to look at the results from individual regressions. Similar to the specification charts for adjusted R$^2$, labels identify characteristics of the results are presented at the bottom of the specification chart. Unlike the adjusted R$^2$ charts, we now present coefficients and confidence intervals for single regressions - 72 per country - and not means of aggregated results and confidence intervals on the mean. In the following discussion, the term significance defines a point estimate with $p>0.95$.

Figure~\ref{fig:v01_sat} presents specification charts for coefficients and confidence intervals on mean daily rainfall by country. As we saw previously, the majority of significant coefficients on mean daily rainfall are positive in each country, with the exception of Tanzania, where a majority of coefficients are not significant. Across countries, there is a consistent pattern in how mean daily rainfall performs based on the remote sensing source of the precipitation data. In Ethiopia, the coefficients on mean rainfall tend to be evenly split between significant and not significant, except for ERA5 in which 10 of 12 coefficients are significant. In contrast, coefficients on MERRA-2 are equally likely to be negative, positive, or not significant. For those coefficients that are significant in Ethiopia, the majority tend to be positive though here MERRA-2 is the outlier, with significant coefficients evenly split between positive and negative values. In Malawi, a majority of coefficients on mean rainfall are significant for each remote sensing source, except MERRA-2, where only four of 12 coefficients are significant. Conditional on being significant, all remote sensing products report a vast majority positive coefficients, with only two of 45 coefficients being negative.

Similar to Ethiopia, Niger and Nigeria coefficients on mean rainfall tend to be evenly split between significant and not significant, with ERA5 reporting 10 of 12 coefficients as significant. However, Niger and Nigeria contrast with Ethiopia by having some remote sensing products report more insignificant coefficients than significant. In Niger, MERRA-2 and TAMSAT report fewer than half of coefficients as significant while in Nigeria CPC reports a minority of coefficients as significant. In both Niger and Nigeria, among significant coefficients, the vast majority are positive. The exceptions being MERRA-2 in Niger, where the sign on significant coefficients is about equally divided, and ERA5 in Nigeria, where a large majority are negative (eight out of 10).

In Tanzania, as has been seen previously, coefficients are more likely to be insignificant than significant. In fact, no remote sensing product produces measures of mean daily rainfall in which a majority of coefficients are significant. Among the small set of significant coefficients in Tanzania, these tend to be negatively correlated with output. Between $75-100\%$ of significant coefficients are negative for CHIRPS, ARC2, and TAMSAT. The outliers here are CPC, with only one third of significant coefficient signs being negative, and ERA5, where all significant coefficients are positive. While mean daily rainfall tends to be uncorrelated or negatively correlated in Tanzania, in Uganda coefficients are more likely to be significant than not significant and those significant coefficients are almost exclusively positive. In fact there is only one negative and significant coefficient in Uganda out of the 72 regressions, which comes from TAMSAT.

Turning to total seasonal rainfall in Figure~\ref{fig:v05_sat}, results are remarkably similar to mean daily rainfall. In fact, other than some variation in results in Uganda, the number of regressions with significant coefficients does not vary within a country if one uses mean daily or total seasonal rainfall. Only the magnitude of the coefficients differs between the results for mean and total rainfall. This result makes intuitive sense, as mean daily rainfall is simply the average of the volume of precipitation every day in the growing seasons while total seasonal rainfall is the sum of the same data. In fact, correlation between coefficients from the same precipitation product tend to be around $0.85$.

Based on these results, we conclude that mean daily rainfall and total seasonal rainfall can be used in the same application and do not produce substantially different results. Results from CHIRPS, ARC2, and TAMSAT tend to be in agreement, producing mostly positive coefficients in all countries, except Tanzania where they all produce mostly negative coefficients. CPC tends to agree with CHIRPS, ARC2, and TAMSAT, though unlike these other three products CPC produces more positive coefficients than negative coefficients in Tanzania. Less consistent, both in comparison to itself and to other remote sensing products, is MERRA-2. MERRA-2 tends to report positive and negative coefficients in equal number in some countries (Ethiopia, Niger, Tanzania) but then exclusively positive coefficients in other countries (Malawi, Nigeria, and Uganda). This inconsistency makes MERRA-2 a hard product to recommend for measuring the volume of precipitation (mean daily or total seasonal) though it does frequently, if inconsistently, agree with CHIRPS, CPC, ARC2, and TAMSAT. By comparison, ERA5 tends to be very consistent with itself across countries but is the notable outlier among the six precipitation products.

ERA5 reports more significant coefficients than the other products. No other product reports significant coefficients in more than $80\%$ of regressions in any country, while ERA5 reports more than $80\%$ of coefficients as significant in Ethiopia, Niger, and Nigeria. Among significant coefficients, ERA5 results in almost exclusively positive signs on mean and total rainfall, even in Tanzania where most of the other products result in negative coefficients. The one exception for ERA5 is Nigeria, where results show $80\%$ of significant coefficients as negative, while all of the other products have a majority of coefficients as positive. Based on these results, MERRA-2 and ERA5 appear to mismeasure the volume of precipitation during the growing season. As shown in Figure~\ref{fig:density_aez_rf}, MERRA-2 reports substantially less rainfall while ERA5 reports substantially more rainfall than the other remote sensing sources. This appears to produce noisy or inconsistent results for mean daily and total seasonal rainfall when using MERRA-2. For ERA5, the large measured volume of precipitation appears to produce results that are in stark contrast to those results produced by other precipitation products.

Next, we consider Figures~\ref{fig:v08_sat} and~\ref{fig:v12_sat} which present specification charts for the number of rainy days and the share of days in the growing season that it rained. As one might expect, given the results of mean and total rain, the two rain day metrics are highly correlated at around $0.80$. The except being for MERRA-2, where the correlation between the number of rainy days and the percentage of rainy days in a season is only $0.60$. Despite the lower correlation within MERRA-2, we will only discuss the number of rainy days in detail.

Results for rainy days are more consistent across country and remote sensing product than mean or total rainfall. Coefficients on rainy day are almost always significant, regardless of country or of remote sensing source. The one exception is MERRA-2, with most coefficients in Ethiopia, Malawi, and Uganda being not significant while $75\%$ of coefficients in Tanzania are significant. These results contrast with most of the other remote sensing products, which tend to report mostly significant coefficients in every country, except Tanzania.

Among significant coefficients, most remote sensing sources produce positive coefficients. In Ethiopia, coefficient signs are divided evenly for CPC but for the other five products a majority of coefficients are positive. For Malawi, all remote sensing products report only positive coefficients. In Niger, MERRA-2 reports negative signs on six of seven significant coefficients, while the other five products report a majority of significant coefficients with positive signs. In Nigeria and Uganda, the vast majority of significant coefficients are positive regardless of remote sensing product. Though unlike Malawi, there are a couple negative and significant coefficients. Finally, in Tanzania, CHIRPS, CPC, and MERRA-2 produce results in which the number of days with rain is positively correlated with output, while ARC2, ERA5, and TAMSAT have the number of days with rain negatively correlated with output.

Synthesizing the evidence from the number and share of rainy days, most remote sensing sources are in agreement. The one exception is MERRA-2, which tends to produce fewer significant coefficients. MERRA-2 is also at odds with the other products in Niger, where the more days with rain reduces crop output while more days with rain, as measured by the other five products, increases crop output. Based on this evidence, MERRA-2 appears to suffer from some measurement error in counting the number of days in which there was precipitation. This result is unsurprising, given that descriptive statistics reveal that MERRA-2 tends to under report precipitation. What is surprising, and difficult to explain, is the sharp division in how the number of days with rain affects output in Tanzania. Data from all six products results in a majority of regression coefficients being significant. But among those significant coefficients, CHIRPS, CPC, and MERRA-2 report mostly positive coefficients, with only $11-14\%$ of coefficients being negative. By contrast, ARC2, ERA5, and TAMSAT all report mostly negative coefficients, with $60-71\%$ of coefficients being negative. While it is unclear what is driving these results, what is clear is that counts or shares of days in which it rained is a poor metric for use in Tanzania.

The results regarding remote sensing temperature products are more ambiguous than those for rainfall. Already we have observed that while mean and median daily temperature produce the most significant coefficients, the signs on these coefficients is nearly evenly split, which violates our assumption about the consistency of sign. As a consequence of this, we also include variance of daily rainfall, which while producing fewer significant coefficients, those coefficients are consistently negative. Figure~\ref{fig:v15_sat} presents results for mean daily temperature and Figure~\ref{fig:v16_sat} presents results for median daily temperature. Results for both metrics are similar, which is expected as the correlation between these two metrics is $0.99$. Results are also similar across country. In general, about half of coefficients are significant. Among significant coefficients, about $60\%$ of coefficients using MERRA-2 data are negative, while between $45$ to $55\%$ of coefficients from ERA5 and CPC are negative. The one exception to this is Niger, where coefficients tend to be either negative or not significant, with very few positive coefficients. 

Examining the specification charts reveals one reason for this near equal division between positive and negative coefficients on mean and median daily temperature. Coefficients from linear specifications tend to be not significant in Ethiopia, Niger, Nigeria, Tanzania, and Uganda. Negative coefficients on mean and median daily temperature in Ethiopia, Nigeria, and Tanzania tend to be from specification~\eqref{eq:quad}, which is the quadratic without household fixed effects or inputs. In this specification, the coefficient on the linear temperature term signifies that higher temperature reduces output. But, when we include household fixed effects, as in specifications~\eqref{eq:quadFE} and~\eqref{eq:quadCOV}, the coefficient on the linear temperature term signifies that higher temperature increases output. In these four countries, including mean or median daily temperature while not accounting for household fixed effects leads one to conclude that higher temperatures reduce crop output while the inclusion of household fixed effects results in the opposite conclusion: higher temperatures increase crop output.

The pattern in Malawi, Niger, and Uganda is slightly different. In Niger and Uganda, the coefficients from the linear specifications tends to be not significant, as in Ethiopia, Nigeria, and Tanzania. But in Niger, all quadratic specifications tend to produce negative coefficients, regardless of whether household fixed effects are included or not. In Uganda, the quadratic specification without household fixed effects tend to be not significant, with only those specifications that include household fixed effects producing significant and positive coefficients. Thus, in Niger, higher temperatures are negatively correlated with output while in Uganda they are positively correlated with output. Finally, in Malawi, linear specifications tend to produce negative coefficients while quadratic specifications tend to produce positive coefficients. Thus, when it comes to mean and median daily temperature, the country and specification are more important in determining what results one gets than which remote sensing product that temperature data comes from.

Figure~\ref{fig:v17_sat} presents results for variance of daily temperature. As we saw before, coefficients on variance are less frequently significant than coefficients on mean and median. For variance, $55\%$ of coefficients are significant while for mean and median $60$ to $63\%$ of coefficients are significant. Here Uganda is an outlier, with only $37\%$ of the coefficients on variance being significant. In terms of sign, significant coefficients on variance are more consistently of the same sign (negative) than coefficients on mean and median. The remote sensing source of the data has little effect on the sign, though as with mean and median, there is variation across countries. In Ethiopia, Malawi, Niger, and Tanzania, a majority of significant coefficients are negative. In Nigeria, however, variance tends to be positively correlated with outcome. In Uganda, among the small number of coefficients that are significant, the sign tends to be split evenly between positive and negative for MERRA-2 and CPC. Variance of daily temperature calculated using ERA5 results in $67\%$ of coefficient being negative.

As with mean and median, which sign one gets on the coefficient of interest is frequently a function of whether household fixed effects are included. For Nigeria, specifications that include fixed effects tend to be show a positive relationship between variance of daily temperature and output, while specifications that lack household fixed effects tend to show a negative relationship. This is regardless of whether the specification is linear or quadratic. By contrast, in Uganda, quadratic specifications tend to result in negative coefficients on variance while linear specifications tend to result in positive coefficients. This is regardless of whether household fixed effects are included or not.

While with rainfall we could conclude that MERRA-2 and ERA5 were poor choices, with temperature all three products produce roughly similar results. But this appears to have as much to do with the lack of a strong relationship between the temperature metrics and outcomes, as with the products themselves. Where we do find differences, it is likely to be due to cross-country variation and with the specific functional form of the regression. To the extent that these differences may reflect the on-the-ground reality in each country, it suggests that the use of variance, or any temperature metric, requires specific knowledge about the country and application where it will be used.


\subsubsection{Selecting a Remote Sensing Product}\label{sec:satelliteselect}

Summarizing our results, mean and total rainfall are clearly substitutes for each other. While these two metrics perform substantially better than most other rainfall metrics, the choice between them seems to be immaterial. The same is true for the number of and share of rainy days. These two metrics outperform other rainfall metrics but produce results extremely similar to one other. In terms of remote sensing products, MERRA-2 is seemingly inferior to the other five precipitation products. Regardless of which of the four rainfall metrics are used, coefficients on these metrics derived from MERRA-2 data tend to be non-significant while coefficients derived from the other remote sensing products are significant. This suggests that MERRA-2 suffers from classical measurement error, with sufficient noise introduced into the data to result in rainfall metrics being uncorrelated with outcomes. ERA5 is also inferior to CHIRPS, CPC, ARC2, and TAMSAT. As is evinced by the descriptive statistics, ERA5 tends to report much higher volumes of rainfall, which suggests the existence of non-classical measurement error. This conclusion is supported by our specification charts for mean and total rainfall, where coefficients derived from ERA5 data tend to be more positive than coefficients from other remote sensing products. This bias in results for ERA5 does not carry over to the number or share of rainy days, since these metrics are not based on the volume of rainfall, which is where ERA5 deviates from the other remote sensing products. Based on this evidence, we conclude that MERRA-2 and ERA5 are poor choices when seeking a remote sensing source for precipitation to match with household data.

Dissimilar from rainfall, for temperature we are unable to draw strong conclusions. In general, results for all three temperature products are broadly alike, suggesting one could use any of the three. But this general similarity is disturbed by differences in highly specific cases. Unlike rainfall, we cannot identify patterns of differences, such as with MERRA-2 and ERA5. Therefore, we caution researchers in using any of the temperature products since there is no apparent pattern of which product works best in what situation. In linear specifications, mean and median daily temperature tend not to be significant, but in quadratic specifications without fixed effects, mean and median tend to be negatively correlated with outcomes, and in quadratic specifications with fixed effects, mean and median tend to be positively correlated with outcomes. For variance, though, even this pattern falls apart, with a single specification producing both positive and negative coefficients. Based on all of this evidence, we conclude that any of the three temperature products are as likely to produce results substantially similar to each other as they are to produce results substantially different from each other. Researchers may need to verify the robustness of their results to the choice of remote sensing temperature data source. Overall, however, temperature seems to be a poor predictor of agricultural outcomes in these Sub-Saharan African countries.


\section{Discussion} \label{sec:discussion}

Having examined results from 129,600 regressions on panel data from six countries with 33,738 total household observations, it is useful to recapitulate our results and provide direction towards a set of best practices when looking to combine remote sensing weather data with household survey data. We also outline future work, both items from our pre-analysis plan that remain unaddressed and exploratory analysis that was not prespecified.


\subsection{Towards a Set of Best Practices} \label{sec:best}

Remotely sensed weather data has become a common component of economic analysis \citep{DellEtAl14, DonaldsonStoreygard16}. Yet there has been little recognition in the economics literature that these data can suffer from measurement error. Nor has there been a convergence on a set of best practices for dealing with measurement error. Few empirical papers today would fail to verify the robustness of the results to different specifications \citep{SimonsohnEtAl20} or different iterations of the data \cite{SteegenEtAl16}. Yet economics papers rarely, if ever, verify the robustness of results to different sources of remote sensing data or the use of different weather metrics. As we have shown in this paper, such a blas\'{e} approach to the use of remote sensing weather data is unjustified.

To investigate the degree to which measurement error is an issue in remote sensing weather data, we tested three hypotheses using a set of heuristics outlined in our pre-analysis plan. Hypothesis 1 considered the degree to which extraction methods used to obfuscate the true location of a household introduced mismeasurement when obfuscated household coordinates are paired with remote sensing data. The preponderance of evidence discussed in section~\ref{sec:ext} leads us to conclude that we cannot reject the null for Hypothesis 1. There is no clear evidence that different obfuscation procedures have different impacts on estimates of agricultural production. Any measurement error which may arise from the use of different extraction methods does not substantially affect estimates. When researchers use publicly available data with obfuscated GPS information, they should feel confident that matching those coordinates with remote sensing data will not introduce measurement error into the analysis.

Our second hypothesis concerned whether the choice of metrics to measure precipitation and temperature are critical to the analysis. Here the concern is that the process economists use to choose one metric, say mean temperature, over another, say growing degree days, is \emph{ad hoc} and lacks the specificity necessary to accurately measure the weather that matters. Certain metrics may mismeasure the relevant meteorological event. If, for example, what matters to human capital development is the number of days with temperatures above a certain threshold, as in \cite{GargEtAl20}, then mean daily temperature mismeasures the relevant event. The results presented in section~\ref{sec:metrics} lead us to reject Hypothesis 2. Based on our heuristics, some weather metrics clearly perform better than others. For precipitation, mean daily rainfall, total seasonal rainfall, the number of rainy days in a growing season, and the share of days in the growing season that saw rain are all strongly correlated with outcomes. For temperature, mean, median, and variance of daily temperature are all frequently correlated with outcomes. Among these seven metrics, mean and total rainfall are highly correlated and could be substituted for each other. Similarly, rainy days and percentage of rainy days are highly correlated, as is mean and median daily temperature. One could substitute these metrics for each other without appreciable differences in results.

Hypothesis 3 considered whether remote sensing products mismeasured the objective facts regarding the volume of precipitation and the temperature in a given location on a given day. The overwhelming evidence presented in section~\ref{sec:satellite} is that some products do, in fact,  mismeasure the weather, though there is general agreement among several products. Specifically, MERRA-2 appears to suffer from classical measurement error, producing noisy data that results in rainfall metrics frequently being uncorrelated with outcomes. ERA5 appears to suffer from non-classical measurement error, producing overestimates of the volume of precipitation and leading to biased coefficients on rainfall metrics. The other four precipitation products all generally report similar data for precipitation. The three temperature products also generally agree with one other. Although there is some evidence that ERA5 reports different temperature data than the other two sources, these differences do not appear to be substantial enough to impact results. However, this lack of impact may be due to the overall poor quality of even the best temperature metrics in predicting outcomes. Whether this is due to temperature actually being uncorrelated with outcomes, or all three products suffering from measurement error is unclear.

An important limitation in our analysis is that we lack data on the objective fact of how much precipitation there actually was or what the temperature actually was in a given location on a given day. While one can easily measure the true size of a field \citep{Carlettoetal17}, the true weight of a harvest \citep{AbayEtAl19}, or the true variety of a seed \citep{KosmowskiEtAl19}, data on the objective facts of the weather frequently do not exist, as it would require rain gauges and thermometers at every location, as well as someone to record the relevant data on a daily (or hourly) basis. This lack of data on the objective facts of the weather is exactly why economists use remote sensing data. Because of this lack of data on the objective facts, we can only compare extraction methods, weather metrics, and remote sensing products relative to each other. For the most part, this does constrain our ability to reject or fail to reject a hypothesis. Extraction methods clearly do not impact results, there are clearly good and bad weather metrics, and some precipitation products clearly report different volumes of precipitation than others. Where we are limited by the lack of objective facts is in drawing conclusions about temperature products. Without knowing the actual temperature experienced by LSMS-ISA households, we cannot determine if the lack of correlation between temperature metrics and outcomes is either a reflection of the true relationship or evidence that all three products mismeasure temperature.

In terms of establishing a set of best practices, we recommend the following:

\begin{itemize}
    \item Researchers need not be concerned about weather data matched to obfuscated household locations. Our results do not change substantially based on which obfuscation method we conduct the analysis on. The current spatial resolution of publicly available remote sensing weather is not fine enough for common extraction methods to result in mismeasurement of the weather that is experienced by a household. Researchers should feel comfortable matching obfuscated coordinates with weather data and need not worry that extraction methods will materially alter their results.
    \item Researchers must take much more care in justifying their choice of weather metric. For the LSMS-ISA data, mean daily rainfall, total seasonal rainfall, the number of rainy days, and percent of rainy days are consistent and positive predictors of outcome in a large number of models and countries. Performance of the other weather metrics are more inconsistent, frequently producing coefficients with opposite signs in similar setting. Since several weather metrics perform equally well, and many weather metrics perform equally poor, one needs to provide a reasonable defense for why one metric (e.g., mean daily) was selected and other metrics (e.g., total seasonal) were not. This is especially true when identification of causal impacts relies on the choice of a weather metric. While specific cases might require a specific weather metric, the reasoning should be made explicit and a justification of why other metrics are inappropriate should be provided.
    \item Researchers should provide evidence that results are robust to the choice of weather metric. There exists substantial cross-country heterogeneity in our results related to rainfall and temperature metrics. This suggests that results are sensitive to specific locations and applications. The current standard in applied economics research is to provide a variety of robustness checks to differences in assumptions about model specification or to changes in the underlying data. When weather is a critical component of the analysis, similar evidence of robustness should be provided regarding the choice of weather metrics. The researcher should be able to answer the question of whether their results hinge on the choice of a particular metric to measure precipitation or temperature, or whether their results reflect a broader relationship between precipitation, temperature, and the outcome of interest.
    \item Researchers should be cautious in using temperature data to predict agricultural outcomes. When looking at agricultural production in the LSMS-ISA data, temperature tends to be a poor predictor. Whether this is because remote sensing products mismeasure temperature or because temperature is not important to agricultural production in tropical Africa is uncertain. Obviously, there are situations that call for the use of temperature in the analysis but simply adding temperature to a regression because the data is available risks overcontrolling. Common temperature metrics for temperate regions, like GDD, are particularly poor predictors of agricultural production in the LSMS-ISA data, and should be avoided.
    \item Researchers must carefully choose which remote sensing source to use in their analysis. Despite the volume of precipitation and the temperature in a given location on a given day being objective facts, remote sensing products differ in how they measure these objective facts. Because of this, remote sensing products can and do disagree on what the weather was. At least for those using the LSMS-ISA data, researchers should not use rainfall data from MERRA-2 or ERA5. Without evidence to the contrary, one cannot assume that the measurement error in MERRA-2 or ERA5 is only limited to the LSMS-ISA, meaning researchers may want to avoid them when using household data from other Sub-Saharan African countries or from countries on other continents.
    \item Researchers may want to demonstrate the robustness of their results to the choice of weather data drawn from different remote sensing products. Again, when weather data is critical to the identification strategy, results should not be sensitive to the choice of remote sensing product. While it is unlikely that the choice between CHIRPS, CPC, ARC2, and TAMSAT will matter, the choice of temperature product could matter in ways that are difficult to predict.
\end{itemize}

\noindent While our analysis is based on agricultural production by LSMS-ISA households in six countries, we believe our conclusions are applicable to and valid in other countries in tropical regions and to other outcomes besides agricultural production. While results are specific to specific contexts, in our case that context is extremely broad. None of our results hinge on a single outcome, a single model, a single year, a single country, or a single region. Our analysis looks at multiple crops using several model specifications. Our data span more than a decade, and come from countries in Eastern, Western, and Southern Africa with different variation in agro-ecolocigal zones and different rainfall patterns. While our two outcomes are both measures of agricultural productivity, many papers that use weather data to investigate non-agricultural outcomes justify the use of weather in identification because of its relationship to agricultural productivity and productivity's relationship with the non-agricultural outcome of interest \citep{Jayachandran06, DescheneGreenstone07, CornoEtAl20}. Therefore, we believe our conclusions, and the best practices built on those conclusions, have broad applicability to the use of remote sensing weather data in economics.


\subsection{Future Work} \label{sec:future}

There are a number of extensions to this work which are currently ongoing. Some of these relate to data and analysis defined in our pre-analysis plan while others are exploratory in that they are not pre-specified.

In terms of pre-specified elements, we are presently working to incorporate data from Mali as well as additional weather metrics. LSMS-ISA data from Mali comes from two rounds but those rounds are not panel data, thus we could not use it as part of the current analysis which includes household fixed effects in four of six model specification. We are also working to incorporate indices of weather variables (e.g. evapotranspiration, water requirement satisfaction index, Palmer drought severity index, standardized precipitation evapotranspiration index). These additional weather metrics are part of the pre-analysis plan but cannot be generated for all remote sensing products. Because of this, and to maintain blinding of the products, we excluded these indices from the current analysis. Additional details on the complete scope of our pre-specified analyses can be found at \href{https://osf.io/8hnz5/wiki/home/}{OSF} and in \cite{MichlerEtAl20}.

With respect to exploratory analyses, at present we have four research questions we are seeking to answer. The first is the degree to which results vary by agro-ecological zone. The analysis in this paper focused on differences across country, which is the level at which data are collected and organized in the LSMS-ISA. Economics researchers is frequently conducted using nationally representative data or regional data from within a single country. However, the relevant organizing structure of weather is not the political boundaries of a country but the geographies that create different agro-ecological zones. We plan to replicate much of the current analysis, ignoring cross-country differences, and instead focusing on differences in agro-ecological zone. Our objective with this question is to determine if some weather metrics or some remote sensing products perform better in one agro-ecological zone versus another. If this is the case, it suggests that researchers using national data may need to use different metrics to measure weather for household in different regions of the country or may need to use data from different remote sensing sources depending on which agro-ecological zone a household lives within.

A second research question which we plan to explore is the degree to which the inclusion of household fixed effects impact results. Here, the integration of cross-sectional data from Mali is critical. Additionally, we will incorporate two waves of cross-sectional data from Malawi, and a single cross section from Tanzania. Our goal is to unpack the result in this paper regarding how little weather matters once one controls for time-invariant household unobservables. While panel data is becoming more common in developing countries, many researchers still rely on cross-sectional data. We are interested in understanding the degree to which results using cross-sectional data may be biased because of the inability to control for household unobservables. If weather matters in cross-sectional contexts but no longer matters once time-invariant household unobservables are controlled for, it indicates that weather may fail the exclusion restrict necessary for its use as an instrumental variable.

Our third research question concerns the extent to which correctly specifying the growing season matters for the use of weather data. The growing seasons differ in northern and southern Nigeria, as do the growing seasons in different regions of Uganda. To account for these differences, we specify different seasons for households in these countries. But, even in a country like Malawi, with a single unimodal growing season, the start and end date of that season can vary depending on where a household is located in the country. Differences in Tanzania can be even more pronounced and our failure to account for this variation may explain why weather appears to matter so little in that country compared to the other five countries. The issue of defining the growing season is another example of potential measurement error in the use of remote sensing weather data. We plan to explore methods for defining the growing season at the household-level based on self-reported planting dates in the LSMS-ISA data along with using remote sensing data to determine the onset of rainfall. If the definition of growing season does matter, it suggest that researchers need to take much more care in defining what days or months to take weather data from.

Our final research question is whether machine learning methods can be effectively used to select weather metrics and remote sensing sources. Our approach has relied on traditional regression techniques to estimate a large contingent of regressions and then determine which metrics matter over a large set of results. However, these research questions also seem well suited to a number of machine learning techniques, in particular feature selection. We are interested to see how the findings of a machine learning algorithm comport with our findings using traditional regression analysis.


\section{Conclusion} \label{sec:conclusion}

In this paper, we investigate the significance and magnitude of the effect of measurement error from various sources on modeling agricultural productivity. Using geospatial weather data from a set of weather products with geo-referenced household survey data from six Sub-Saharan African countries, we are able to provide systematic evidence of measurement error, as related to spatial measurement of weather metrics over a large set of crops and countries. We find that while extraction method does not introduce substantial measurement error into the analysis, results vary depending on the choice of weather metric and the choice of remote sensing data source. Based on these results, we propose a set of best practices for economists seeking to use remote sensing weather in conjunction with household survey data.

This work contributes to an ongoing body of research which explores the implications of measurement error for estimates of agricultural production and agricultural productivity. Previous work has focused on measurement error in self-reported data such as land area, harvest quantity, and seed variety. We contribute to this literature by reporting on the extent of measurement error in remote sensing weather data, inaccuracies introduced by the obfuscation of farm or household GPS coordinates, and inaccuracies introduced by the use of weather metrics that are not fit for the purpose.

The use of remote sensing weather data has become common in economics research. Weather data has been used as an explanatory variable in studying agricultural production and productivity, and also as part of identification strategies to explain everything from migration to human capital development to economic growth to technology adoption. Our results demonstrate that care must be taken when using weather data as measurement error in these data can substantially alter results.


\newpage
\singlespacing
\bibliographystyle{chicago}
\bibliography{LSMSref}


\newpage 
\FloatBarrier


\begin{landscape}
\begin{table}[htbp]	\centering
    \caption{Sources of Weather Data}  \label{tab:weather}
	\scalebox{0.9}
	{ \setlength{\linewidth}{.1cm}\newcommand{\contents}
		{\begin{tabular}{p{0.65\linewidth} lllll}
            \\[-1.8ex]\hline 
			\hline \\[-1.8ex]
            Data set & Length of record & Resolution & Time step & Data & Units \\
            \midrule
            \multicolumn{6}{l}{\textbf{Precipitation}} \\
            -Africa Rainfall Climatology version 2 (ARC2) & 1983-current & 0.1 deg & daily & total precip & mm \\
            -Climate Hazards group InfraRed Precipitation with Station data (CHIRPS) & 1981-current & 0.05 deg & daily & total precip & mm \\
            -CPC Global Unified Gauge-Based Analysis of Daily Precipitation & 1979-current & 0.5 deg & daily & total precip & mm \\
            -European Centre for Medium-Range Weather Forecasts (ECMWF) ERA5 & 1979-current & 0.28 deg & hourly & total precip & m \\
            -Modern-Era Retrospective analysis for Research and Applications, version 2 (MERRA-2) Surface Flux Diagnostics & 1980-current & 0.625x0.5 deg & hourly & rain rate & kg m$^2$ s$^1$ \\
            -Tropical Applications of Meteorology using SATellite data  and ground-based observations (TAMSAT) & 1983-current & 0.0375 deg & daily & total precip & mm \\
            \midrule
            \multicolumn{6}{l}{\textbf{Temperature}} \\
            -CPC Global Unified Gauge-Based Analysis of Daily Temperature & 1979-current & 0.5 deg & daily & min, max temp & C \\
            -European Centre for Medium-Range Weather Forecasts (ECMWF) ERA5 & 1979-current & 0.28 deg & hourly & mean temp & K \\
            -Modern-Era Retrospective analysis for Research and Applications, version 2 (MERRA-2) statD & 1980-current & 0.625x0.5 deg & daily & mean temp & K \\
			\\[-1.8ex]\hline 
			\hline \\[-1.8ex]
    		\multicolumn{6}{l}{\footnotesize \textit{Note}: The table summarizes the remote sensing sources and related details for precipitation and temperature data.}
    	\end{tabular}}
	\setbox0=\hbox{\contents}
    \setlength{\linewidth}{\wd0-2\tabcolsep-.25em}
    \contents}
\end{table}
\end{landscape}


\begin{table}[htbp]	\centering
    \caption{Sources of Household Data}  \label{tab:lsms}
	\scalebox{0.9}
	{ \setlength{\linewidth}{.1cm}\newcommand{\contents}
		{\begin{tabular}{llrrr}
            \\[-1.8ex]\hline 
			\hline \\[-1.8ex]
            Country & Survey Name & Years & Original $n$ & Final $n$  \\
            \midrule
            Ethiopia & Ethiopia Socioeconomic Survey (ERSS) & 2011/2012 & 3,969 & 1,689 \\
                        & & 2013/2014 & 5,262 & 2,865 \\
                        & & 2015/2016 & 4,954 & 2,718 \\
            Malawi & Integrated Household Panel Survey (IHPS)  & 2010/2011 & 3,246 & 1,241 \\
                        & & 2013 & 4,000 & 968 \\
                        & & 2016/2017 & 2,508 & 1,041 \\
            Niger & Enqu\^{e}te Nationale sur les Conditions de Vie des & 2011 & 3,968 & 2,223 \\
                        & $\:$ M\'{e}nages et l'Agriculture (ECVMA) & 2014 & 3,617 & 1,690 \\
            Nigeria & General Household Survey (GHS) & 2010/2011 & 5,000 & 2,833 \\
                        & & 2012/2013 & 4,802 & 2,768 \\
                        & & 2015/2016 & 4,613 & 2,783 \\
            Tanzania & Tanzania National Panel Survey (TZNPS) & 2008/2009 & 3,280 & 1,907 \\
                        & & 2010/2011 & 3,924 & 1,914 \\
                        & & 2012/2013 & 3,924 & 1,848 \\
            Uganda & Uganda National Panel Survey (UNPS) & 2009/2010 & 2,975 & 1,704 \\
                        & & 2010/2011 & 2,716 & 1,741 \\
                        & & 2011/2012 & 2,850 & 1,805 \\
            \midrule
            Total & 6 countries & 17 waves & 65,608 & 33,738 \\
			\\[-1.8ex]\hline 
			\hline \\[-1.8ex]
    		\multicolumn{5}{l}{\footnotesize \textit{Note}: The table summarizes the household data details for each country, per LSMS Basic Information Documents.}
    	\end{tabular}}
	\setbox0=\hbox{\contents}
    \setlength{\linewidth}{\wd0-2\tabcolsep-.25em}
    \contents}
\end{table}


\begin{table}[htbp]	\centering
	\caption{Household Variables and Definitions} \label{tab:HHvar}
	\scalebox{0.9}
	{ \setlength{\linewidth}{.1cm}\newcommand{\contents}
		{\begin{tabular}{ll}
			\\[-1.8ex]\hline 
			\hline \\[-1.8ex]
			\multicolumn{2}{l}{\emph{\textbf{Panel A}: Outcome Variables}} \\
			\multicolumn{1}{l}{Yield} & \multicolumn{1}{p{11cm}}{Output in kilograms per hectare for the primary cereal crop in each country data set} \\
			\multicolumn{1}{l}{Value} & \multicolumn{1}{p{11cm}}{Output in real USD per hectare for all seasonal farm crop production} \\
			\midrule
			& \\
			\multicolumn{2}{l}{\emph{\textbf{Panel B}: Input Variables}} \\
			\multicolumn{1}{l}{Labor use rate} & \multicolumn{1}{p{11cm}}{Number of days per hectare (combined or divided by age and gender)} \\
			\multicolumn{1}{l}{Fertilizer application rate} & \multicolumn{1}{p{11cm}}{Kilograms per hectare (combined or divided into Urea and DAP)} \\
			\multicolumn{1}{l}{Seed application rate} & \multicolumn{1}{l}{Value in USD per hectare} \\
			\multicolumn{1}{l}{Pesticide use} & \multicolumn{1}{l}{Equal to 1 if yes, 0 if no} \\
			\multicolumn{1}{l}{Herbicide use} & \multicolumn{1}{l}{Equal to 1 if yes, 0 if no} \\
			\multicolumn{1}{l}{Irrigation use} & \multicolumn{1}{l}{Equal to 1 if yes, 0 if no} \\
			\\[-1.8ex]\hline 
			\hline \\[-1.8ex]
			\multicolumn{2}{p{\linewidth}}{\footnotesize  \textit{Note}: The table presents definitions for included outcome and input variables from LSMS sources defined in Table~\ref{tab:lsms}.} \\
		\end{tabular}}
	\setbox0=\hbox{\contents}
    \setlength{\linewidth}{\wd0-2\tabcolsep-.25em}
    \contents}
\end{table}


\begin{table}[htbp]	\centering
	\caption{Weather Variables \& Transformations} \label{tab:Wvar}
	\scalebox{0.9}
	{ \setlength{\linewidth}{.1cm}\newcommand{\contents}
		{\begin{tabular}{ll}
			\\[-1.8ex]\hline 
			\hline \\[-1.8ex]
			\multicolumn{2}{l}{\emph{\textbf{Panel A}: Rainfall}} \\
			\multicolumn{1}{l}{Daily rainfall} & \multicolumn{1}{l}{In millimeters} \\
			\multicolumn{1}{l}{Mean} & \multicolumn{1}{p{11cm}}{The first moment of the daily rainfall distribution for the growing season$^\dagger$} \\
			\multicolumn{1}{l}{Median} & \multicolumn{1}{p{11cm}}{The median daily rainfall for the growing season$^\dagger$} \\
			\multicolumn{1}{l}{Variance} & \multicolumn{1}{p{11cm}}{The second moment of the daily rainfall distribution for the growing season$^\dagger$} \\
			\multicolumn{1}{l}{Skew} & \multicolumn{1}{p{11cm}}{The third moment of the daily rainfall distribution for the growing season$^\dagger$} \\
			\multicolumn{1}{l}{Total} & \multicolumn{1}{p{11cm}}{Cumulative daily rainfall for the growing season$^\dagger$} \\
			\multicolumn{1}{l}{Deviations in total rainfall} & \multicolumn{1}{p{11cm}}{The z-score for cumulative daily rainfall for the growing season$^\dagger$} \\
			\multicolumn{1}{l}{Scaled deviations in total rainfall} & \multicolumn{1}{p{11cm}}{The z-score for cumulative daily rainfall for the growing season$^\dagger$} \\
			\multicolumn{1}{l}{Rainfall days} & \multicolumn{1}{p{11cm}}{The number of days with at least 1 mm of rain for the growing season$^\dagger$} \\
			\multicolumn{1}{l}{Deviation in rainfall days} & \multicolumn{1}{p{11cm}}{The number of days with rain for the growing season minus the long run average$^*$} \\
			\multicolumn{1}{l}{No rain days} & \multicolumn{1}{p{11cm}}{The number of days with less than 1 mm of rain for the growing season$^\dagger$} \\
			\multicolumn{1}{l}{Deviation in no rain days} & \multicolumn{1}{p{11cm}}{The number of days without rain for the growing season minus the long run average$^*$} \\
			\multicolumn{1}{l}{Share of rainy days} & \multicolumn{1}{p{11cm}}{The percent of growing season days with rain$^\dagger$} \\
			\multicolumn{1}{l}{Deviation in share of rainy days} & \multicolumn{1}{p{11cm}}{The percent of growing season days with rain minus the long run average$^\dagger$ $^*$} \\
			\multicolumn{1}{l}{Intra-season dry spells} & \multicolumn{1}{p{11cm}}{The maximum length of time (measured in days) without rain during the growing season$^\dagger$} \\
			\midrule
			& \\
			\multicolumn{2}{l}{\emph{\textbf{Panel B}: Temperature}} \\
			\multicolumn{1}{l}{Daily average temperature} & \multicolumn{1}{l}{In Celsius} \\
			\multicolumn{1}{l}{Daily maximum temperature} & \multicolumn{1}{l}{In Celsius} \\	
			\multicolumn{1}{l}{Mean} & \multicolumn{1}{p{11cm}}{The first moment of the daily temperature distribution for the growing season$^\dagger$} \\
			\multicolumn{1}{l}{Median} & \multicolumn{1}{p{11cm}}{The median daily temperature for the growing season$^\dagger$} \\
			\multicolumn{1}{l}{Variance} & \multicolumn{1}{p{11cm}}{The second moment of the daily temperature distribution for the growing season$^\dagger$} \\
			\multicolumn{1}{l}{Skew} & \multicolumn{1}{p{11cm}}{The third moment of the daily temperature distribution for the growing season$^\dagger$} \\
			\multicolumn{1}{l}{Growing degree days (GDD)} & \multicolumn{1}{p{11cm}}{The number of days within bound temperature for the growing season, following \cite{RS1991}$^\dagger$} \\
			\multicolumn{1}{l}{Deviation in GDD} & \multicolumn{1}{p{11cm}}{GDD for the growing season minus the long run average$^\dagger$ $^*$} \\
			\multicolumn{1}{l}{Scaled deviation in GDD} & \multicolumn{1}{p{11cm}}{The z-score for GDD} \\
			\multicolumn{1}{l}{Maximum temperature} & \multicolumn{1}{p{11cm}}{The average maximum daily temperature} \\
			\\[-1.8ex]\hline 
			\hline \\[-1.8ex]
			\multicolumn{2}{p{\linewidth}}{\footnotesize  \textit{Note}: The table presents definitions for included weather variables and transformations from weather sources defined in Table~\ref{tab:weather}. $^\dagger$Growing season determined for each country following \href{http://www.fao.org/agriculture/seed/cropcalendar/welcome.do}{FAO crop calendar} (see Table~\ref{tab:growseason}). $^*$For variables when ``long run'' is referenced, long run is defined as the entire length of the weather data set. While each weather source has a different start date, to ensure blinding all data sets were shortened to 1983, which is the latest start date of the data sources.} \\
		\end{tabular}}
	\setbox0=\hbox{\contents}
    \setlength{\linewidth}{\wd0-2\tabcolsep-.25em}
    \contents}
\end{table}


\begin{table}[htbp]	
	\begin{center}\caption{Data Scope \label{tab:sourcesetc}}
		\resizebox{6in}{!}
		{\setlength{\linewidth}{.1cm}\newcommand{\contents}
			{\begin{tabular}{cc}
		        	\\[-1.8ex]\hline 
    		    	\hline \\[-1.8ex]
					\multicolumn{1}{l}{Countries (6)}& \multicolumn{1}{l}{Ethiopia, Malawi, Niger, Nigeria, Tanzania, Uganda}\\
					\midrule
					\multicolumn{1}{l}{Weather Products (9)}& \multicolumn{1}{l}{Precipitation}\\
					\multicolumn{1}{l}{}& \multicolumn{1}{l}{ \ \ \ \ \  \ ARC2, CHIRPS, CPC, ERA5, MERRA-2, TAMSAT}\\
					\multicolumn{1}{l}{}& \multicolumn{1}{l}{Temperature}\\
					\multicolumn{1}{l}{}& \multicolumn{1}{l}{ \ \ \ \ \  \ CPC, ERA5, MERRA-2}\\
					\midrule
					\multicolumn{1}{l}{Extractions (10)}& \multicolumn{1}{l}{Points (simple)}\\
					\multicolumn{1}{l}{}& \multicolumn{1}{l}{ \ \ \ \ \  \ household, EA centerpoint, modified EA centerpoint, administrative centroid}\\
					\multicolumn{1}{l}{}& \multicolumn{1}{l}{Points (bilinear)}\\
					\multicolumn{1}{l}{}& \multicolumn{1}{l}{ \ \ \ \ \  \ household, EA centerpoint, modified EA centerpoint, administrative centroid}\\
					\multicolumn{1}{l}{}& \multicolumn{1}{l}{Area (zonal mean)}\\
					\multicolumn{1}{l}{}& \multicolumn{1}{l}{ \ \ \ \ \  \ EA range of uncertainty, administrative unit}\\
					\midrule
					\multicolumn{1}{l}{Weather metrics (22)}& \multicolumn{1}{l}{14 rainfall}\\
					\multicolumn{1}{l}{}& \multicolumn{1}{l}{8 temperature}\\
					\midrule
					\multicolumn{1}{l}{Dependent variables (2)}& \multicolumn{1}{l}{value, quantity}\\
					\midrule
					\multicolumn{1}{l}{Specifications (6)}& \multicolumn{1}{l}{linear, quadratic: with/out controls, with/out fixed effects}\\
		        	\\[-1.8ex]\hline 
    		    	\hline \\[-1.8ex]
    		    	\multicolumn{2}{p{\linewidth}}{\footnotesize  \textit{Note}: The table summarizes the scope of the data across country, weather products, extraction method, weather metric, dependent variable, and econometric specification.} \\
			\end{tabular}}
			\setbox0=\hbox{\contents}
			\setlength{\linewidth}{\wd0-2\tabcolsep-.25em}
			\contents}
	\end{center}
\end{table}


\begin{table}[htbp]
	\begin{center}\caption{Household Data Summary Statistics \label{tab:sumstattab}}
		\resizebox{6in}{!}
		{\setlength{\linewidth}{.1cm}\newcommand{\contents}
			{\begin{tabular}{lcccccc}
		        	\\[-1.8ex]\hline 
    		    	\hline \\[-1.8ex]
				                    &\multicolumn{1}{c}{Ethiopia}&\multicolumn{1}{c}{Malawi}&\multicolumn{1}{c}{Niger}&\multicolumn{1}{c}{Nigeria}&\multicolumn{1}{c}{Tanzania}&\multicolumn{1}{c}{Uganda}\\
\midrule
&&&&&& \\ 
\multicolumn{7}{l}{\emph{\textbf{Panel A}: Total Farm Production}} \\ 
Total farm production (2010 USD)&       245.4&       351.2&       188.5&       663.5&       207.3&       169.0\\
                    &     (322.8)&     (727.5)&     (201.7)&     (704.1)&     (269.4)&     (226.0)\\
Total farmed area (ha)&       1.055&       0.719&       11.14&       2.197&       1.598&       5.553\\
                    &     (5.670)&     (0.659)&     (15.59)&     (3.598)&     (2.985)&     (12.32)\\
Total farm yield (2010 USD/ha)&       440.6&       525.4&       60.93&       680.9&       235.2&       93.64\\
                    &     (607.2)&     (560.3)&     (169.0)&     (850.7)&     (286.8)&     (145.0)\\
Total farm labor rate (days/ha)&       434.4&       247.8&       94.94&       156.3&       311.1&       282.6\\
                    &     (834.9)&     (193.2)&     (219.1)&     (210.3)&     (353.9)&     (353.4)\\
Total farm fertilizer rate (kg/ha)&       56.26&       0.691&       5.120&       99.02&       20.89&       0.160\\
                    &     (94.98)&     (0.393)&     (23.83)&     (219.4)&     (68.17)&     (1.136)\\
Total farm pesticide use (\%)&      0.062&      0.042&      0.080&       0.212&      0.033&      0.066\\
                    &     (0.240)&     (0.200)&     (0.271)&     (0.408)&     (0.177)&     (0.248)\\
Total farm herbicide use (\%)&       0.235&      0.013&      0.037&       0.292&      0.098&      0.030\\
                    &     (0.424)&     (0.112)&     (0.190)&     (0.455)&     (0.297)&     (0.169)\\
Total farm irrigation use (\%)&      0.068&      0.015&           0.000&      0.027&      0.042&      0.020\\
                    &     (0.252)&     (0.123)&         (0.000)&     (0.162)&     (0.201)&     (0.141)\\
\midrule 
Observations&        7272&        3250&        3913&        8384&        5669&        5250\\
\midrule
&&&&&& \\ 
\multicolumn{7}{l}{\emph{\textbf{Panel B}: Primary Crop Production}} \\  
Primary crop production (kg)&       314.6&       636.8&       471.5&       763.8&       608.5&       51.50\\
                    &     (476.6)&    (1035)&     (488.6)&    (1151)&     (793.5)&     (88.16)\\
Primary crop farmed area (ha)&       0.322&       0.661&       4.930&       0.720&       0.934&       1.501\\
                    &     (1.310)&     (0.622)&     (6.451)&     (1.115)&     (1.417)&     (2.669)\\
Primary crop yield (kg/ha)&      1593&      1192&       219.8&      1620&      1059&       87.89\\
                    &    (1783)&    (1242)&     (363.5)&    (2242)&    (1193)&     (146.7)\\
Primary crop labor rate (days/ha)&       466.1&       249.6&       70.72&       161.8&       263.2&       306.4\\
                    &     (666.6)&     (198.8)&     (121.1)&     (235.9)&     (295.9)&     (387.4)\\
Primary crop fertilizer rate (kg/ha)&       71.28&       0.722&       1.945&       146.3&       25.75&       0.196\\
                    &     (133.7)&     (0.399)&     (7.754)&     (302.0)&     (80.67)&     (1.638)\\
Primary crop pesticide use (\%)&      0.026&      0.039&      0.055&       0.189&      0.020&      0.034\\
                    &     (0.159)&     (0.193)&     (0.228)&     (0.392)&     (0.139)&     (0.180)\\
Primary crop herbicide use (\%)&      0.042&      0.011&      0.028&       0.378&      0.095&      0.022\\
                    &     (0.201)&     (0.106)&     (0.164)&     (0.485)&     (0.292)&     (0.146)\\
Primary crop irrigation use (\%)&      0.052&      0.014&           0&      0.019&      0.027&      0.020\\
                    &     (0.221)&     (0.118)&         (0)&     (0.135)&     (0.161)&     (0.138)\\
\midrule
Observations&        3906&        3198&        3492&        3590&        4076&        2509\\   		    		\\[-1.8ex]\hline 
    		    	\hline \\[-1.8ex]
    		    	\multicolumn{7}{p{\linewidth}}{\footnotesize  \textit{Note}: The table presents the mean  and (standard deviation) of total farm production and primary crop production. Statistics are calculated for each country, aggregated across all waves used.} \\
			\end{tabular}}
			\setbox0=\hbox{\contents}
			\setlength{\linewidth}{\wd0-2\tabcolsep-.25em}
			\contents}
	\end{center}
\end{table}


\newpage 
\FloatBarrier 

\begin{figure}[!htbp]
	\begin{minipage}{\linewidth}
		
		\caption{Varying Resolution of Rainfall Measurement}
    	\label{fig:rain_res}
		\begin{center}
			\includegraphics[width=.9\linewidth,keepaspectratio]{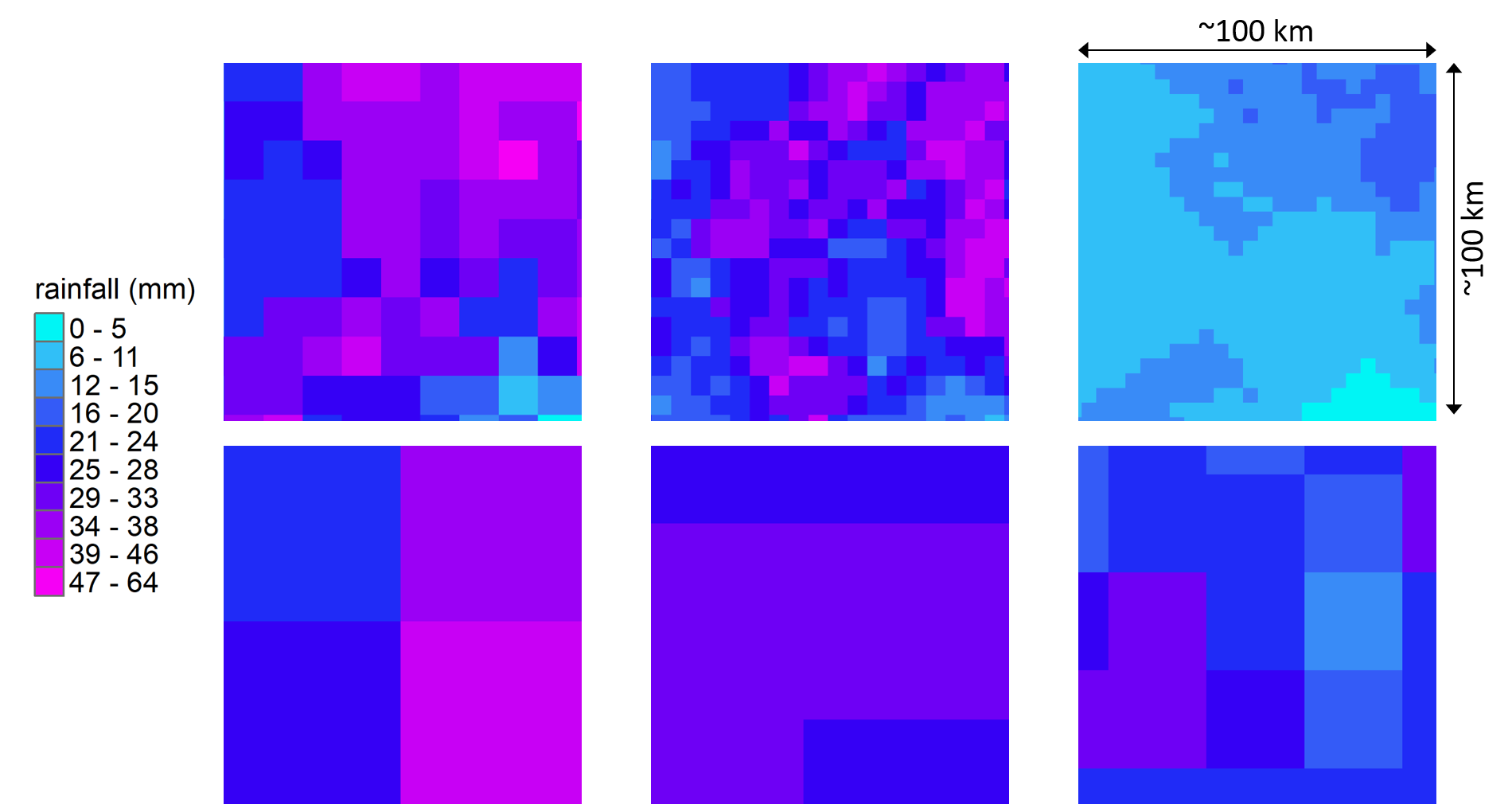}
		\end{center}
		\footnotesize  \textit{Note}: The figure captures rainfall as measured by all six precipitation products for the same 100km x 100km area on a single day (7 January 2010).

	    \vspace{2cm}
	
		\caption{Varying Resolution of Temperature Measurement}
    	\label{fig:temp_res}
		\begin{center}
			\includegraphics[width=.9\linewidth,keepaspectratio]{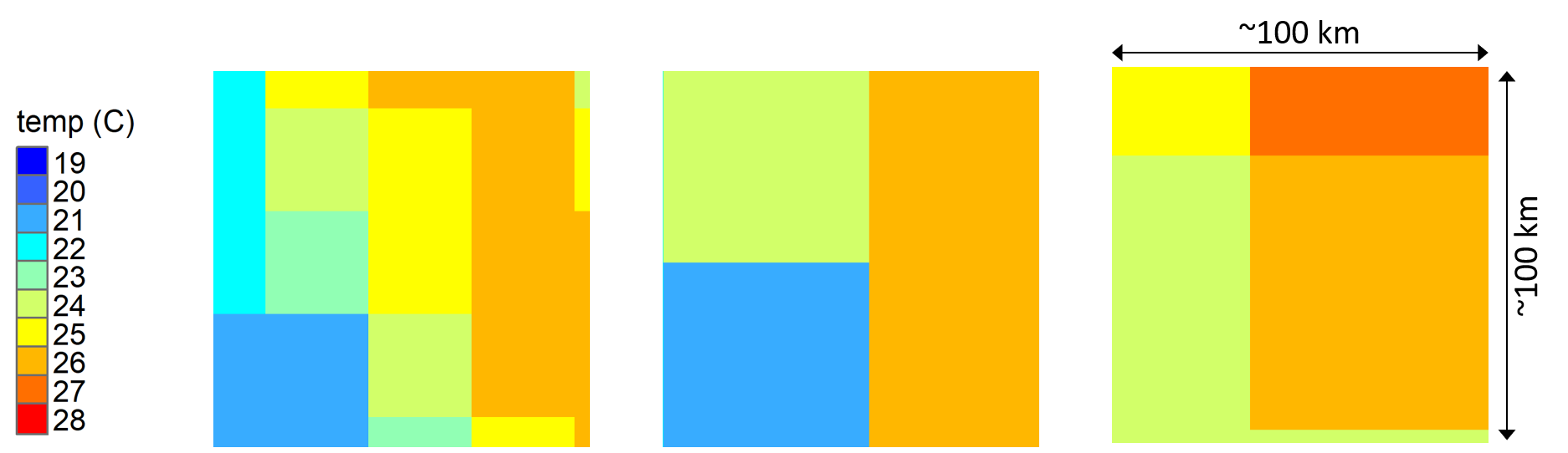}
		\end{center}
		\footnotesize  \textit{Note}: The figure captures temperature as measured by all three temperature products for the same 100km x 100km area on a single day (7 January 2010).
	\end{minipage}
\end{figure}


\begin{figure}[!htbp]
	\begin{minipage}{\linewidth}
		\caption{Visualization of Extraction Methods}
    	\label{fig:features}
		\begin{center}
			\includegraphics[width=.5\linewidth,keepaspectratio]{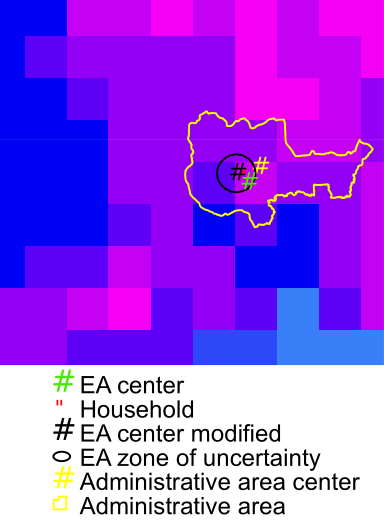}
		\end{center}
		\footnotesize  \textit{Note}: The figure presents the different extraction methods (see Table~\ref{tab:sourcesetc}) and how the measurement of extraction method would vary across a particular precipitation product (from Figure~\ref{fig:rain_res}).
	\end{minipage}

\end{figure}


\newpage 

\begin{landscape}
\begin{figure}[!htbp]
	\begin{minipage}{\linewidth}	
		\caption{Distribution of Total Season Rainfall, by Remote Sensing Source and Agro-ecological Zone}
    	\label{fig:density_aez_rf}
		\begin{center}
			\includegraphics[width=.95\linewidth,keepaspectratio]{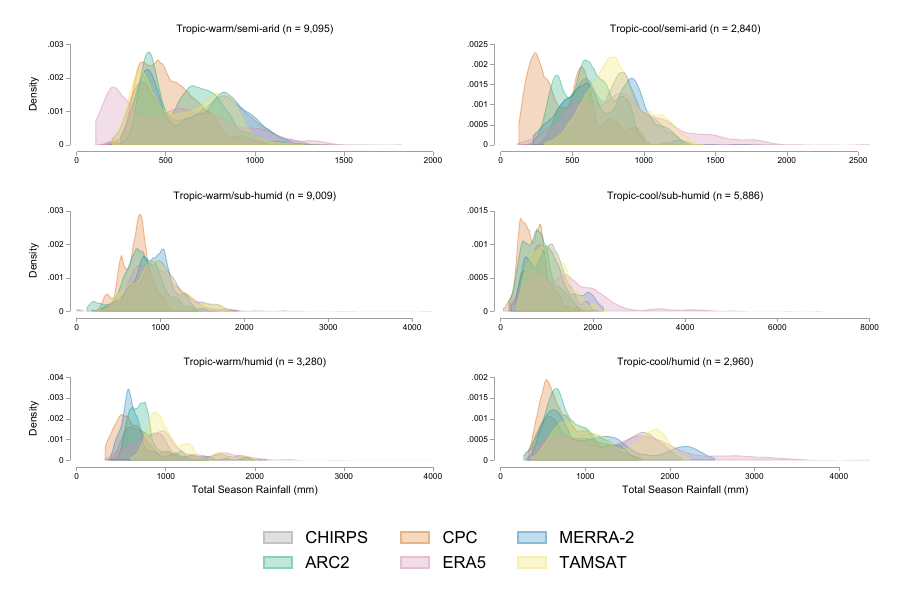}
		\end{center}
		\footnotesize  \textit{Note}: The figure presents rainfall distributions pooled across all countries and years, disaggregated by agro-ecological zones. 
	\end{minipage}
\end{figure}
\end{landscape}	

\begin{landscape}
\begin{figure}[!htbp]
	\begin{minipage}{\linewidth}		
		\caption{Prediction of Mean Number of No Rain Days, by Remote Sensing Source and Agro-ecological Zone}
		\label{fig:norain_aez_rf}
		\begin{center}
			\includegraphics[width=.9\linewidth,keepaspectratio]{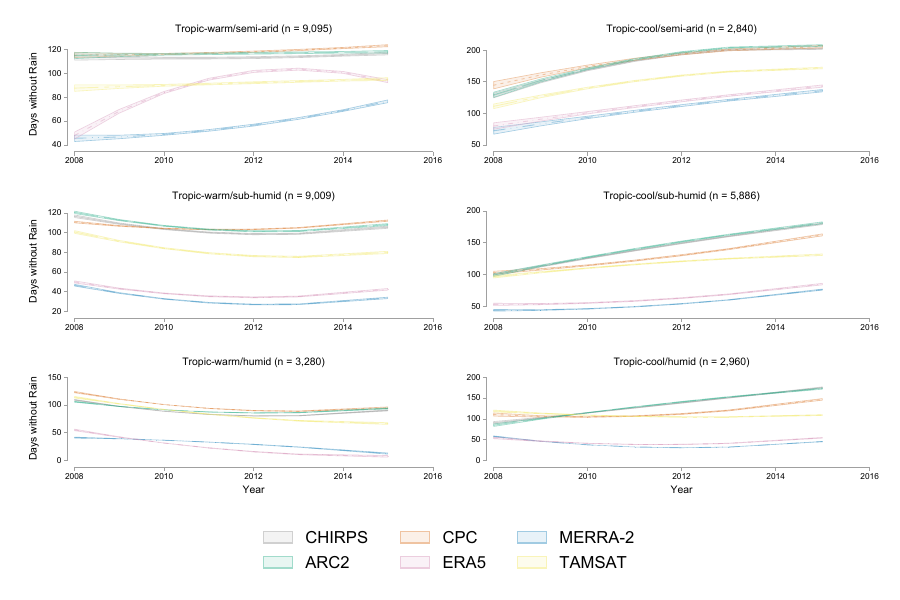}
		\end{center}
		\footnotesize  \textit{Note}: The figure presents the mean number of no rain days, pooled across all countries and years, disaggregated by agro-ecological zones. Prediction made via Fractional-Polynomial. 
	\end{minipage}
\end{figure}
\end{landscape}	

\begin{landscape}
\begin{figure}[!htbp]
	\begin{minipage}{\linewidth}
		\caption{Distribution of Mean Seasonal Temperature, by Remote Sensing Source and Agro-ecological Zone}
    	\label{fig:density_aez_tp}
		\begin{center}
			\includegraphics[width=.95\linewidth,keepaspectratio]{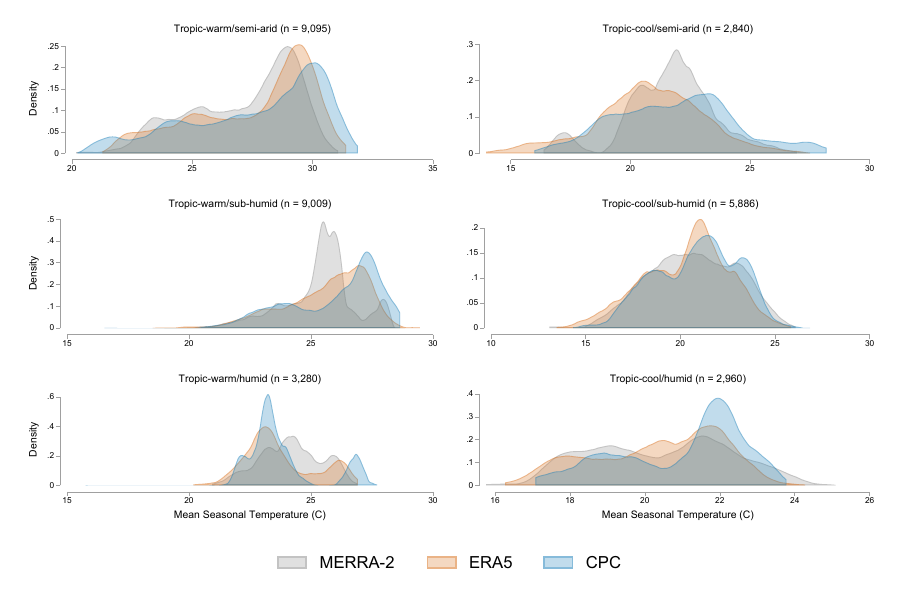}
		\end{center}
		\footnotesize  \textit{Note}:  The figure presents temperature distributions pooled across all countries and years, disaggregated by agro-ecological zones.
	\end{minipage}
\end{figure}
\end{landscape}	

\begin{landscape}
\begin{figure}[!htbp]
	\begin{minipage}{\linewidth}
		\caption{Prediction of Mean Number of Mean Growing Degree Days, by Remote Sensing Source and Agro-ecological Zone}
		\label{fig:gdd_aez_tp}
		\begin{center}
			\includegraphics[width=.9\linewidth,keepaspectratio]{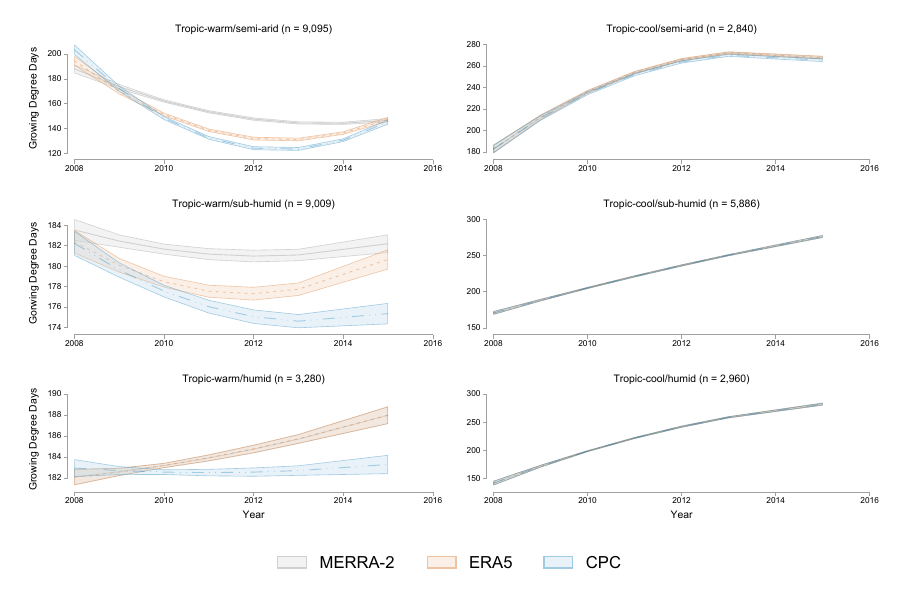}
		\end{center}
		\footnotesize  \textit{Note}: The figure presents the mean number of growing degree days pooled across all countries and years, disaggregated by agro-ecological zones. Prediction made via Fractional-Polynomial. 
	\end{minipage}
\end{figure}
\end{landscape}	


\newpage 


\begin{figure}[!htbp]
	\begin{minipage}{\linewidth}		
		\caption{Mean Adjusted R$^2$, by Extraction and Model}
		\label{fig:r2_ext}
		\begin{center}
			\includegraphics[width=\linewidth,keepaspectratio]{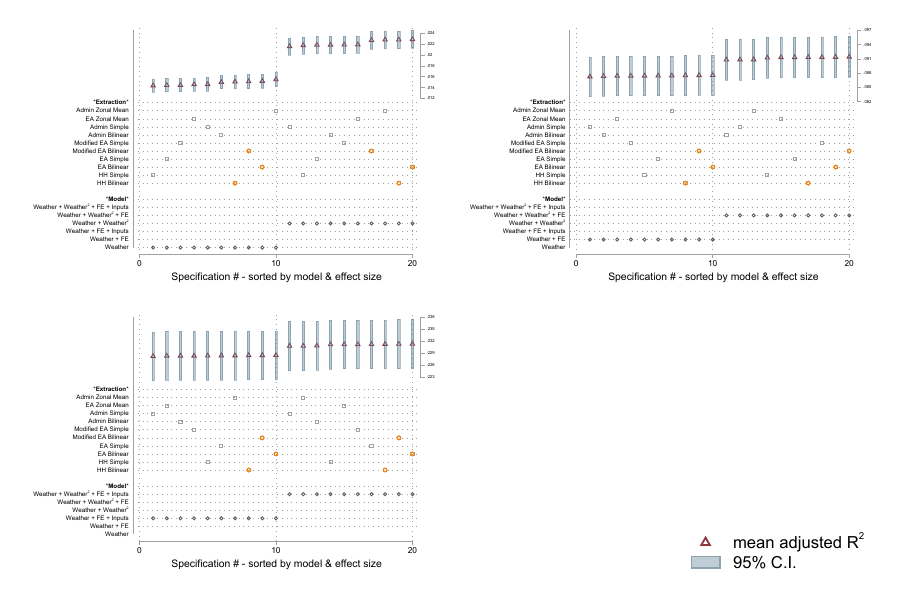}
		\end{center}
		\footnotesize  \textit{Note}: The figure presents the mean adjusted R$^2$, by extraction method and model specification, aggregated over country, weather metric, remote sensing source, and outcome variable. Across panels, the figure includes 77,760 regressions, with each panel including 25,920 regressions. Each column includes 1,296 regressions, which are for each specification model and each extraction method. The northwest panel presents results from the linear and quadratic model specifications. The northeast panel presents results from the linear and quadratic model specifications with fixed effects. The southwest panel presents results from the linear and quadratic model specifications with fixed effects and input controls. Orange diamonds identify bilinear extract methods that tend to perform particularly well (household, EA, and modified EA).
	\end{minipage}	
\end{figure}


\begin{landscape}
\begin{figure}[!htbp]
	\begin{minipage}{\linewidth}		
		\caption{$p$-values of Rainfall and Temperature, by Extraction Method}
		\label{fig:pval_ext}
		\begin{center}
			\includegraphics[width=.9\linewidth,keepaspectratio]{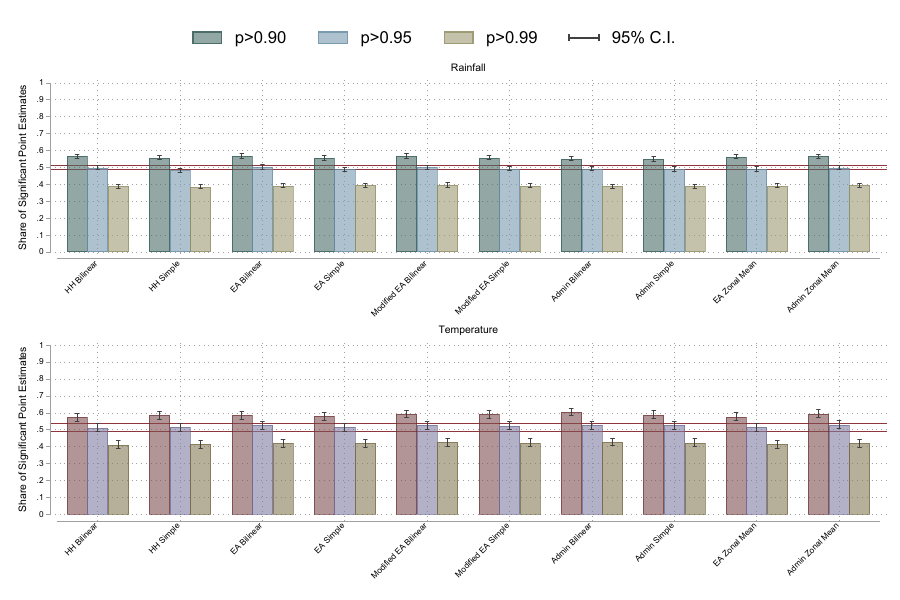}
		\end{center}
		\footnotesize  \textit{Note}: The figure displays the share of coefficients on the rainfall and temperature variables that are statistically significant from each extraction method, aggregated over country, weather metric, remote sensing source, outcome variable, and specification. The northern panel presents rainfall while the southern panel presents temperature. The data summarized in the northern panel includes 60,480 regressions, with each column including 6,048 regressions. The data summarized in the southern panel includes 17,280 regressions, with each column including 1,728 regressions.  
	\end{minipage}	
\end{figure}
\end{landscape}

\begin{landscape}
\begin{center}
\begin{figure}[!htbp]
	\begin{minipage}{\linewidth}
		\caption{$p$-values of Rainfall, by Country and Extraction Method}
		\label{fig:pval_ext_rf}
		\begin{center}
			\includegraphics[width=.95\linewidth,keepaspectratio]{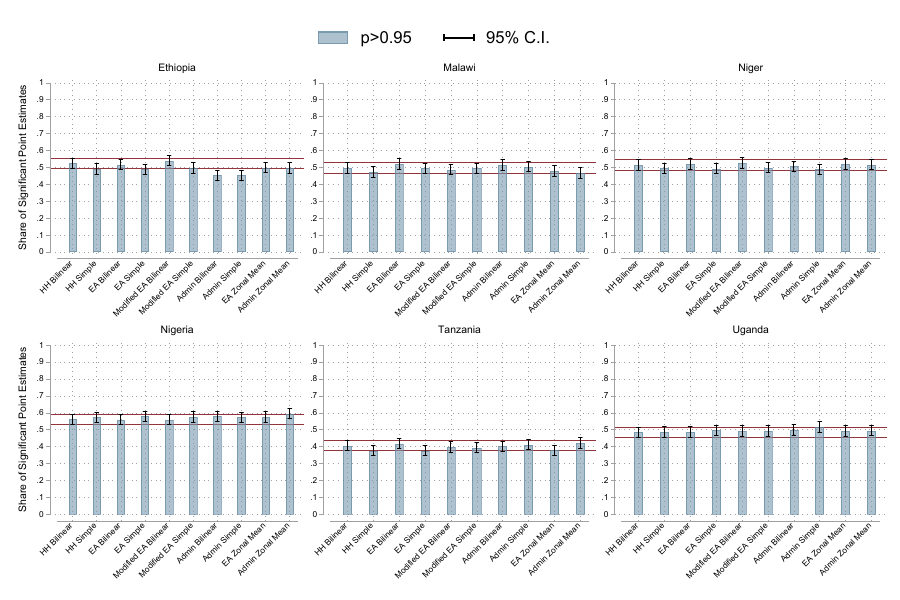}
		\end{center}
		\footnotesize  \textit{Note}: The figure displays the share of coefficients on the rainfall variables that are statistically significant from each extraction method for each country, aggregated over weather metric, remote sensing source, outcome variable, and specification. The figure presents results from a total of 60,480 regressions. Each country includes results from 10,080 regressions and thus each column is based on 1,008 regressions. 
	\end{minipage}	
\end{figure}
\end{center}
\end{landscape}

\begin{landscape}
\begin{center}
\begin{figure}[!htbp]
	\begin{minipage}{\linewidth}		
		\caption{$p$-values of Temperature, by Country and Extraction Method}
		\label{fig:pval_ext_tp}
		\begin{center}
			\includegraphics[width=.95\linewidth,keepaspectratio]{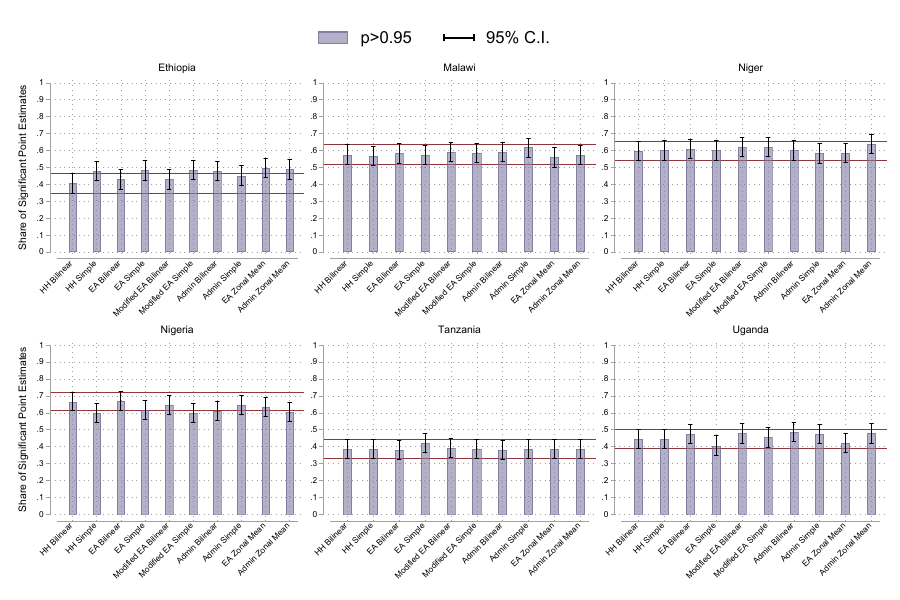}
		\end{center}
		\footnotesize  \textit{Note}: The figure displays the share of coefficients on the temperature variables that are statistically significant from each extraction method for each country, aggregated over weather metric, remote sensing source, outcome variable, and specification. The figure presents results from a total of 17,280 regressions. Each country includes results from 2,880 regressions and thus each column is based on 288 regressions. 
	\end{minipage}	
\end{figure}
\end{center}
\end{landscape}


\newpage 


\begin{figure}[!htbp]
	\begin{minipage}{\linewidth}		
		\caption{Mean Adjusted R$^2$, by Rainfall Metric and Model}
		\label{fig:r2_rf}
		\begin{center}
			\includegraphics[width=\linewidth,keepaspectratio]{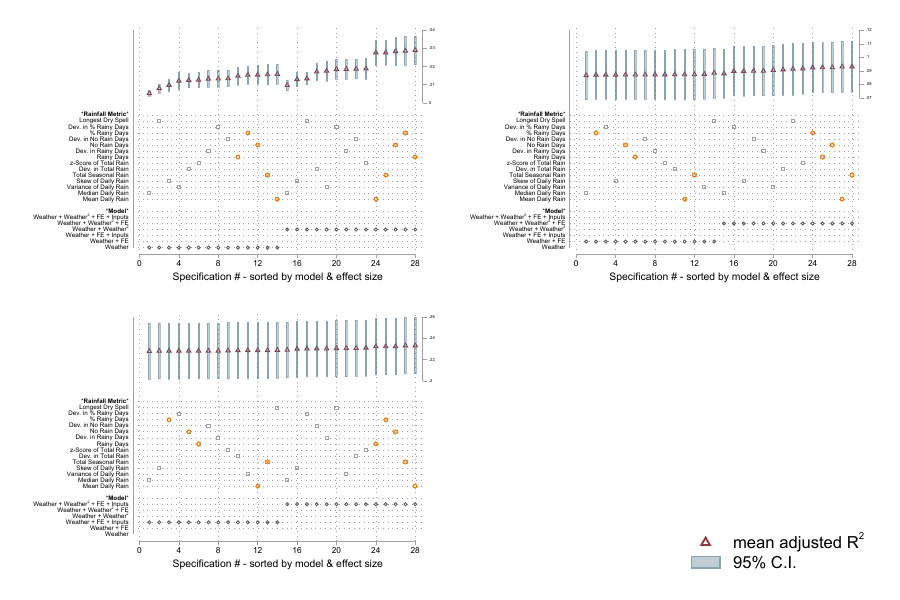}
		\end{center}
		\footnotesize  \textit{Note}: The figure presents the mean adjusted R$^2$, by rainfall metric and model specification, aggregated over country, remote sensing source, and outcome variable. Only one extraction method is included. Across panels, the figure includes 6,048 regressions, with each panel including 2,016 regressions. Each column includes 72 regressions, which are for each specification model and each rainfall metric. The northwest panel presents results from the linear and quadratic model specifications. The northeast panel presents results from the linear and quadratic model specifications with fixed effects. The southwest panel presents results from the linear and quadratic model specifications with fixed effects and input controls. Orange diamonds identify rainfall metrics that tend to perform particularly well.
	\end{minipage}	
\end{figure}

\begin{figure}[!htbp]
	\begin{minipage}{\linewidth}		
		\caption{Mean Adjusted R$^2$, by Temperature Metric and Model}
		\label{fig:r2_tp}
		\begin{center}
			\includegraphics[width=\linewidth,keepaspectratio]{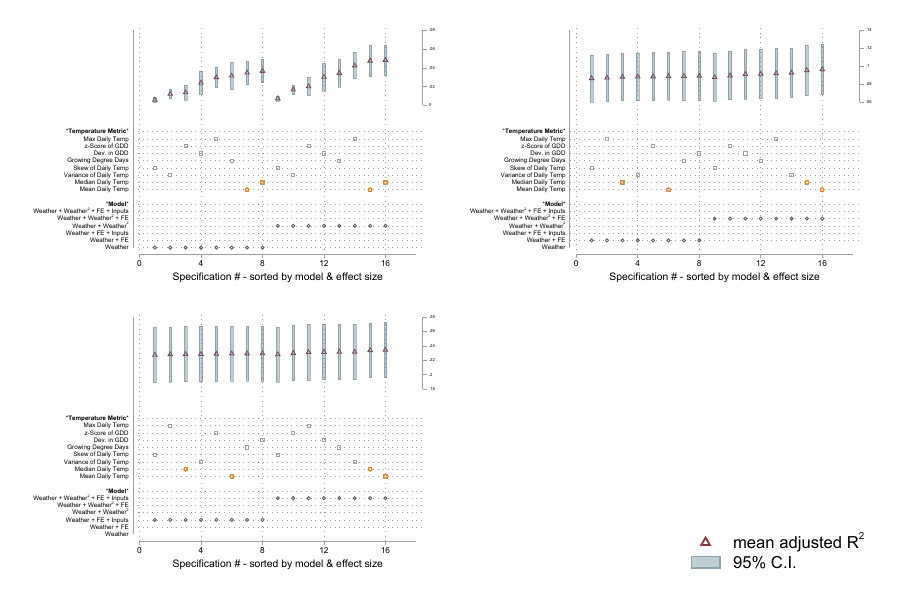}
		\end{center}
		\footnotesize  \textit{Note}: The figure presents mean adjusted R$^2$, by temperature metric and model specification, aggregated over country, remote sensing source, and outcome variable. Only one extraction method is included. Across panels, the figure includes 1,728 regressions, with each panel including 576 regressions. Each column includes 36 regressions, which are for each specification model and each temperature metric. The northwest panel presents results from the linear and quadratic model specifications. The northeast panel presents results from the linear and quadratic model specifications with fixed effects. The southwest panel presents results from the linear and quadratic model specifications with fixed effects and input controls. Orange diamonds identify temperature metrics that tend to perform particularly well.
	\end{minipage}	
\end{figure}


\begin{landscape}
\begin{figure}[!htbp]
	\begin{minipage}{\linewidth}		
		\caption{$p$-values of Rainfall and Temperature, by Weather Metric}
		\label{fig:pval_varname}
		\begin{center}
			\includegraphics[width=.9\linewidth,keepaspectratio]{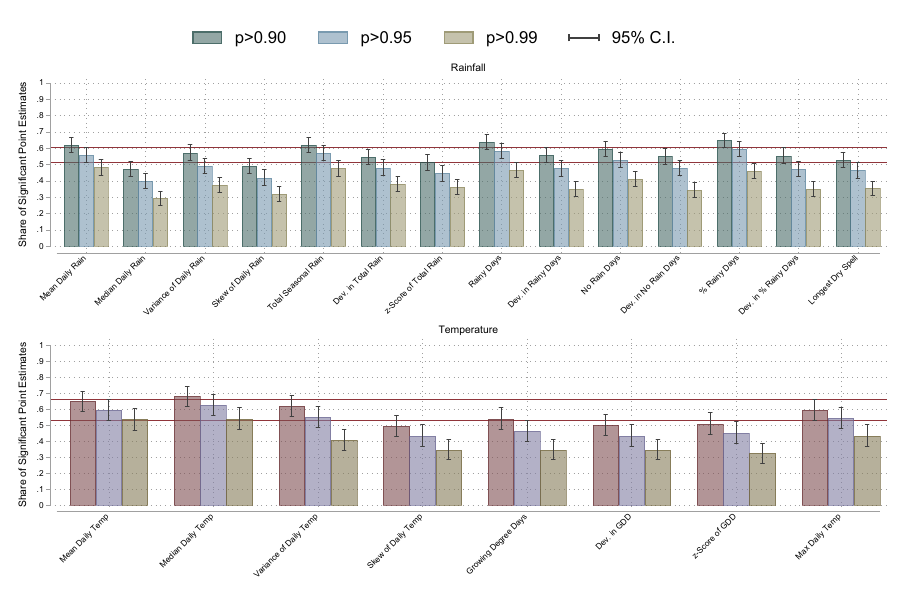}
		\end{center}
		\footnotesize  \textit{Note}: The figure displays the share of coefficients on the rainfall and temperature variables that are statistically significant from each weather metric, aggregated over country, remote sensing source, outcome variable, and specification. Only one extraction method is included. The data summarized in the northern panel is for rainfall and includes 6,048 regressions, with each column including 432 regressions. The data summarized in the southern panel is for temperature and includes 1,728 regressions, with each column including 216 regressions.
	\end{minipage}	
\end{figure}
\end{landscape}

\begin{landscape}
	\begin{center}
		\begin{figure}[!htbp]
			\begin{minipage}{\linewidth}
				\caption{$p$-values, by Country and Rainfall Metric}
				\label{fig:pval_varname_rf}
				\begin{center}
					\includegraphics[width=.95\linewidth,keepaspectratio]{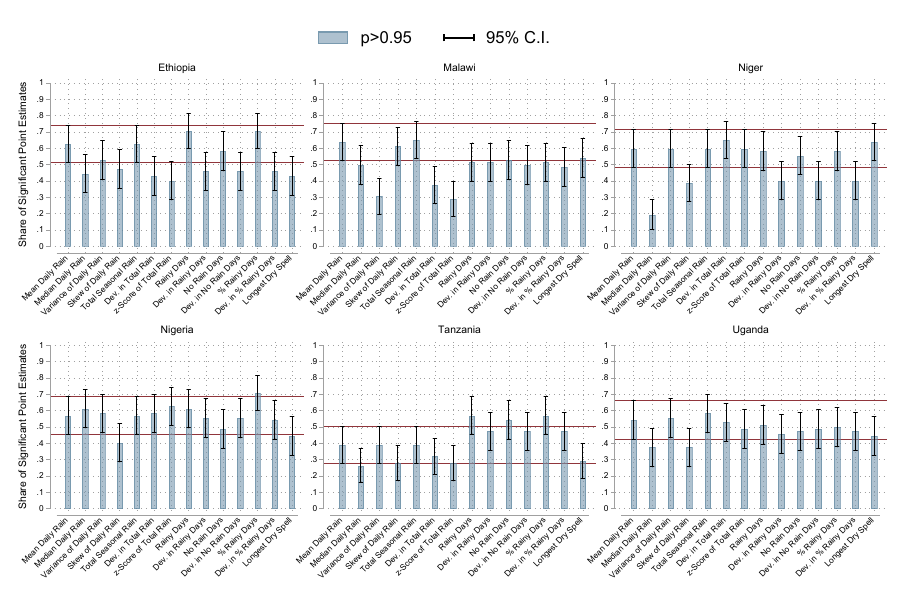}
				\end{center}
				\footnotesize  \textit{Note}: The figure displays the share of coefficients on the rainfall variables that are statistically significant from each weather metric for each country, aggregated over remote sensing source, outcome variable, and specification. Only one extraction method is included. The figure presents results from a total of 6,048 regressions. Each country includes results from 1,008 regressions and thus each column is based on 72 regressions. 
			\end{minipage}	
		\end{figure}
	\end{center}
\end{landscape}

\begin{landscape}
	\begin{center}
		\begin{figure}[!htbp]
			\begin{minipage}{\linewidth}		
				\caption{$p$-values, by Country and Temperature Metric}
				\label{fig:pval_varname_tp}
				\begin{center}
					\includegraphics[width=.95\linewidth,keepaspectratio]{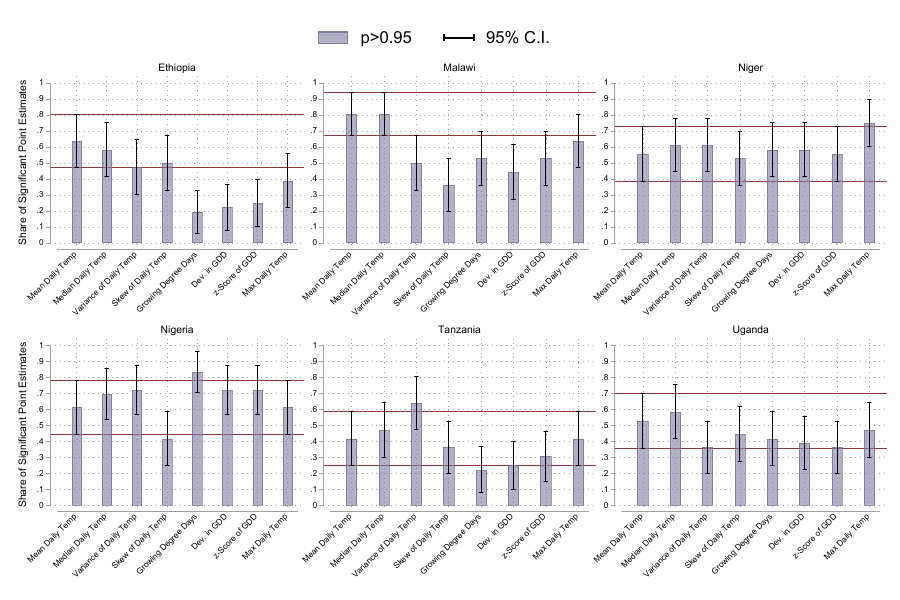}
				\end{center}
				\footnotesize  \textit{Note}: The figure displays the share of coefficients on the temperature variables that are statistically significant from each weather metric for each country, aggregated over remote sensing source, outcome variable, and specification. Only one extraction method is included. The figure presents results from a total of 1,728 regressions. Each country includes results from 288 regressions and thus each column is based on 36 regressions. 
			\end{minipage}	
		\end{figure}
	\end{center}
\end{landscape}


\begin{figure}[!htbp]
	\begin{minipage}{\linewidth}		
		\caption{Coefficients and Confidence Intervals for Mean Daily Rainfall, by Country}
		\label{fig:v01_cty}
		\begin{center}
			\includegraphics[width=\linewidth,keepaspectratio]{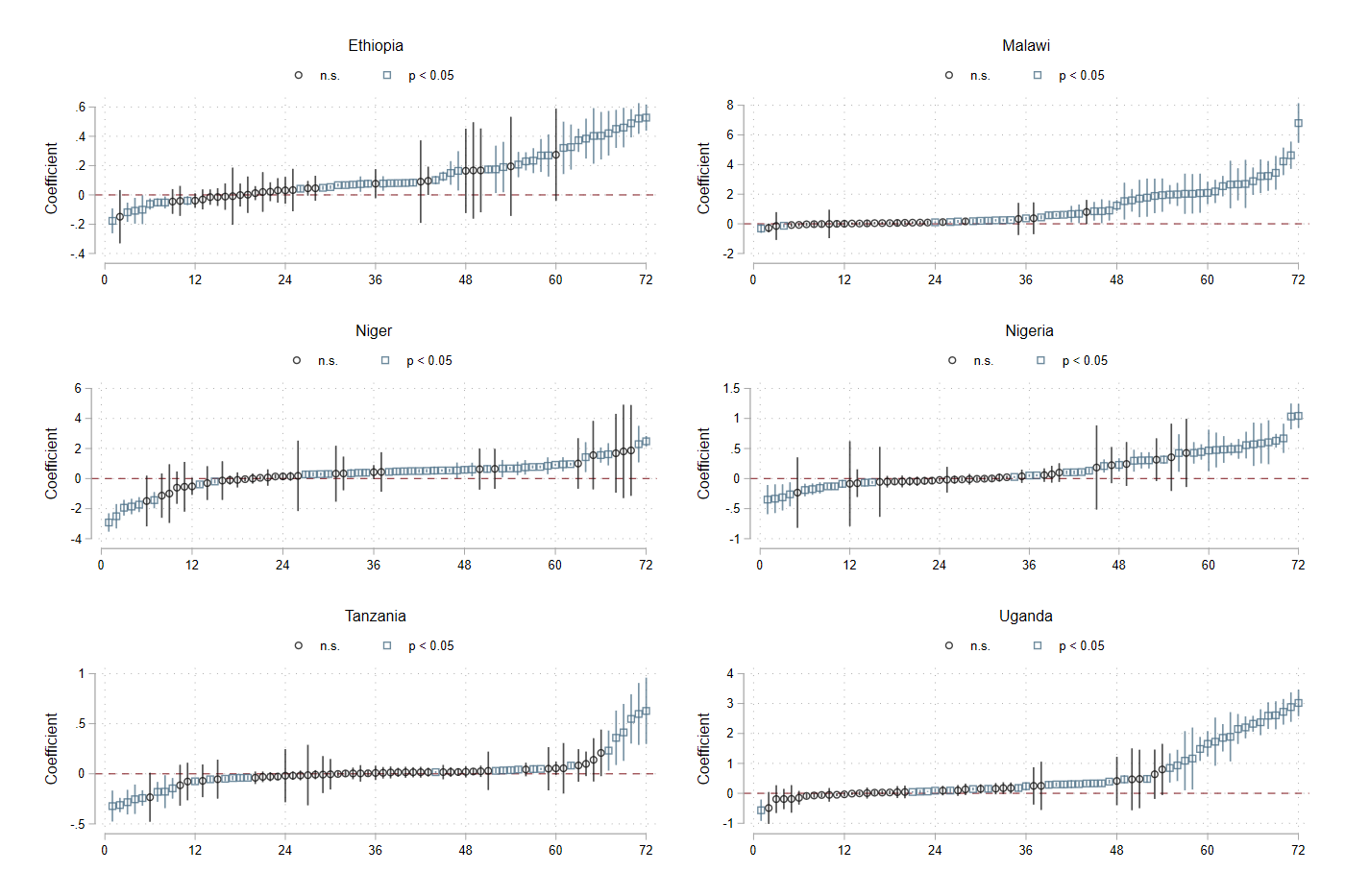}
		\end{center}
		\footnotesize  \textit{Note}: The figure presents the coefficients and confidence intervals for mean daily rainfall, by country. Only one extraction method is included. Each panel includes 72 coefficients and confidence intervals (designated on the $x$-axis). Each column, as such, represents the findings of a single regressions, e.g. the coefficient and $95\%$ confidence interval itself. The significance of these are denoted by color and shape of the identifier in the figure, as designated at the top of each panel. 
	\end{minipage}	
\end{figure}

\begin{figure}[!htbp]
	\begin{minipage}{\linewidth}		
		\caption{Coefficients and Confidence Intervals for Median Daily Rainfall, by Country}
		\label{fig:v02_cty}
		\begin{center}
			\includegraphics[width=\linewidth,keepaspectratio]{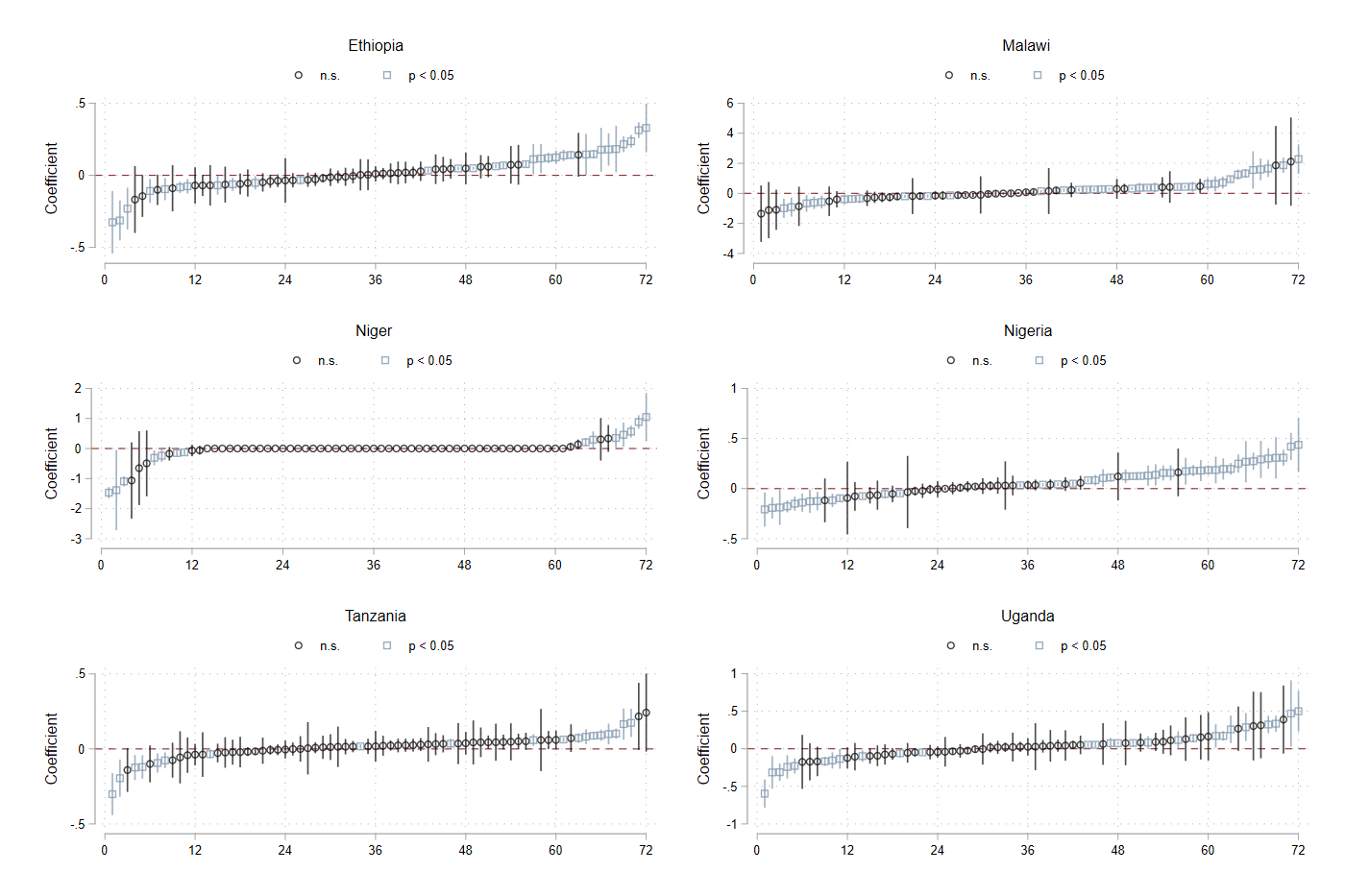}
		\end{center}
		\footnotesize  \textit{Note}: The figure presents the coefficients and confidence intervals for median daily rainfall, by country. Only one extraction method is included. Each panel includes 72 coefficients and confidence intervals (designated on the $x$-axis). Each column, as such, represents the findings of a single regressions, e.g. the coefficient and $95\%$ confidence interval itself. The significance level of these are denoted by color and shape of the identifier in the figure, as designated at the top of each panel. 
	\end{minipage}	
\end{figure}

\begin{figure}[!htbp]
	\begin{minipage}{\linewidth}		
		\caption{Coefficients and Confidence Intervals for Total Seasonal Rainfall, by Country}
		\label{fig:v05_cty}
		\begin{center}
			\includegraphics[width=\linewidth,keepaspectratio]{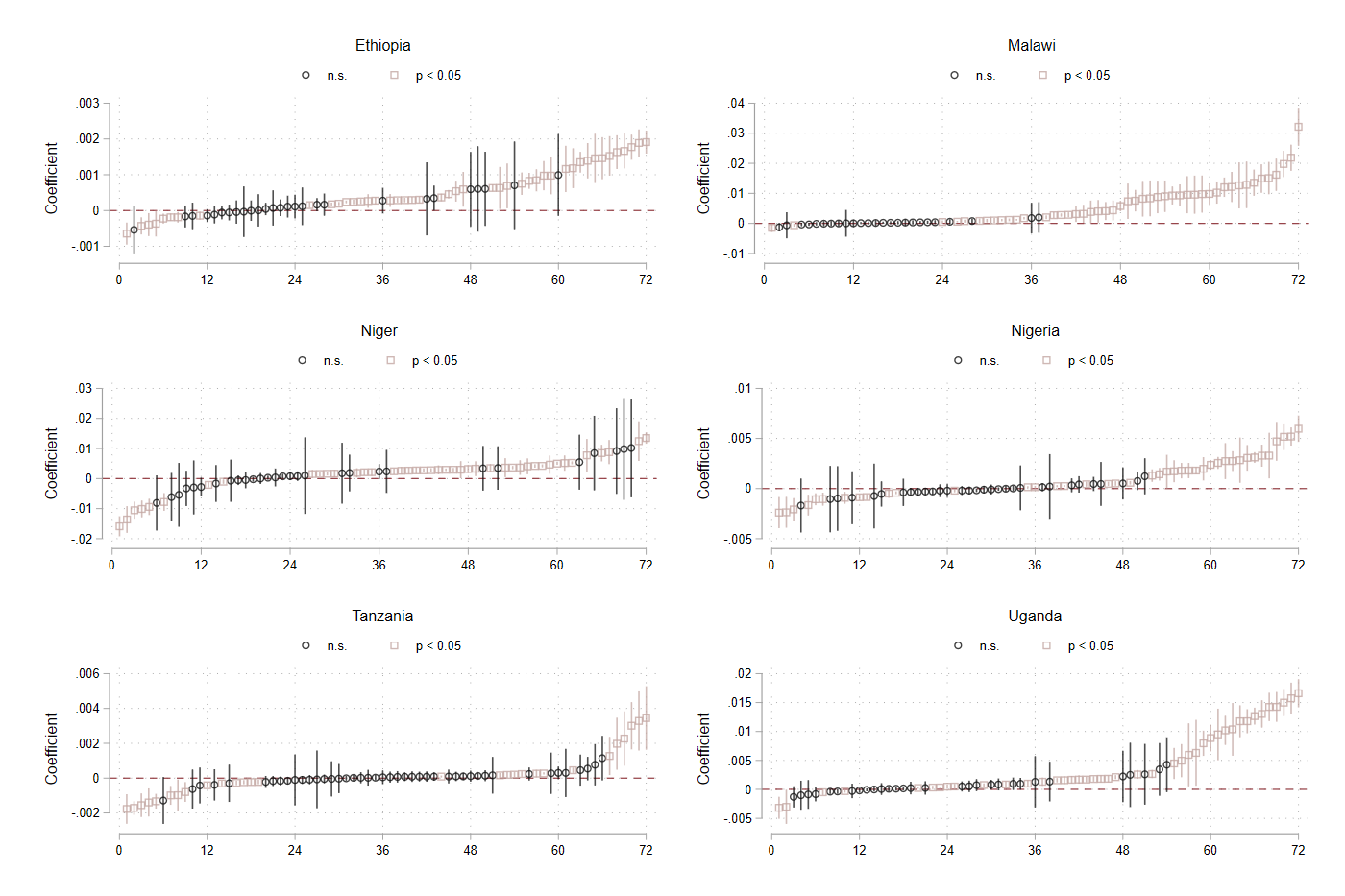}
		\end{center}
		\footnotesize  \textit{Note}: The figure presents the coefficients and confidence intervals for total seasonal rainfall, by country. Only one extraction method is included. Each panel includes 72 coefficients and confidence intervals (designated on the $x$-axis). Each column, as such, represents the findings of a single regressions, e.g. the coefficient and $95\%$ confidence interval itself. The significance level of these are denoted by color and shape of the identifier in the figure, as designated at the top of each panel. 
	\end{minipage}	
\end{figure}

\begin{figure}[!htbp]
	\begin{minipage}{\linewidth}		
		\caption{Coefficients and Confidence Intervals for Number of Days with Rain, by Country}
		\label{fig:v08_cty}
		\begin{center}
			\includegraphics[width=\linewidth,keepaspectratio]{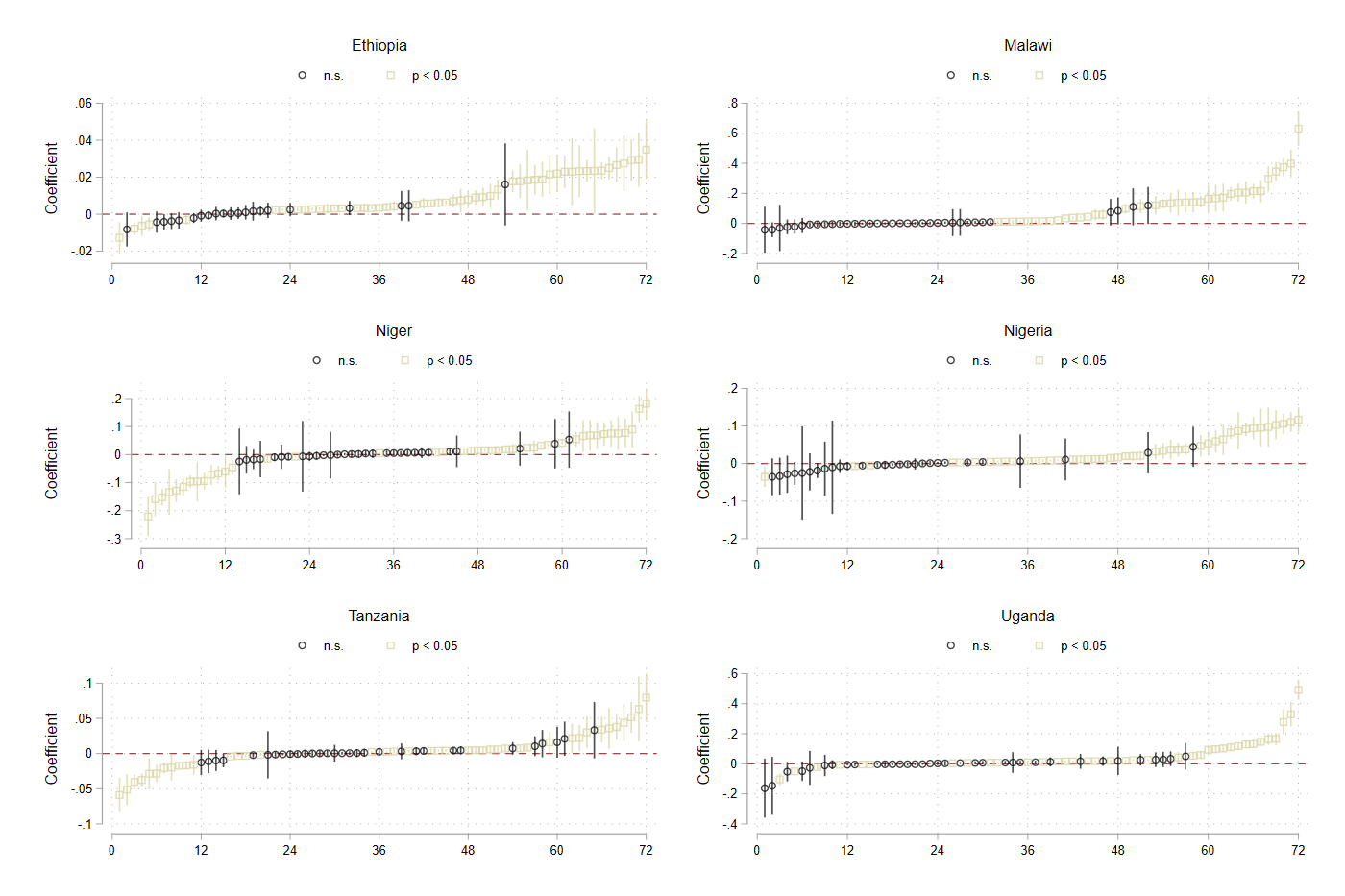}
		\end{center}
		\footnotesize  \textit{Note}: The figure presents the coefficients and confidence intervals for the number of days with rain, by country. Only one extraction method is included. Each panel includes 72 coefficients and confidence intervals (designated on the $x$-axis). Each column, as such, represents the findings of a single regressions, e.g. the coefficient and $95\%$ confidence interval itself. The significance level of these are denoted by color and shape of the identifier in the figure, as designated at the top of each panel. 
	\end{minipage}	
\end{figure}

\begin{figure}[!htbp]
	\begin{minipage}{\linewidth}		
		\caption{Coefficients and Confidence Intervals for Percentage of Days with Rain, by Country}
		\label{fig:v12_cty}
		\begin{center}
			\includegraphics[width=\linewidth,keepaspectratio]{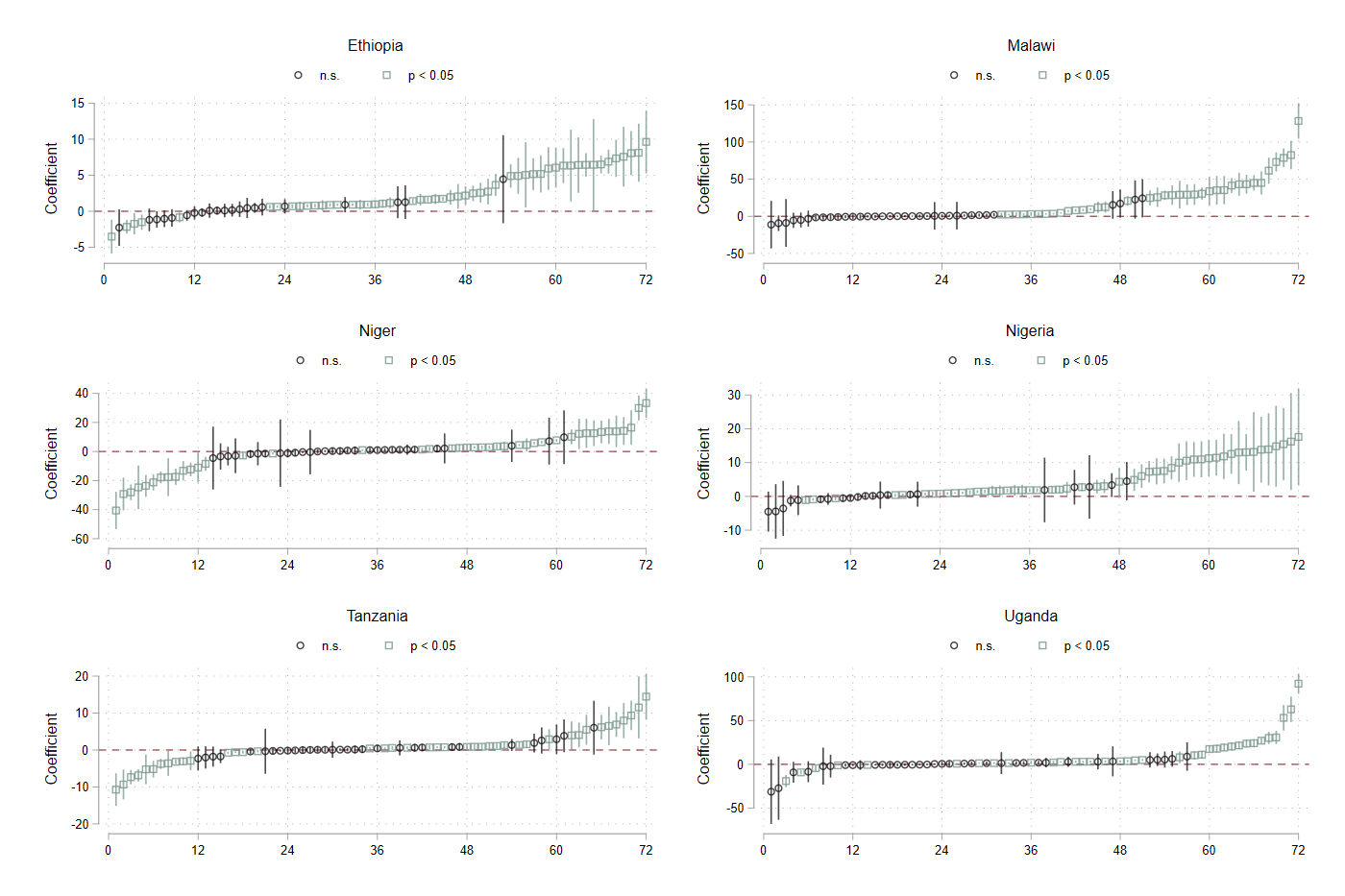}
		\end{center}
		\footnotesize  \textit{Note}: The figure presents the coefficients and confidence intervals for the percentage of days with rain, by country. Only one extraction method is included. Each panel includes 72 coefficients and confidence intervals (designated on the $x$-axis). Each column, as such, represents the findings of a single regressions, e.g. the coefficient and $95\%$ confidence interval itself. The significance level of these are denoted by color and shape of the identifier in the figure, as designated at the top of each panel. 
	\end{minipage}	
\end{figure}

\begin{figure}[!htbp]
	\begin{minipage}{\linewidth}		
		\caption{Coefficients and Confidence Intervals for Longest Dry Spell, by Country}
		\label{fig:v14_cty}
		\begin{center}
			\includegraphics[width=\linewidth,keepaspectratio]{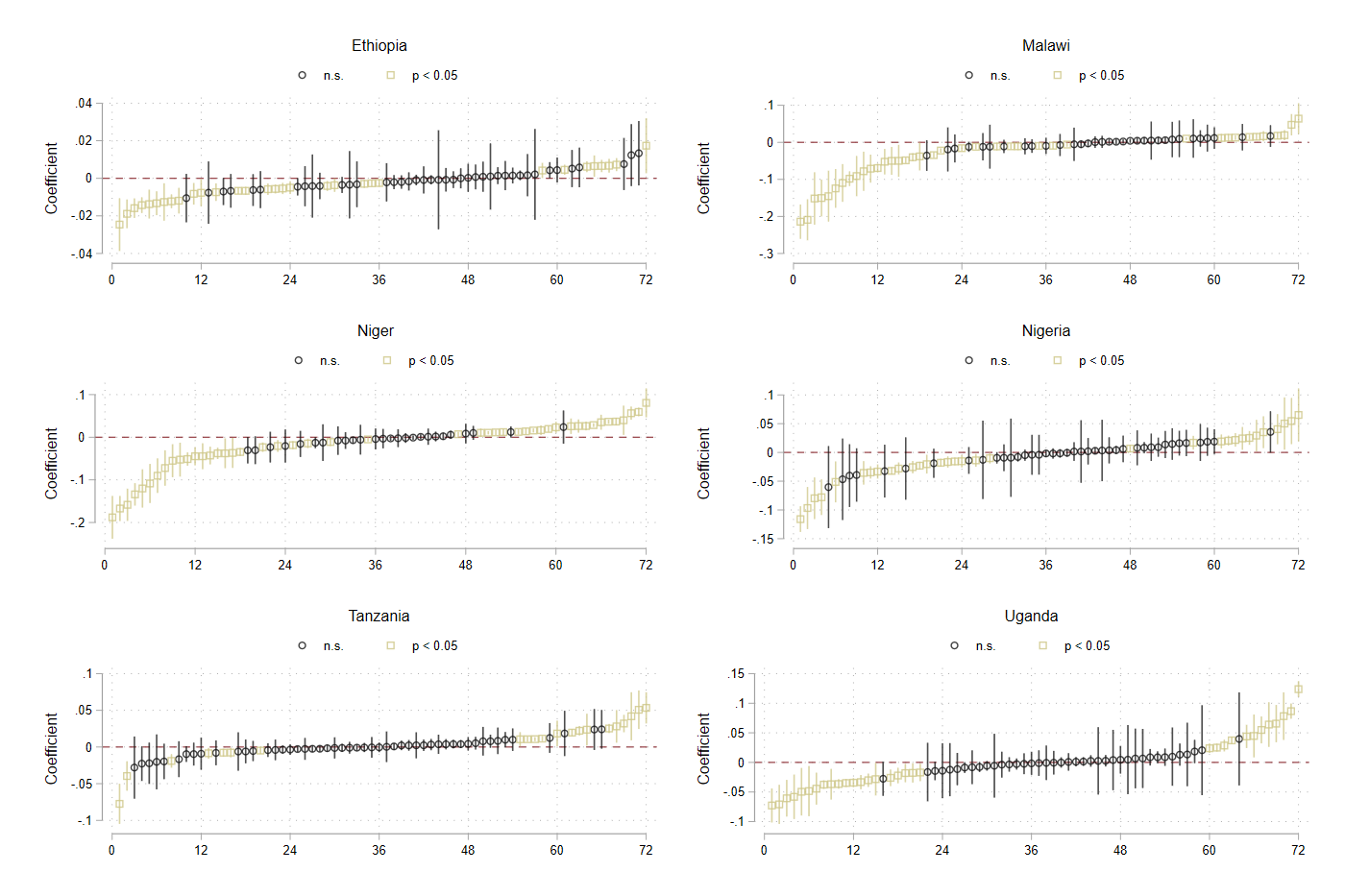}
		\end{center}
		\footnotesize  \textit{Note}: The figure presents the coefficients and confidence intervals for the longest dry spell, by country. Only one extraction method is included. Each panel includes 72 coefficients and confidence intervals (designated on the $x$-axis). Each column, as such, represents the findings of a single regressions, e.g. the coefficient and $95\%$ confidence interval itself. The significance level of these are denoted by color and shape of the identifier in the figure, as designated at the top of each panel. 
	\end{minipage}	
\end{figure}

\begin{figure}[!htbp]
	\begin{minipage}{\linewidth}		
		\caption{Coefficients and Confidence Intervals for Mean Daily Temperature, by Country}
		\label{fig:v15_cty}
		\begin{center}
			\includegraphics[width=\linewidth,keepaspectratio]{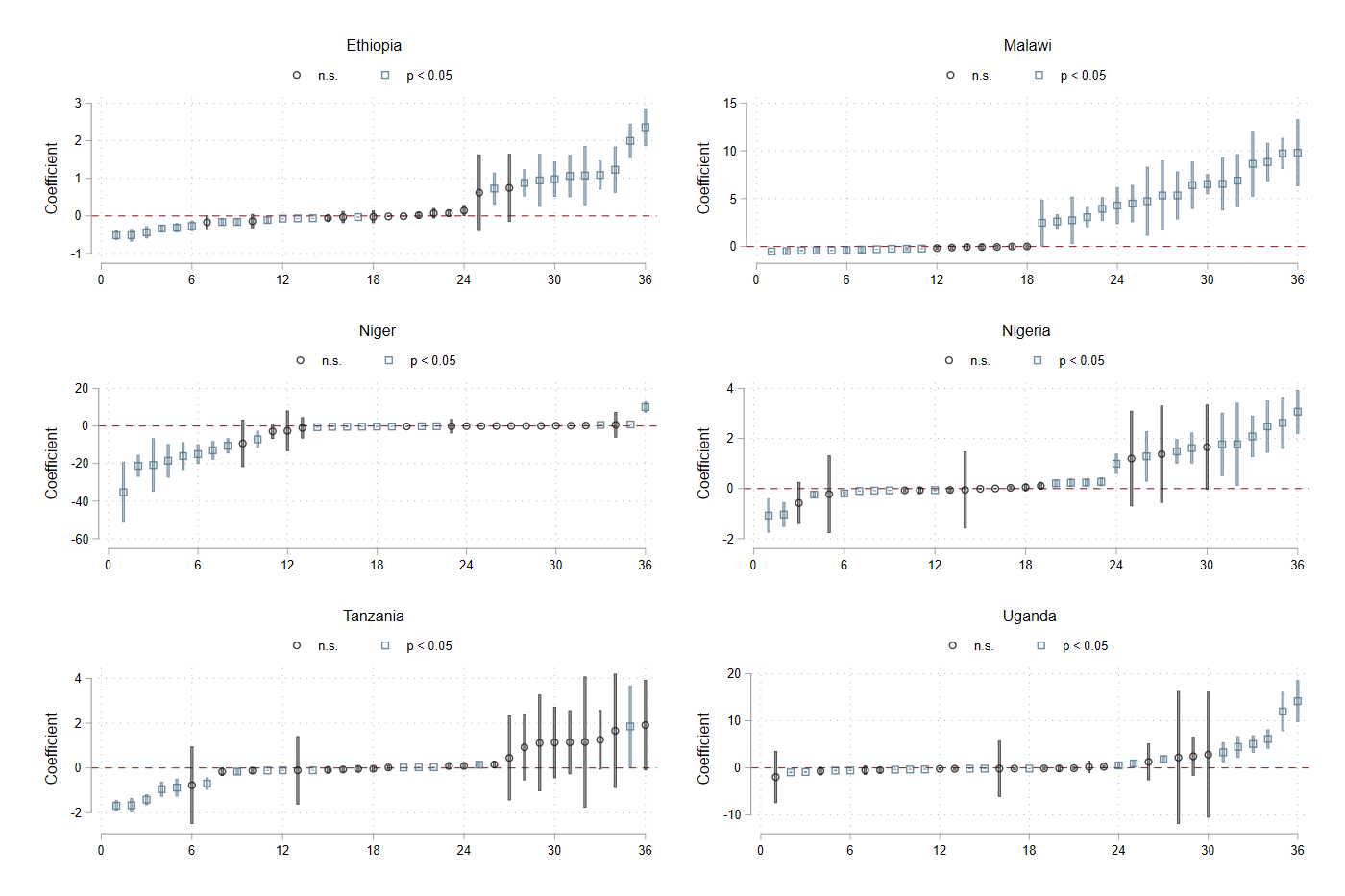}
		\end{center}
		\footnotesize  \textit{Note}: The figure presents the coefficients and confidence intervals for mean daily temperature, by country. Only one extraction method is included. Each panel includes 72 coefficients and confidence intervals (designated on the $x$-axis). Each column, as such, represents the findings of a single regressions, e.g. the coefficient and $95\%$ confidence interval itself. The significance level of these are denoted by color and shape of the identifier in the figure, as designated at the top of each panel. 
	\end{minipage}	
\end{figure}

\begin{figure}[!htbp]
	\begin{minipage}{\linewidth}		
		\caption{Coefficients and Confidence Intervals for Median Daily Temperature, by Country}
		\label{fig:v16_cty}
		\begin{center}
			\includegraphics[width=\linewidth,keepaspectratio]{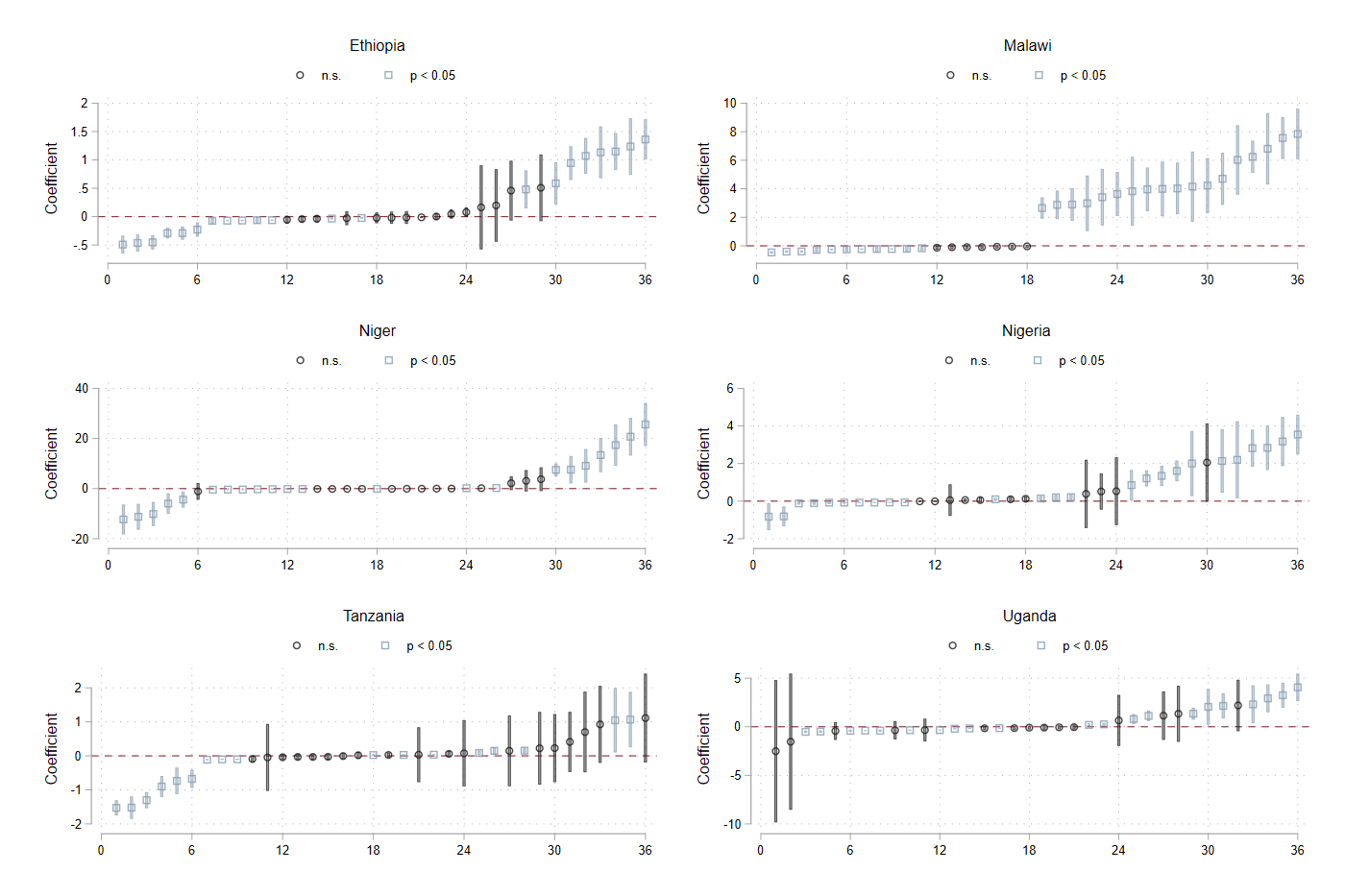}
		\end{center}
		\footnotesize  \textit{Note}: The figure presents the coefficients and confidence intervals for median daily temperature, by country. Only one extraction method is included. Each panel includes 72 coefficients and confidence intervals (designated on the $x$-axis). Each column, as such, represents the findings of a single regressions, e.g. the coefficient and $95\%$ confidence interval itself. The significance level of these are denoted by color and shape of the identifier in the figure, as designated at the top of each panel. 
	\end{minipage}	
\end{figure}

\begin{figure}[!htbp]
	\begin{minipage}{\linewidth}		
		\caption{Coefficients and Confidence Intervals for Variance of Daily Temperature, by Country}
		\label{fig:v17_cty}
		\begin{center}
			\includegraphics[width=\linewidth,keepaspectratio]{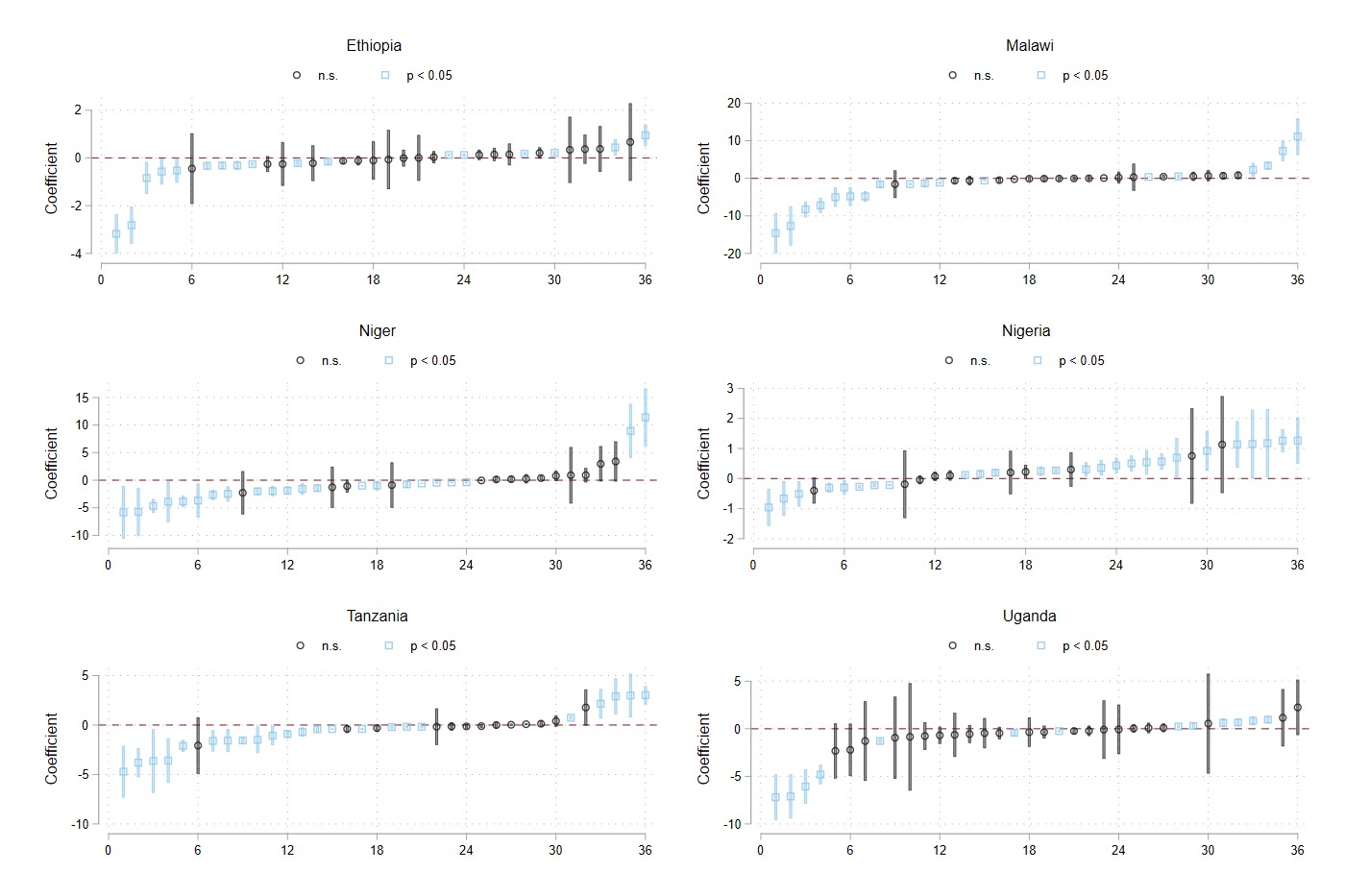}
		\end{center}
		\footnotesize  \textit{Note}: The figure presents the coefficients and confidence intervals for the variance in daily temperature, by country. Only one extraction method is included. Each panel includes 72 coefficients and confidence intervals (designated on the $x$-axis). Each column, as such, represents the findings of a single regressions, e.g. the coefficient and $95\%$ confidence interval itself. The significance level of these are denoted by color and shape of the identifier in the figure, as designated at the top of each panel. 
	\end{minipage}	
\end{figure}

\begin{figure}[!htbp]
	\begin{minipage}{\linewidth}		
		\caption{Coefficients and Confidence Intervals for Growing Degree Days (GDD), by Country}
		\label{fig:v19_cty}
		\begin{center}
			\includegraphics[width=\linewidth,keepaspectratio]{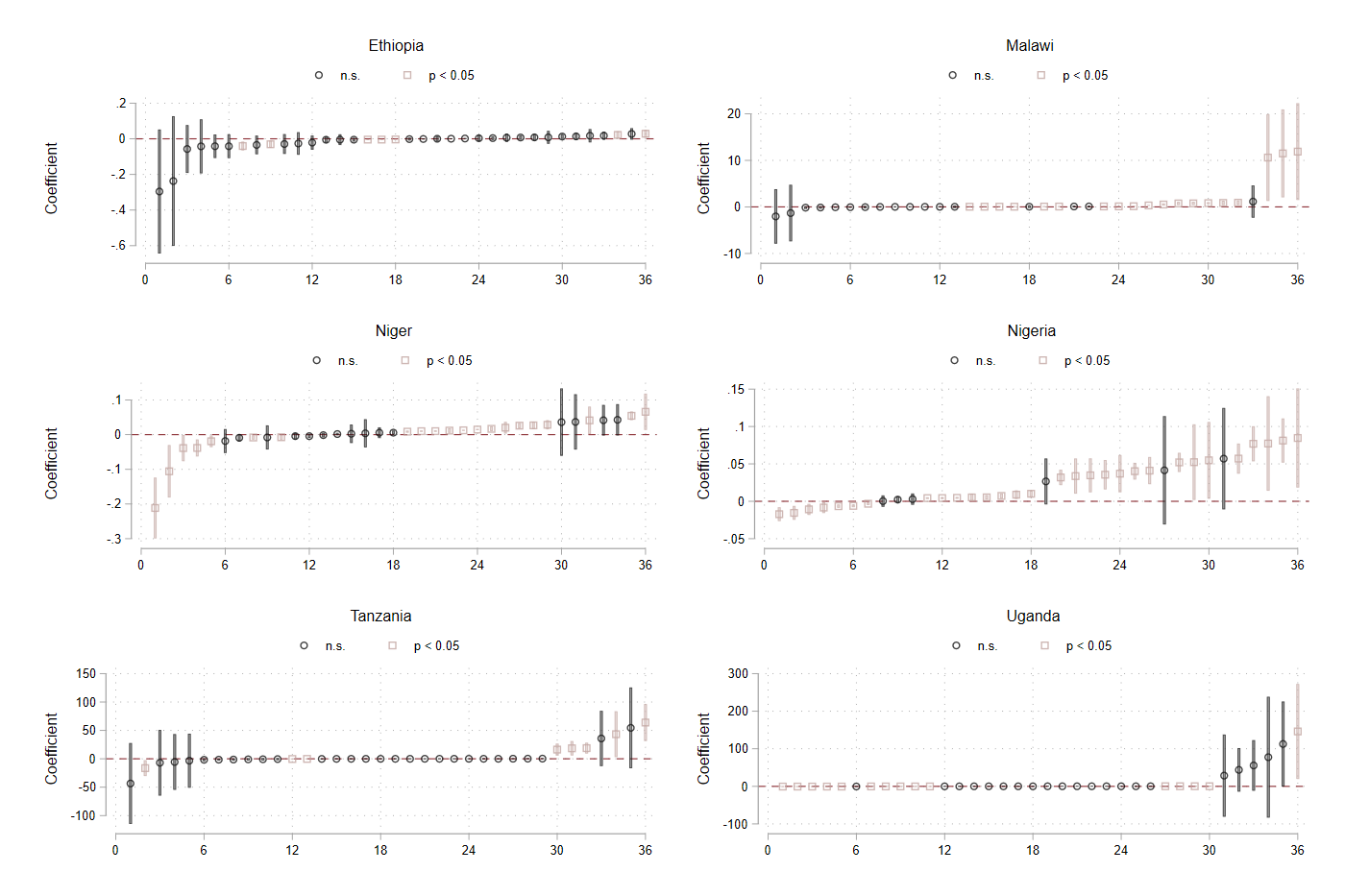}
		\end{center}
		\footnotesize  \textit{Note}: The figure presents the coefficients and confidence intervals for growing degree days, by country. Only one extraction method is included. Each panel includes 72 coefficients and confidence intervals (designated on the $x$-axis). Each column, as such, represents the findings of a single regressions, e.g. the coefficient and $95\%$ confidence interval itself. The significance level of these are denoted by color and shape of the identifier in the figure, as designated at the top of each panel. 
	\end{minipage}	
\end{figure}


\newpage 


\begin{figure}[!htbp]
	\begin{minipage}{\linewidth}		
		\caption{Mean Adjusted R$^2$, by Weather Product, Rainfall Metric, and Model}
		\label{fig:r2_sat_rf}
		\begin{center}
			\includegraphics[width=\linewidth,keepaspectratio]{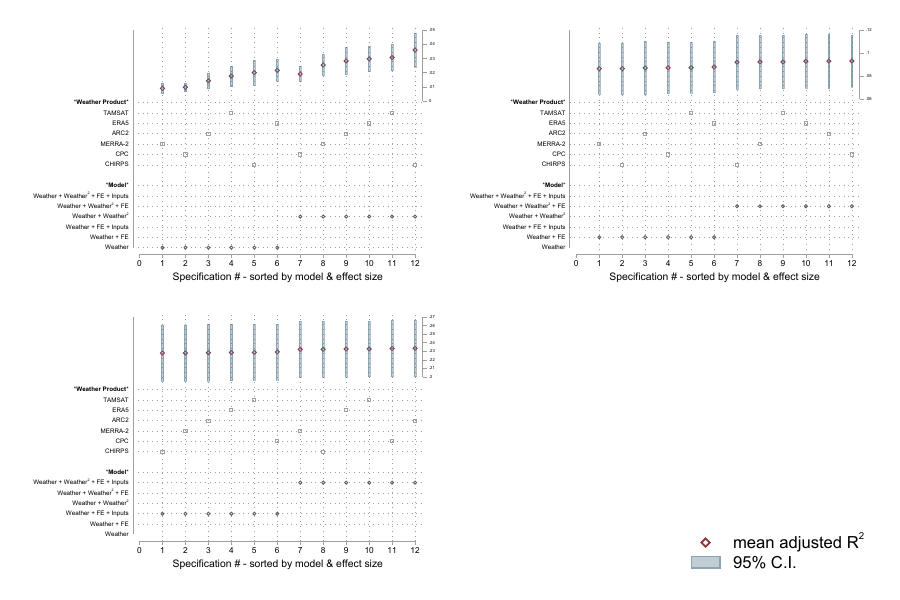}
		\end{center}
		\footnotesize  \textit{Note}: The figure presents the mean adjusted R$^2$, by weather product and model specification, aggregated over rainfall metric, country, and outcome variable. Only one extraction method is included. Across panels, the figure includes 1,728 regressions, with each panel including 576 regressions. Each column includes 48 regressions, which are for each specification model and each rainfall metric. The northwest panel presents results from the linear and quadratic model specifications. The northeast panel presents results from the linear and quadratic model specifications with fixed effects. The southwest panel presents results from the linear and quadratic model specifications with fixed effects and input controls. 
	\end{minipage}	
\end{figure}

\begin{figure}[!htbp]
	\begin{minipage}{\linewidth}		
		\caption{Mean Adjusted R$^2$, by Weather Product, Temperature Metric, and Model}
		\label{fig:r2_sat_tp}
		\begin{center}
			\includegraphics[width=\linewidth,keepaspectratio]{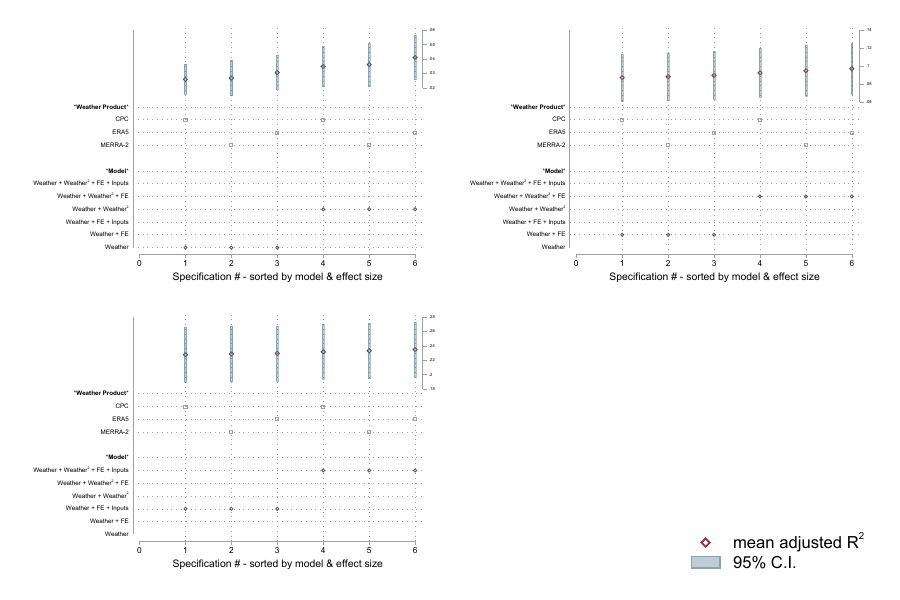}
		\end{center}
		\footnotesize  \textit{Note}: The figure presents the mean adjusted R$^2$, by weather product and model specification, aggregated over temperature metric, country, and outcome variable. Only one extraction method is included. Across panels, the figure includes 864 regressions, with each panel including 288 regressions. Each column includes 48 regressions, which are for each specification model and each temperature metric. The northwest panel presents results from the linear and quadratic model specifications. The northeast panel presents results from the linear and quadratic model specifications with fixed effects. The southwest panel presents results from the linear and quadratic model specifications with fixed effects and input controls.  
	\end{minipage}	
\end{figure}


\begin{figure}[!htbp]
	\begin{minipage}{\linewidth}		
		\caption{$p$-values for Rainfall and Temperature by Weather Product}
		\label{fig:pval_sat}
		\begin{center}
			\includegraphics[width=\linewidth,keepaspectratio]{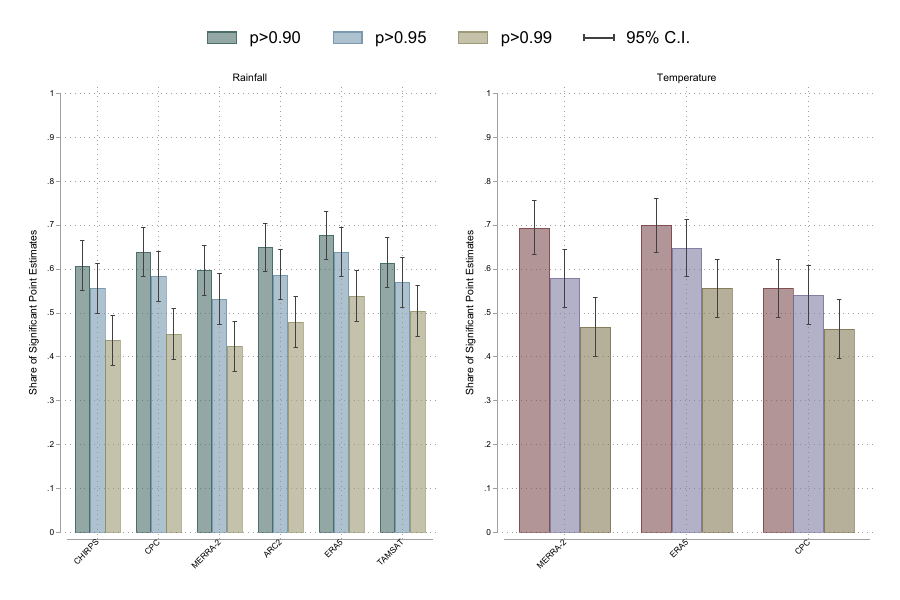}
		\end{center}
		\footnotesize  \textit{Note}: The figure displays the share of coefficients for rainfall and temperature variables, by weather product. These are aggregated over weather metric, country, outcome variable, and specification. Only one extraction method is included. The rainfall panel includes 1,728 regressions, with each column representing 288 regressions. The temperature panel includes 648 regressions, with each column representing 216 regressions.
	\end{minipage}	
\end{figure}

\begin{figure}[!htbp]
	\begin{minipage}{\linewidth}		
		\caption{$p$-values for Rainfall Metrics by Weather Product}
		\label{fig:pval_rf}
		\begin{center}
			\includegraphics[width=\linewidth,keepaspectratio]{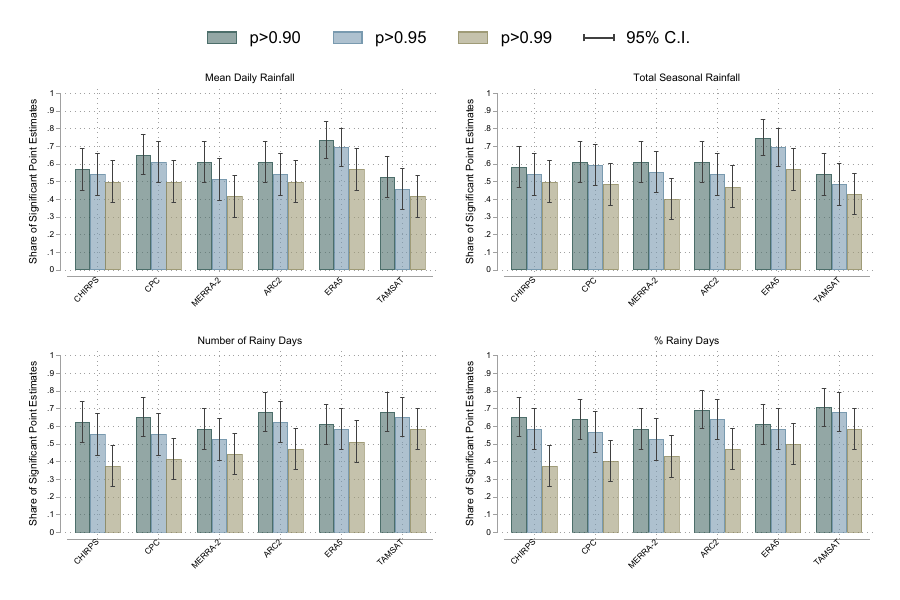}
		\end{center}
		\footnotesize  \textit{Note}: The figure displays the share of coefficients on the rainfall variables that are statistically significant for three rainfall metrics for each product, aggregated over country, outcome variable, and specification. Only one extraction method is included. Each panel includes 432 regressions, with each column representing 72 regressions. 
	\end{minipage}	
\end{figure}

\begin{figure}[!htbp]
	\begin{minipage}{\linewidth}		
		\caption{$p$-values for Temperature Metrics by Weather Product}
		\label{fig:pval_tp}
		\begin{center}
			\includegraphics[width=\linewidth,keepaspectratio]{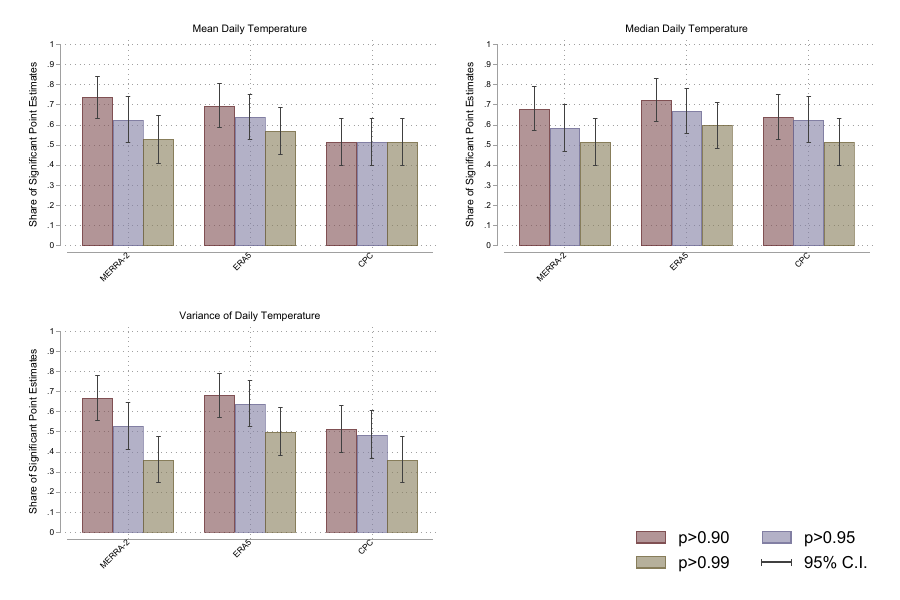}
		\end{center}
		\footnotesize  \textit{Note}: The figure displays the share of coefficients on the temperature variables that are statistically significant for three temperature metrics for each product, aggregated over country, outcome variable, and specification. Only one extraction method is included. Each panel includes 216 regressions, with each column representing 72 regressions.  
	\end{minipage}	
\end{figure}


\begin{figure}[!htbp]
	\begin{minipage}{\linewidth}		
		\caption{Specification Curve for Mean Daily Rainfall by Country}
		\label{fig:v01_sat}
		\begin{center}
			\includegraphics[width=.49\linewidth,keepaspectratio]{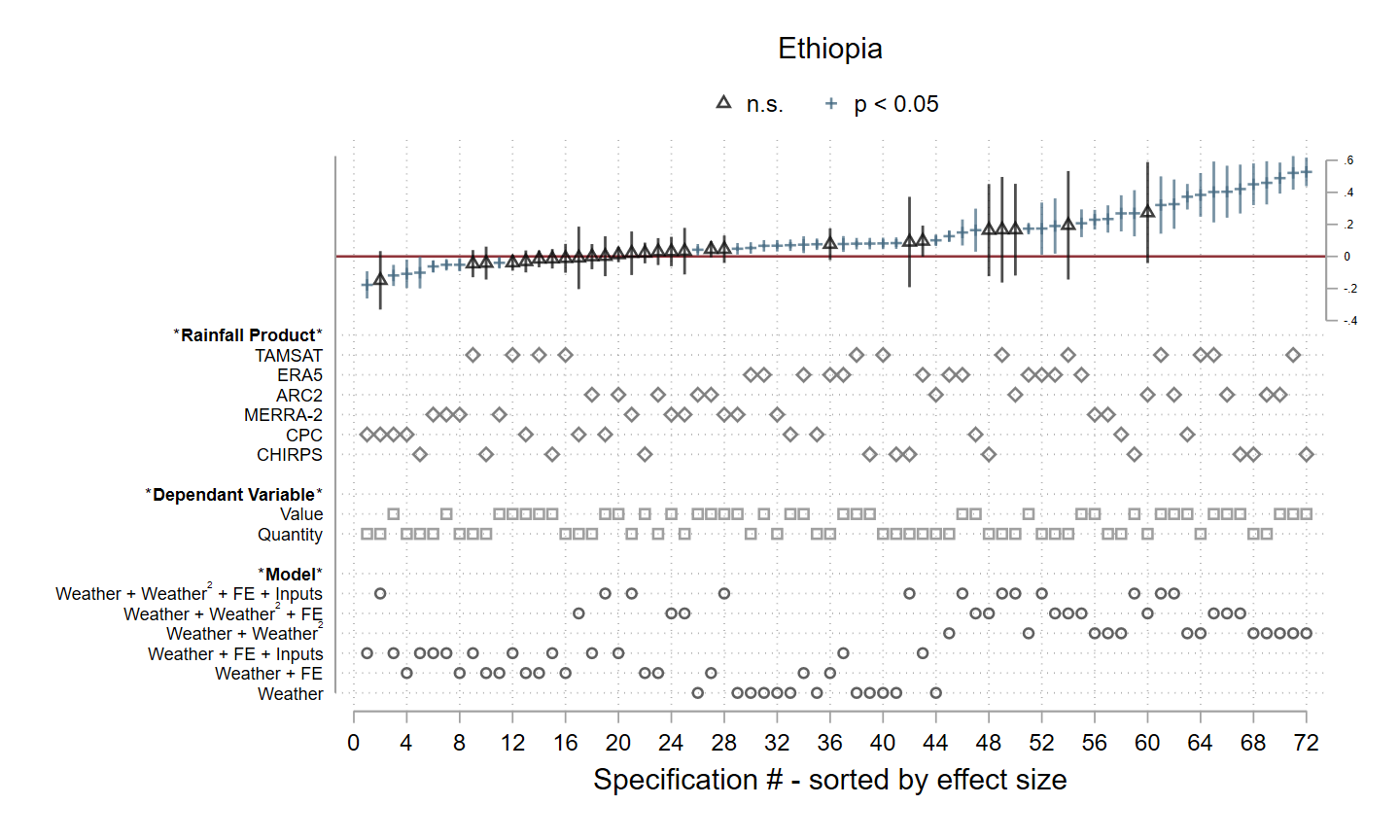}
			\includegraphics[width=.49\linewidth,keepaspectratio]{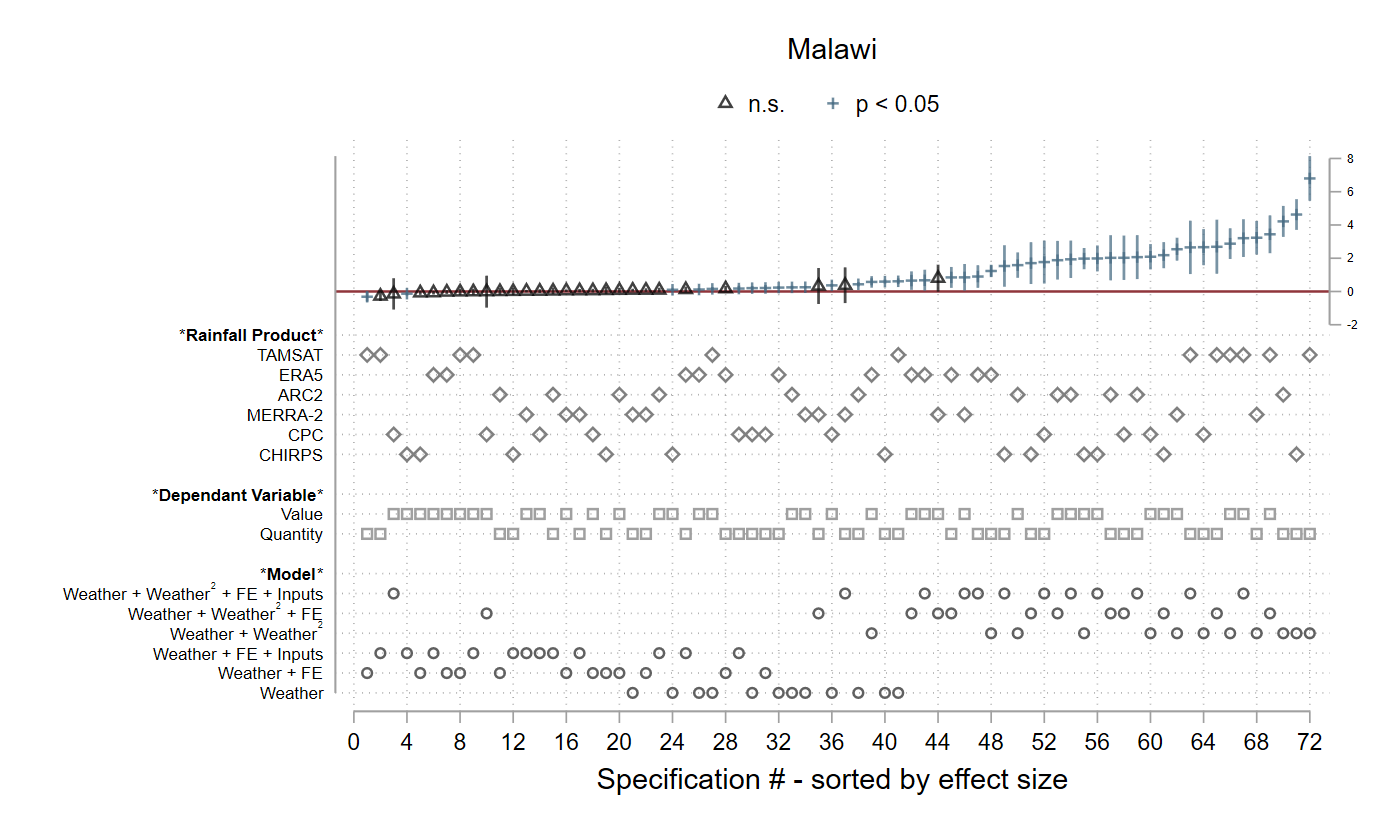}
			\includegraphics[width=.49\linewidth,keepaspectratio]{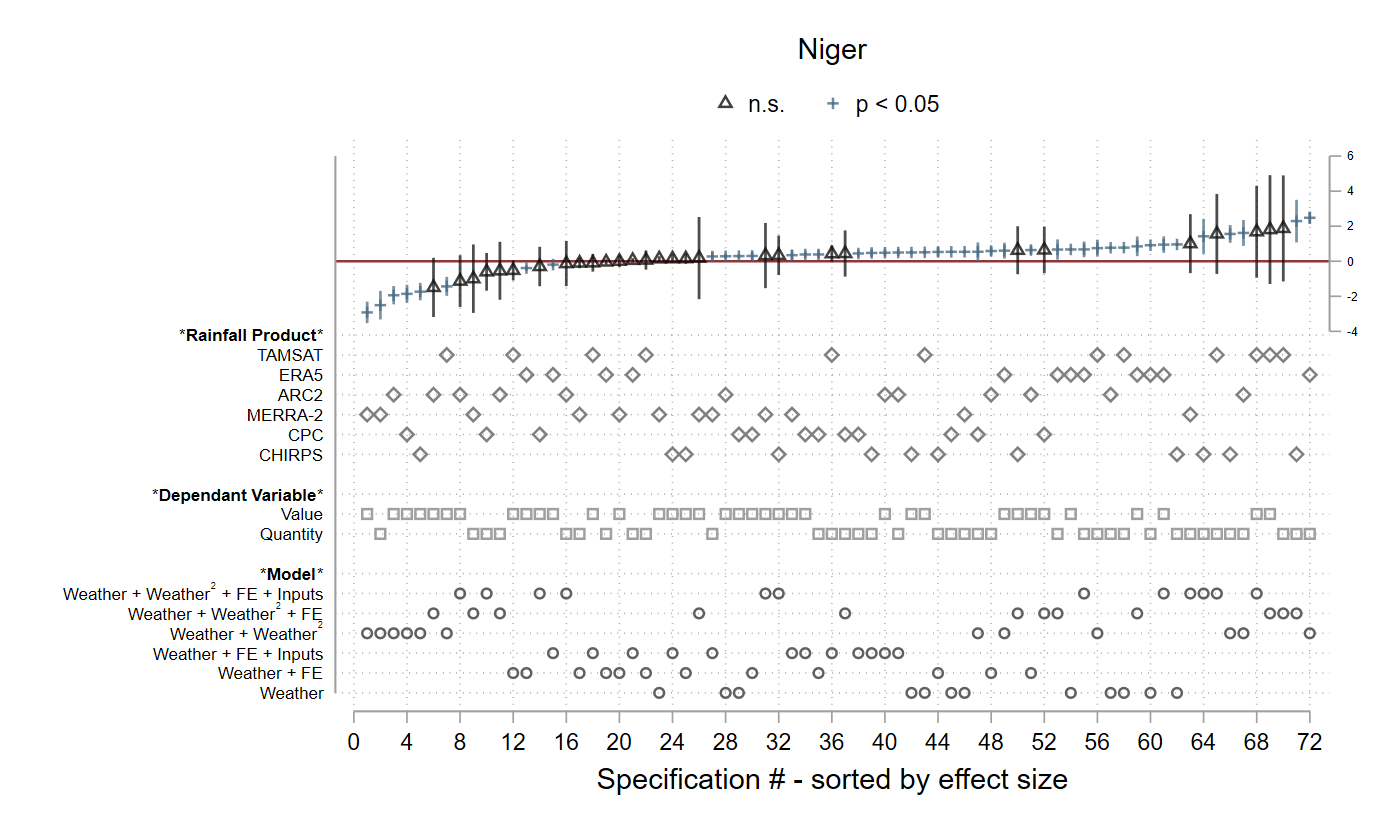}
			\includegraphics[width=.49\linewidth,keepaspectratio]{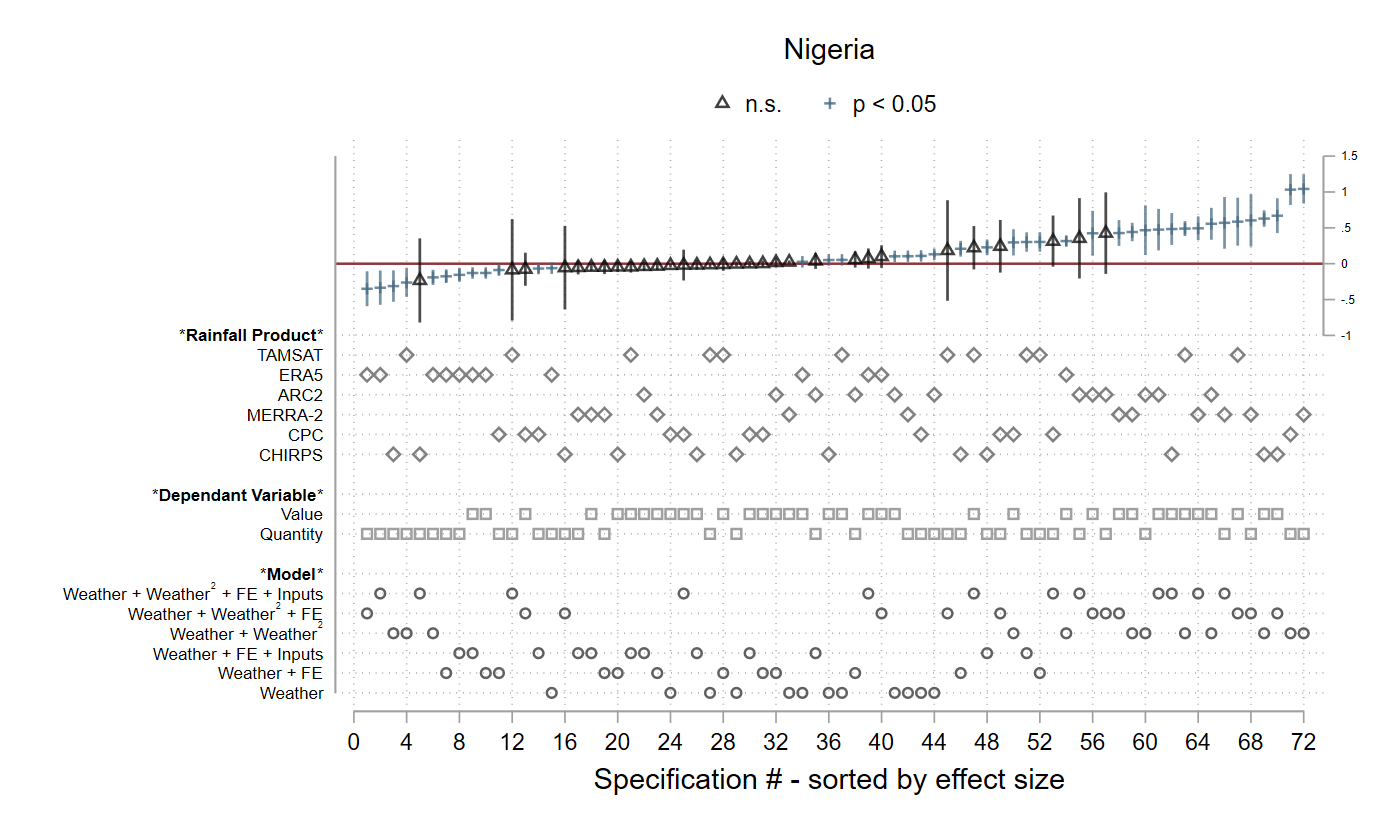}
			\includegraphics[width=.49\linewidth,keepaspectratio]{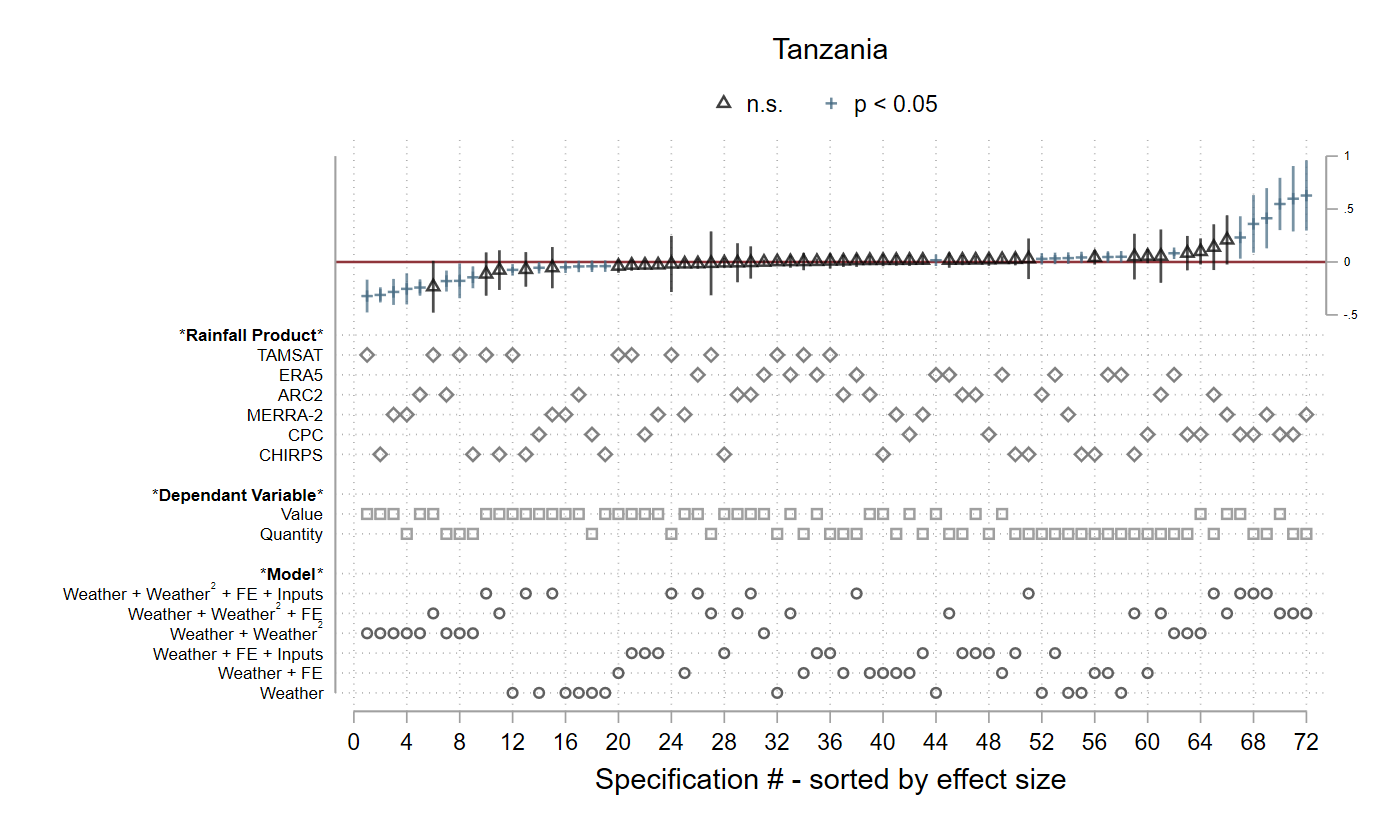}
			\includegraphics[width=.49\linewidth,keepaspectratio]{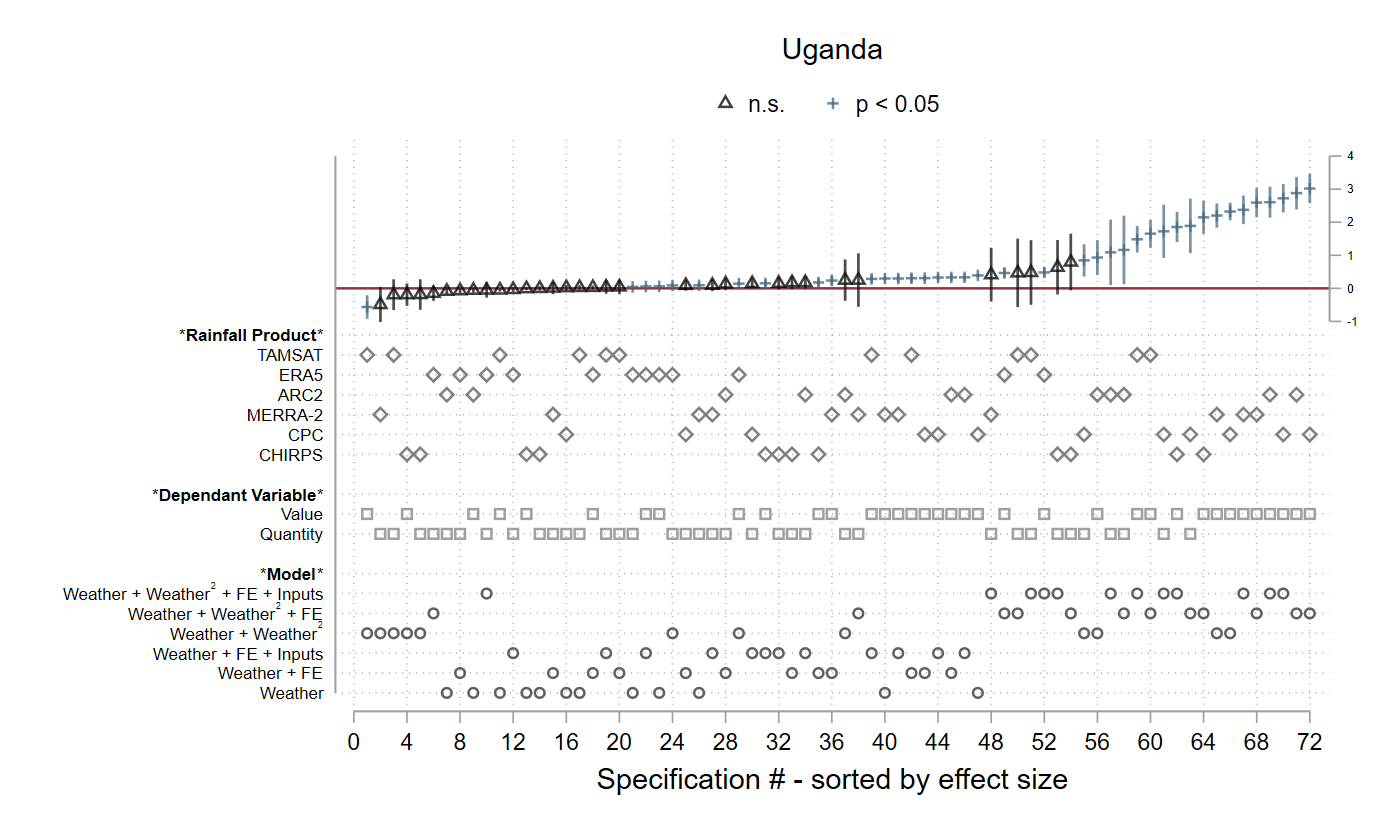}
		\end{center}
		\footnotesize  \textit{Note}: The figure presents specification curves for mean daily rainfall, where each country is represented in a different panel. Each panel includes 72 regressions, where each column represents a single regression. Significant and non-significant coefficients are designated at the top of each panel. 
	\end{minipage}	
\end{figure}

\begin{figure}[!htbp]
	\begin{minipage}{\linewidth}		
		\caption{Specification Curve for Total Seasonal Rainfall by Country}
		\label{fig:v05_sat}
		\begin{center}
			\includegraphics[width=.49\linewidth,keepaspectratio]{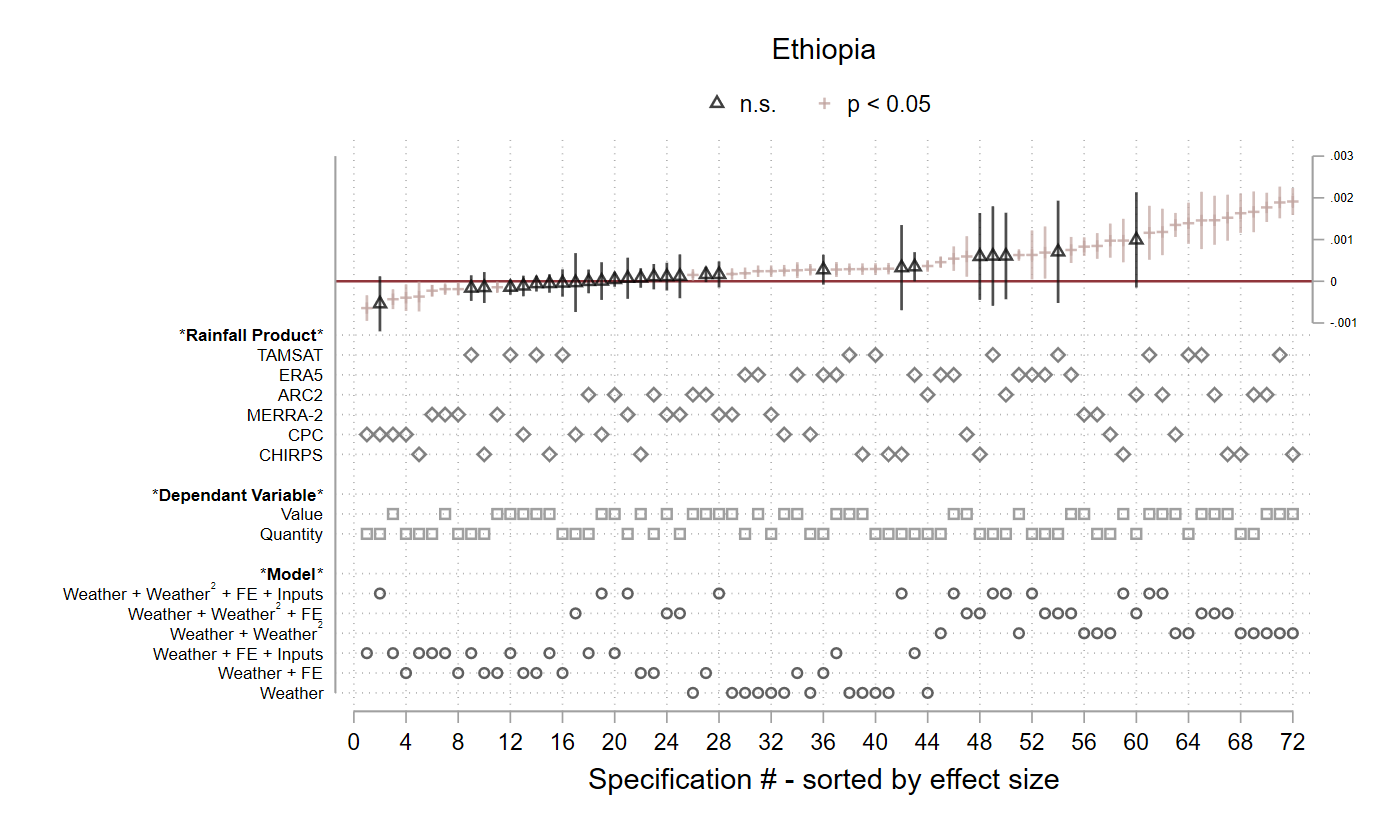}
			\includegraphics[width=.49\linewidth,keepaspectratio]{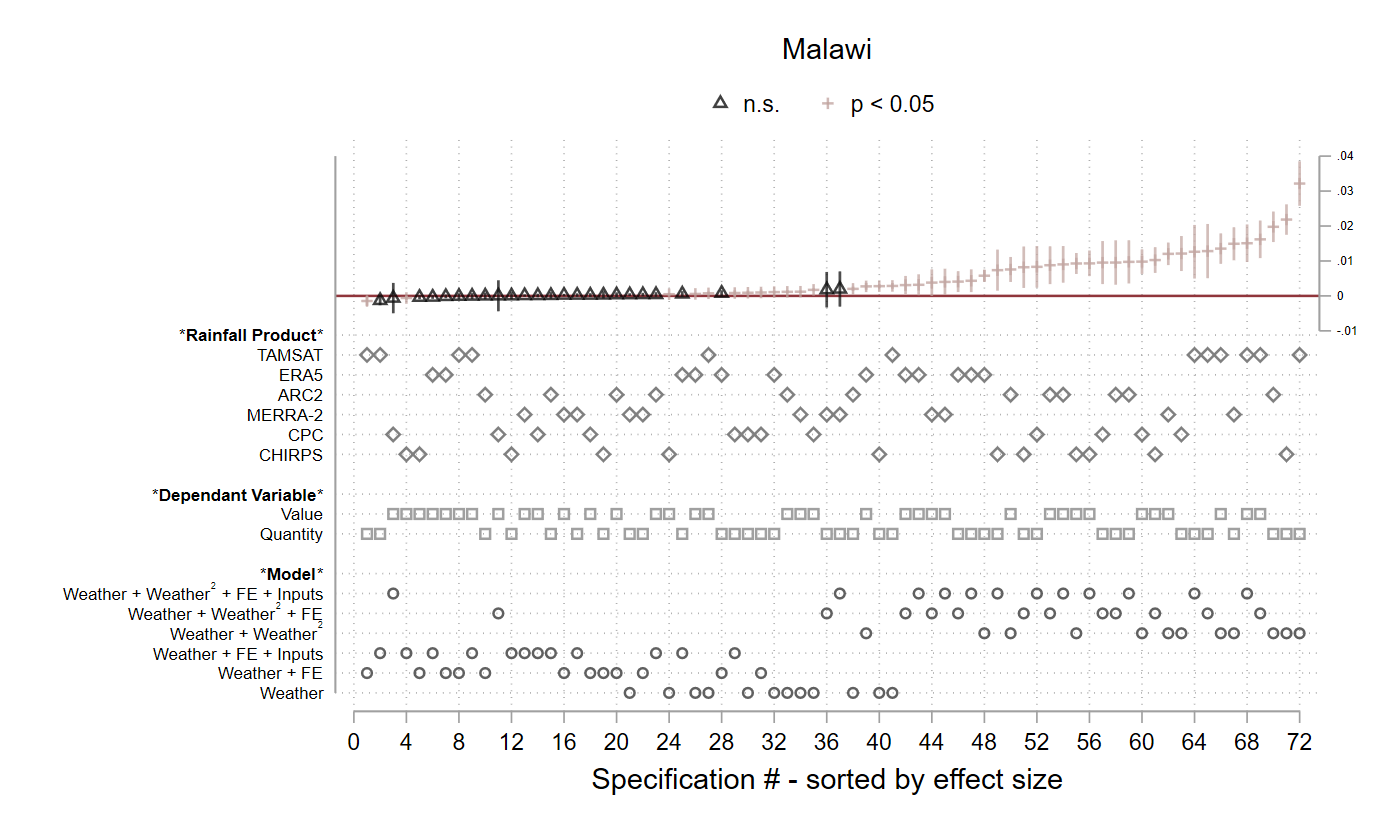}
			\includegraphics[width=.49\linewidth,keepaspectratio]{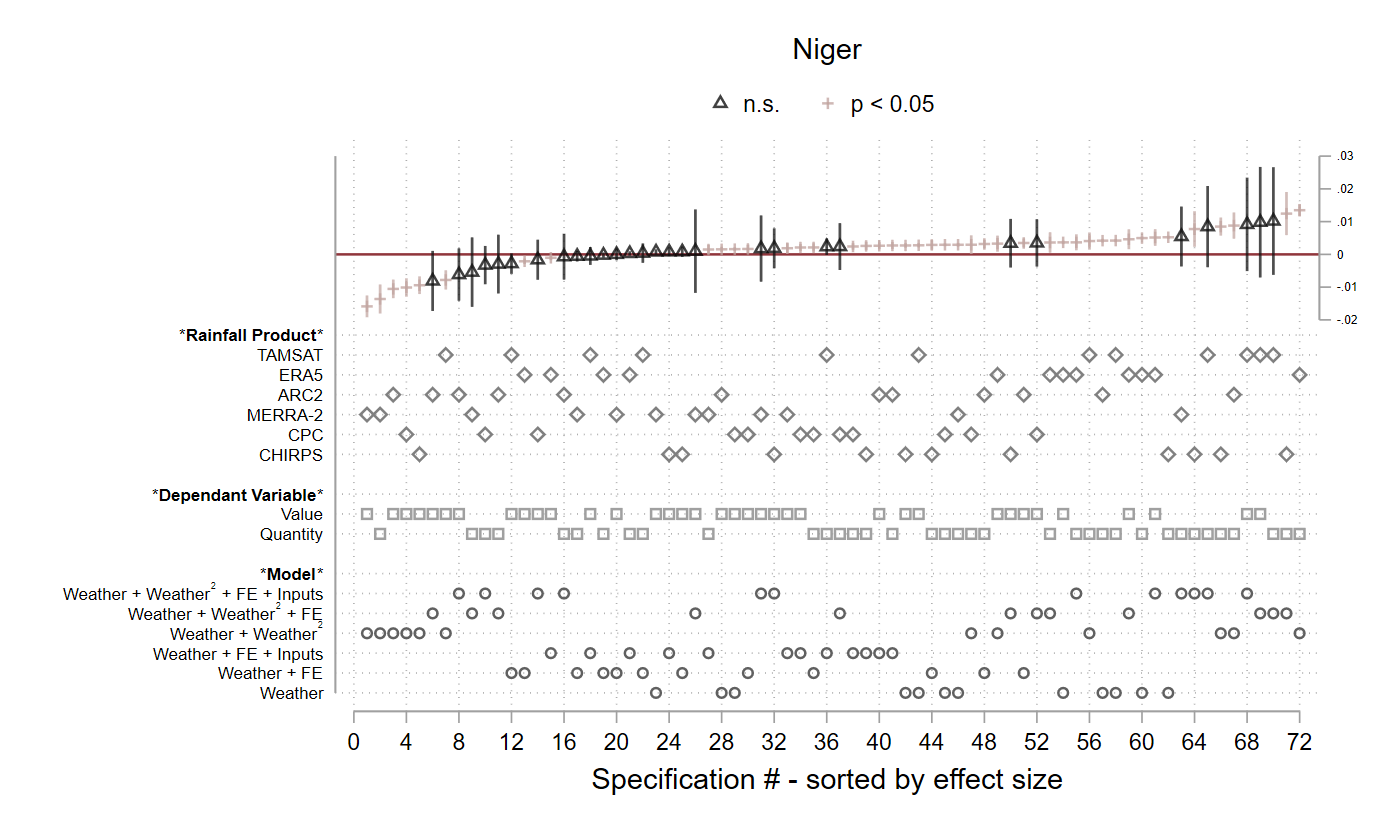}
			\includegraphics[width=.49\linewidth,keepaspectratio]{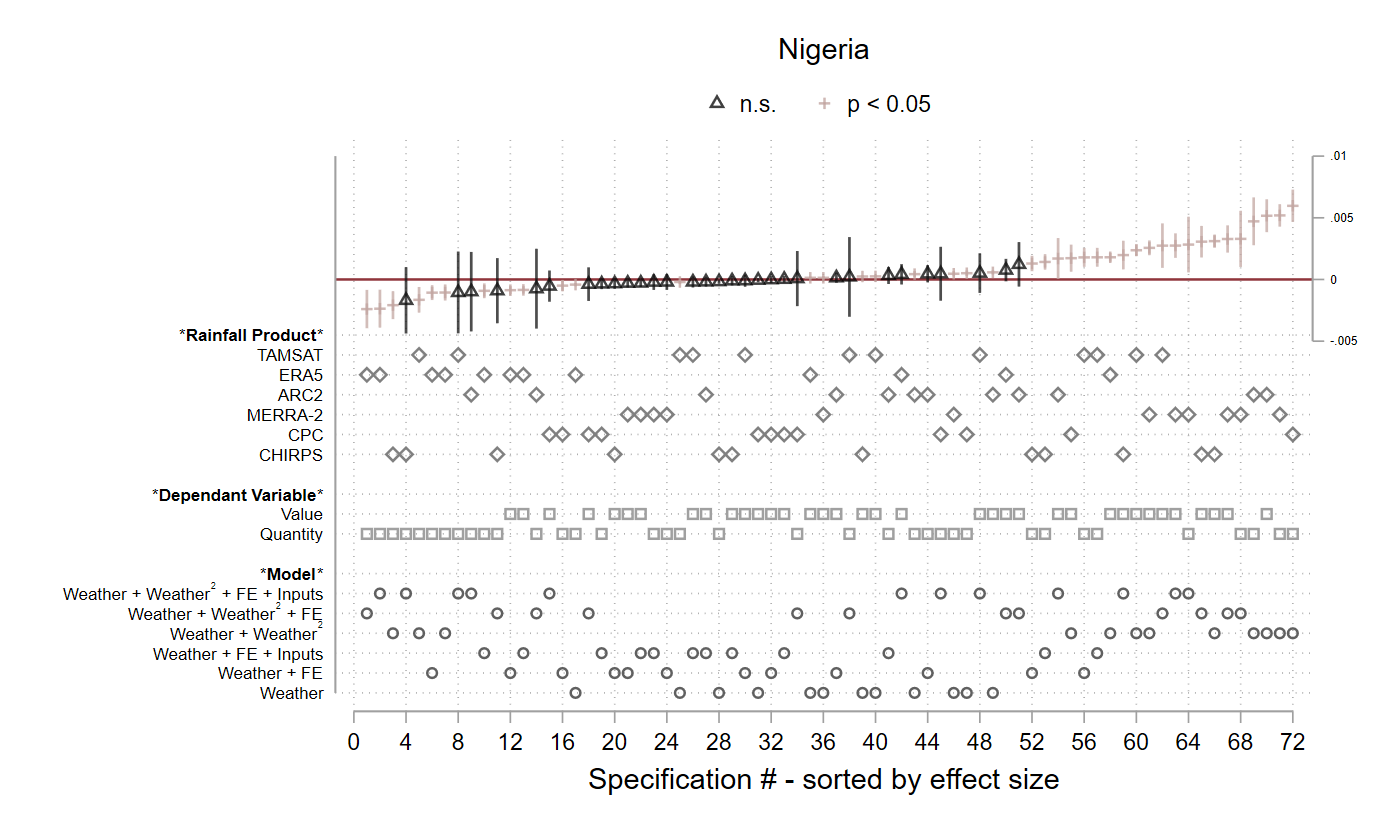}
			\includegraphics[width=.49\linewidth,keepaspectratio]{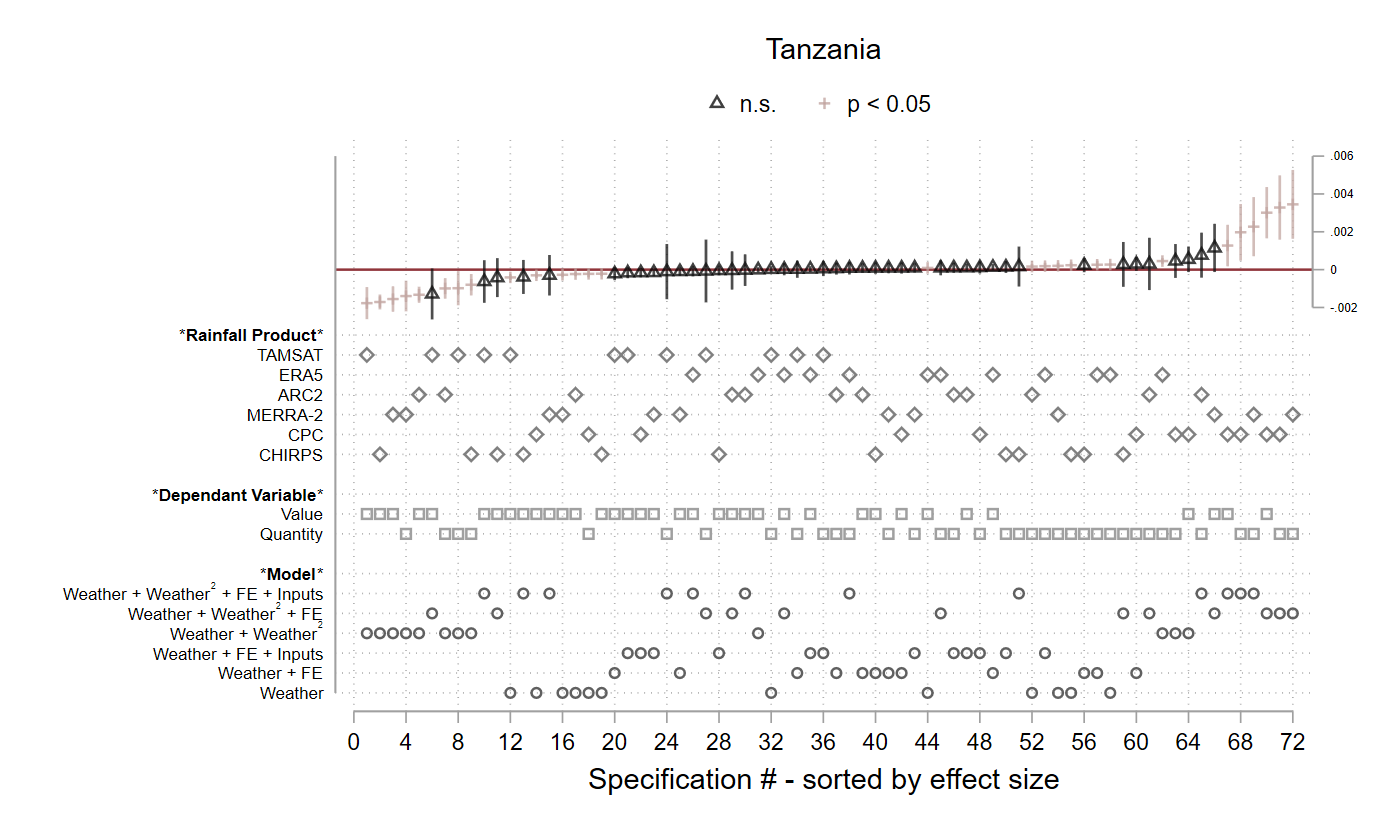}
			\includegraphics[width=.49\linewidth,keepaspectratio]{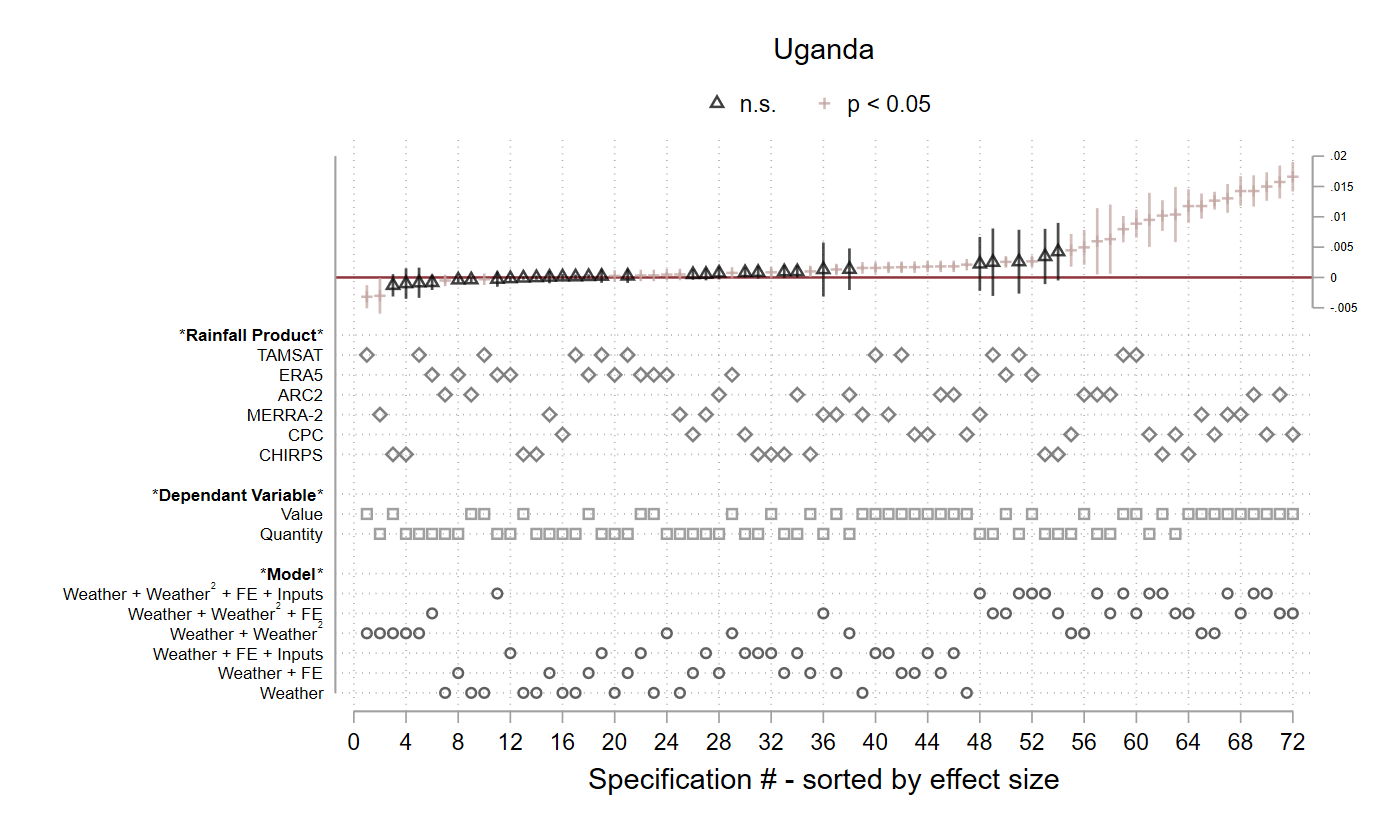}
		\end{center}
		\footnotesize  \textit{Note}: The figure presents specification curves for total season rainfall, where each country is represented in a different panel. Each panel includes 72 regressions, where each column represents a single regression. Significant and non-significant coefficients are designated at the top of each panel. 
	\end{minipage}	
\end{figure}

\begin{figure}[!htbp]
	\begin{minipage}{\linewidth}		
		\caption{Specification Curve for Number of Rainy Days by Country}
		\label{fig:v08_sat}
		\begin{center}
			\includegraphics[width=.49\linewidth,keepaspectratio]{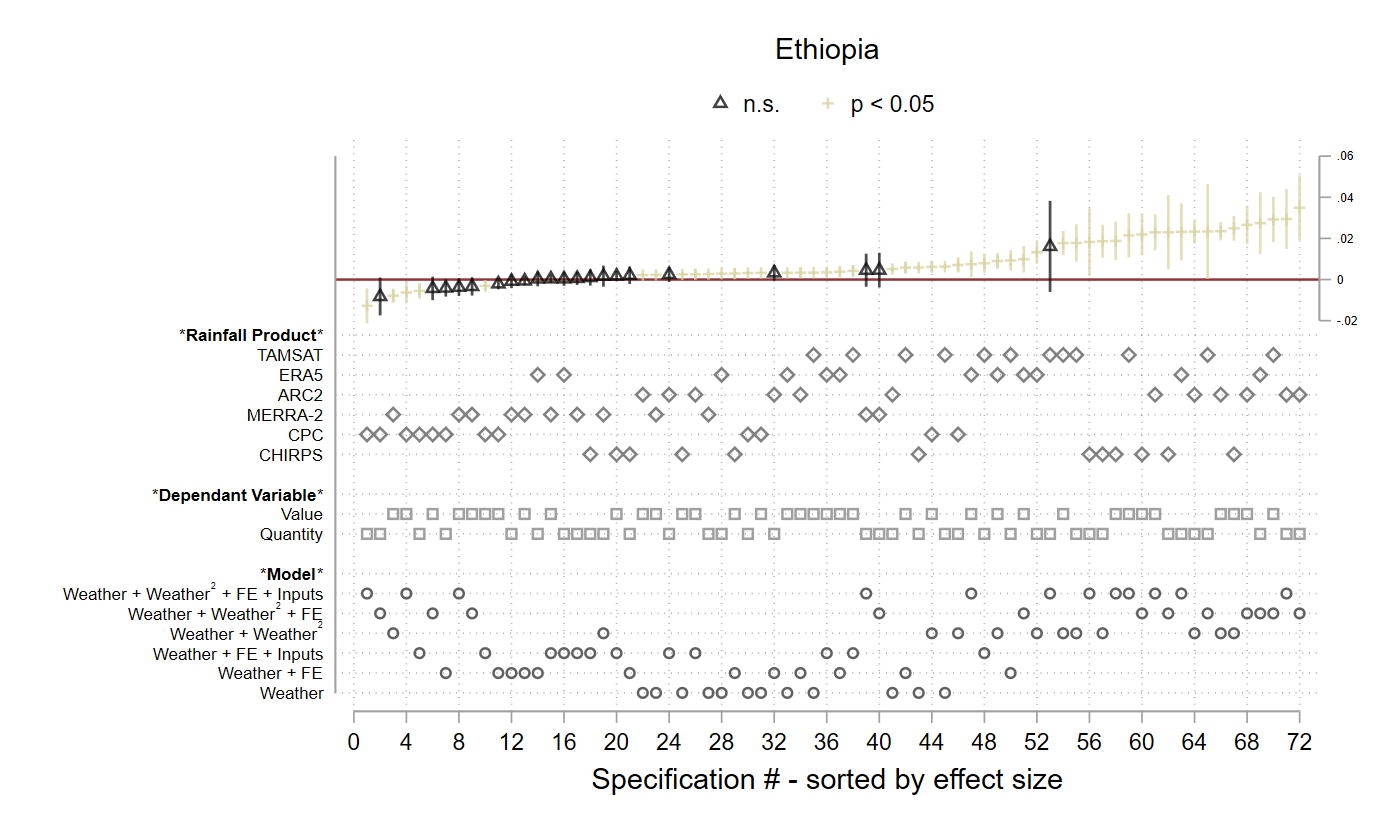}
			\includegraphics[width=.49\linewidth,keepaspectratio]{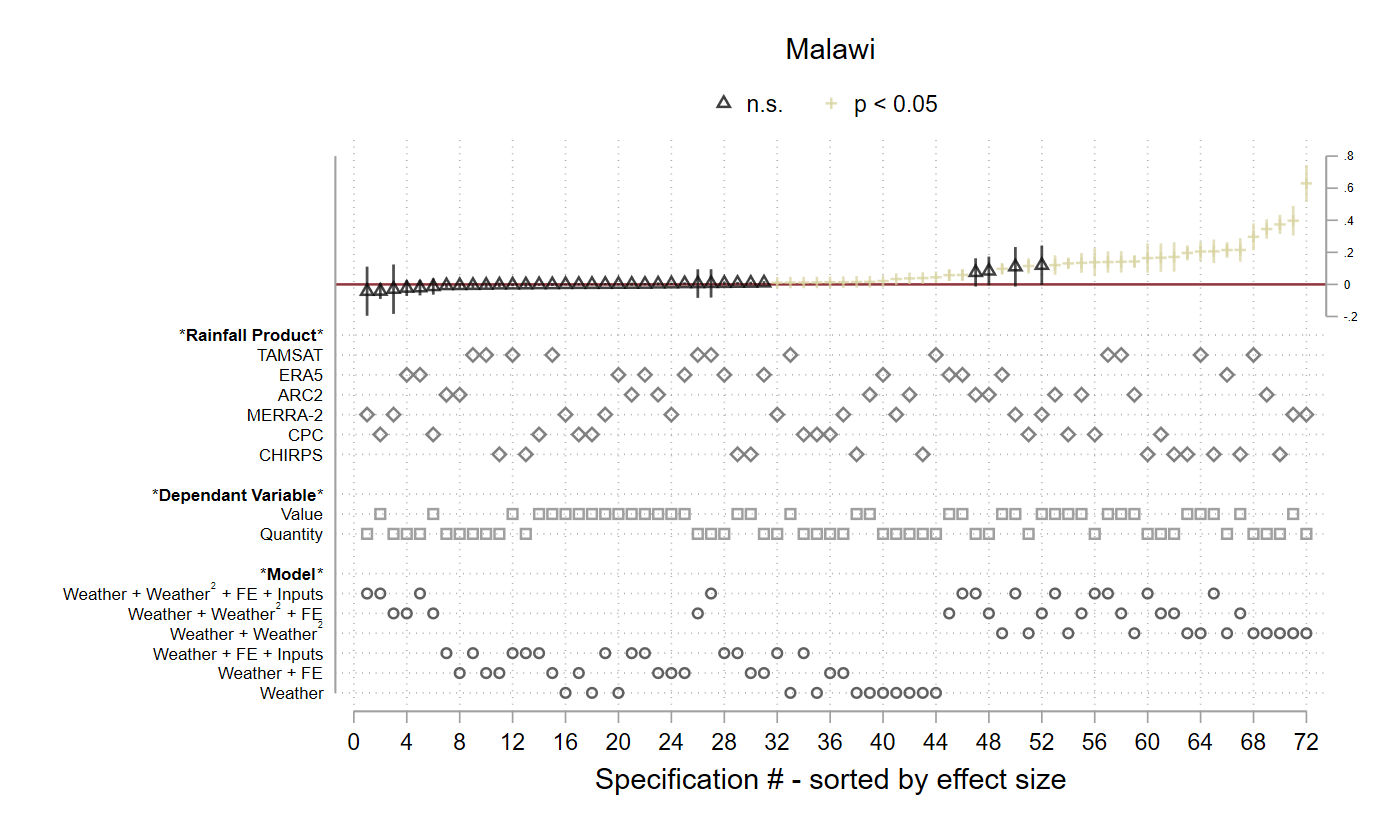}
			\includegraphics[width=.49\linewidth,keepaspectratio]{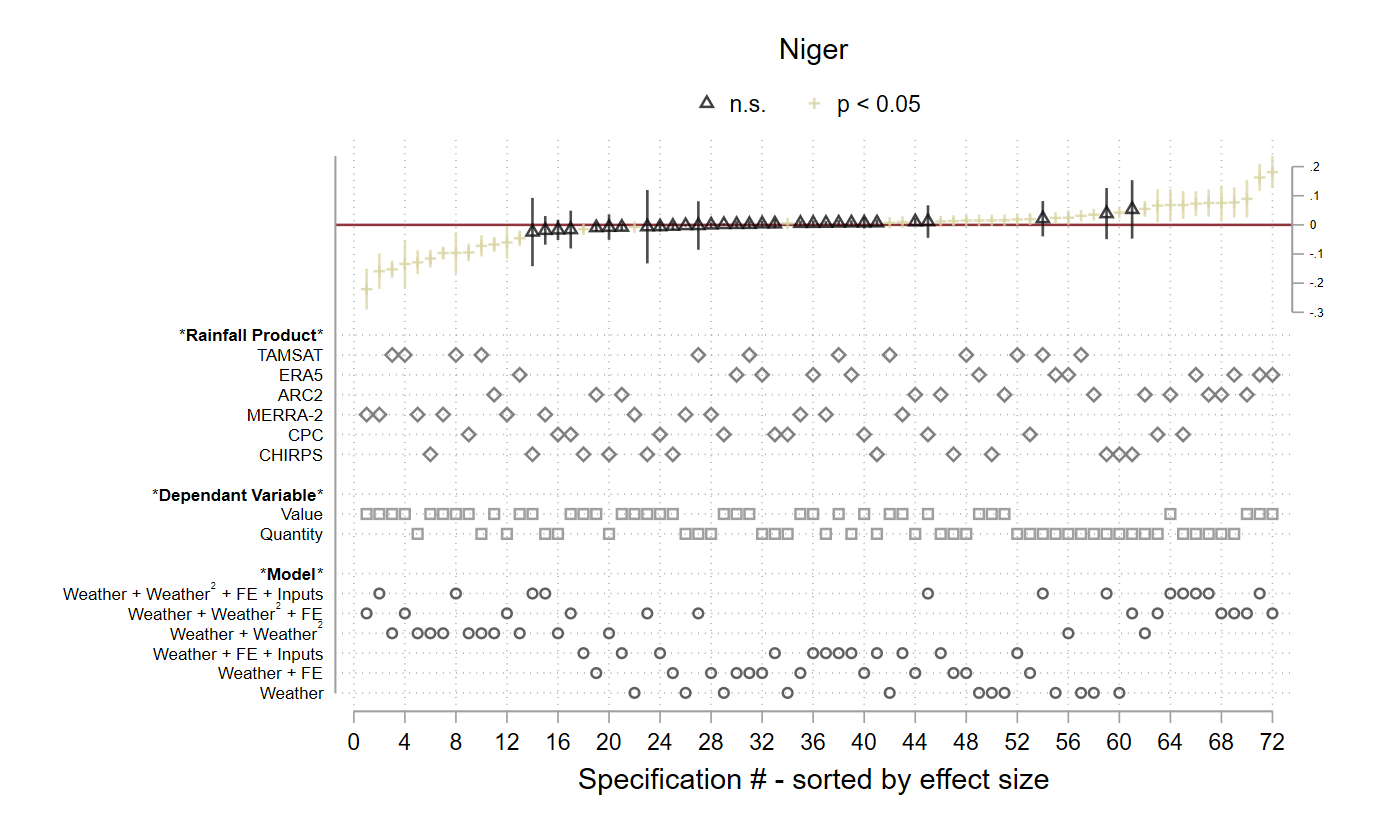}
			\includegraphics[width=.49\linewidth,keepaspectratio]{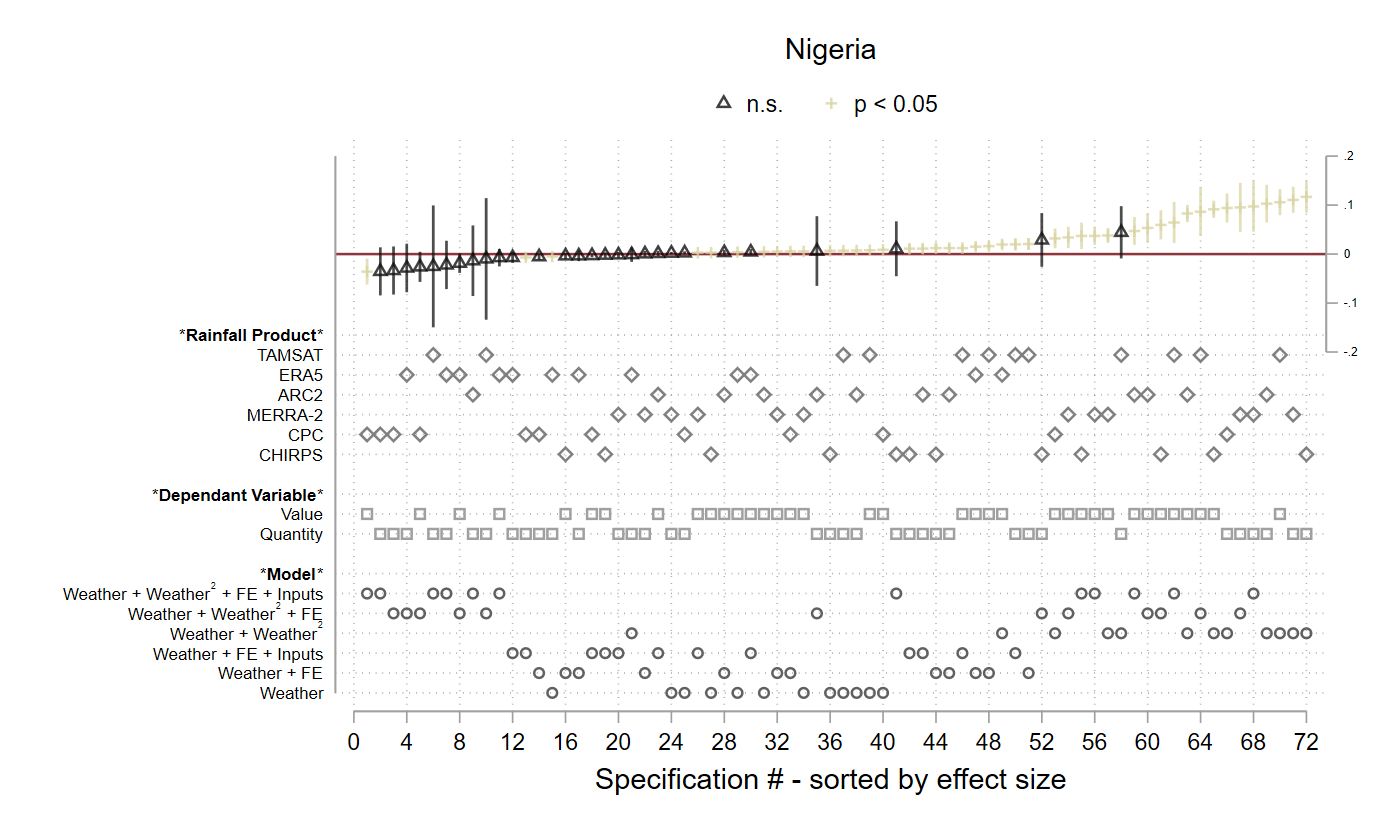}
			\includegraphics[width=.49\linewidth,keepaspectratio]{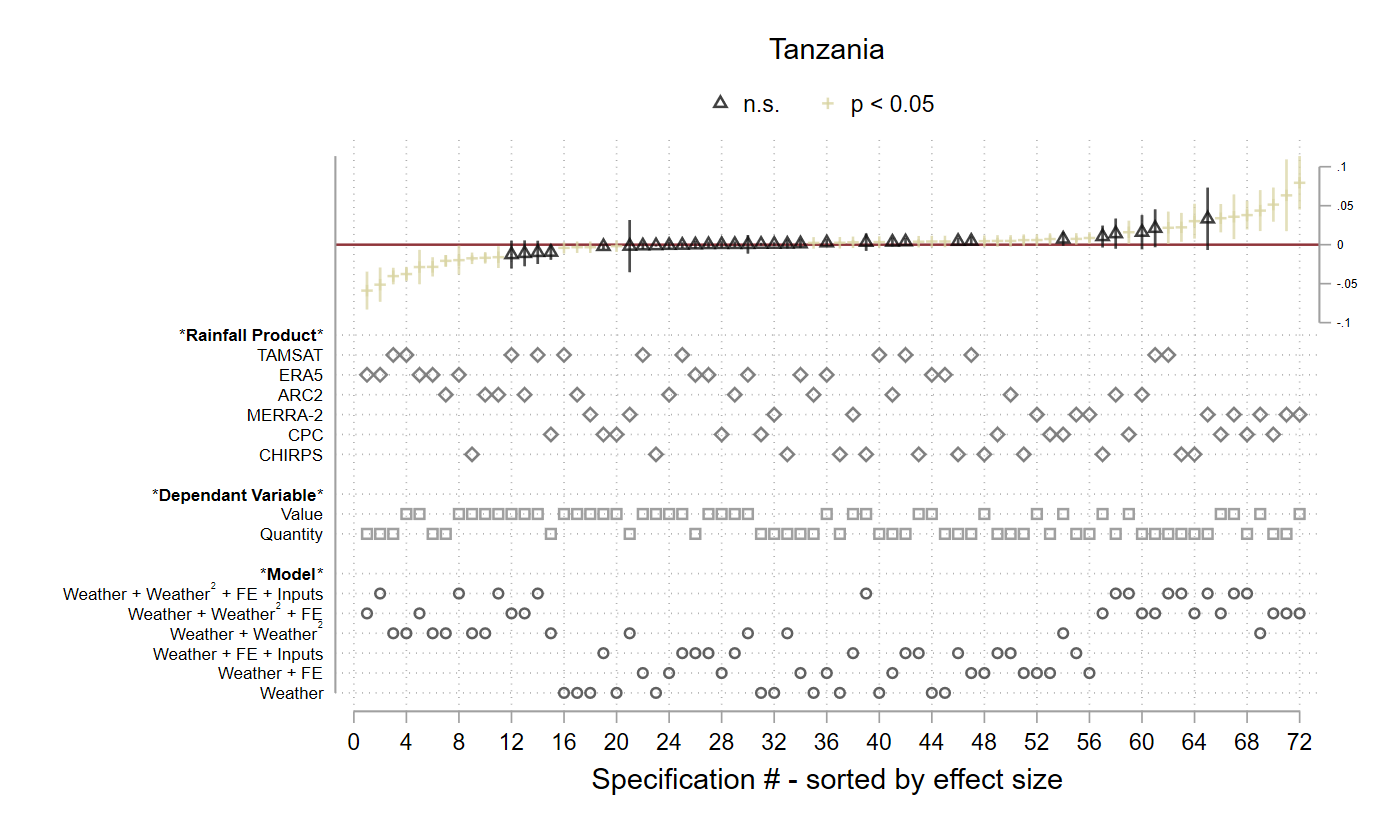}
			\includegraphics[width=.49\linewidth,keepaspectratio]{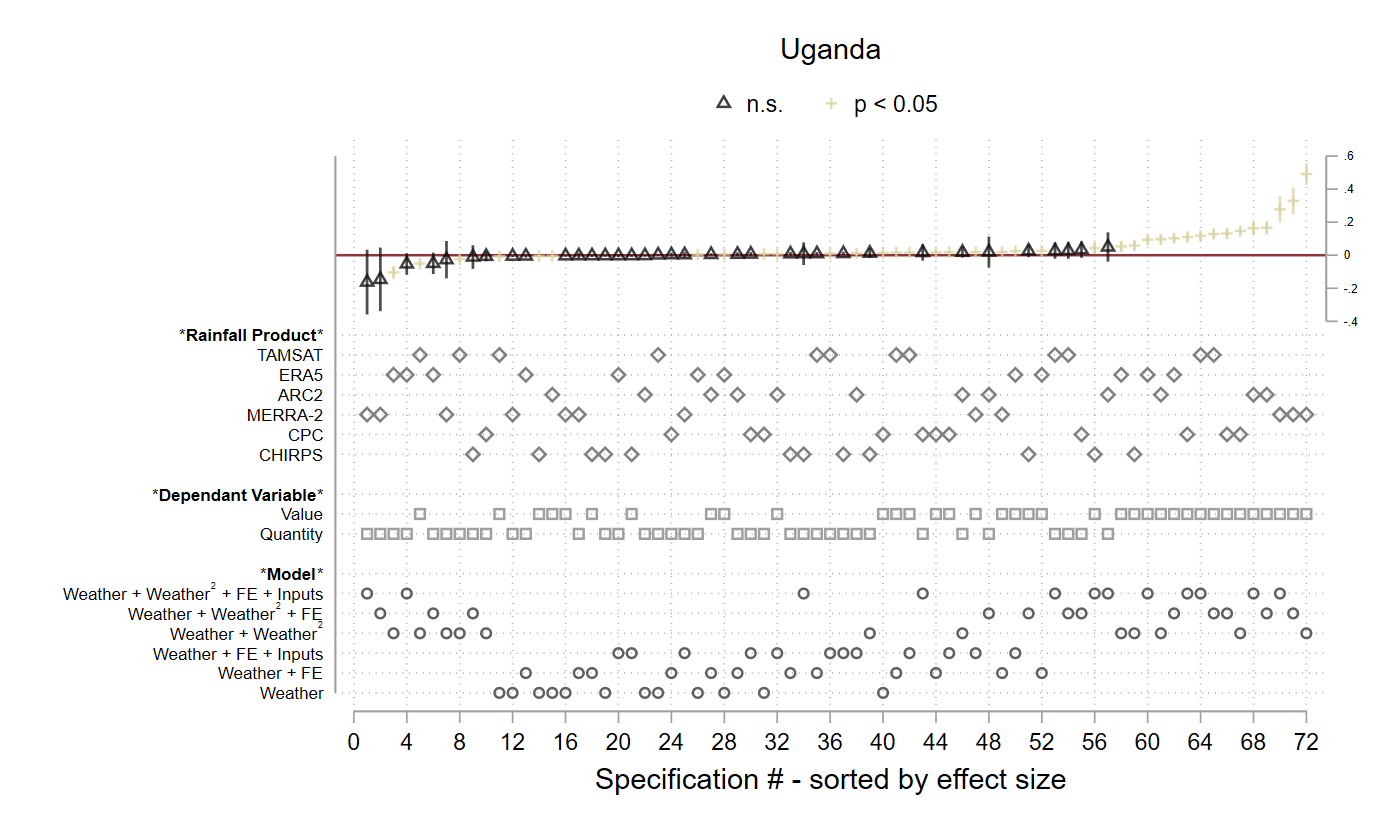}
		\end{center}
		\footnotesize  \textit{Note}: The figure presents specification curves for the number of rainy days, where each country is represented in a different panel. Each panel includes 72 regressions, where each column represents a single regression. Significant and non-significant coefficients are designated at the top of each panel.   
	\end{minipage}	
\end{figure}

\begin{figure}[!htbp]
	\begin{minipage}{\linewidth}		
		\caption{Specification Curve for Percentage of Rainy Days by Country}
		\label{fig:v12_sat}
		\begin{center}
			\includegraphics[width=.49\linewidth,keepaspectratio]{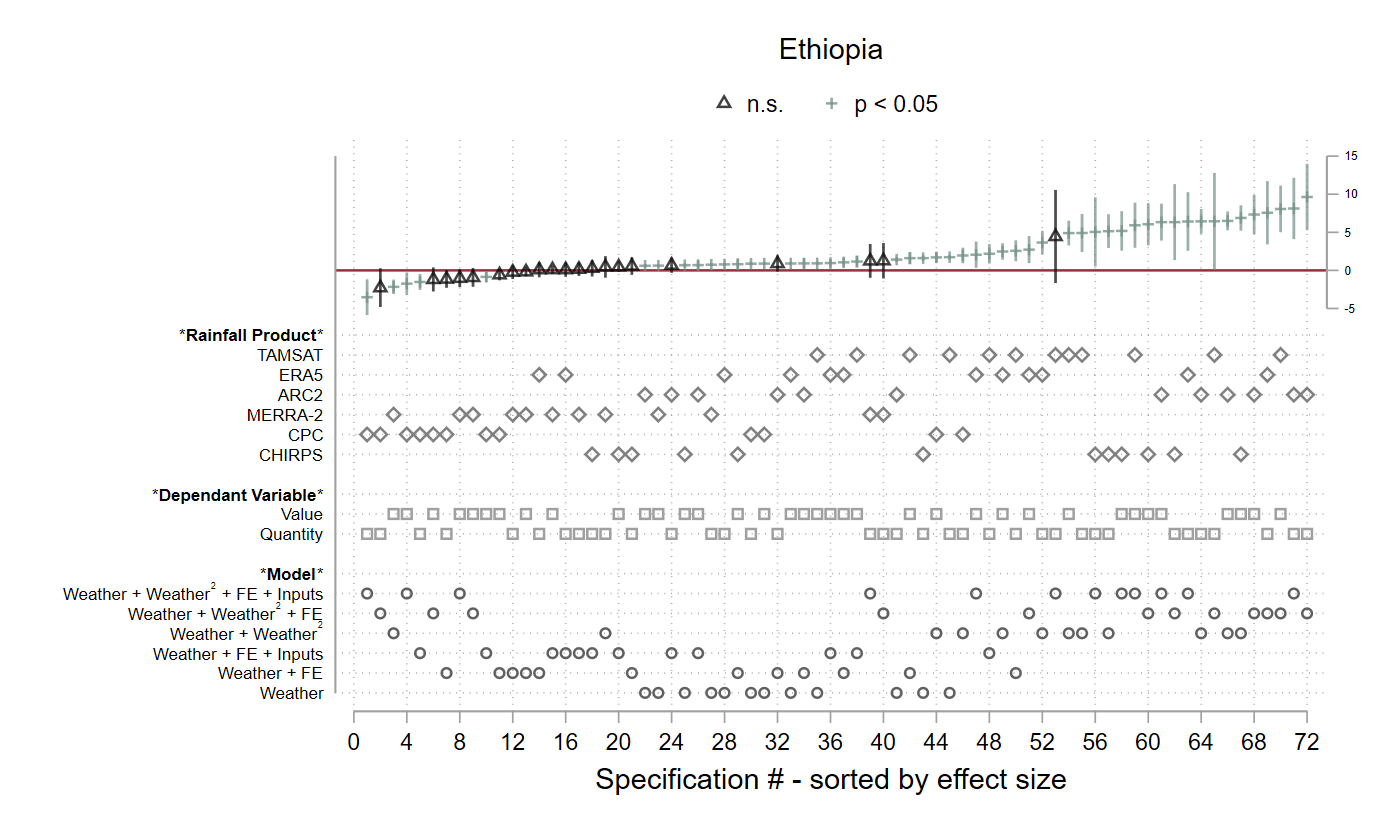}
			\includegraphics[width=.49\linewidth,keepaspectratio]{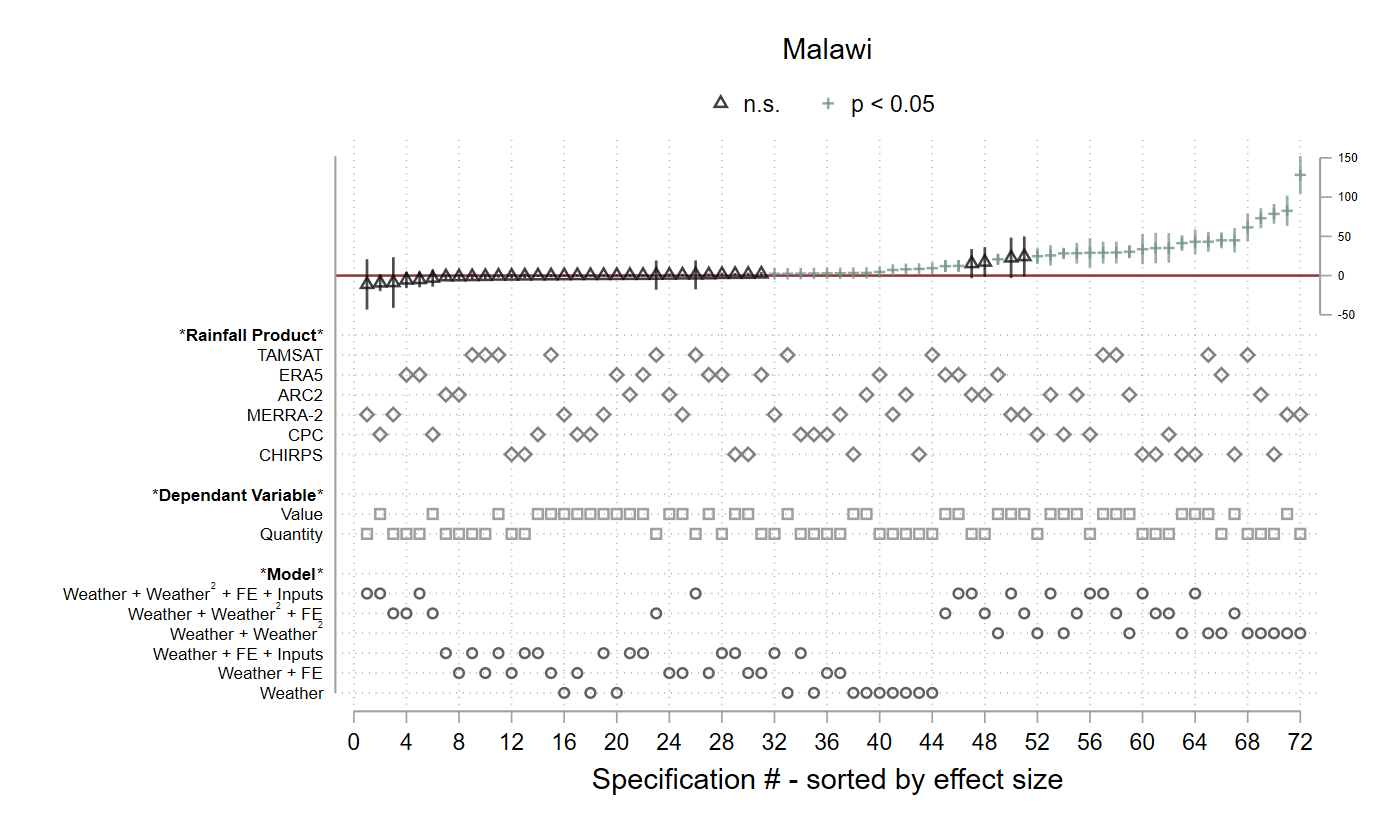}
			\includegraphics[width=.49\linewidth,keepaspectratio]{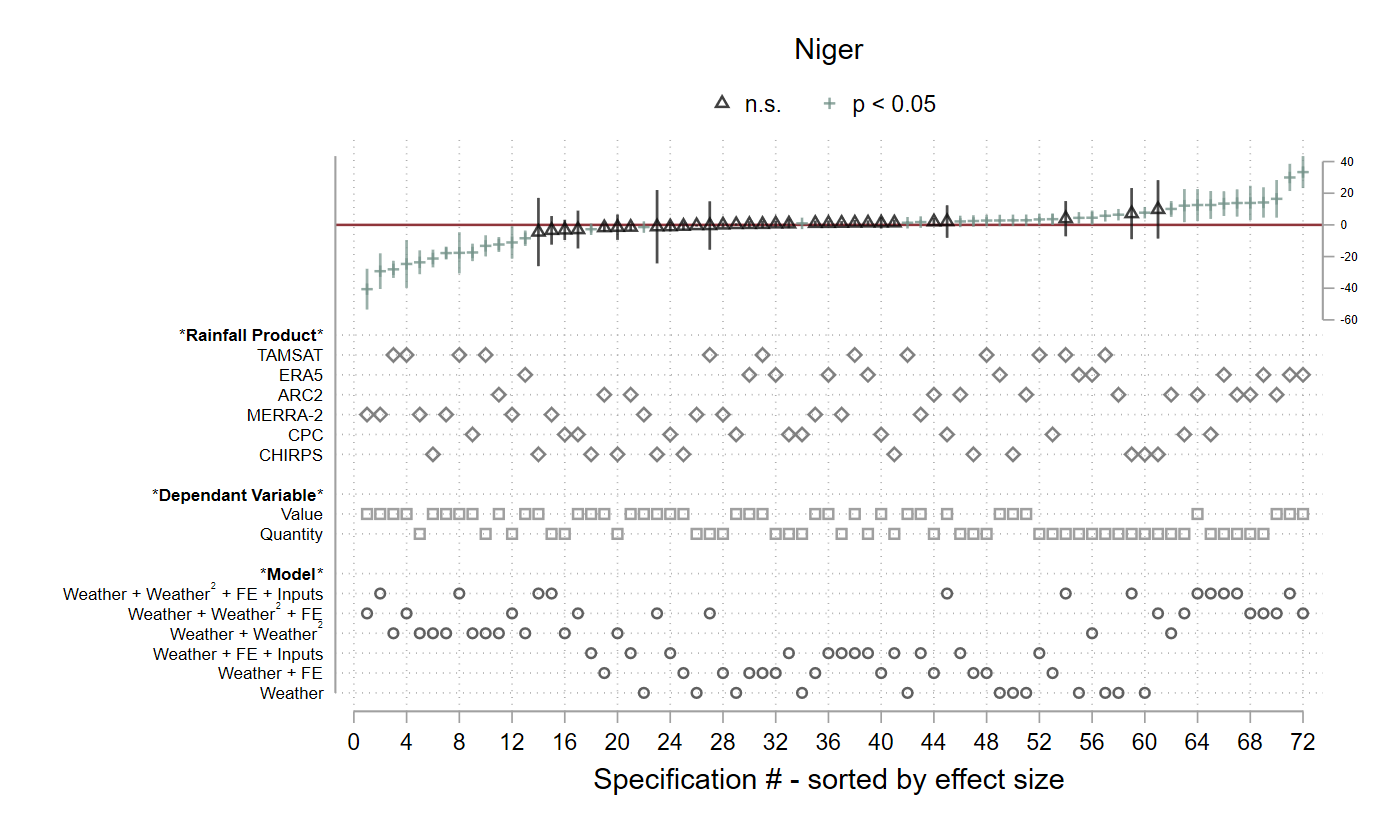}
			\includegraphics[width=.49\linewidth,keepaspectratio]{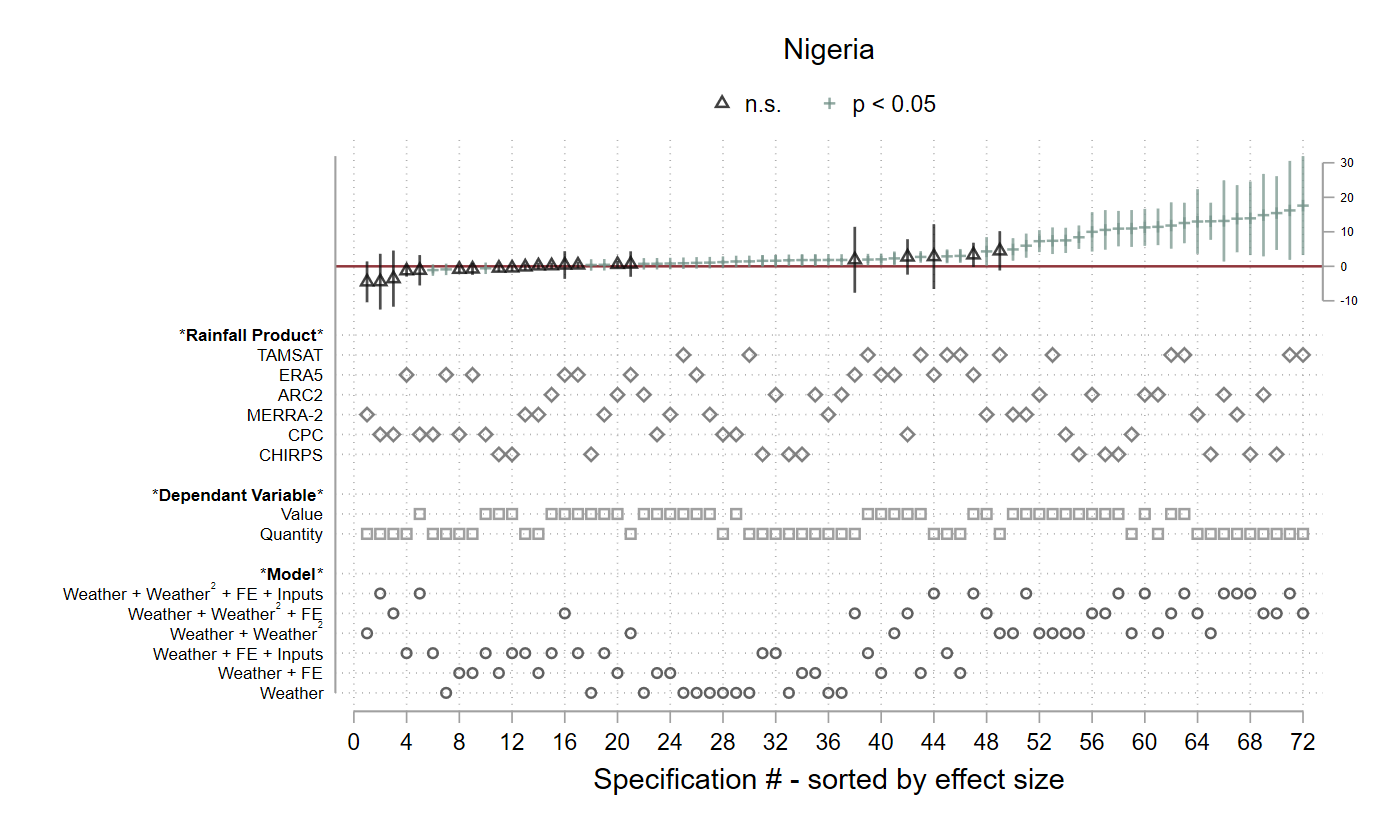}
			\includegraphics[width=.49\linewidth,keepaspectratio]{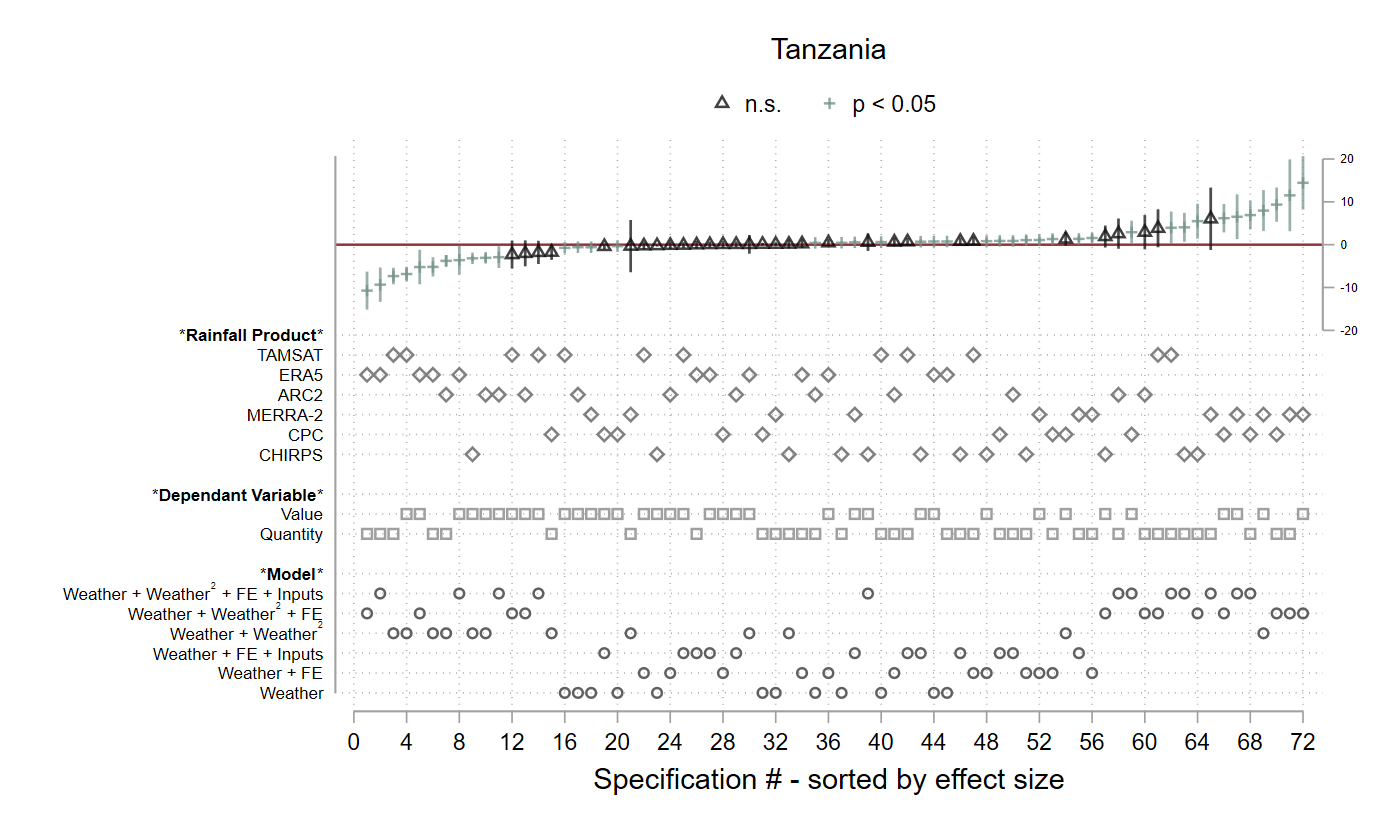}
			\includegraphics[width=.49\linewidth,keepaspectratio]{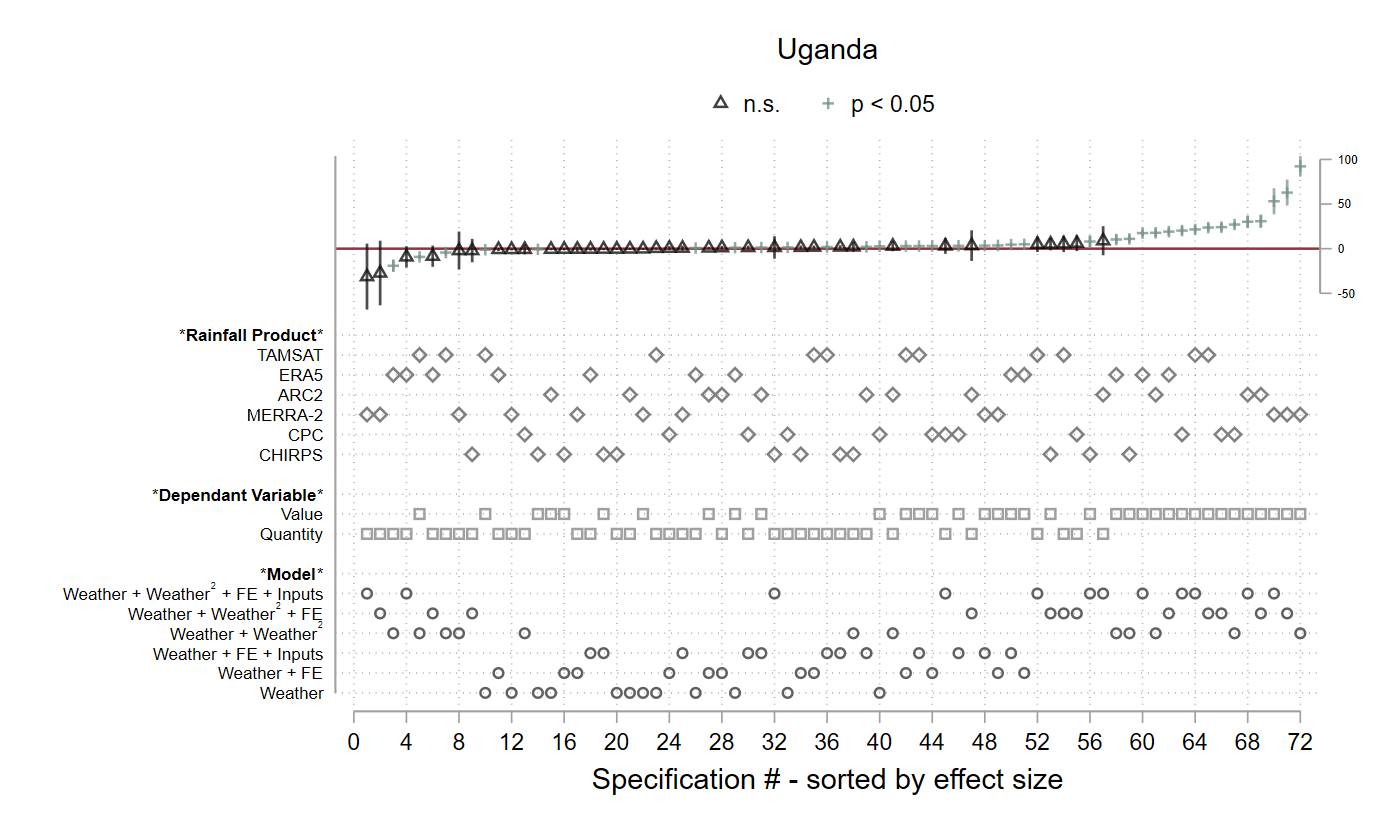}
		\end{center}
		\footnotesize  \textit{Note}: The figure presents specification curves for the percentage of rainy days, where each country is represented in a different panel. Each panel includes 72 regressions, where each column represents a single regression. Significant and non-significant coefficients are designated at the top of each panel.  
	\end{minipage}	
\end{figure}

\begin{figure}[!htbp]
	\begin{minipage}{\linewidth}		
		\caption{Specification Curve for Mean Daily Temperature by Country}
		\label{fig:v15_sat}
		\begin{center}
			\includegraphics[width=.49\linewidth,keepaspectratio]{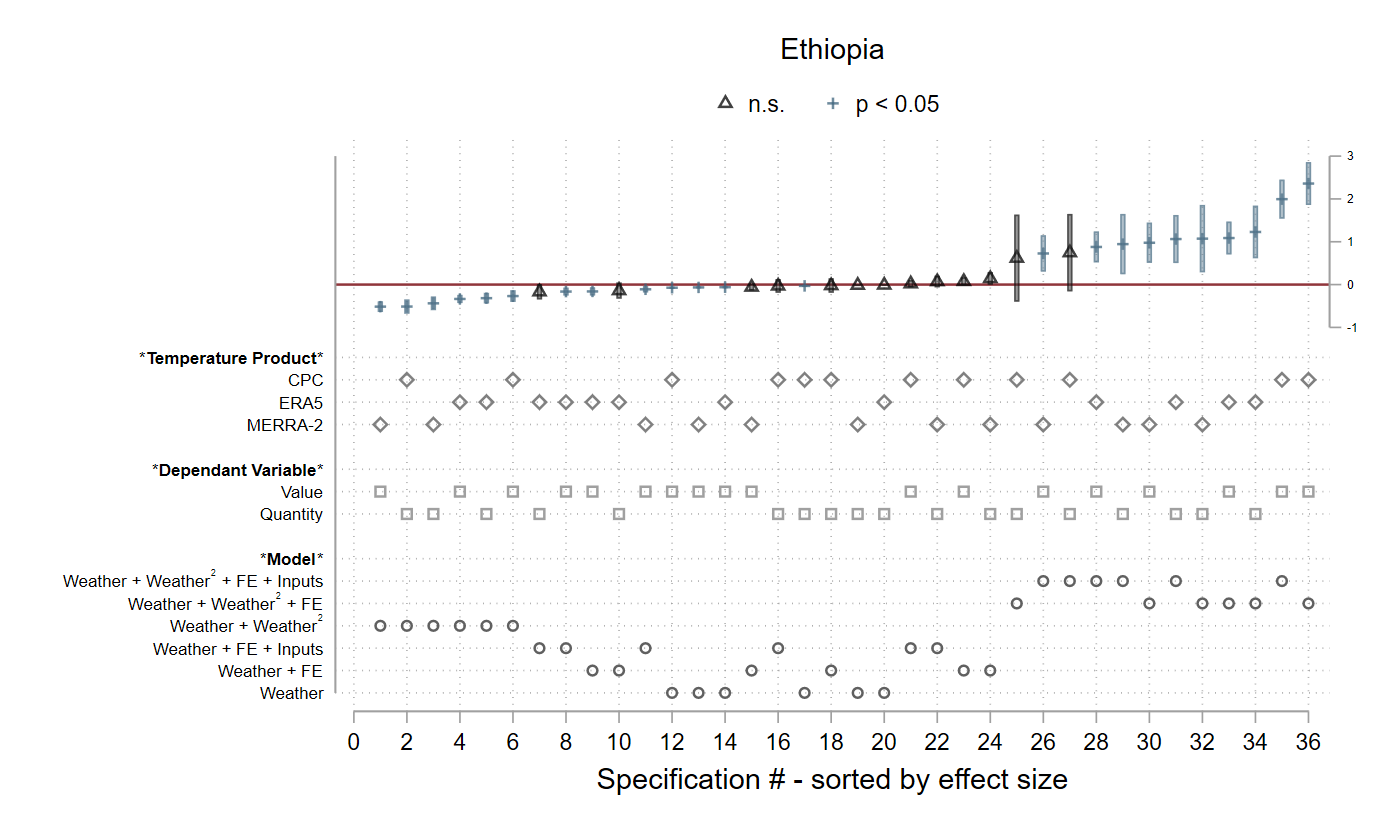}
			\includegraphics[width=.49\linewidth,keepaspectratio]{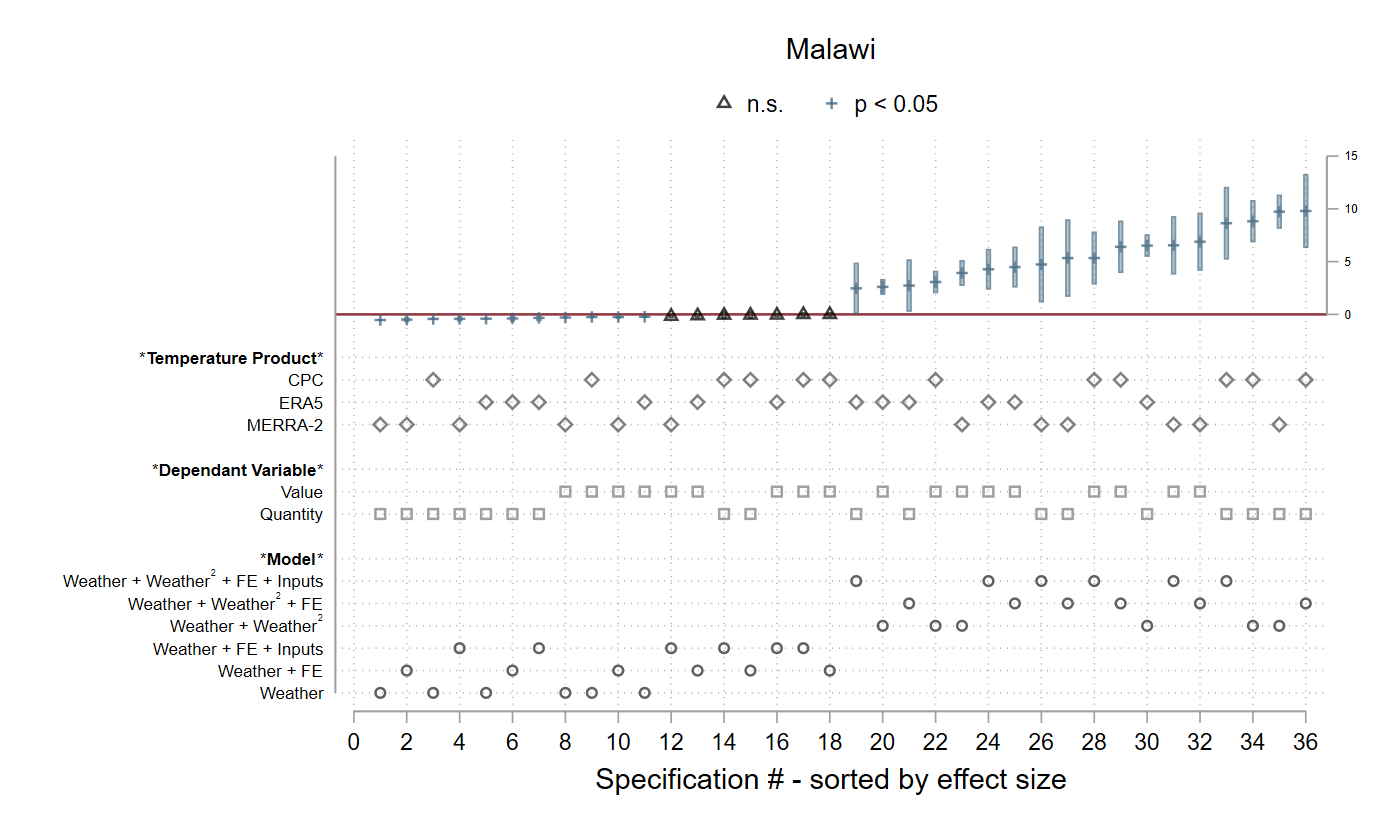}
			\includegraphics[width=.49\linewidth,keepaspectratio]{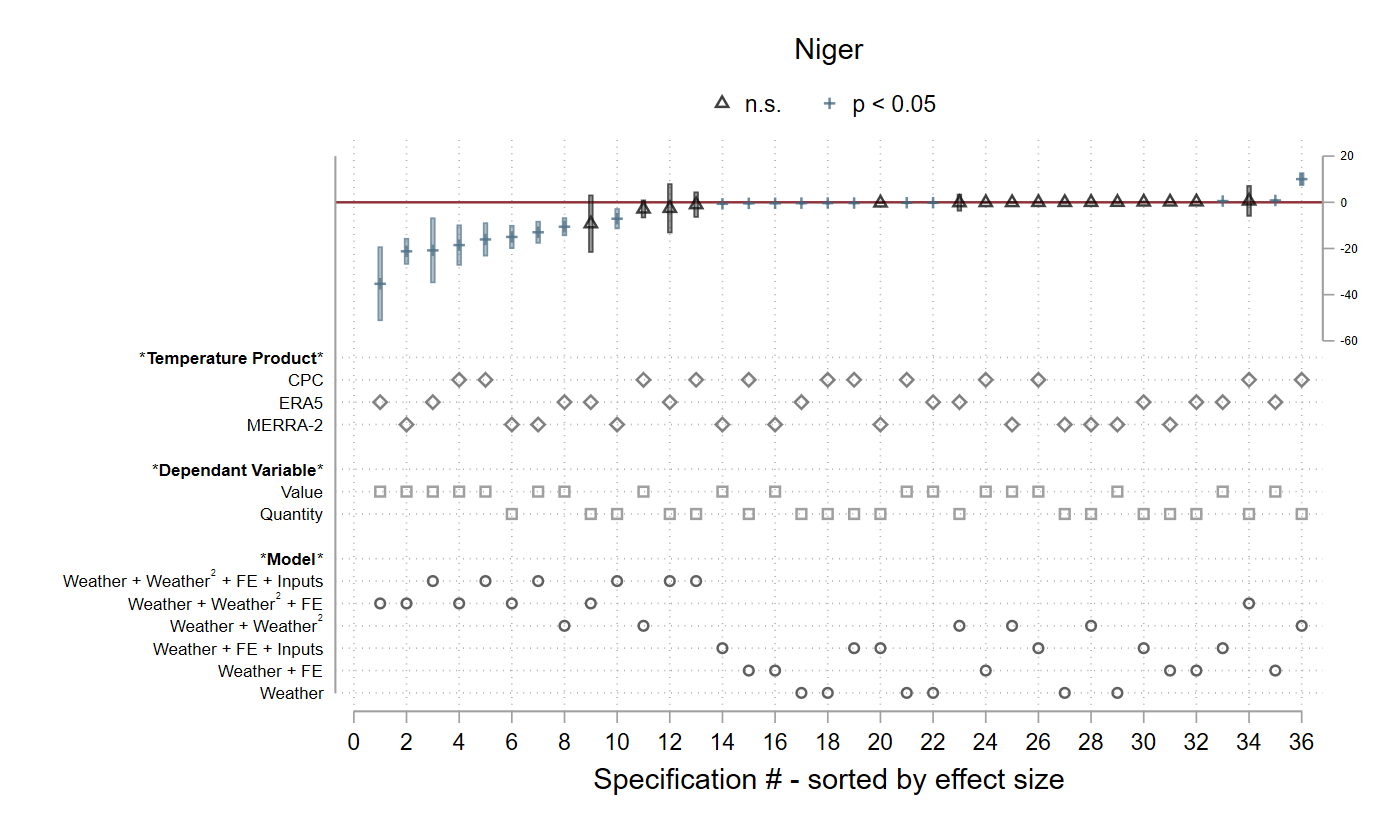}
			\includegraphics[width=.49\linewidth,keepaspectratio]{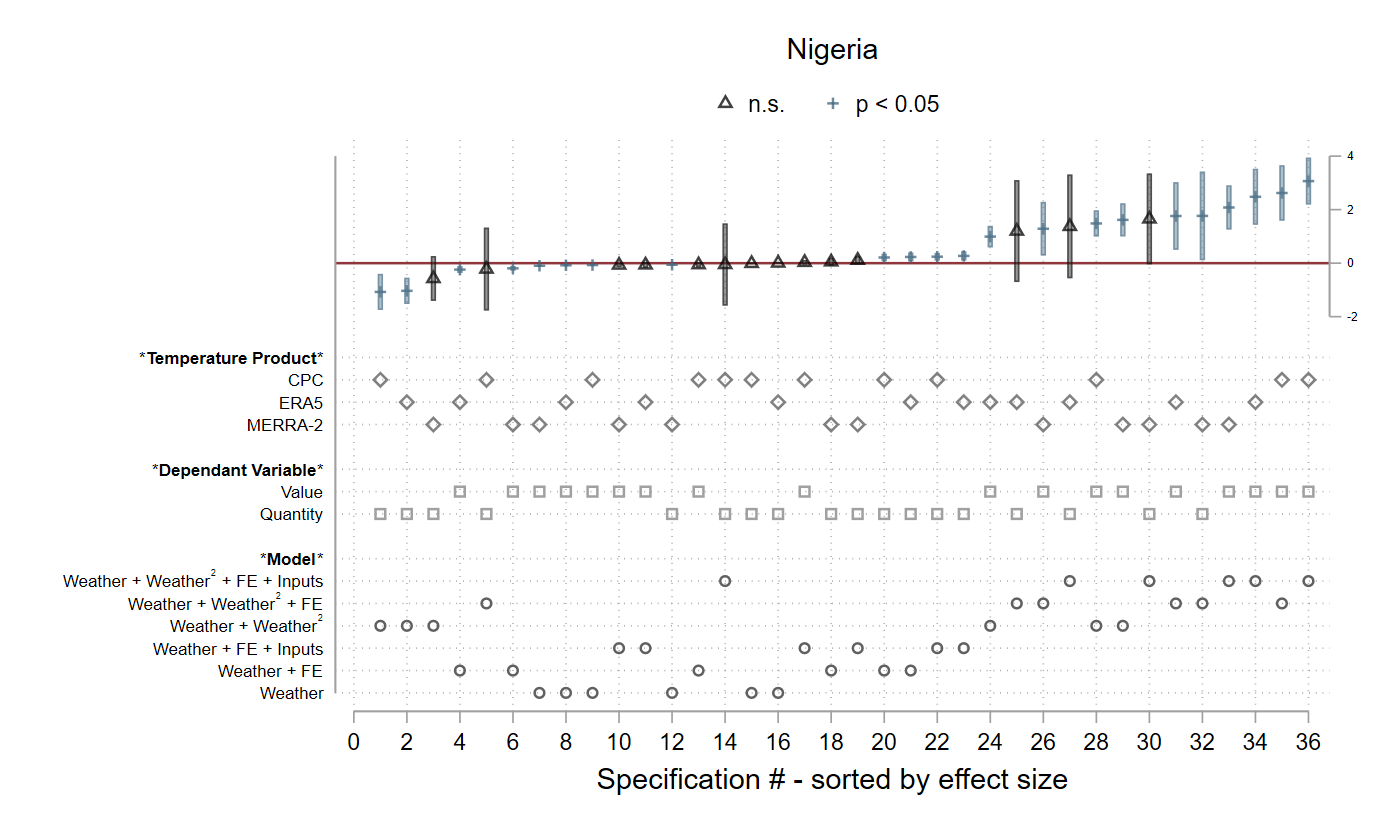}
			\includegraphics[width=.49\linewidth,keepaspectratio]{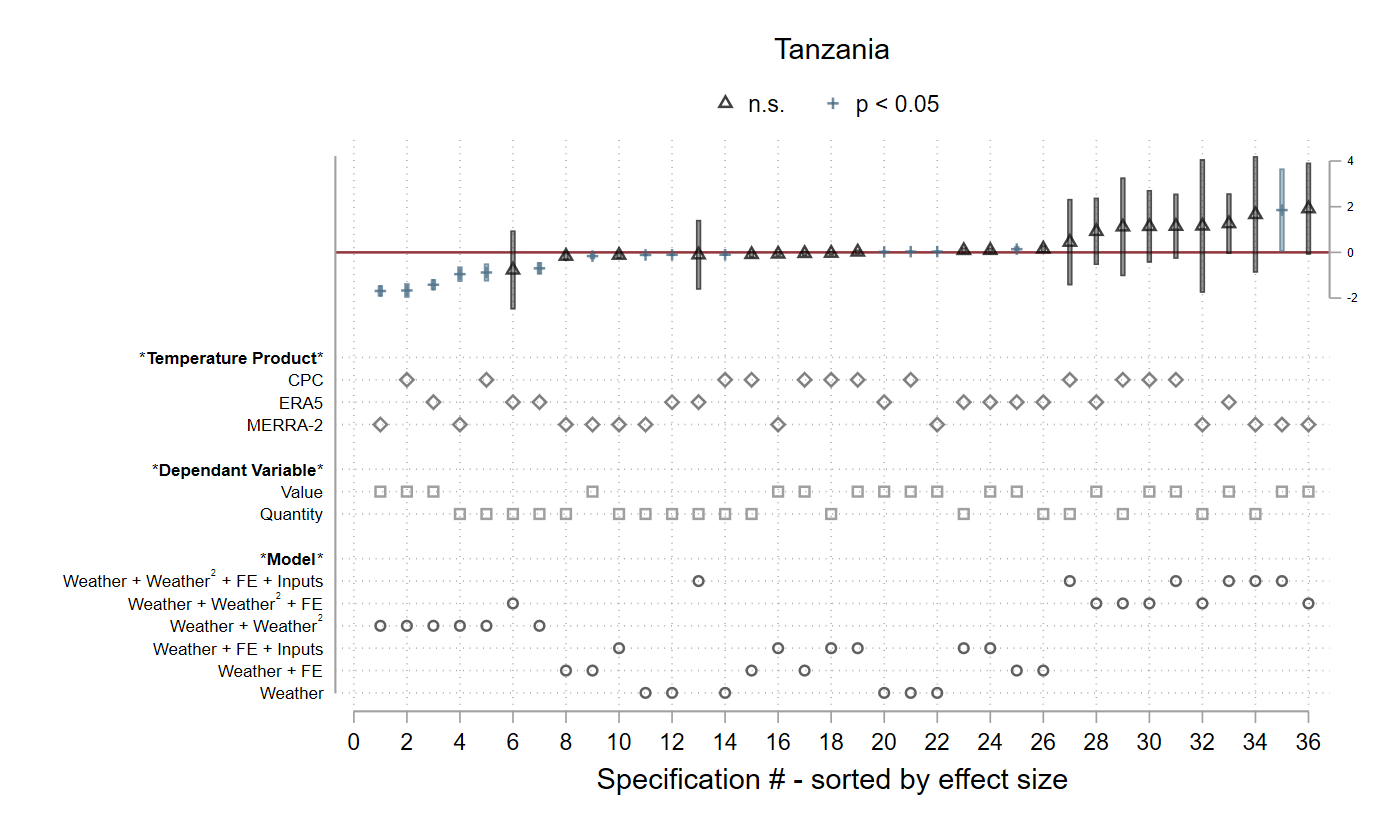}
			\includegraphics[width=.49\linewidth,keepaspectratio]{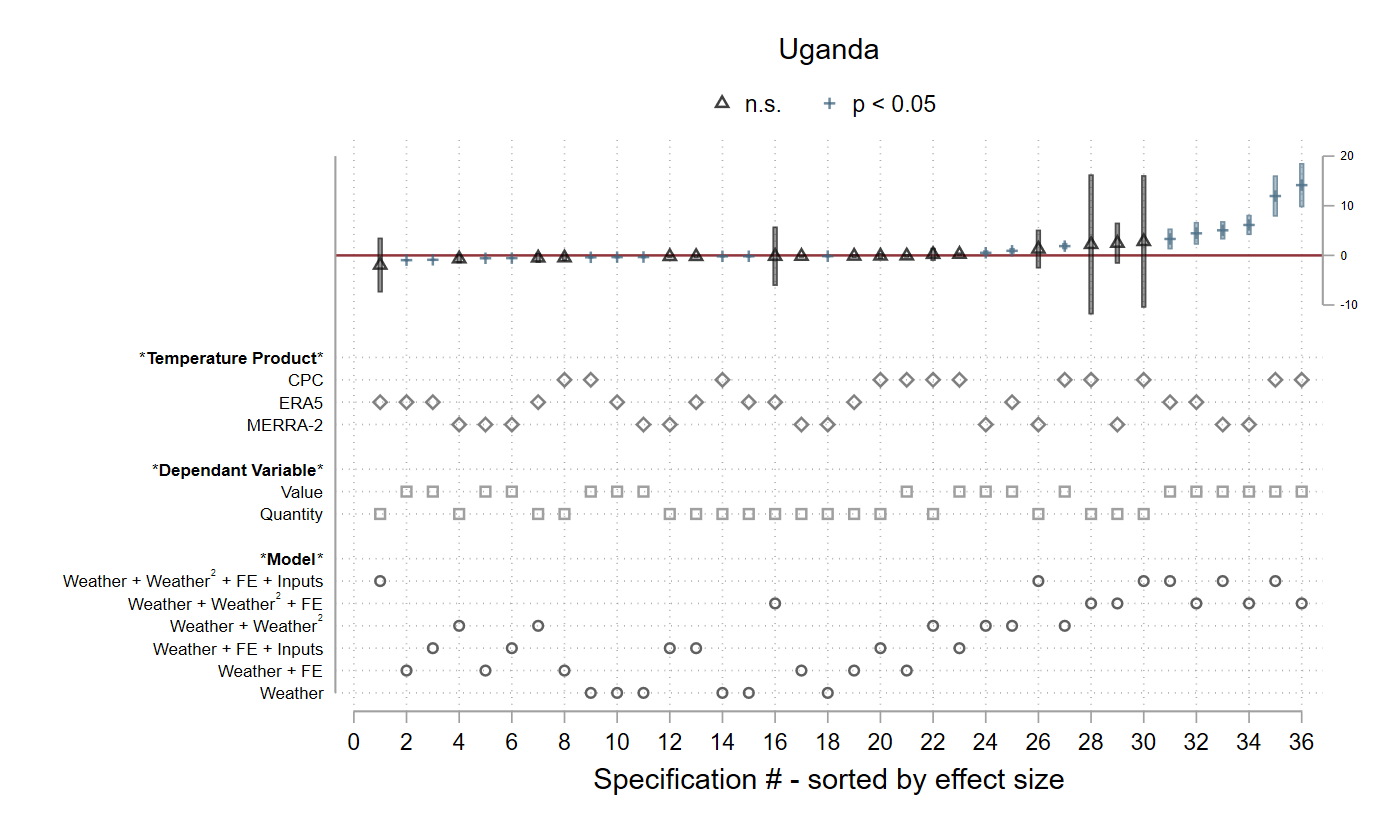}
		\end{center}
		\footnotesize  \textit{Note}: The figure presents specification curves for mean daily temperature, where each country is represented in a different panel. Each panel includes 36 regressions, where each column represents a single regression. Significant and non-significant coefficients are designated at the top of each panel.  
	\end{minipage}	
\end{figure}

\begin{figure}[!htbp]
	\begin{minipage}{\linewidth}		
		\caption{Specification Curve for Median Daily Temperature by Country}
		\label{fig:v16_sat}
		\begin{center}
			\includegraphics[width=.49\linewidth,keepaspectratio]{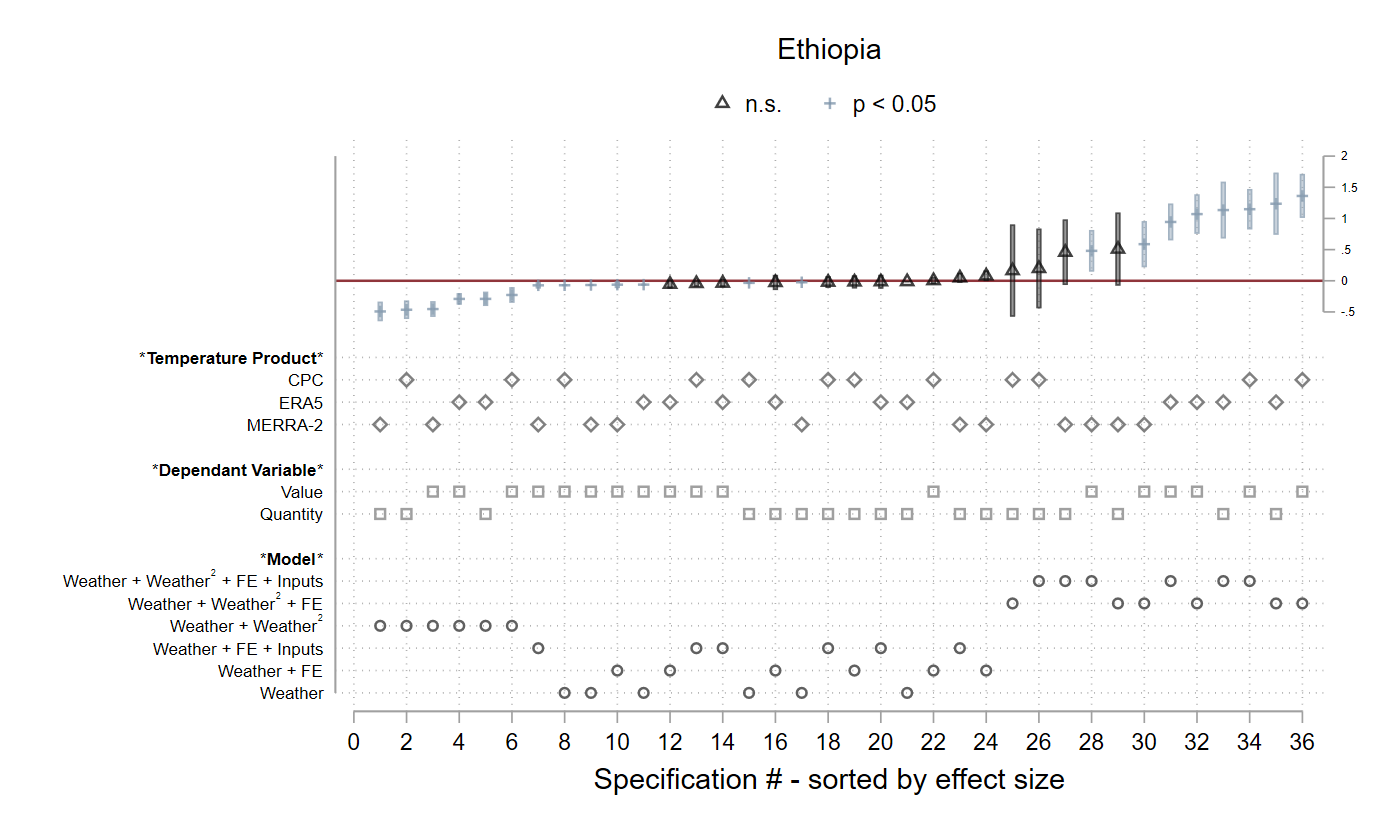}
			\includegraphics[width=.49\linewidth,keepaspectratio]{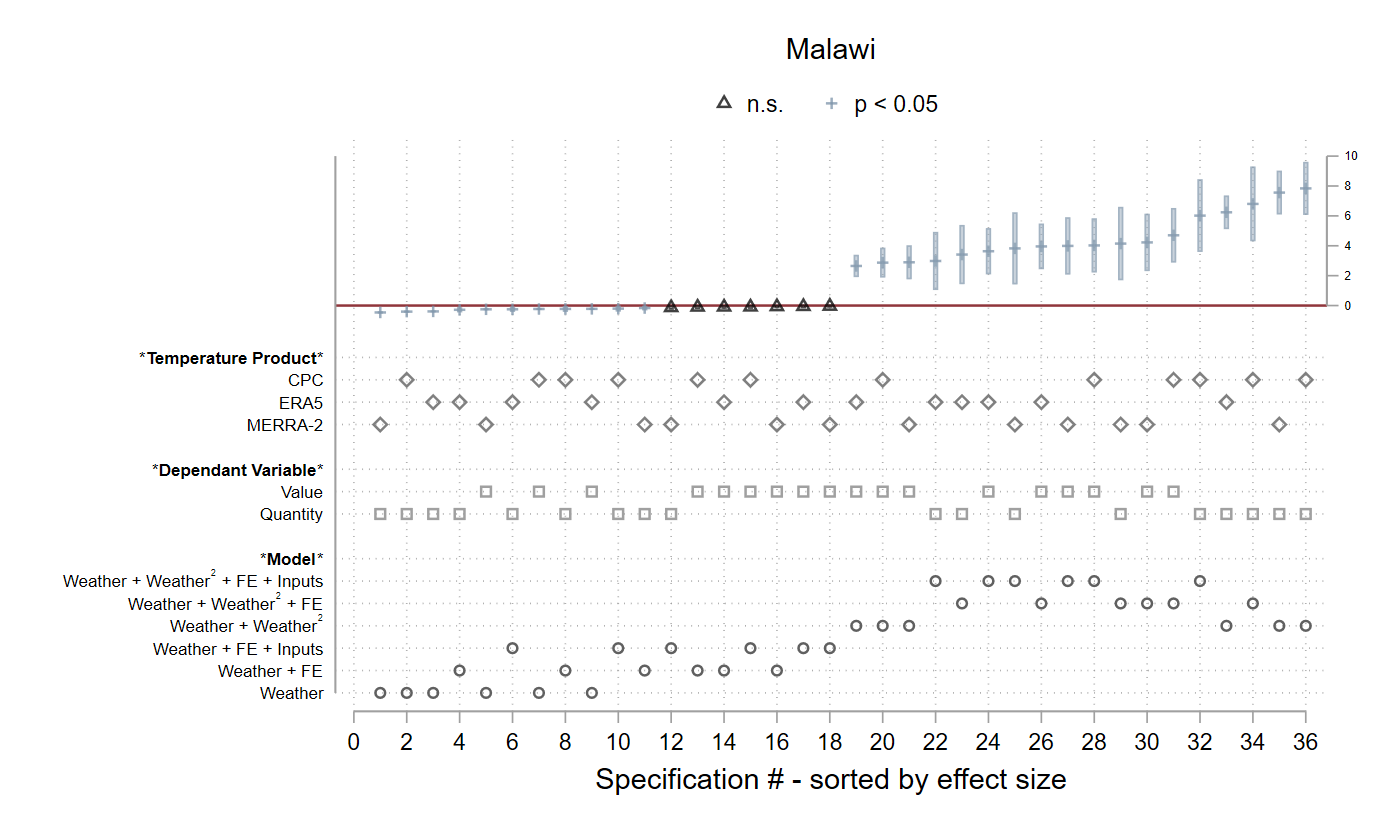}
			\includegraphics[width=.49\linewidth,keepaspectratio]{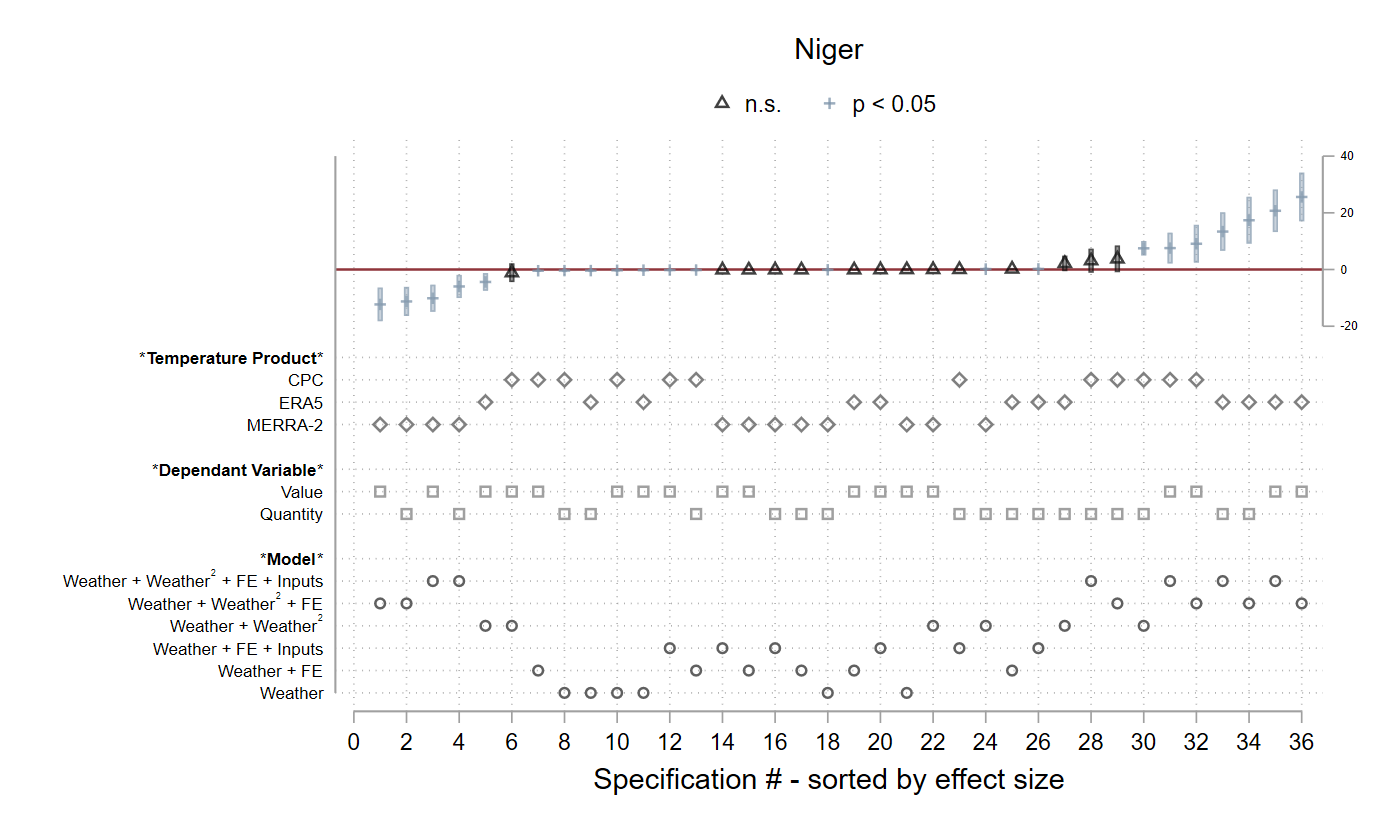}
			\includegraphics[width=.49\linewidth,keepaspectratio]{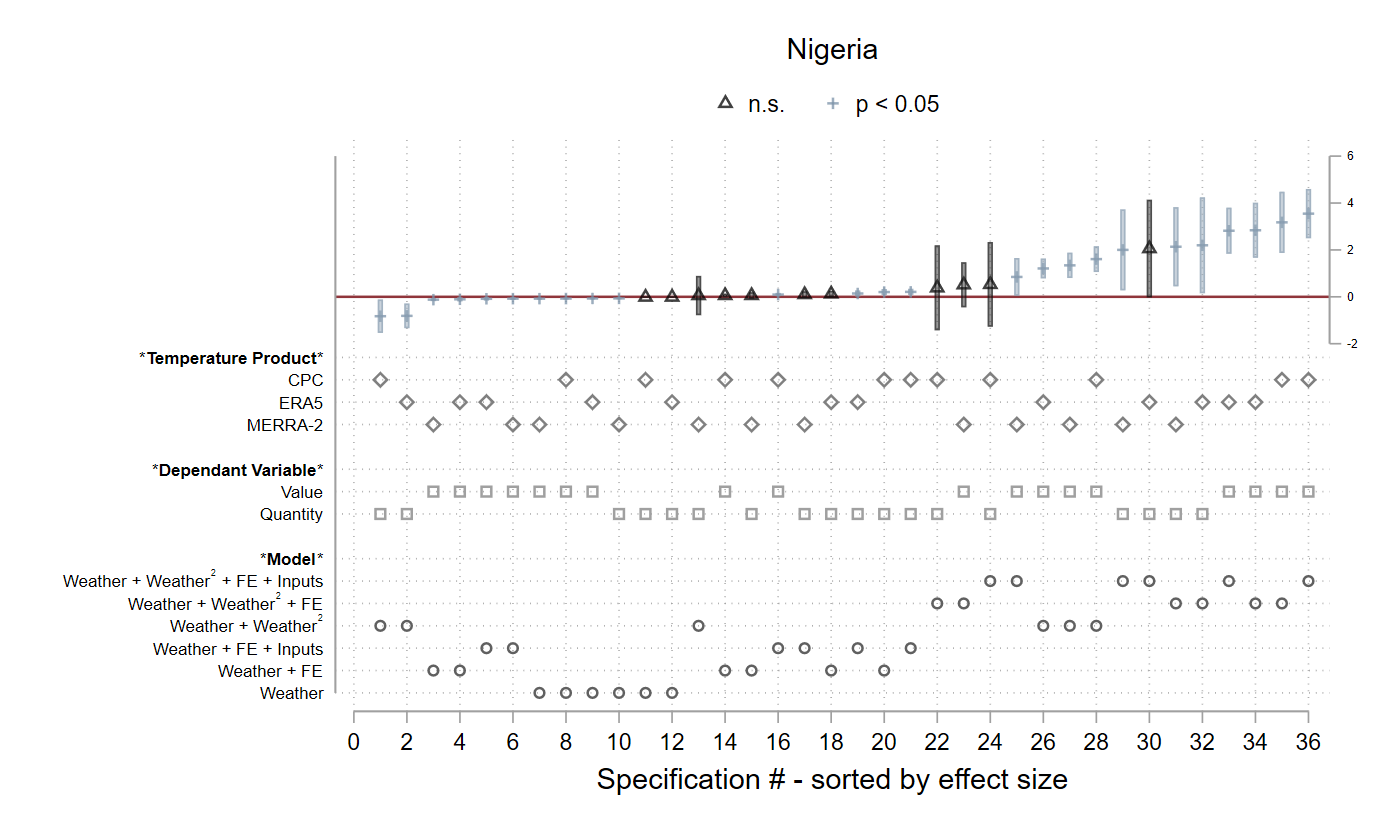}
			\includegraphics[width=.49\linewidth,keepaspectratio]{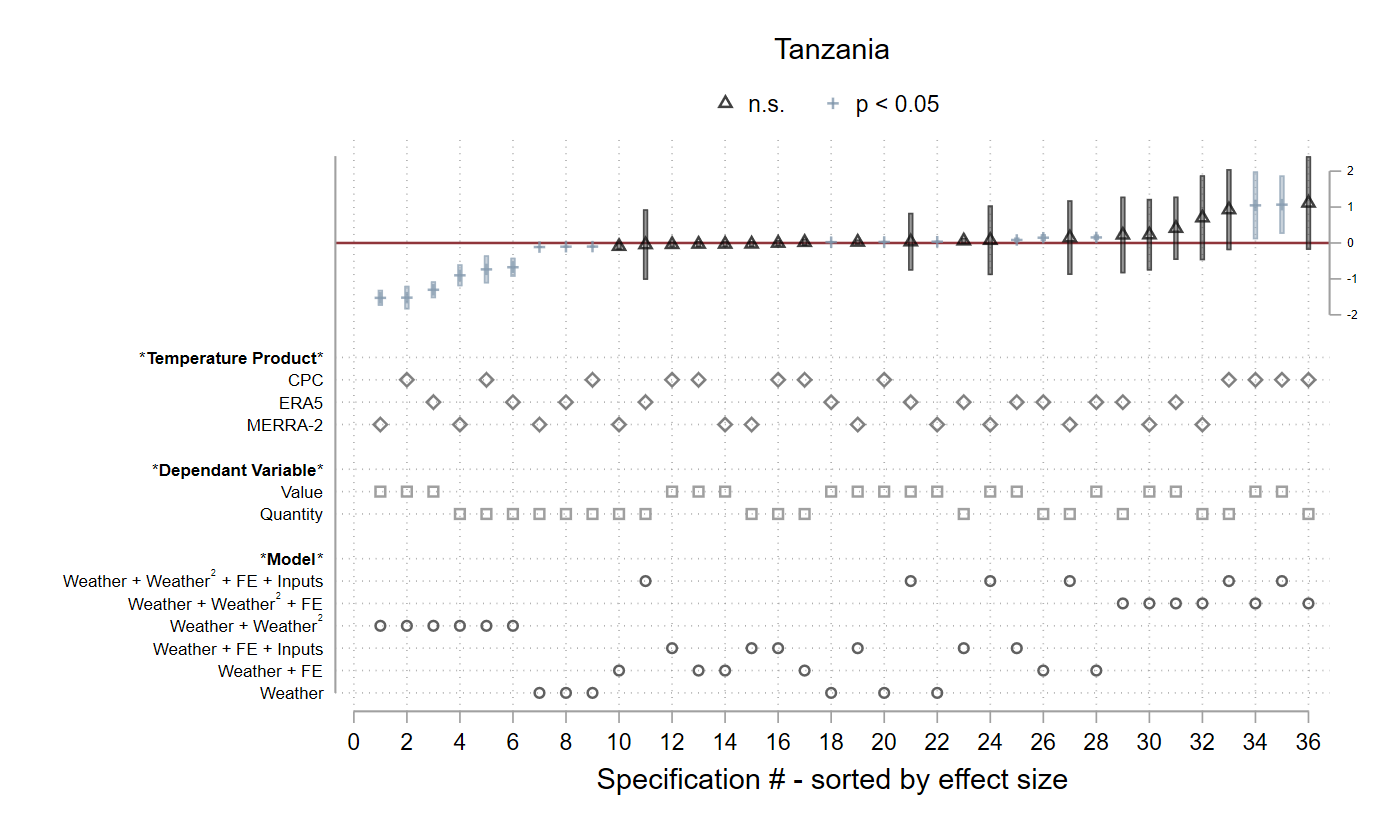}
			\includegraphics[width=.49\linewidth,keepaspectratio]{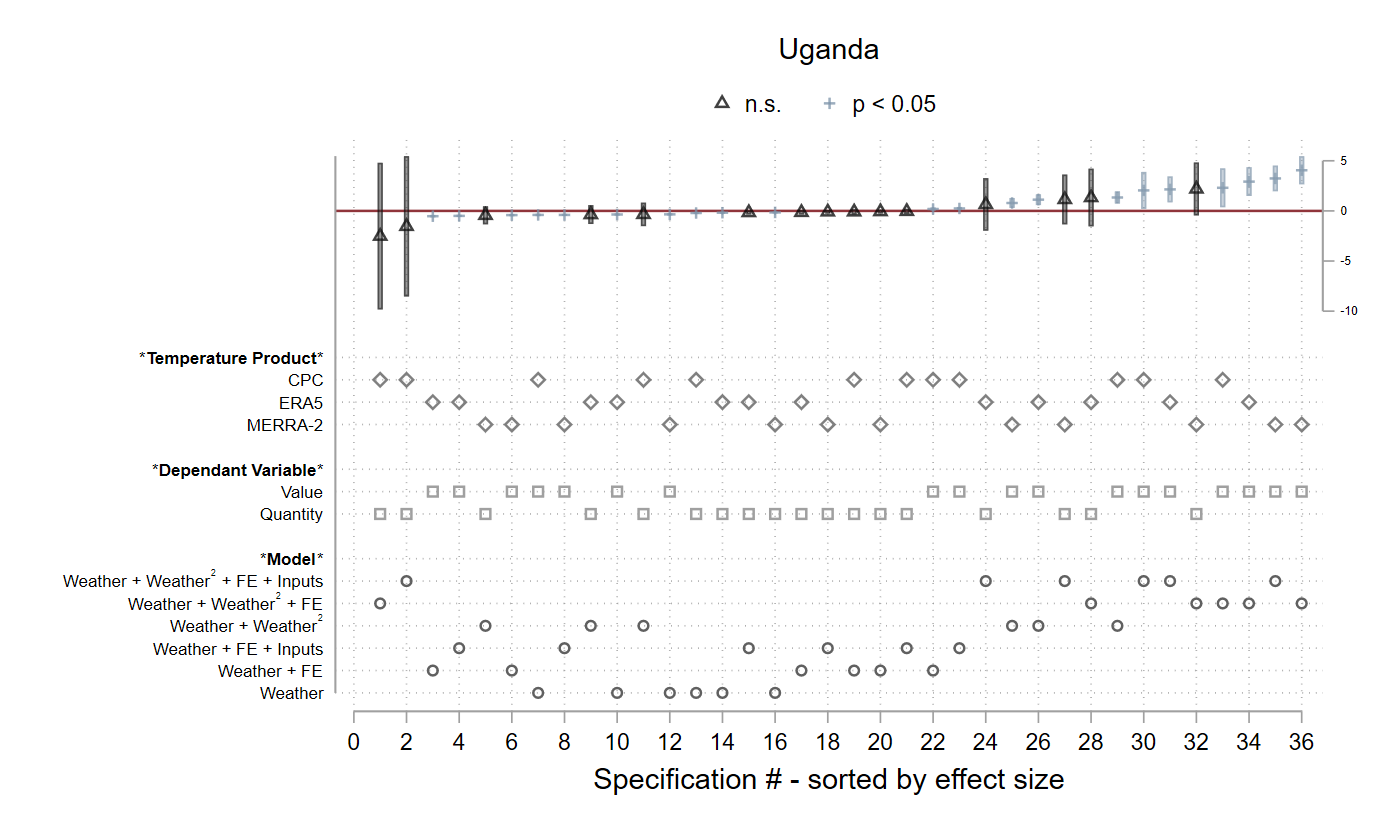}
		\end{center}
		\footnotesize  \textit{Note}: The figure presents specification curves for median daily rainfall, where each country is represented in a different panel. Each panel includes 36 regressions, where each column represents a single regression. Significant and non-significant coefficients are designated at the top of each panel.  
	\end{minipage}	
\end{figure}

\begin{figure}[!htbp]
	\begin{minipage}{\linewidth}		
		\caption{Specification Curve for Variance of Daily Temperature by Country}
		\label{fig:v17_sat}
		\begin{center}
			\includegraphics[width=.49\linewidth,keepaspectratio]{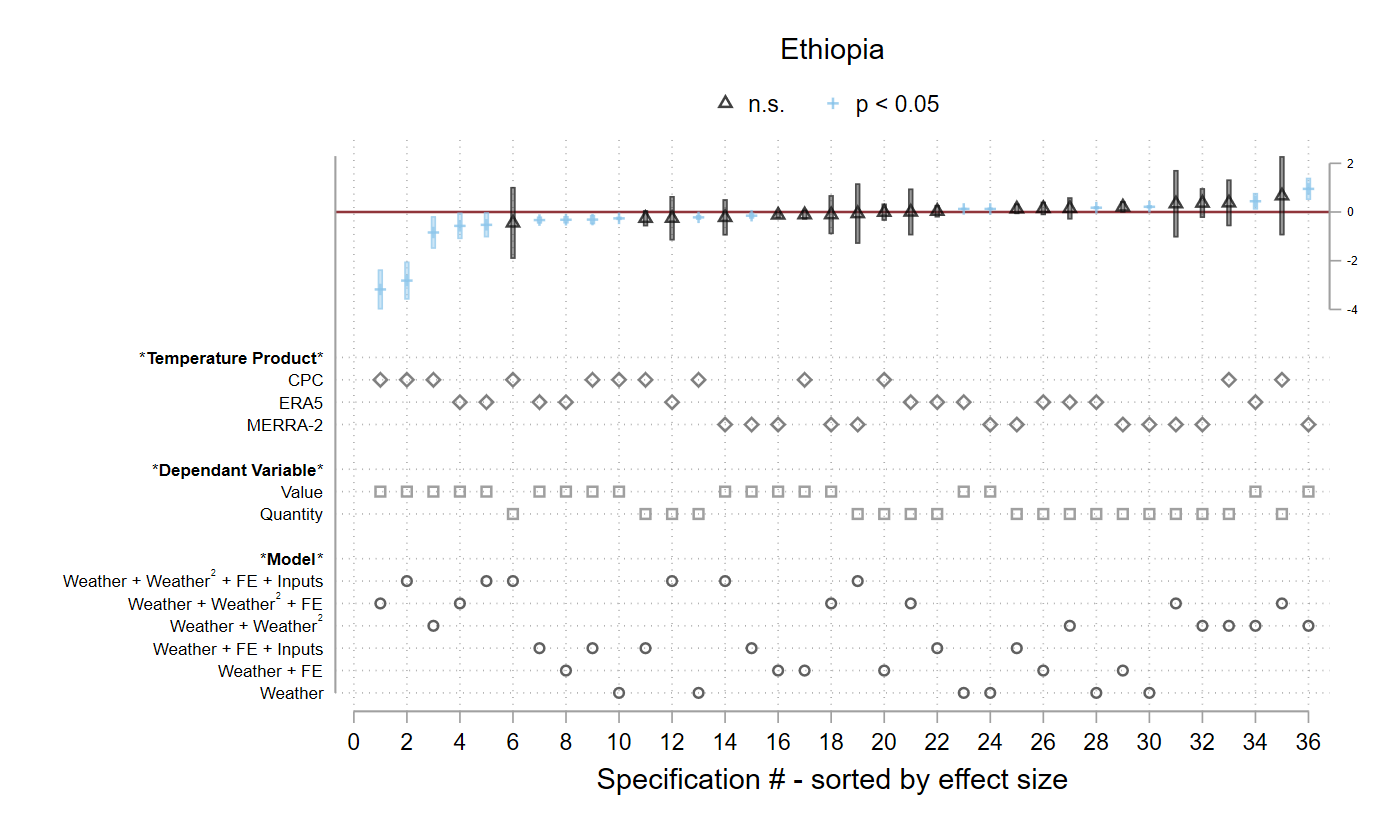}
			\includegraphics[width=.49\linewidth,keepaspectratio]{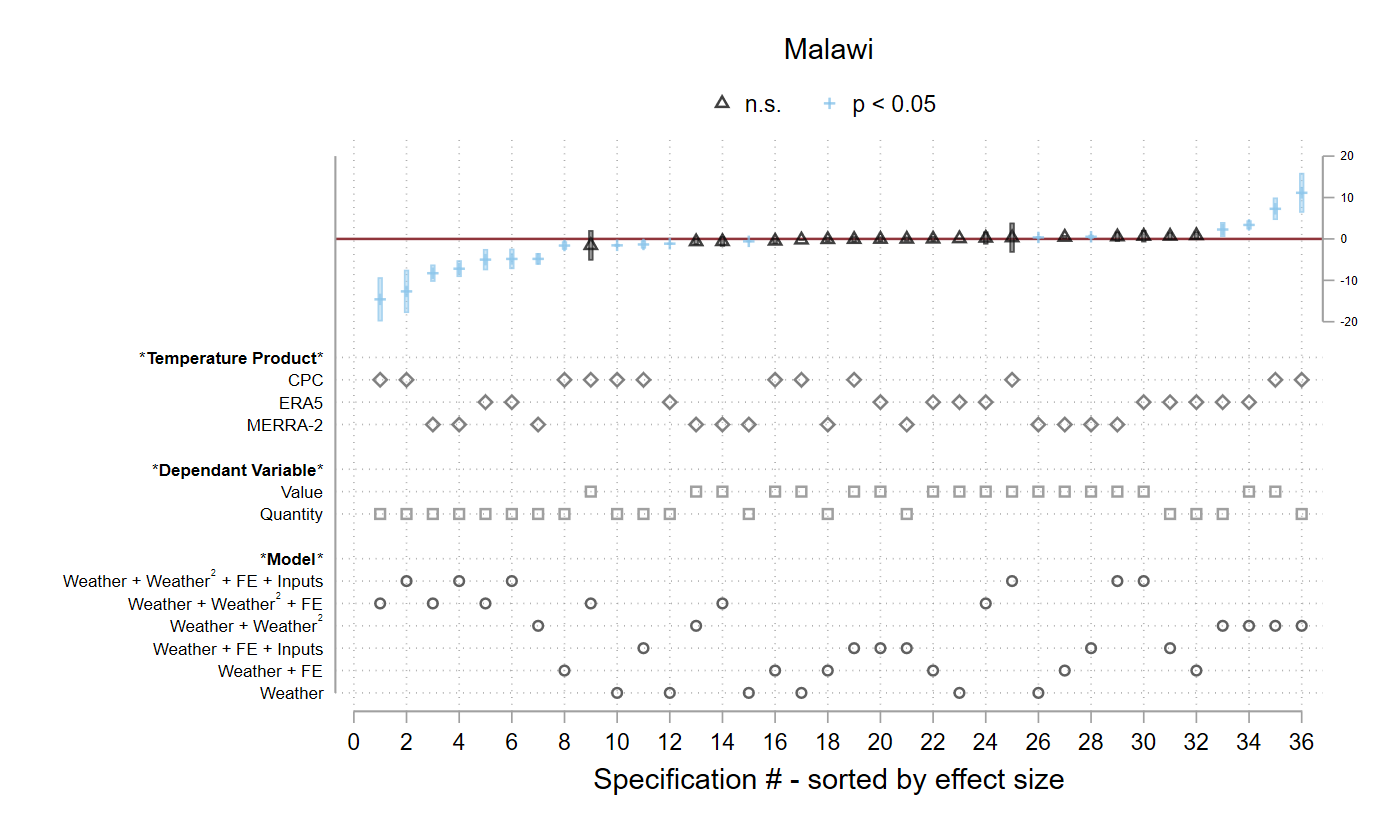}
			\includegraphics[width=.49\linewidth,keepaspectratio]{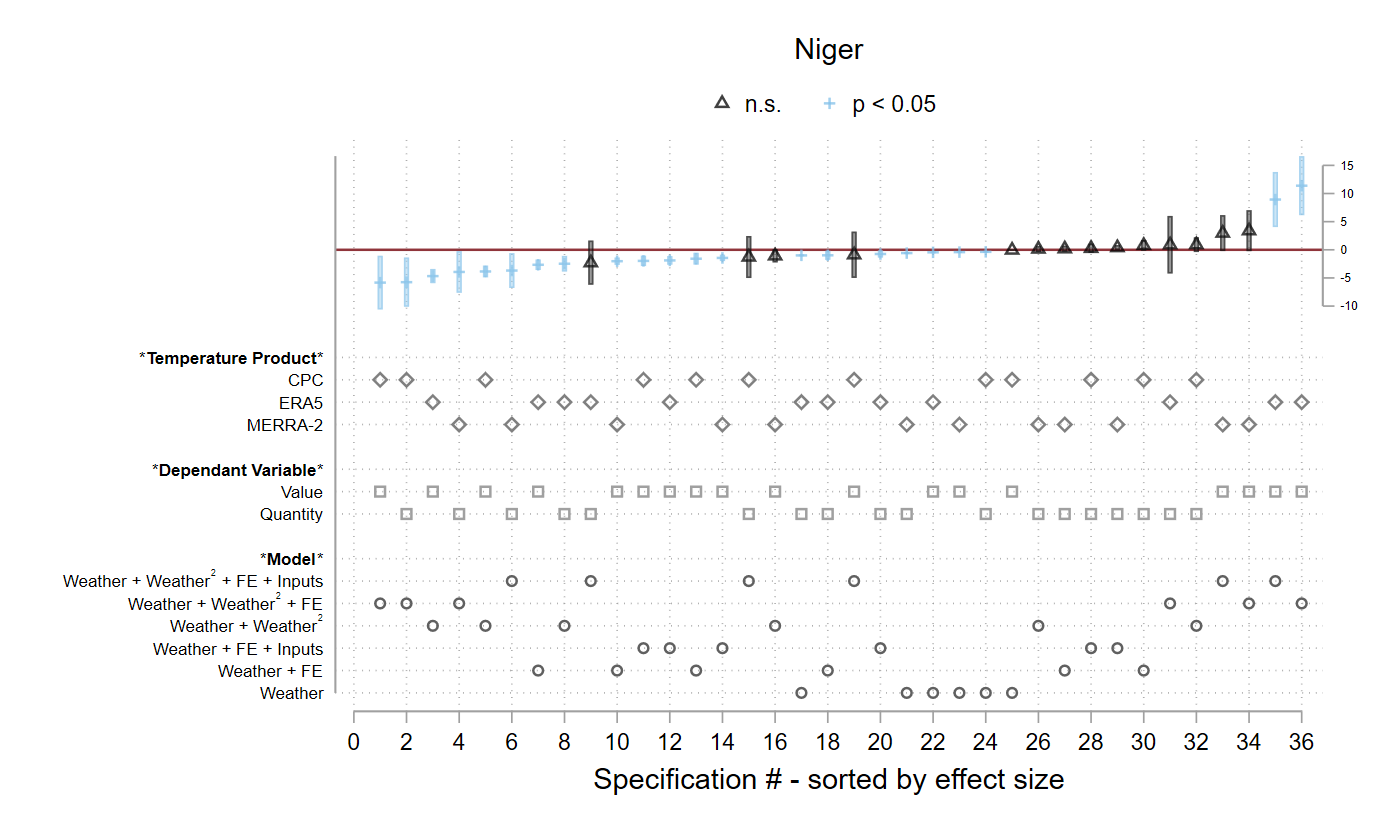}
			\includegraphics[width=.49\linewidth,keepaspectratio]{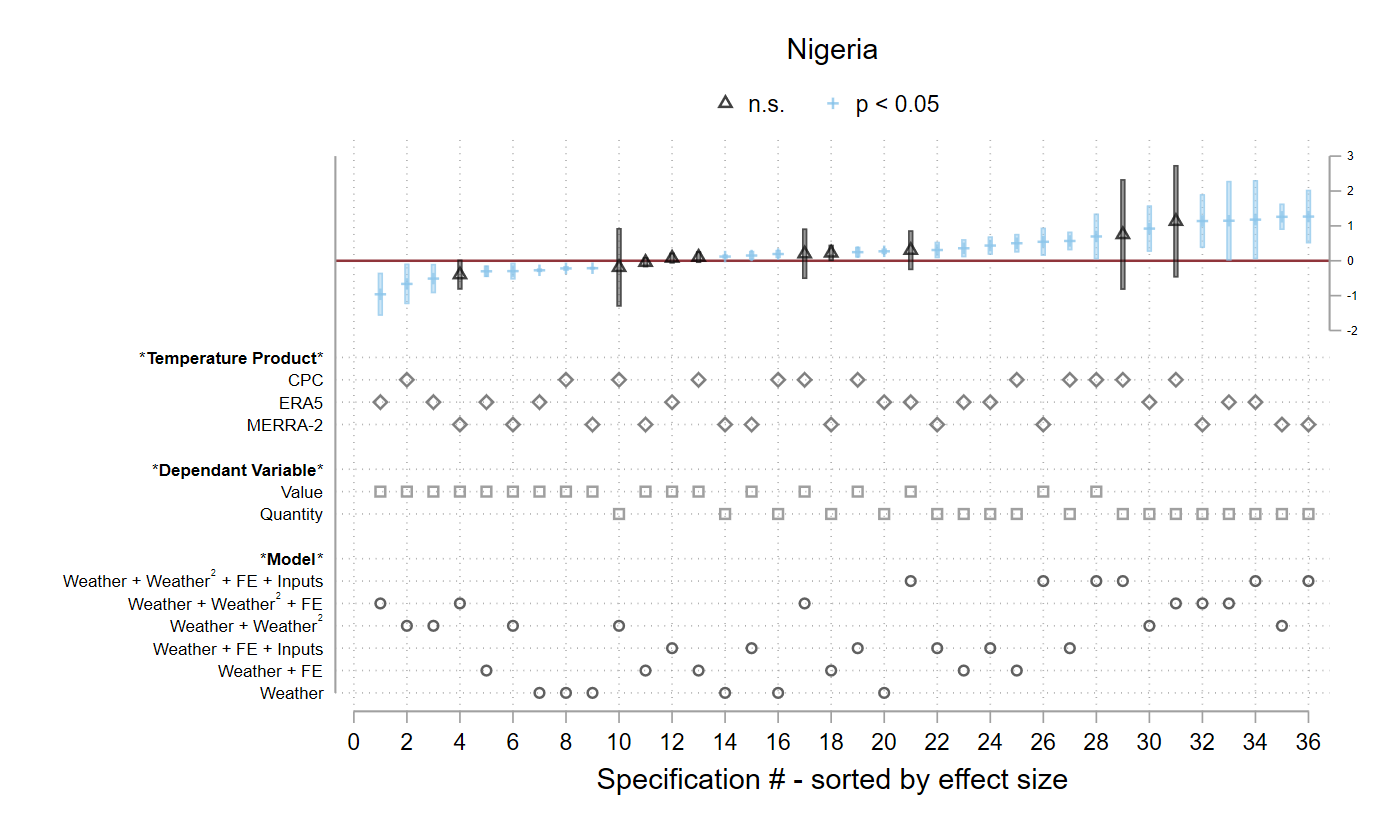}
			\includegraphics[width=.49\linewidth,keepaspectratio]{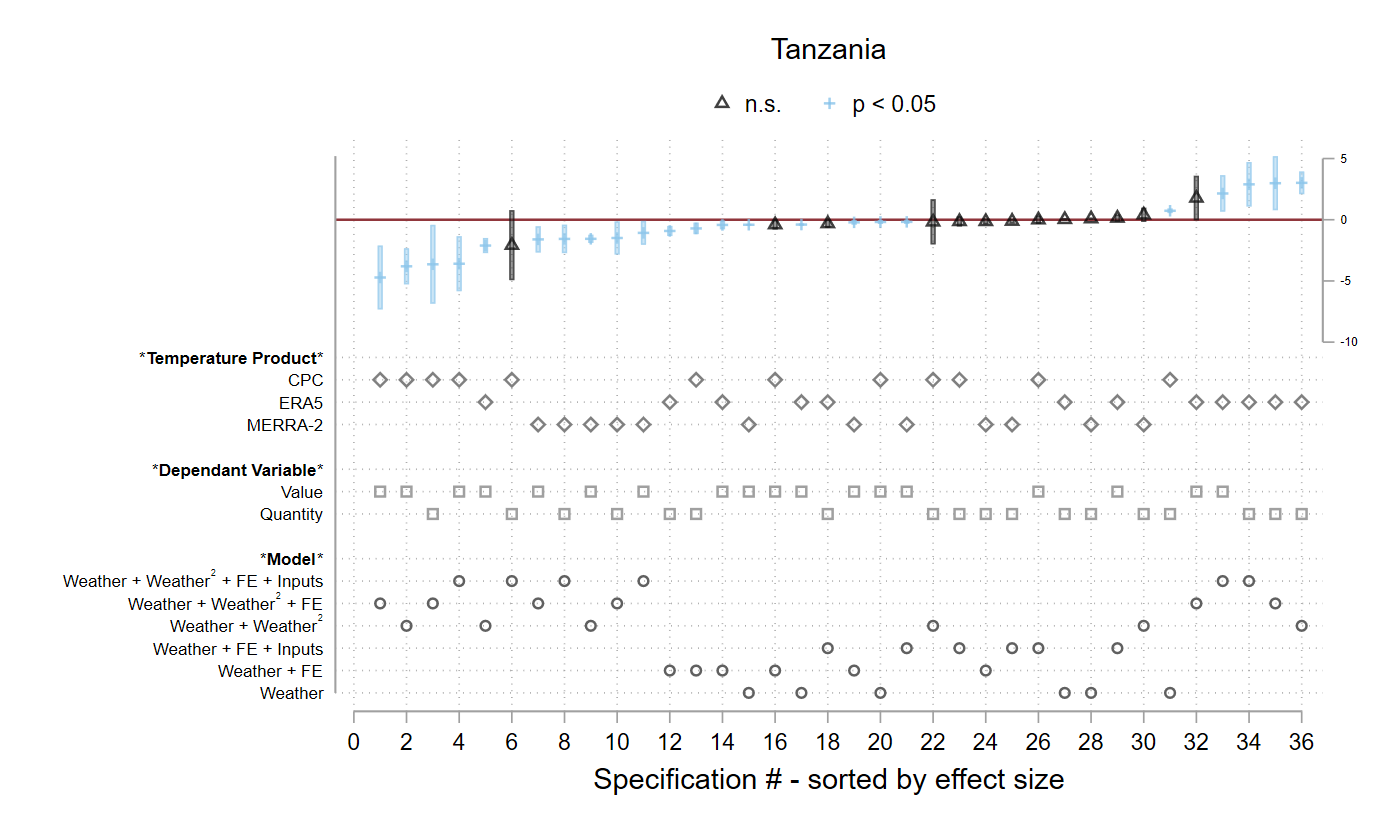}
			\includegraphics[width=.49\linewidth,keepaspectratio]{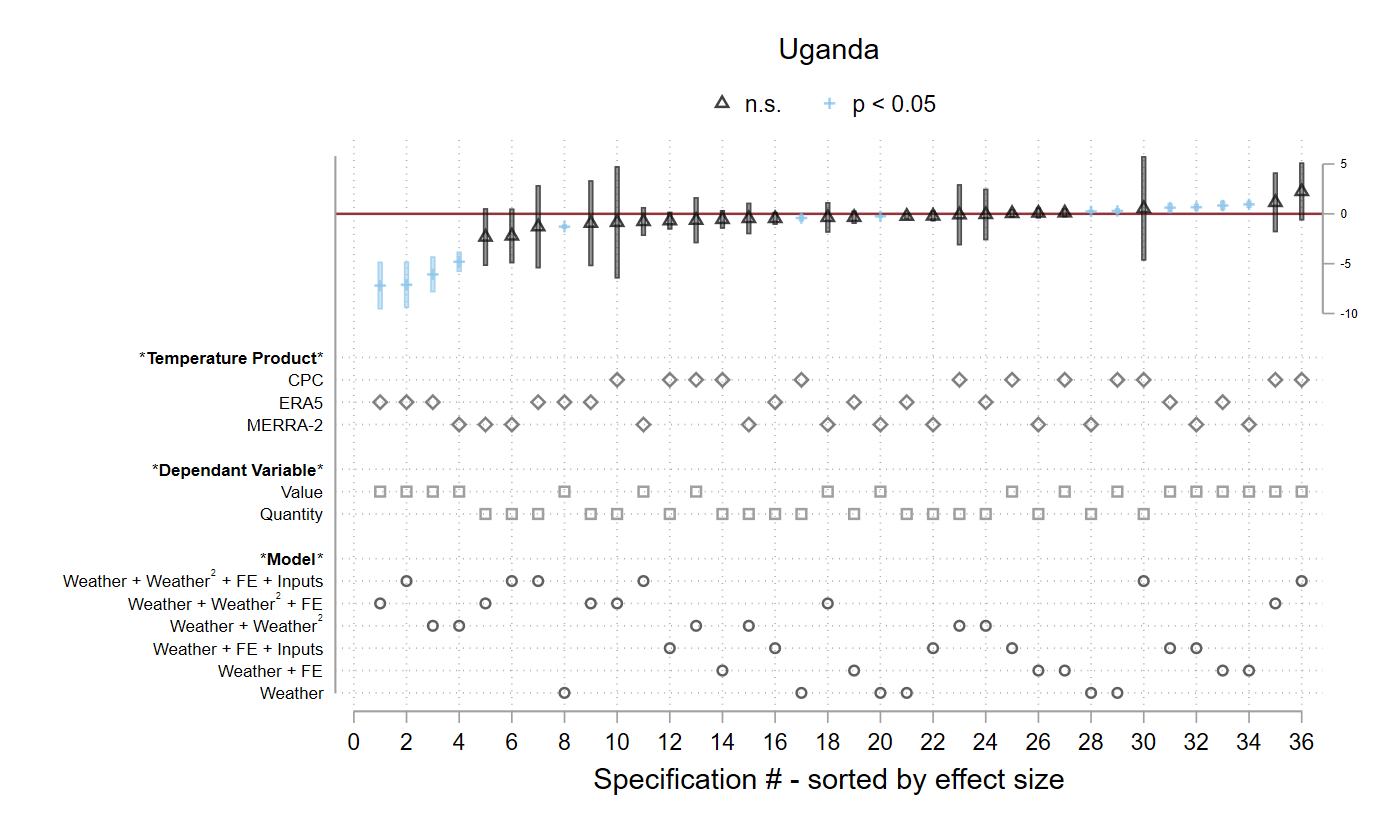}
		\end{center}
		\footnotesize  \textit{Note}: The figure presents specification curves for variance of daily temperature, where each country is represented in a different panel. Each panel includes 36 regressions, where each column represents a single regression. Significant and non-significant coefficients are designated at the top of each panel. 
	\end{minipage}	
\end{figure}


\clearpage
\newpage
\appendix
\onehalfspacing

\begin{center}
	\section*{Appendix to ``Estimating the Impact of Weather on Agriculture''} \label{sec:app}
\end{center}


\section{Details on Remote Sensing Weather Data} \label{sec:appRS}

\setcounter{table}{0}
\renewcommand{\thetable}{A\arabic{table}}
\setcounter{figure}{0}
\renewcommand{\thefigure}{A\arabic{figure}}

\subsection{On Using Remote Sensing Products for Economics} \label{sec:appRS_sat}

Uncertainty is present in all model outputs, and weather data sets are no exception. Spatial data sets of weather variables, like precipitation and temperature, that are produced using remotely sensed data, are not direct measurements of the variable of interest. Satellite sensors provide spatially continuous observation of reflectance from the earth's surface in different parts of the magnetic spectrum. These values are used to estimate related phenomena, such as cloud presence, cloud top temperature or earth surface temperature. The continuous data sets are then used in combination with directly observed, but often sparsely distributed, gauge data to produce weather variables. Some inputs are common across products, but there are differences in other inputs as well as modeling techniques.

The type of analysis matters in assessing weather data sets for use in economic research. Is the goal to understand climate trends, capture characteristics of a particular agricultural season, or identify extreme weather events occurring in near real-time? This can help determine the relative importance of different data set characteristics, such as spatial detail, temporal frequency and length of record, with respect to the intended analysis. The data sets used in this analysis were constrained by certain minimum criteria, leading to elimination of some commonly used data sets, such as the product from the Center for Climatic Research at the University of Delaware. We also did not consider proprietary data sets, preferring to use sources currently in the public domain. Despite the exclusions imposed by our minimum criteria, the data sets summarized in Table~\ref{tab:weather} represent a range of spatial resolutions and model types commonly used by economists. Further details on the specifics of each remote sensing product are provided below with the goal of providing economists with direction to a data set that meets the requirements of their analysis.


\subsubsection{Africa Rainfall Climatology version 2 (ARC2)}

ARC2 is a merged gauge data and remote sensing product that provides daily rainfall outputs for the African continent. The data set, produced by the National Oceanic and Atmospheric Administration (NOAA) Climate Prediction Center (CPC) provides improvements over ARC1 and a longer length of record compared to the rainfall estimate (RFE), the operational data set of USAID's Famine Early Warning Systems Network (FEWSNET) program. Inputs are Global Telecommunications System (GTS) rain gauge data over Africa, geostationary Meteosat infrared (IR) imagery, and polar-orbiting microwave Special Sensor Microwave Image (SSM/I) and Advanced Microwave Sounding Unit (AMSU-B).

Validation efforts by \cite{ARC2} found that low reporting rates for some GTS stations degrades model performance in those regions. Other findings are a general tendency to underestimate rainfall, which is enhanced in areas of high relief or complex topography.

Data and technical documentation are available for download from \url{https://www.cpc.ncep.noaa.gov/products/international/data.shtml}.


\subsubsection{Climate Hazards group InfraRed Precipitation with Station data (CHIRPS)}

Like ARC2, the CHIRPS rainfall data set builds on established techniques for merging gauge and remote sensing data. Produced by the Climate Hazards Group at University of California, Santa Barbara this data set is designed for monitoring of drought and environmental change at a global level. To minimize latency, there are two products, a preliminary version with two day lag, and final output available at three weeks. Outputs are available at time-steps from six hours to three months. As inputs, CHIRPS makes use of a monthly climatology CHPclim, Tropical Rainfall Measuring Mission Multi-satellite Precipitation Analysis version 7 (TMPA 3B42 v7) and global Thermal Infrared Cold Cloud Duration (TIR CCD) from two NOAA archives. The remote sensing data are then merged with gauge data from five public archives, including the Global Historical Climatology Network (GHCN) and GTS, several private sources, and meteorological agencies. While targeted gauge data collection efforts resulted in a greater number of input stations for years prior to 2010, the number of stations going forward is more limited, particularly in Sub-Saharan Africa. Detailed metadata by country is available and may be a useful reference to determine if coverage for a region of interest is sufficient for the analysis.

Validation for select countries found that the climatology input CHPclim outperformed other climatology data sets in data sparse regions and complex terrain \citep{CHIRPS}. Furthermore, in an assessment of wet season statistics CHIRPS showed less bias than other rainfall sources and good correspondence with Global Precipitation Climatology Centre (GPCC) estimates. 

Data and technical documentation are available for download from \url{https://data.chc.ucsb.edu/products/CHIRPS-2.0/}.


\subsubsection{CPC Global Unified Gauge-Based Analysis of Daily Precipitation and Temperature}

NOAA's Climate Prediction Center (CPC) Unified Gauge-based (CPC-U) data sets for daily temperature and precipitation do not incorporate remote sensing data in the estimation of weather variables. Instead, an optimal interpolation (OI) technique is used on gauge data for precipitation, and Shepard's algorithm for temperature. CPC-U provides systematic global data sets for validation and climate monitoring. GTS is a primary input data source, with some national collections, but density is most sparse over Africa.

As to be expected, even though the OI interpolation performs better than other techniques, a cross-validation exercise shows performance to degrade significantly with increasing distance to nearest station \citep{CPC}. As a result, this data set may not be suitable for analysis in some parts of Africa, with high spatial variation and low density of stations.

Data and technical documentation are available for download from \url{https://psl.noaa.gov/data/gridded/data.cpc.globalprecip.html} for precipitation and
\url{https://psl.noaa.gov/data/gridded/data.cpc.globaltemp.html} for temperature.


\subsubsection{European Centre for Medium-Range Weather Forecasts (ECMWF) ERA5}

ERA5, based on the European forecasting model ECMWF, is one of two assimilation model data sets used in this paper. The inputs are far too numerous to mention but include a range of satellite inputs as well as gauge data sets. There are a wide range of outputs as well, including 2-meter air temperature and rainfall, available at sub-daily intervals and differentiated vertically. ERA5 is coarser spatial resolution than the global and regional merged rainfall data sets, but more detailed than MERRA2. 

The sheer number and complexity of outputs can be a deterrent to the use of weather variables from assimilation models. Uncertainty or lack of understanding about inaccuracies associated with individual output variables of assimilation models, compared to other types of models, is another reason to carefully consider their suitability for particular research \citep{Parker16}. Nevertheless, reanalysis data sets are used in a broad range of applications and even outperform other gridded climate data sets in some settings \citep{ZandlerEtAl20}.

Data and technical documentation are available for download from \url{https://cds.climate.copernicus.eu}.


\subsubsection{Modern-Era Retrospective analysis for Research and Applications, version 2 (MERRA-2)}

The second reanalysis data set used in this analysis is MERRA-2, a product of NASA's Goddard Earth Observing System, version 5 (GEOS-5) assimilation model. Specifically we make use of the variables T2MMEAN from the statD daily statistics collection, and PRECTOTLAND from the Land Surface Diagnostics collection. 
  
Data and technical documentation are available for download from \url{https://disc.gsfc.nasa.gov/}.


\subsubsection{Tropical Applications of Meteorology using SATellite data (TAMSAT)}

The TAMSAT rainfall data set is the highest spatial resolution gridded data set used in this analysis. Inputs are similar to other merged gauge and remote sensing products: Meteosat TIR imagery, purposefully collected archival (1983-2010) rain gauge data from meteorological agencies and other sources and GTS gauge data. Rainfall estimation is based on cold cloud duration (CCD) inferred from TIR and calibrated using gauge data within discrete calibration zones.

Validation of TAMSAT found a mean underestimation of rainfall of approximately four millimeters per dekad, though the bias was not always negative \citep{TAMSAT}. Due to differences in methodology from CHIRPS and ARC2 precipitation products, TAMSAT is not affected by inconsistency in gauge data inputs. This makes it suitable for placing rainfall variability in the context of a long-term climatology and thus detecting unusually wet or dry conditions.

Data and technical documentation are available for download from
\url{http://www.tamsat.org.uk/data/}.


\subsection{Defining Growing Season} \label{sec:appRS_gs}

We define growing season following the FAO \href{http://www.fao.org/agriculture/seed/cropcalendar/welcome.do}{crop calendar} for each country. Table~\ref{tab:growseason} presents details for each country on the growing season used, as well as whether that season spans years and whether it is unimodal or bimodal. Remote sensing data used in our analysis follows the defined growing season in each respective country.

Of the six countries, two (Malawi and Tanzania) span calendar years, which means that the growing season begins in one year and stretches into the year that follows. Take, for example, Malawi. The growing season in that country begins on 1 October and ends on 30 April. This means that it would begin 1 October 2021 and would end 30 April 2022. 

Similarly, of the six countries, two (Nigeria and Uganda) are bimodal. The season modality designates whether different regions within the countries have different growing seasons. In both Nigeria and Uganda, the northern part of the country has a different growing season from the southern part of the country. In these cases we designate the modality of the season, and also provide the growing season dates for both regions. 


\newpage
\begin{table}[htbp]	\centering
	\caption{Growing Seasons} \label{tab:growseason}
	\scalebox{0.9}
	{ \setlength{\linewidth}{.1cm}\newcommand{\contents}
		{\begin{tabular}{llll}
			\\[-1.8ex]\hline 
			\hline \\[-1.8ex]
			& \multicolumn{1}{c}{Growing Season} & \multicolumn{1}{c}{Span Calendar Years} & \multicolumn{1}{c}{Season Modality}  \\
			\multicolumn{1}{l}{\href{http://www.fao.org/giews/countrybrief/country.jsp?code=ETH}{Ethiopia}} & \multicolumn{1}{l}{1 March - 30 November} & \multicolumn{1}{c}{no} & \multicolumn{1}{c}{unimodal}  \\
			\multicolumn{1}{l}{\href{http://www.fao.org/giews/countrybrief/country.jsp?code=MWI}{Malawi}} & \multicolumn{1}{l}{1 October - 30 April} & \multicolumn{1}{c}{yes} & \multicolumn{1}{c}{unimodal}  \\
			\multicolumn{1}{l}{\href{http://www.fao.org/giews/countrybrief/country.jsp?code=NER}{Niger}} & \multicolumn{1}{l}{1 June - 30 November} & \multicolumn{1}{c}{no} & \multicolumn{1}{c}{unimodal}  \\
			\multicolumn{1}{l}{\href{http://www.fao.org/giews/countrybrief/country.jsp?code=NGA}{Nigeria}} & \multicolumn{1}{l}{\emph{North}: 1 May - 30 September} & \multicolumn{1}{c}{no} & \multicolumn{1}{c}{bimodal}  \\
			 & \multicolumn{1}{l}{\emph{South}: 1 March - 31 August} &  &  \\
			\multicolumn{1}{l}{\href{http://www.fao.org/giews/countrybrief/country.jsp?code=TZA}{Tanzania}} & \multicolumn{1}{l}{1 November - 30 April} & \multicolumn{1}{c}{yes} & \multicolumn{1}{c}{unimodal}  \\
			\multicolumn{1}{l}{\href{http://www.fao.org/giews/countrybrief/country.jsp?code=UGA}{Uganda}} & \multicolumn{1}{l}{\emph{North}: 1 April - 30 September}  & \multicolumn{1}{c}{no} & \multicolumn{1}{c}{bimodal} \\
	    	& \multicolumn{1}{l}{\emph{South}: 1 February - 31 July} & &  \\
			\\[-1.8ex]\hline 
			\hline \\[-1.8ex]
			\multicolumn{4}{p{\linewidth}}{\footnotesize  \textit{Note}: \footnotesize The table presents the growing season ranges, as defined by following FAO \href{http://www.fao.org/agriculture/seed/cropcalendar/welcome.do}{crop calendar} for each country, respectively.} \\
		\end{tabular}}
	\setbox0=\hbox{\contents}
    \setlength{\linewidth}{\wd0-2\tabcolsep-.25em}
    \contents}
\end{table}


\clearpage
\newpage
\section{Details on Household Data from the LSMS-ISA} \label{sec:appHH}

The World Bank Living Standards Measurement Study - Integrated Surveys on Agriculture (LSMS-ISA) is a household survey program that provides financial and technical assistance to national statistical offices in Sub-Saharan Africa for the design and implementation of national, multi-topic longitudinal household surveys with a focus on agriculture. The LSMS-ISA-supported countries include Burkina Faso, Ethiopia, Malawi, Mali, Niger, Nigeria, Uganda and Tanzania. We use the data sets from Ethiopia, Malawi, Niger, Nigeria, Uganda, and Tanzania in this work.\footnote{We intend to extend our analysis to include Mali. We do not intend to include Burkina Faso, due to issues with geo-reference locations which make its use incompatible with the project methodology.} More details on each country are included in the following sub-sections and details on samples are provided in Table~\ref{tab:lsms}. 

A common feature of the LSMS-ISA-supported surveys is that each sample household receives a multi-topic Household Questionnaire that elicit comprehensive socioeconomic information that also allows for the construction of consumption and income aggregates. Households engaged in agricultural activities additionally receive an Agriculture Questionnaire that elicits comprehensive information on smallholder crop, livestock and fishery activities and that allows for the construction of plot-level indicators of land and labor productivity and input use, among others. Finally, while the key variables that drive each survey's sampling design is household consumption and income, each survey provides a large sample of agricultural households in each round.

In our analysis, we only include households which did not move. Although the LSMS-ISA surveys follow individuals who ``split off'' and create new households, we do not include these movers in our analysis. 

\subsection{Ethiopia}

The LSMS-ISA data from Ethiopia includes three waves. Wave 1 (2011/12) includes 4,000 households in rural and small towns across the country \citepalias{ETH1}. This initial sample was followed in 2013/14 and 2015/16 \citepalias{ETH2, ETH3}. Beginning in Wave 2 (2013/14) the survey was also expanded to include 1,500 households in urban areas.

The Wave 1 data is representative at the regional level for the most populous regions (Amhara, Oromiya, Southern Nations, Nationalities, and People's Region, and Tigray). In Wave 2, in order to align with the existing Wave 1 design while ensuring that all urban areas were included, the population frame was stratified to provide population inferences for the same five domains as in Wave 1 as well as an additional domain for the city state of Addis Ababa. However, the sample size in both waves, is not sufficient to support region-specific estimates for each of the small regions (Afar, Benshangul Gumuz, Dire Dawa, Gambella, Harari, and Somalie). 

\subsection{Malawi}

The LSMS-ISA data from Malawi includes two separate surveys: (1) Integrated Household Survey, from which we include the first wave and (2) Integrated Household Panel Survey which includes three waves \citepalias{MWI1, MWI2, MWI3}. The two surveys are different in their representation of various households within the country. In this analysis, we rely only on the Integrated Household Panel Survey. 

The Integrated Household Panel Survey begins with Wave 1 in 2010 and includes 3,247 households from 204 enumeration areas that were visited as part of the Third Integrated Household Survey 2010/11 and that were designated as ``panel'' for follow-up, starting again in 2013. The sample was designed to be representative at the national-, urban/rural-, and regional-level at baseline. Wave 2 from 2013 aimed to track all panel households from Wave 1, including all individuals that changed locations between the waves. The Wave 2 household sample size was 4,000, including new households that were formed by split off individuals that were tracked. Finally, Wave 3 from 2016 aimed to track all households and split off individuals that were ever associated with a random half of 204 original enumeration areas that had been visited in 2010. The Wave 3 household sample was 2,500 households, including again new households that were formed by split off individuals that were tracked from previous rounds.




\subsection{Niger}

The LSMS-ISA data from Niger includes two rounds. In Wave 1. approximately 4,000 households in 270 Zones de Dénombrement \citepalias{NGR1}. The sample is nationally representative, as well as representative of Niamey, other urban, and rural areas. Households visited in Wave 1 were re-visited in Wave 2, including households and individuals who moved after the 2011 survey \citepalias{NGR2}. When the entire household moved within Niger, the household was found and re-interviewed in the second wave. When individuals from the household moved, one individual per household was selected to follow. This forms a sample of approximately 3,600 households in Wave 2. 

\subsection{Nigeria}

The LSMS-ISA data from Nigeria includes three waves \citepalias{NGA1, NGA2, NGA3}. The total sample consists of 5,000 panel households and is representative at the national level. Households are visited twice per wave of the Panel, both post-planting and post-harvest. The post-harvest visit is implemented jointly with a larger General Household Survey of 22,000 households (5,000 panel and 17,000 non-panel households). The sample is representative at the national level and provides reliable estimates of key socio-economic variables for the six zones in the country.

\subsection{Tanzania}

Three waves of the LSMS-ISA data from Tanzania are included in our analysis. The first wave includes 3,265 households and the sample is representative for the nation, and provides reliable estimates of key socioeconomic variables for mainland rural areas, Dar es Salaam, other mainland urban areas, and Zanzibar \citepalias{TZA1}. In Wave 2, all original households were targeted for revisit \citepalias{TZA2}. For those household members still residing in their original location, they were simply re-interviewed. For adults who had relocated, these individuals were tracked and re-interviewed in their new location with their new households. As a result of this, the sample size for the second round expanded to 3,924 households. Wave 3 adhered to the same tracking protocol as Wave 2, resulting in a final sample size of 5,015 households \citepalias{TZA3}.

\subsection{Uganda}

The LSMS-ISA from Uganda includes five waves, of which we use three in this analysis. Wave 1 (2009/10) includes approximately 3,200 households that were previously interviewed by the Uganda National Household Survey (UNHS) in 2005/06  \citepalias{UGA1}. The sample was designed to be representative at the national-, urban/rural- and regional-level. For subsequent waves, the Wave 1 sample was followed, including tracking of shifted and split-off households, for two additional rounds: 2010/11 and 2011/12 \citepalias{UGA2, UGA3}. Each round includes nearly 3,000 households. 

\setcounter{table}{0}
\renewcommand{\thetable}{A\arabic{table}}
\setcounter{figure}{0}
\renewcommand{\thefigure}{A\arabic{figure}}


\clearpage
\newpage
\section{Extraction Methods}\label{sec:apextraction}

\setcounter{table}{0}
\renewcommand{\thetable}{C\arabic{table}}
\setcounter{figure}{0}
\renewcommand{\thefigure}{C\arabic{figure}}

Extending section~\ref{sec:ext}, in this section, we present further evidence on Hypothesis $1$ ($H_0^1$ - different obfuscation procedures implemented to preserve privacy of farms or households have no impact on estimates of agricultural productivity). The following figures (Figures~\ref{fig:ext_moment_rf} through \ref{fig:ext_total_tp}) pool the results from 77,760 regressions and then divide the pool into ten bins, one for each extraction method for 7,776 regression results. Extending the results presented in the main text, these figures examine the differences in coefficients $(\beta_{1})$ and their relative significance by extraction method. 

Reviewing the coefficients and confidence intervals for rainfall and temperature measures, disaggregated by extraction methods, as presented in Figures~\ref{fig:ext_moment_rf} through \ref{fig:ext_total_tp} overwhelmingly supports the conclusions of section~\ref{sec:ext}, specifically that we cannot reject the null for Hypothesis 1. 

\subsection{Selecting a Single Extraction Method}

Prior to the unblinding of the Data Analysis Group, all analysis was conducted using a randomly selected extraction method. The following text is what was contained in section 4.1.3 ``Selecting a Single Extraction Method'' in version 2 of the paper:

Concluding that different extraction method do not produce substantially different results, we narrow the set of results used from this point forward. As that the Data Analysis Group is still blinded to the identity of each extraction method, we cannot simply select the ``true'' coordinates and proceed. Instead, we us a random number generator in Stata to select one extraction method to continue the analysis with. Once the Data Analysis Group is unblinded, the analysis will be updated using the true coordinates. If our conclusion that extraction method does not introduce substantial mismeasurement, then the results and analysis using a randomly selected extraction method should be qualitatively the same as results and analysis using the true GPS coordinates.

The code for the random selection process is: 

\begin{lstlisting}[
    style = myListingStyle,
    caption = {}
    ]
* choose one extraction method at random
preserve
	clear			all
	set obs			1
	set seed		3317230
	gen double 		u = (10-1) * runiform() + 1
	gen 			i = round(u)
	sum		 	u i 
restore
\end{lstlisting}

\noindent Following the results of this random number generation, the findings presented from this point on only rely on extraction method. Time-stamped code for this process is available in the \href{https://github.com/jdavidm/weather_project}{Github repository}.


\newpage

\begin{figure}[!htbp]
	\begin{minipage}{\linewidth}		
		\caption{Coefficients and Confidence Intervals for (1) Daily Mean, (2) Daily Median, (3) Daily Variance, and (4) Daily Skew, for Rainfall by Extraction Method}
		\label{fig:ext_moment_rf}
		\begin{center}
			\includegraphics[width=\linewidth,keepaspectratio]{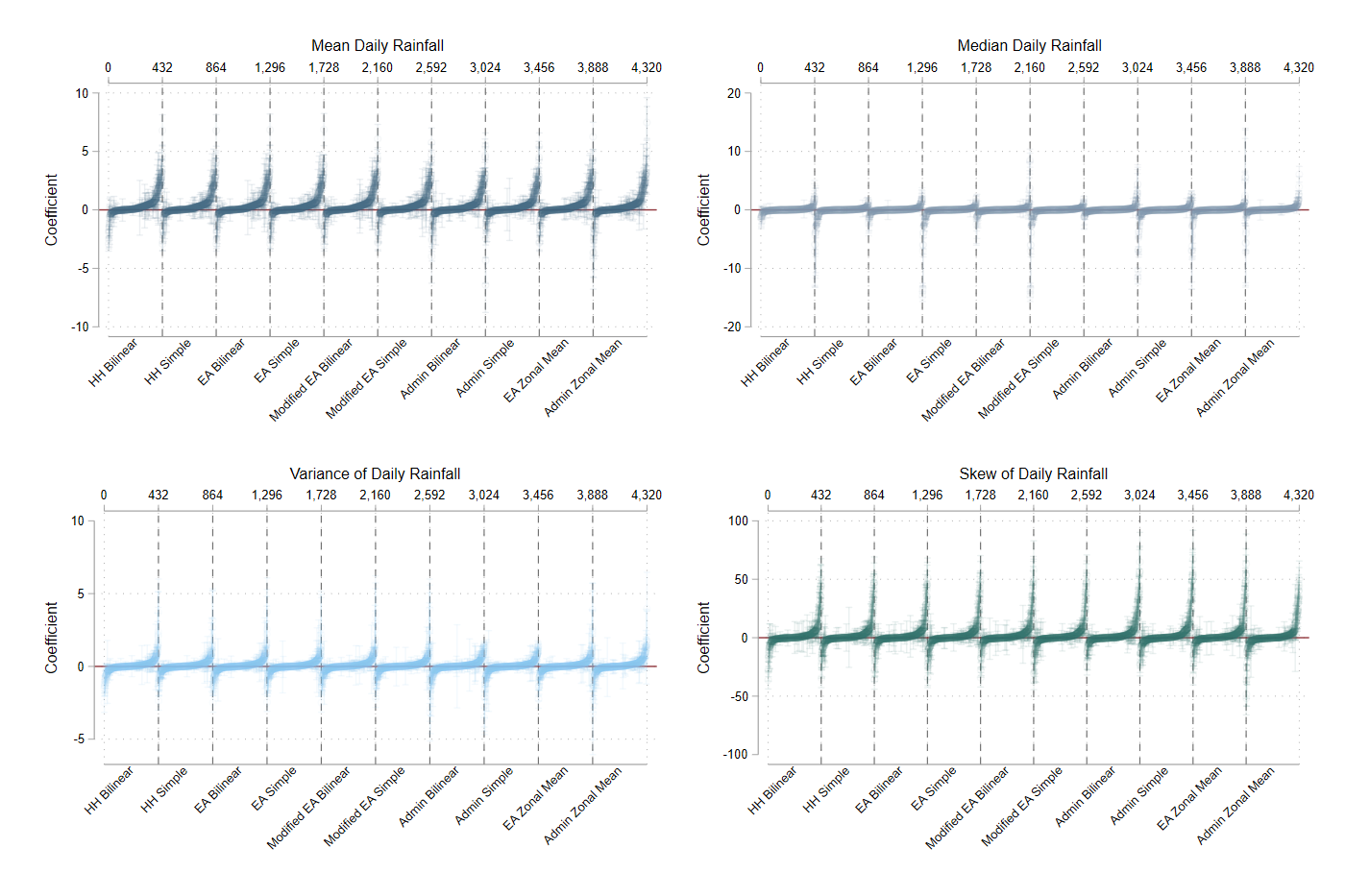}
		\end{center}
		\footnotesize  \textit{Note}: The figure presents the coefficients and confidence intervals for four rainfall metrics, by extraction method. These are aggregated across country, remote sensing source, outcome variable, and specification. Each panel includes 4,320 coefficients and confidence intervals (designated on the $x$-axis), with each bin (extraction method) including 432 coefficients and confidence intervals. Each column, as such, represents the findings of a single regressions, e.g. the coefficient and confidence interval itself. 
	\end{minipage}	
\end{figure}

\begin{figure}[!htbp]
	\begin{minipage}{\linewidth}		
		\caption{Coefficients and Confidence Intervals for (1) Total Season, (2) Deviation in Total Season, and (3) z-score of Total Season, for Rainfall by Extraction Method}
		\label{fig:ext_total_rf}
		\begin{center}
			\includegraphics[width=\linewidth,keepaspectratio]{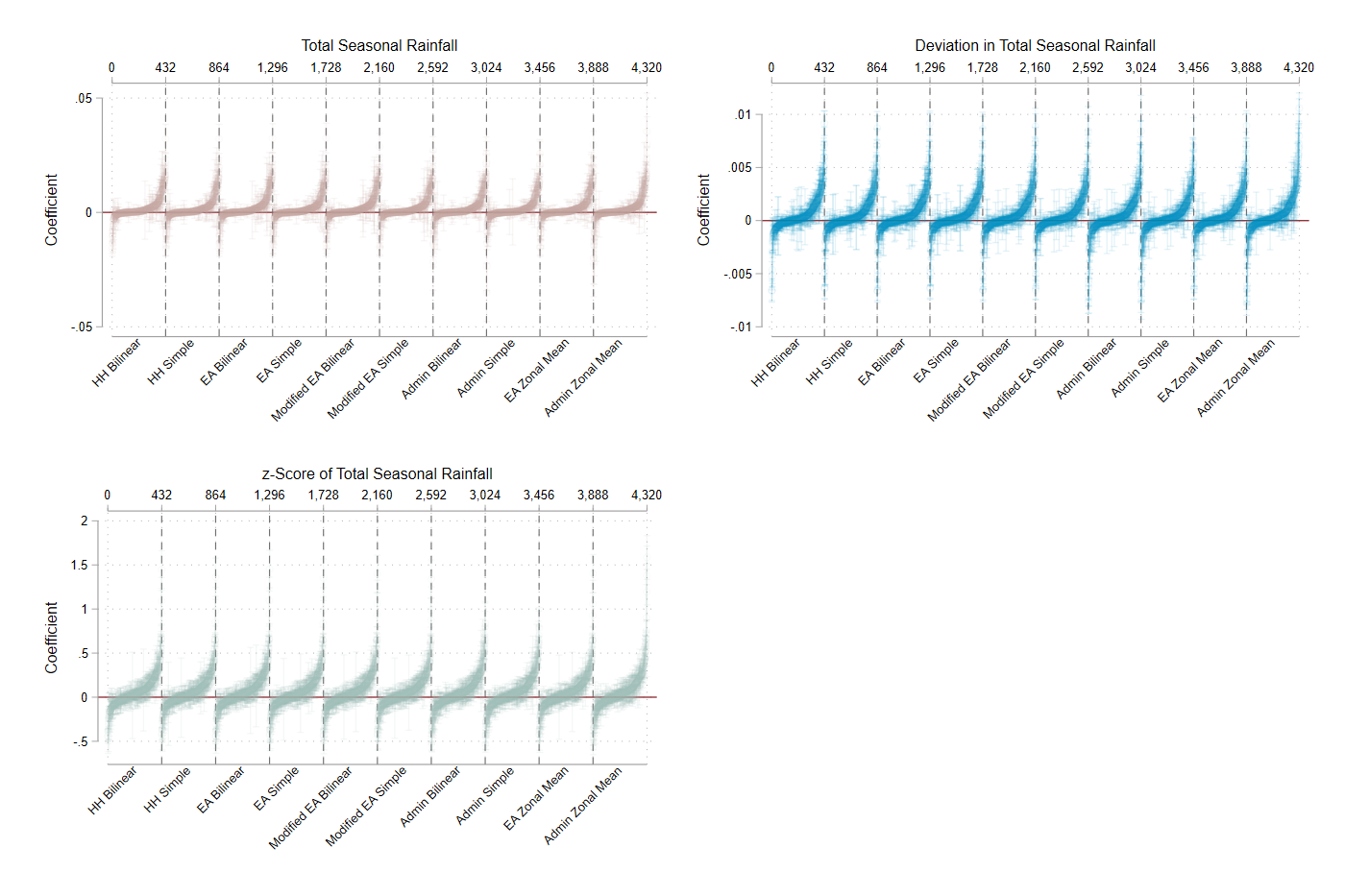}
		\end{center}
		\footnotesize  \textit{Note}: The figure presents the coefficients and confidence intervals for four rainfall metrics, by extraction method. These are aggregated across country, remote sensing source, outcome variable, and specification. Each panel includes 4,320 coefficients and confidence intervals (designated on the $x$-axis), with each bin (extraction method) including 432 coefficients and confidence intervals. Each column, as such, represents the findings of a single regressions, e.g. the coefficient and confidence interval itself.  
	\end{minipage}
\end{figure}

\begin{figure}[!htbp]
	\begin{minipage}{\linewidth}
		\caption{Coefficients and Confidence Intervals for (1) Number of Days with Rain, (2) Deviation in Number of Days with Rain, (3) Percentage of Days with Rain, and (4) Deviation in Percentage of Days with Rain, for Rainfall by Extraction Method}
		\label{fig:ext_rain_rf}
		\begin{center}
			\includegraphics[width=\linewidth,keepaspectratio]{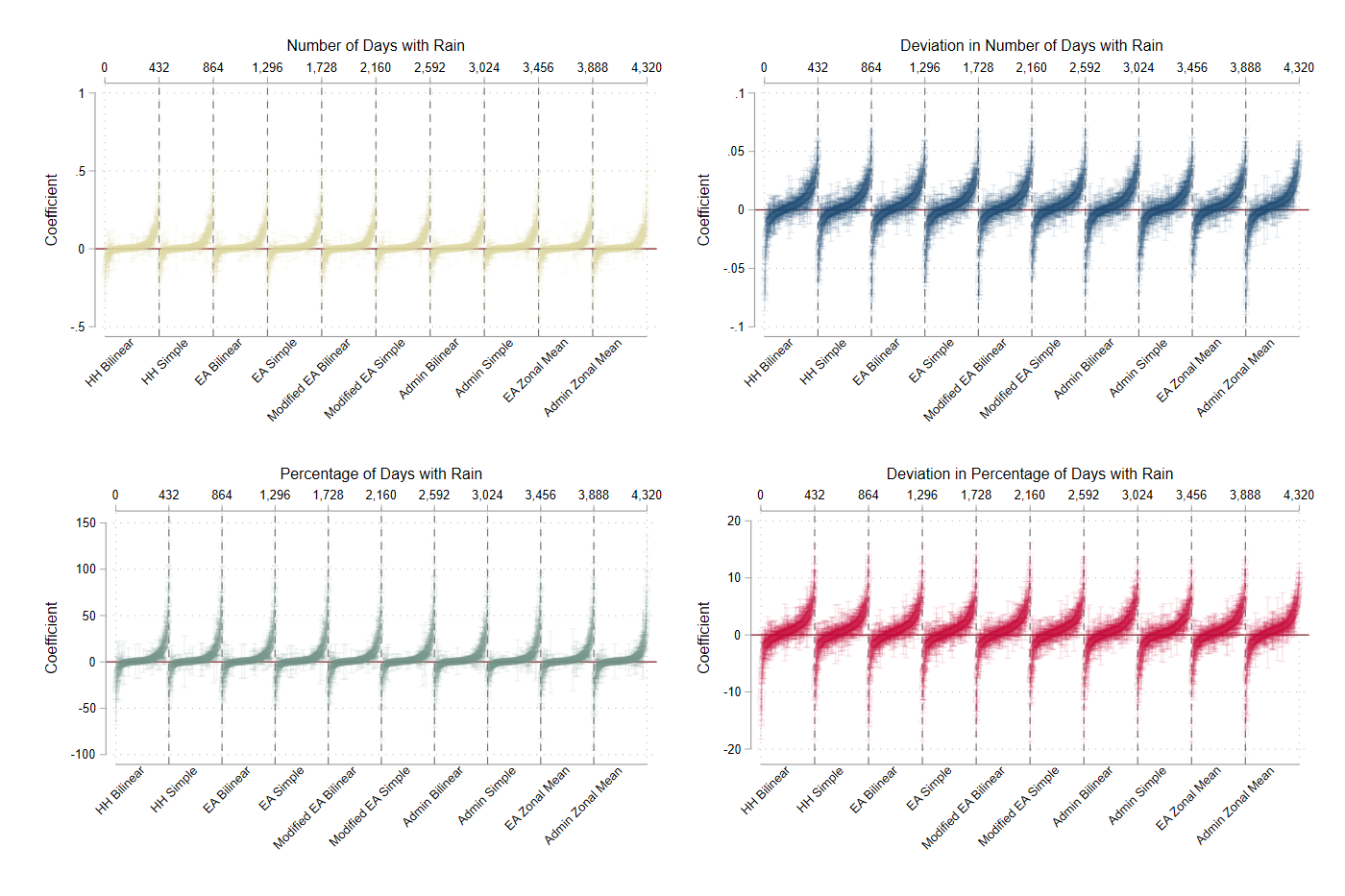}
		\end{center}
		\footnotesize  \textit{Note}: The figure presents the coefficients and confidence intervals for four rainfall metrics, by extraction method. These are aggregated across country, remote sensing source, outcome variable, and specification. Each panel includes 4,320 coefficients and confidence intervals (designated on the $x$-axis), with each bin (extraction method) including 432 coefficients and confidence intervals. Each column, as such, represents the findings of a single regressions, e.g. the coefficient and confidence interval itself. 
	\end{minipage}	
\end{figure}

\begin{figure}[!htbp]
	\begin{minipage}{\linewidth}		
		\caption{Coefficients and Confidence Intervals for (1) Number of Days without Rain, (2) Deviation in Number of Days without Rain, and (3) Longest Dry Spell, for Rainfall by Extraction Method}
		\label{fig:ext_none_rf}
		\begin{center}
			\includegraphics[width=\linewidth,keepaspectratio]{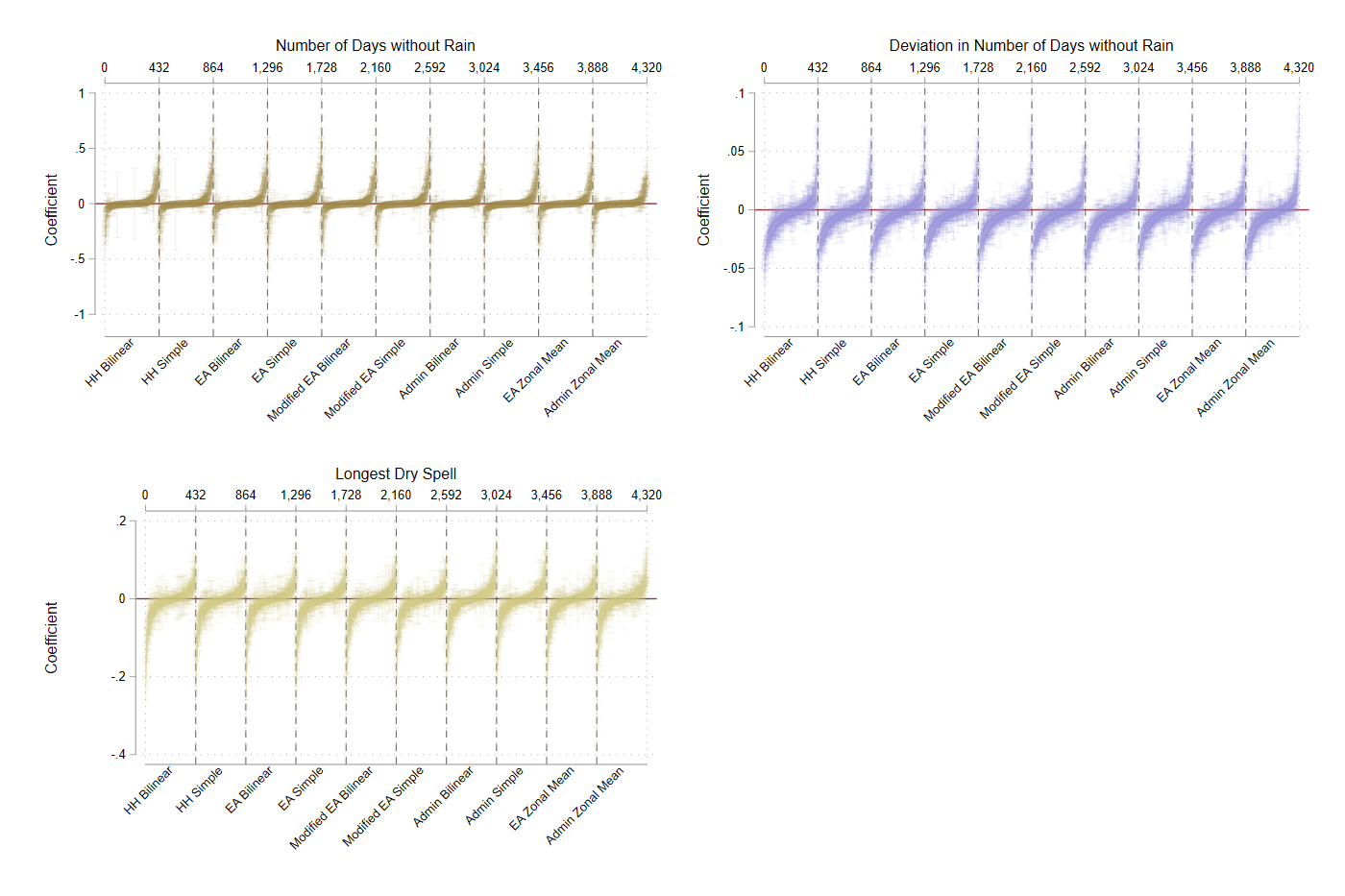}
		\end{center}
		\footnotesize  \textit{Note}: The figure presents the coefficients and confidence intervals for four rainfall metrics, by extraction method. These are aggregated across country, remote sensing source, outcome variable, and specification. Each panel includes 4,320 coefficients and confidence intervals (designated on the $x$-axis), with each bin (extraction method) including 432 coefficients and confidence intervals. Each column, as such, represents the findings of a single regressions, e.g. the coefficient and confidence interval itself. 
	\end{minipage}	
\end{figure}

\begin{figure}[!htbp]
	\begin{minipage}{\linewidth}		
		\caption{Coefficients and Confidence Intervals for (1) Daily Mean, (2) Daily Median, (3) Daily Variance, and (4) Daily Skew, for Temperature by Extraction Method}
		\label{fig:ext_moment_tp}
		\begin{center}
			\includegraphics[width=\linewidth,keepaspectratio]{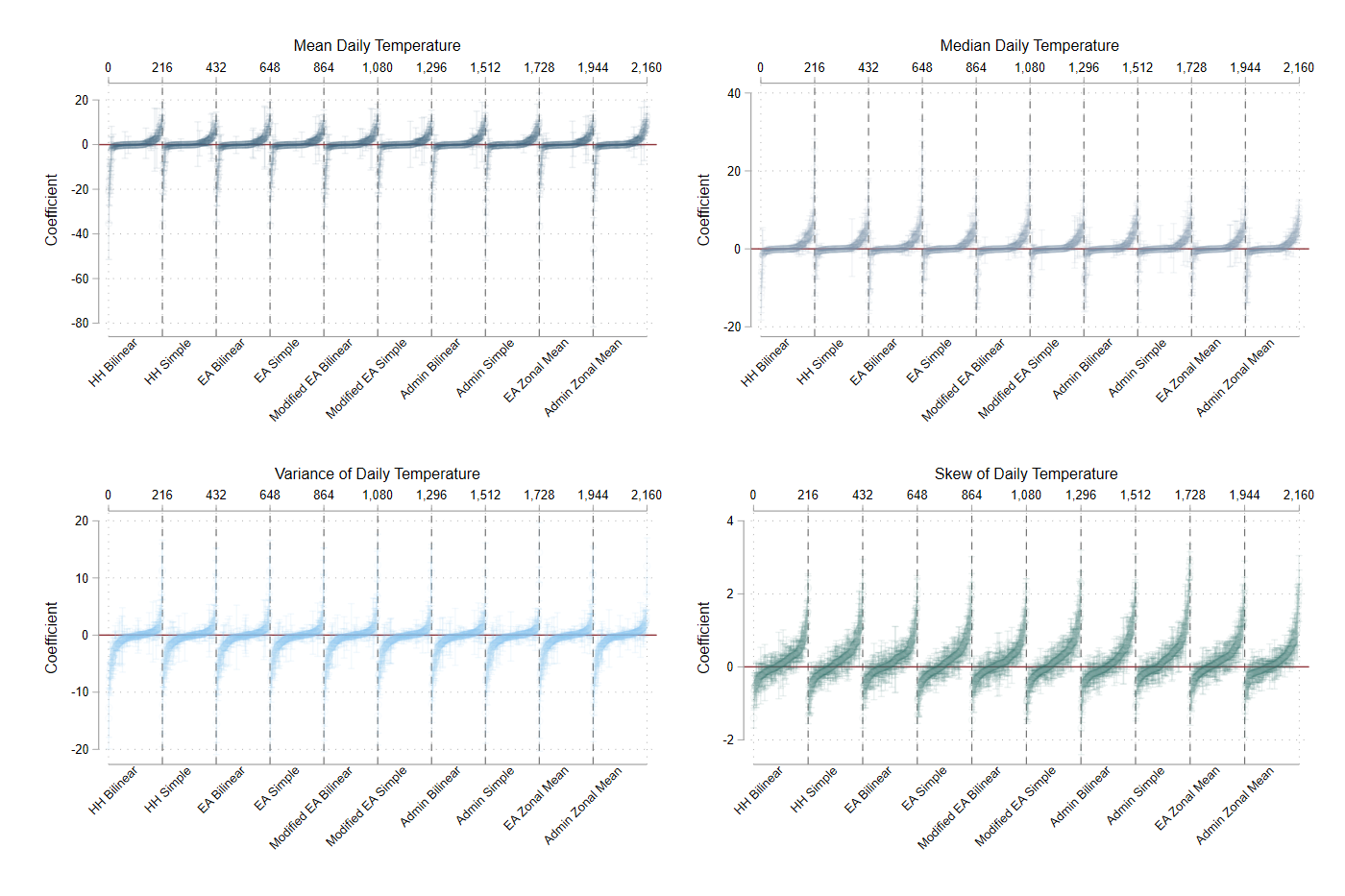}
		\end{center}
		\footnotesize  \textit{Note}: The figure presents the coefficients and confidence intervals for four temperature metrics, by extraction method. These are aggregated across country, remote sensing source, outcome variable, and specification. Each panel includes 2,160 coefficients and confidence intervals (designated on the $x$-axis), with each bin (extraction method) including 216 coefficients and confidence intervals. Each column, as such, represents the findings of a single regressions, e.g. the coefficient and confidence interval itself. 
	\end{minipage}	
\end{figure}

\begin{figure}[!htbp]
	\begin{minipage}{\linewidth}		
		\caption{Coefficients and Confidence Intervals for (1) Total Season, (2) Deviation in Total Season, and (3) z-score of Total Season, for Temperature by Extraction Method}
		\label{fig:ext_total_tp}
		\begin{center}
			\includegraphics[width=\linewidth,keepaspectratio]{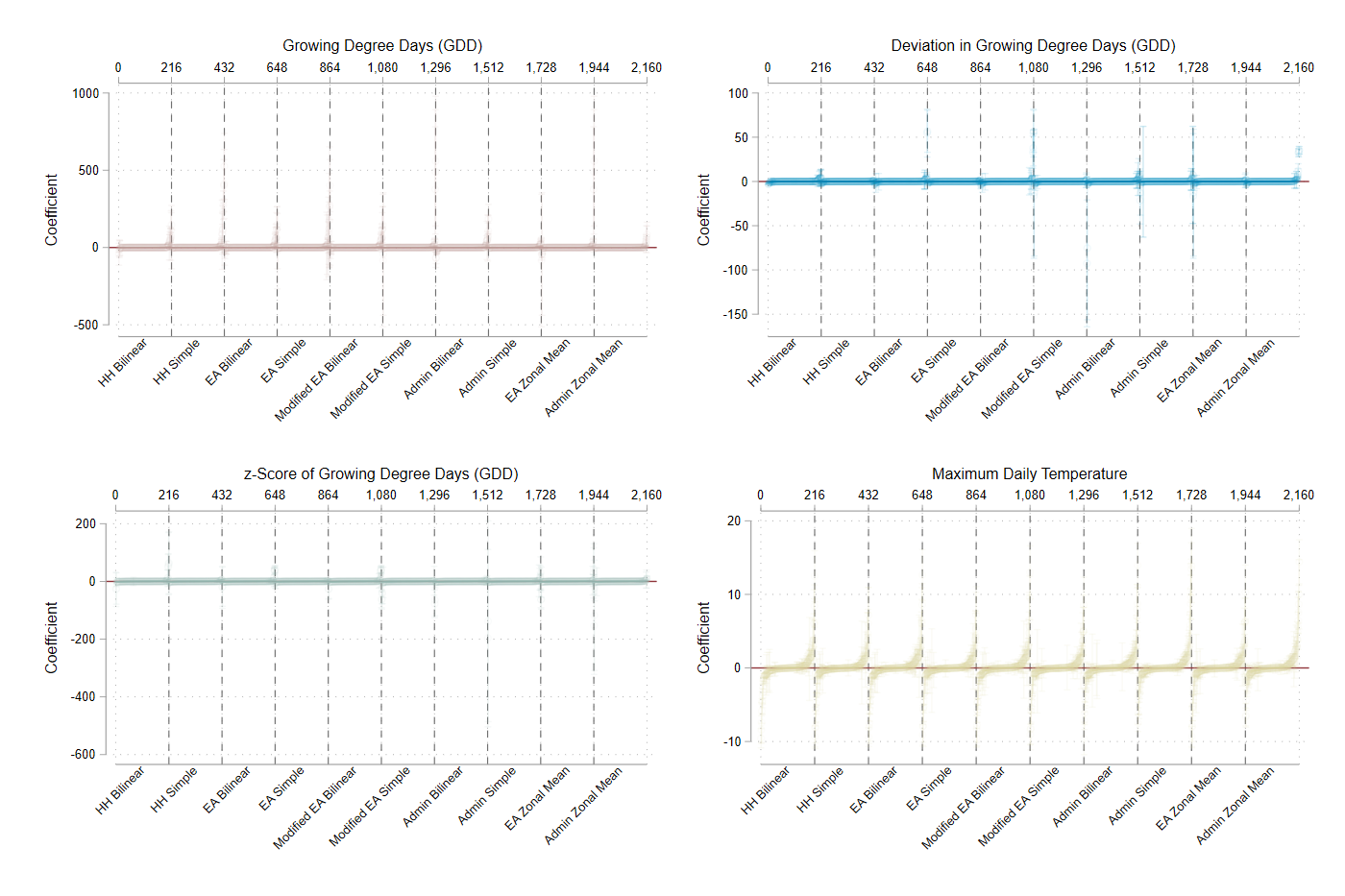}
		\end{center}
		\footnotesize  \textit{Note}: The figure presents the coefficients and confidence intervals for four temperature metrics, by extraction method. These are aggregated across country, remote sensing source, outcome variable, and specification. Each panel includes 2,160 coefficients and confidence intervals (designated on the $x$-axis), with each bin (extraction method) including 216 coefficients and confidence intervals. Each column, as such, represents the findings of a single regressions, e.g. the coefficient and confidence interval itself. 
	\end{minipage}
\end{figure}


\clearpage
\newpage
\section{Weather Metrics}\label{sec:apmetrics}

\setcounter{table}{0}
\renewcommand{\thetable}{D\arabic{table}}
\setcounter{figure}{0}
\renewcommand{\thefigure}{D\arabic{figure}}

Expanding on section~\ref{sec:metriccoef}, we present plots of coefficients and confidence intervals for individual regressions by weather metrics and country. As the objective is to determine which weather metrics conform to our priors about consistency in sign, we classify a weather metric as more accurate if a larger share of its significant coefficients share the same sign. We also consider how these coefficients comport with our priors. Table~\ref{tab:Wvarsign} in the appendix defines our assumption for the expected sign for each of the 22 metrics. As before, we propose that a weather metric is classified as suffering from mismeasurement if a nearly equal proportion of its significant coefficients are negative and positive. As such, we interpret a non-significant coefficient as a zero, neither positive nor negative. As stated previously, because we lack data on the objective facts regarding rainfall and temperature, we seek to draw broad conclusions about which weather metrics adhere to the assumptions on consistency in sign. 

We first consider the remaining rainfall metrics not discussed in section~\ref{sec:metriccoef}. These include: the variance in daily rainfall (Figure~\ref{fig:v03_cty}); the skew in daily rainfall (Figure~\ref{fig:v04_cty}); the deviation in total seasonal rainfall (Figure~\ref{fig:v06_cty}); the z-score in total seasonal rainfall (Figure~\ref{fig:v07_cty}); the deviation in number of days with rain (Figure~\ref{fig:v09_cty}); the number of days without rain (Figure~\ref{fig:v10_cty}); the deviation in no rain days (Figure~\ref{fig:v11_cty}); and the deviation in percentage of days with rain (Figure~\ref{fig:v13_cty}). Reviewing these figures, significance is mixed across them and thus no clear trend emerges in that regard. However, there are some disparities with respect to how closely these variables comport with our expectations about signs. We see that most variables are about the same as a coin flip about whether they will be significant or not. This includes the variance of daily rainfall (significant $49\%$ of the time), the deviation in total seasonal rainfall ($48\%$), the deviation in the days with rain  ($48\%$), the number of days without rain ($53\%$), the deviation in days without rain ($48\%$), and the deviation in percentage days of rain ($47\%$). Other variables are more frequently insignificant than significant: skew of daily rainfall (significant $42\%$ of the time) and z-score of total seasonal rainfall ($45\%$). 

Among the subset of coefficients that are significant, we find that most coefficient do not comport with our expectations for the directionality of their signs: for the variance in daily rainfall, $62\%$ of coefficients are positive; for the deviation in total season rainfall, $68\%$ of coefficients are positive; for the z-score of total seasonal rainfall, $68\%$ of coefficients are positive; for the deviation in the number of days with rain, $75\%$ of coefficients are positive; for the deviation in number of days without rain, $24\%$ of coefficients are positive; for the deviation in percentage of days with rain, $75\%$ are positive. Only the number of days without rain comport with our expectations, with only $40\%$ of coefficients positive. The skew of daily rainfall is somewhat ambiguous as both signs may be expected: we see that  $69\%$ of coefficients are positive and $31\%$ are negative for these coefficients.

We next consider the remaining four temperature metrics. These include: the skew of daily temperature (Figure~\ref{fig:v18_cty}); the deviation in growing degree days (Figure~\ref{fig:v20_cty}); the z-score of growing degree days (Figure~\ref{fig:v21_cty}); and the maximum daily temperature (Figure~\ref{fig:v22_cty}). As with the remaining rainfall coefficients, significance is mixed across, around $50\%$. This includes the largest share of significant coefficients at $55\%$ for maximum daily temperature, followed by the z-score of growing degree days ($45\%$), deviation in growing degree days ($44\%$), and the skew of daily temperature ($44\%$). Of these, only the maximum daily temperature comport with expectations of sign: $43\%$ of coefficients are positive. The remaining metrics have unexpected signs: for skew of daily temperature, $75\%$ of coefficients are positive; for the deviation in growing degree days, $65\%$ of coefficients are positive; for the z-score of growing degree days, $60\%$ of coefficients are positive.


\newpage

\begin{table}[htbp]	\centering
	\caption{Expected Sign of Weather Metrics} \label{tab:Wvarsign}
	\scalebox{0.9}
	{ \setlength{\linewidth}{.1cm}\newcommand{\contents}
		{\begin{tabular}{ll}
			\\[-1.8ex]\hline 
			\hline \\[-1.8ex]
			& \multicolumn{1}{c}{Expected Sign} \\
			\multicolumn{2}{l}{\emph{\textbf{Panel A}: Rainfall}} \\
			\multicolumn{1}{l}{Mean Daily Rain} & \multicolumn{1}{c}{$+$} \\
			\multicolumn{1}{l}{Median Daily Rain} & \multicolumn{1}{c}{$+$} \\
			\multicolumn{1}{l}{Variance of Daily Rain} & \multicolumn{1}{c}{$-$} \\
			\multicolumn{1}{l}{Skew of Daily Rain} & \multicolumn{1}{c}{$+/-$} \\
			\multicolumn{1}{l}{Total Seasonal Rain} & \multicolumn{1}{c}{$+$} \\
			\multicolumn{1}{l}{Deviation in Total Rain} & \multicolumn{1}{c}{$-$} \\
			\multicolumn{1}{l}{z-score of Total Rain} & \multicolumn{1}{c}{$-$} \\
			\multicolumn{1}{l}{Rainy Days} & \multicolumn{1}{c}{$+$} \\
			\multicolumn{1}{l}{Deviation in Rainy Days} & \multicolumn{1}{c}{$-$} \\
			\multicolumn{1}{l}{No Rain Days} & \multicolumn{1}{c}{$-$} \\
			\multicolumn{1}{l}{Deviation in No Rain Days} & \multicolumn{1}{c}{$+$} \\
			\multicolumn{1}{l}{$\%$ of Rainy Days} & \multicolumn{1}{c}{$+$} \\
			\multicolumn{1}{l}{Deviation in $\%$ of Rainy Days} & \multicolumn{1}{c}{$-$} \\
			\multicolumn{1}{l}{Longest Dry Spell} & \multicolumn{1}{c}{$-$} \\
			\midrule
			& \\
			\multicolumn{2}{l}{\emph{\textbf{Panel B}: Temperature}} \\
			\multicolumn{1}{l}{Mean Daily Temp} & \multicolumn{1}{c}{$+$} \\
			\multicolumn{1}{l}{Median Daily Temp} & \multicolumn{1}{c}{$+$} \\
			\multicolumn{1}{l}{Variance of Daily Temp} & \multicolumn{1}{c}{$-$} \\
			\multicolumn{1}{l}{Skew of Daily Temp} & \multicolumn{1}{c}{$+/-$} \\
			\multicolumn{1}{l}{Growing degree days} & \multicolumn{1}{c}{$+$} \\
			\multicolumn{1}{l}{Deviation in GDD} & \multicolumn{1}{c}{$-$} \\
			\multicolumn{1}{l}{z-score of GDD} & \multicolumn{1}{c}{$-$} \\
			\multicolumn{1}{l}{Max Daily Temp} & \multicolumn{1}{c}{$-$} \\
			\\[-1.8ex]\hline 
			\hline \\[-1.8ex]
			\multicolumn{2}{p{\linewidth}}{\footnotesize  \textit{Note}: \footnotesize The table presents weather metrics and the expected directionality of coefficients.} \\
		\end{tabular}}
	\setbox0=\hbox{\contents}
    \setlength{\linewidth}{\wd0-2\tabcolsep-.25em}
    \contents}
\end{table}


\begin{figure}[!htbp]
	\begin{minipage}{\linewidth}		
		\caption{Coefficients and Confidence Intervals for Variance of Daily Rainfall, by Country}
		\label{fig:v03_cty}
		\begin{center}
			\includegraphics[width=\linewidth,keepaspectratio]{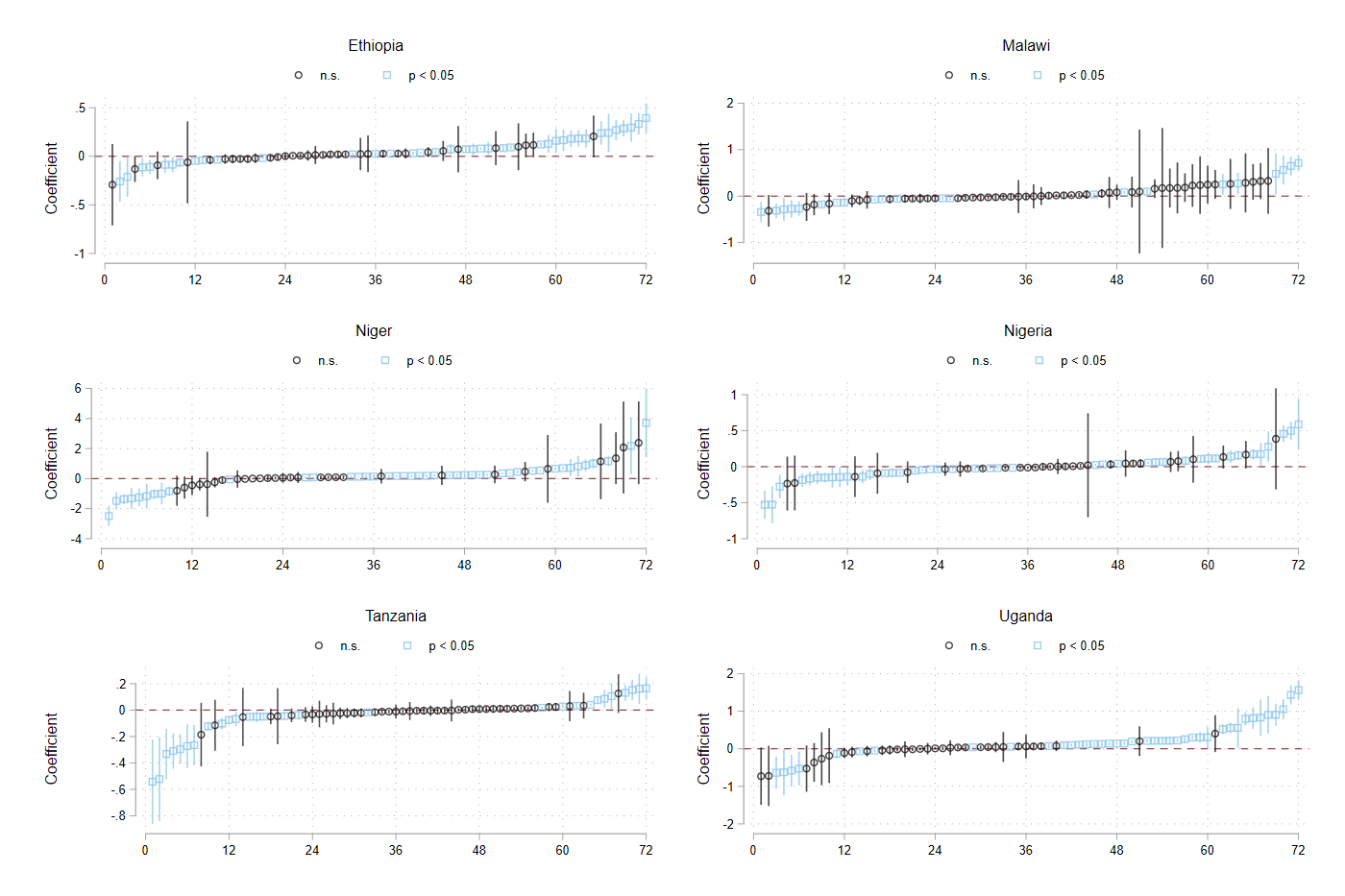}
		\end{center}
		\footnotesize  \textit{Note}: The figure presents the coefficients and confidence intervals for variance of daily rainfall, by country. Only one extraction method is included. Each panel includes 72 coefficients and confidence intervals (designated on the $x$-axis). Each column, as such, represents the findings of a single regressions, e.g. the coefficient and confidence interval itself. The significance level of these are denoted by color and shape of the identifier in the figure, as designated at the top of each panel. 
	\end{minipage}	
\end{figure}

\begin{figure}[!htbp]
	\begin{minipage}{\linewidth}		
		\caption{Coefficients and Confidence Intervals for Skew of Daily Rainfall, by Country}
		\label{fig:v04_cty}
		\begin{center}
			\includegraphics[width=\linewidth,keepaspectratio]{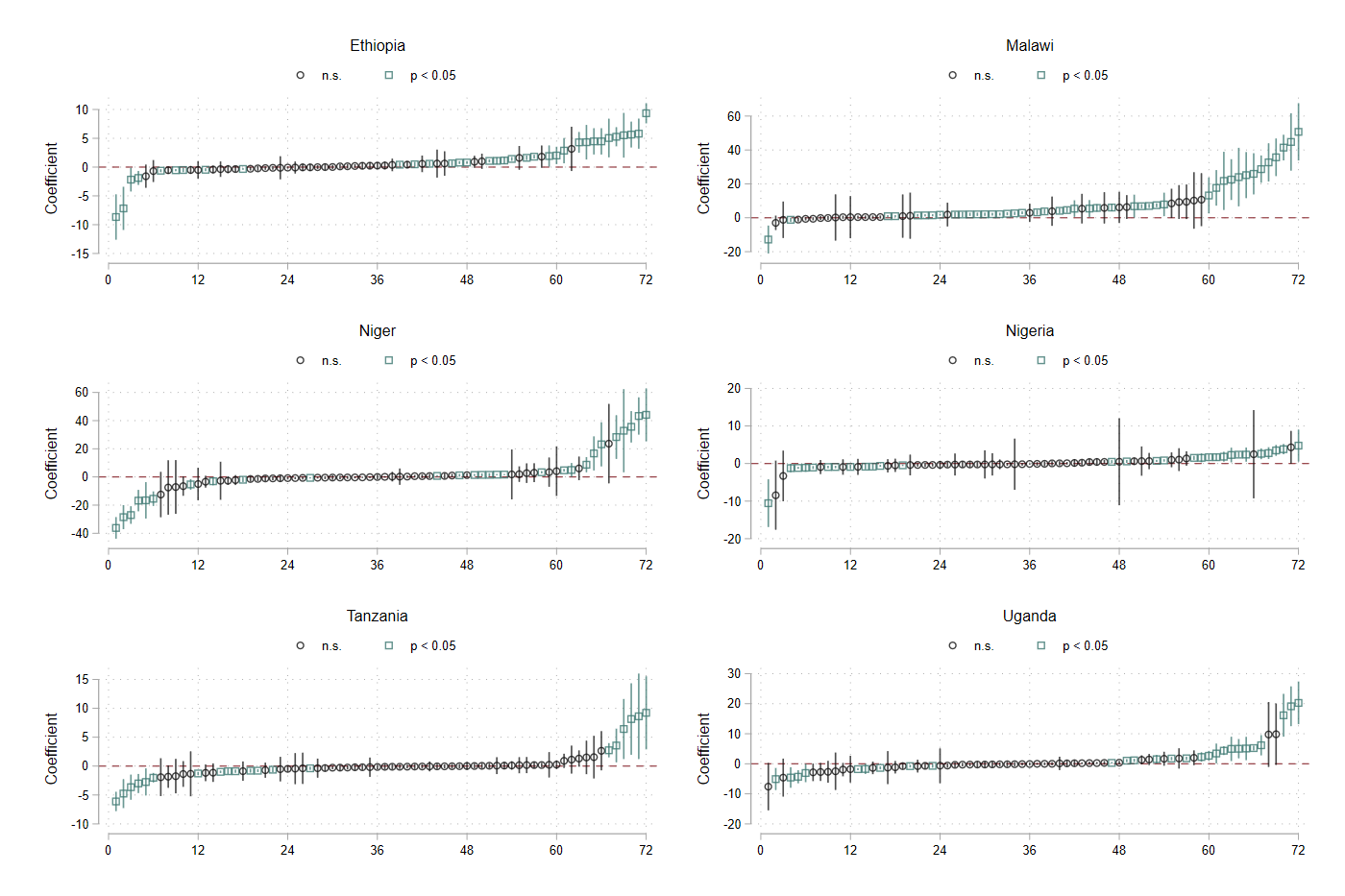}
		\end{center}
		\footnotesize  \textit{Note}: The figure presents the coefficients and confidence intervals for skew of daily rainfall, by country. Only one extraction method is included. Each panel includes 72 coefficients and confidence intervals (designated on the $x$-axis). Each column, as such, represents the findings of a single regressions, e.g. the coefficient and confidence interval itself. The significance level of these are denoted by color and shape of the identifier in the figure, as designated at the top of each panel. 
	\end{minipage}	
\end{figure}

\begin{figure}[!htbp]
	\begin{minipage}{\linewidth}		
		\caption{Coefficients and Confidence Intervals for Deviation in Total Seasonal Rainfall, by Country}
		\label{fig:v06_cty}
		\begin{center}
			\includegraphics[width=\linewidth,keepaspectratio]{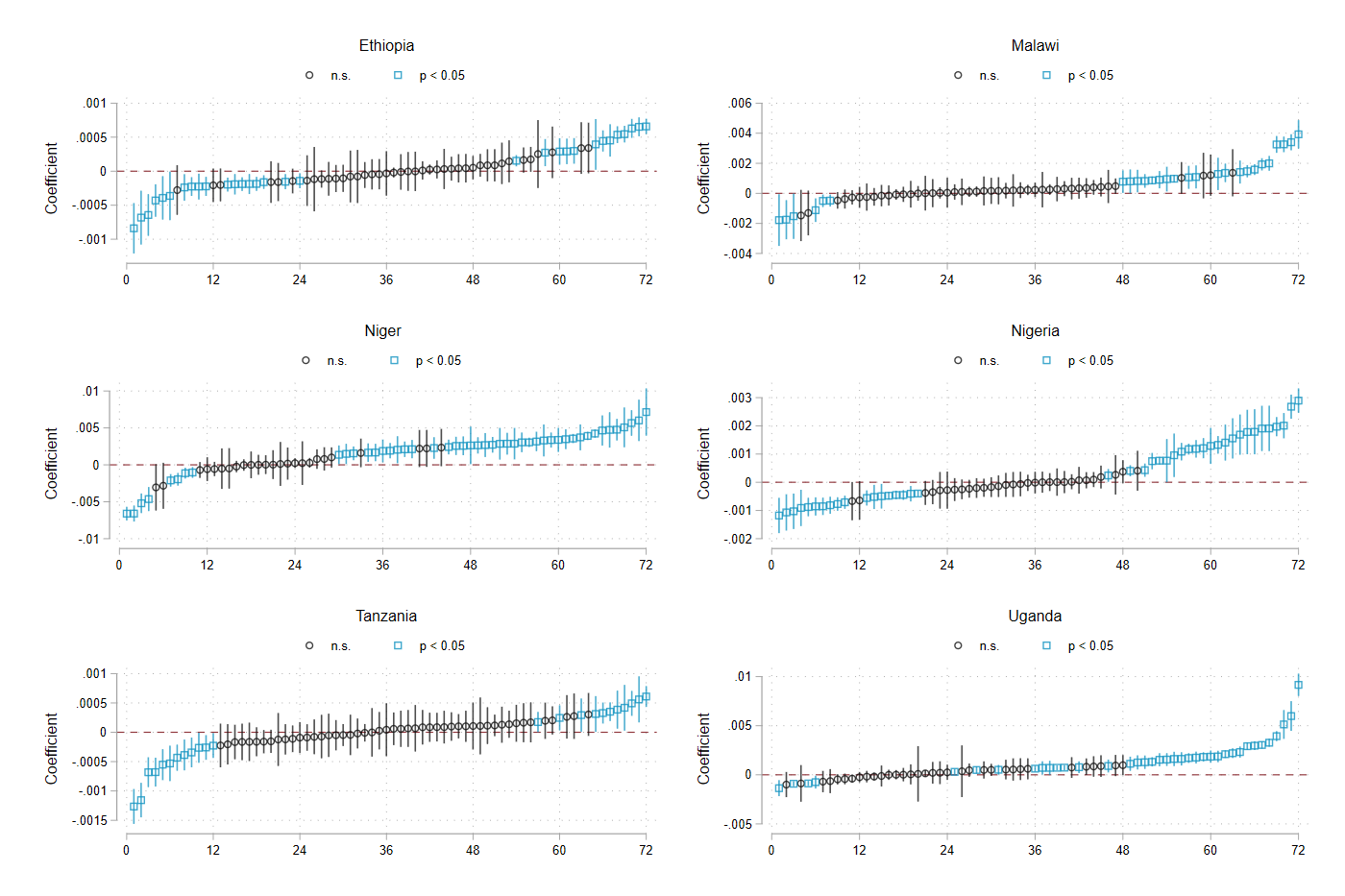}
		\end{center}
		\footnotesize  \textit{Note}: The figure presents the coefficients and confidence intervals for the deviation in total seasonal rainfall, by country. Only one extraction method is included. Each panel includes 72 coefficients and confidence intervals (designated on the $x$-axis). Each column, as such, represents the findings of a single regressions, e.g. the coefficient and confidence interval itself. The significance level of these are denoted by color and shape of the identifier in the figure, as designated at the top of each panel. 
	\end{minipage}	
\end{figure}

\begin{figure}[!htbp]
	\begin{minipage}{\linewidth}		
		\caption{Coefficients and Confidence Intervals for z-score of Total Seasonal Rainfall, by Country}
		\label{fig:v07_cty}
		\begin{center}
			\includegraphics[width=\linewidth,keepaspectratio]{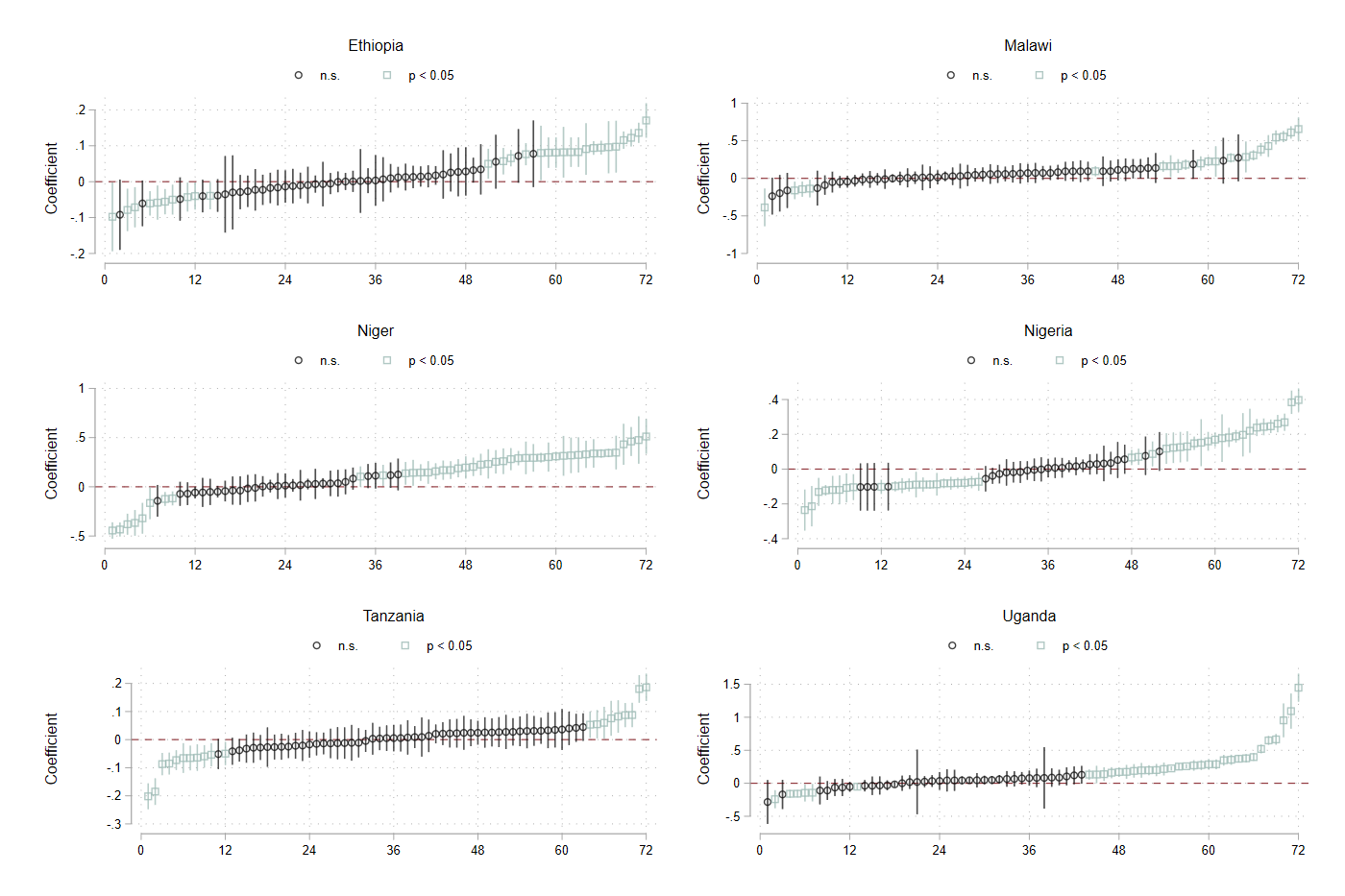}
		\end{center}
		\footnotesize  \textit{Note}: The figure presents the coefficients and confidence intervals for the z-score of total seasonal rainfall, by country. Only one extraction method is included. Each panel includes 72 coefficients and confidence intervals (designated on the $x$-axis). Each column, as such, represents the findings of a single regressions, e.g. the coefficient and confidence interval itself. The significance level of these are denoted by color and shape of the identifier in the figure, as designated at the top of each panel. 
	\end{minipage}	
\end{figure}

\begin{figure}[!htbp]
	\begin{minipage}{\linewidth}		
		\caption{Coefficients and Confidence Intervals for Deviation in Number of Days with Rain, by Country}
		\label{fig:v09_cty}
		\begin{center}
			\includegraphics[width=\linewidth,keepaspectratio]{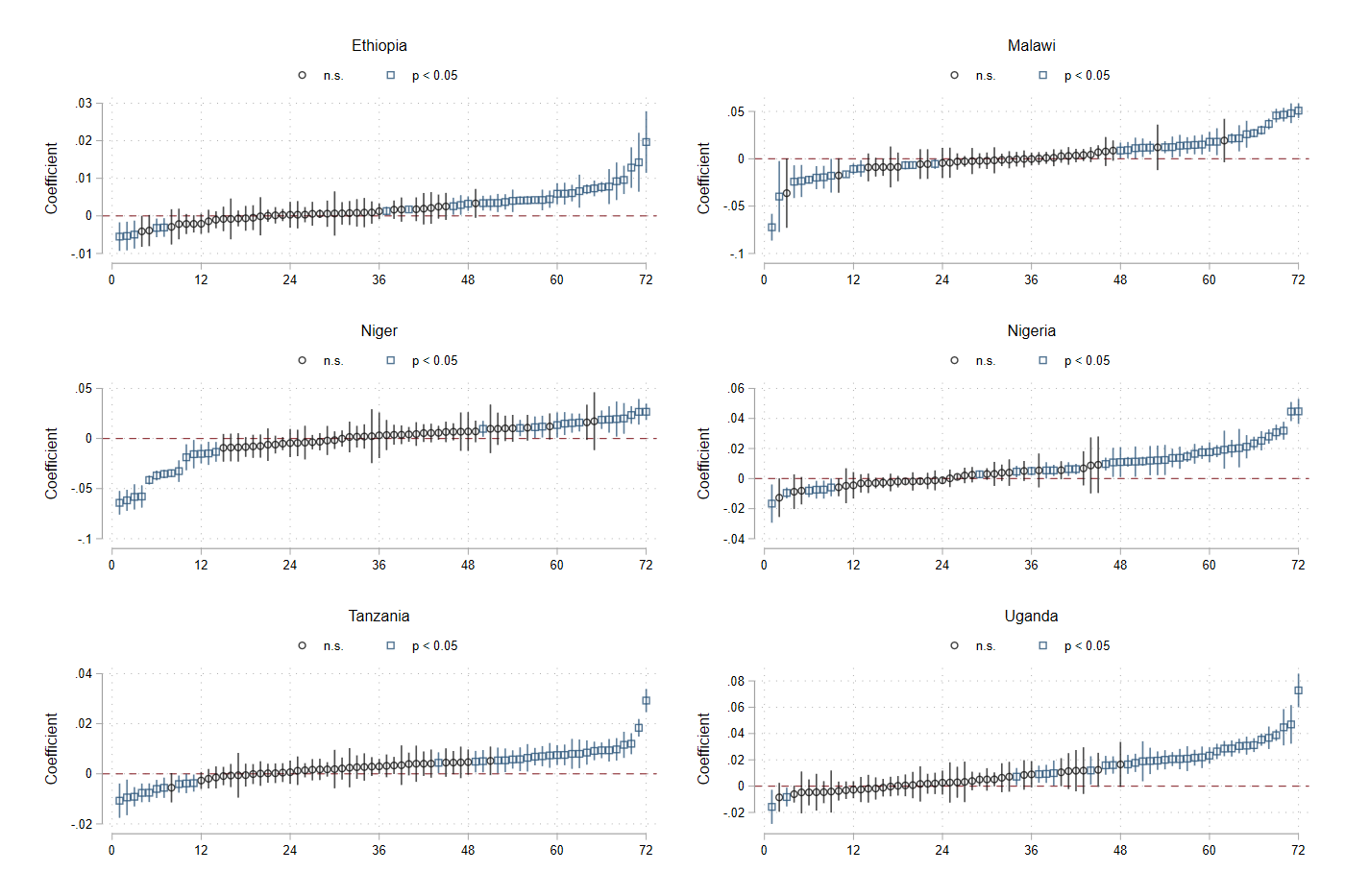}
		\end{center}
		\footnotesize  \textit{Note}: The figure presents the coefficients and confidence intervals for the deviation in number of days with rain, by country. Only one extraction method is included. Each panel includes 72 coefficients and confidence intervals (designated on the $x$-axis). Each column, as such, represents the findings of a single regressions, e.g. the coefficient and confidence interval itself. The significance level of these are denoted by color and shape of the identifier in the figure, as designated at the top of each panel. 
	\end{minipage}	
\end{figure}

\begin{figure}[!htbp]
	\begin{minipage}{\linewidth}		
		\caption{Coefficients and Confidence Intervals for Number of Days without Rain, by Country}
		\label{fig:v10_cty}
		\begin{center}
			\includegraphics[width=\linewidth,keepaspectratio]{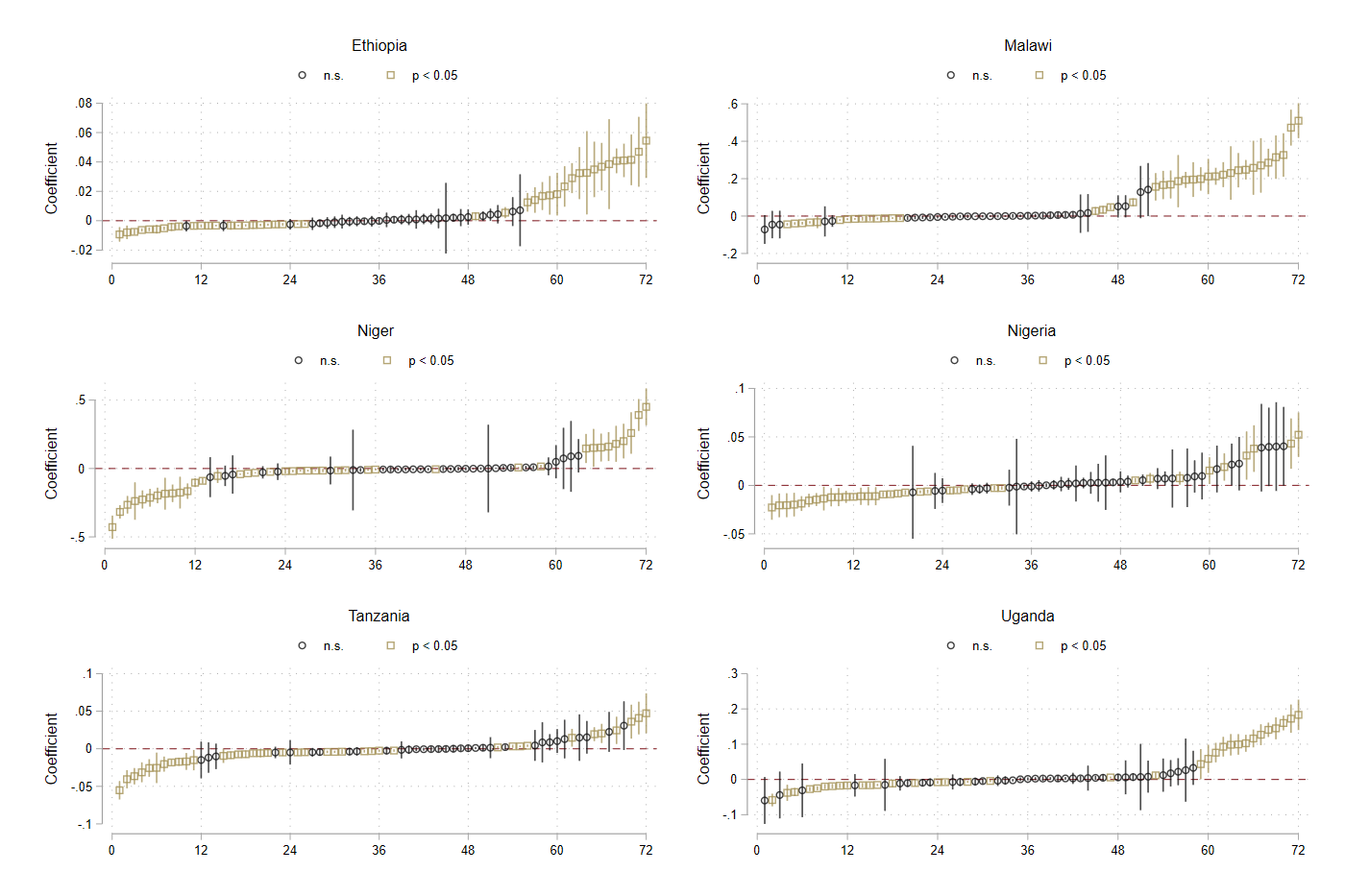}
		\end{center}
		\footnotesize  \textit{Note}: The figure presents the coefficients and confidence intervals for the number of days without rain, by country. These are aggregated across remote sensing source, outcome variable, and specification. Only one extraction method is included. Each panel includes 72 coefficients and confidence intervals (designated on the $x$-axis). Each column, as such, represents the findings of a single regressions, e.g. the coefficient and confidence interval itself. The significance level of these are denoted by color and shape of the identifier in the figure, as designated at the top of each panel. 
	\end{minipage}	
\end{figure}

\begin{figure}[!htbp]
	\begin{minipage}{\linewidth}		
		\caption{Coefficients and Confidence Intervals for Deviation in No Rain Days, by Country}
		\label{fig:v11_cty}
		\begin{center}
			\includegraphics[width=\linewidth,keepaspectratio]{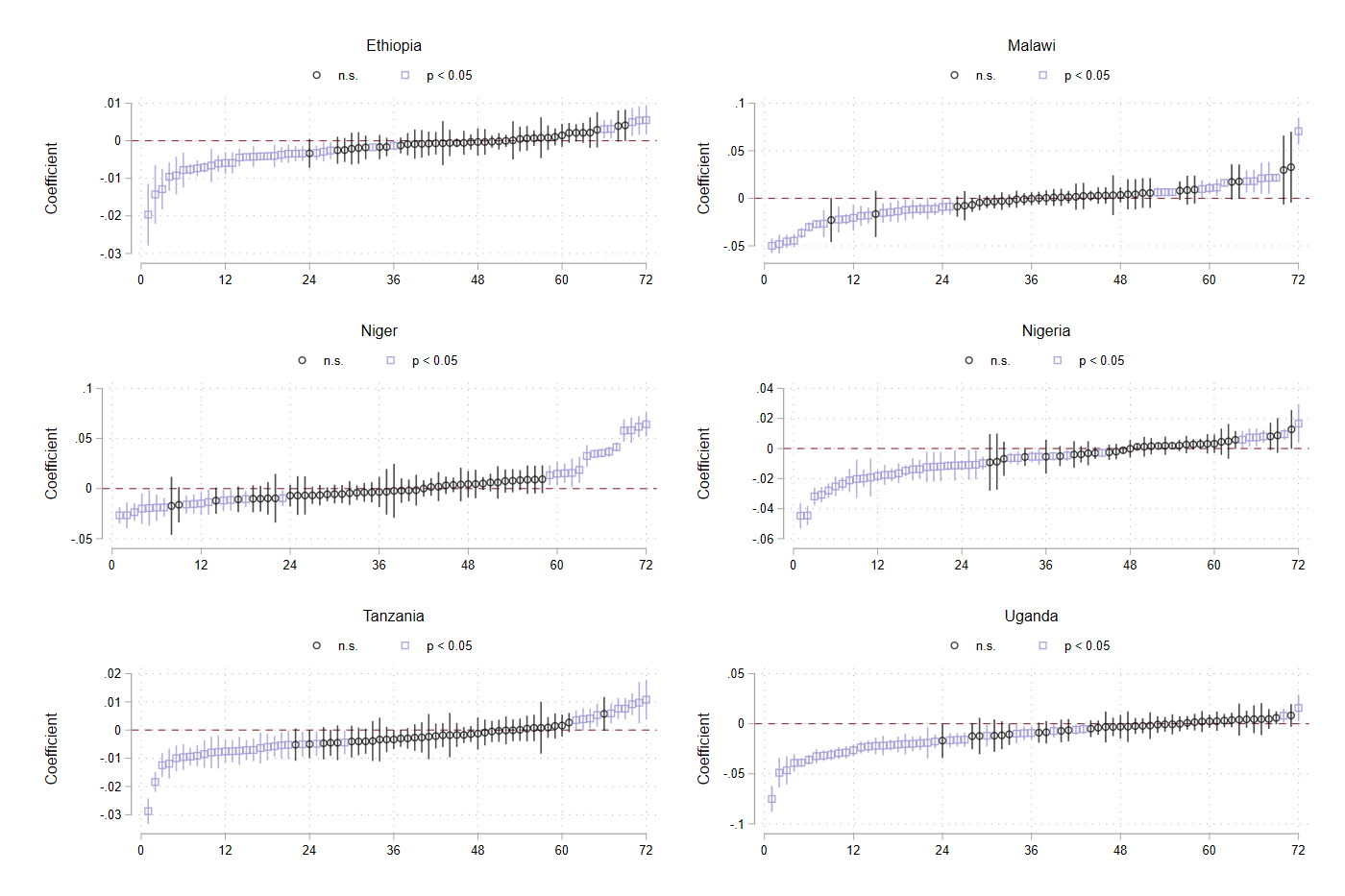}
		\end{center}
		\footnotesize  \textit{Note}: The figure presents the coefficients and confidence intervals for the deviation in number of days without rain, by country. These are aggregated across remote sensing source, outcome variable, and specification. Only one extraction method is included. Each panel includes 72 coefficients and confidence intervals (designated on the $x$-axis). Each column, as such, represents the findings of a single regressions, e.g. the coefficient and confidence interval itself. The significance level of these are denoted by color and shape of the identifier in the figure, as designated at the top of each panel. 
	\end{minipage}	
\end{figure}

\begin{figure}[!htbp]
	\begin{minipage}{\linewidth}		
		\caption{Coefficients and Confidence Intervals for Deviation in Percentage of Days with Rain, by Country}
		\label{fig:v13_cty}
		\begin{center}
			\includegraphics[width=\linewidth,keepaspectratio]{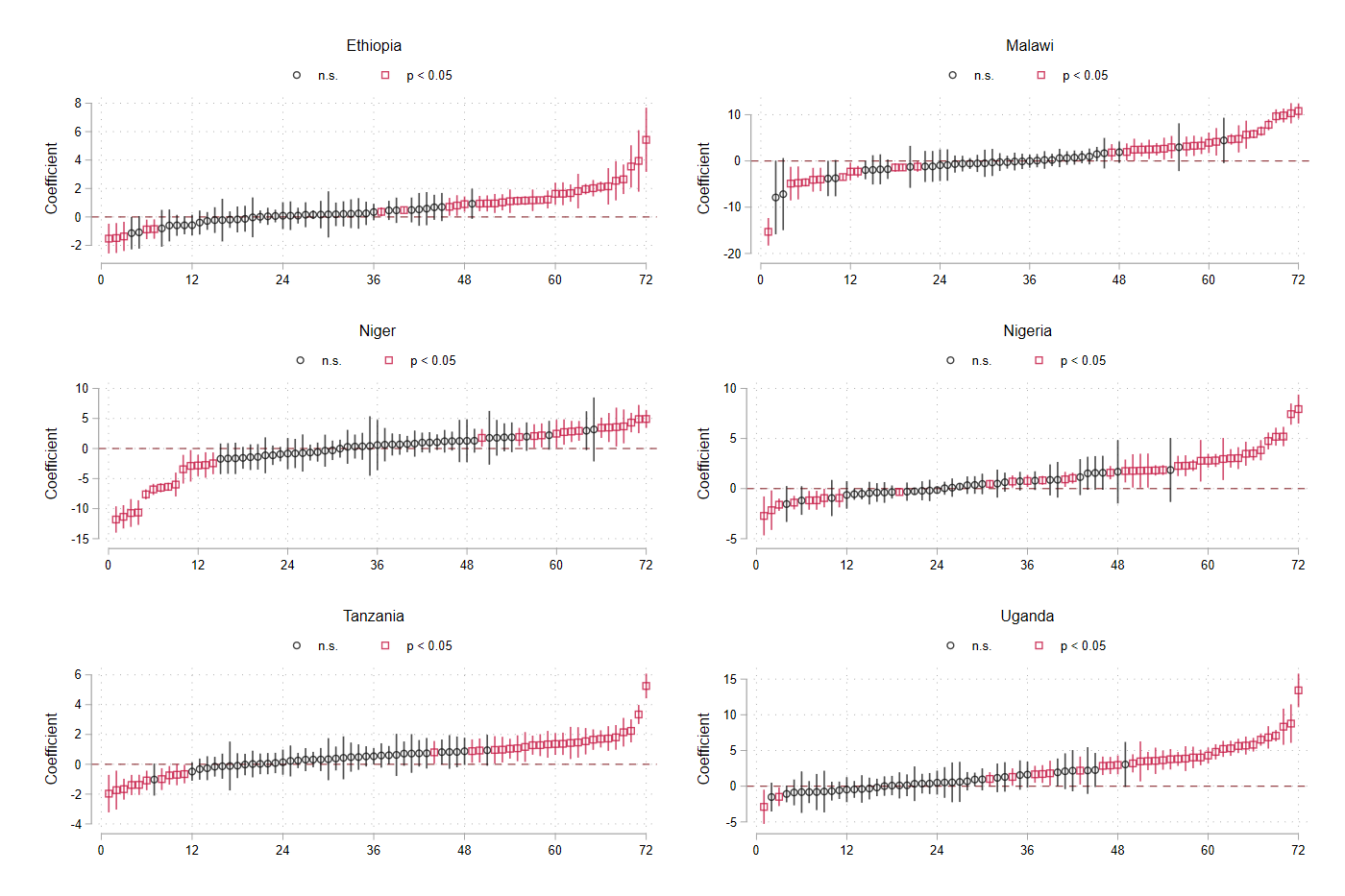}
		\end{center}
		\footnotesize  \textit{Note}: The figure presents the coefficients and confidence intervals for the deviation in percentage of days with rain, by country. Only one extraction method is included. Each panel includes 72 coefficients and confidence intervals (designated on the $x$-axis). Each column, as such, represents the findings of a single regressions, e.g. the coefficient and confidence interval itself. The significance level of these are denoted by color and shape of the identifier in the figure, as designated at the top of each panel. 
	\end{minipage}	
\end{figure}

\begin{figure}[!htbp]
	\begin{minipage}{\linewidth}		
		\caption{Coefficients and Confidence Intervals for Skew of Daily Temperature, by Country}
		\label{fig:v18_cty}
		\begin{center}
			\includegraphics[width=\linewidth,keepaspectratio]{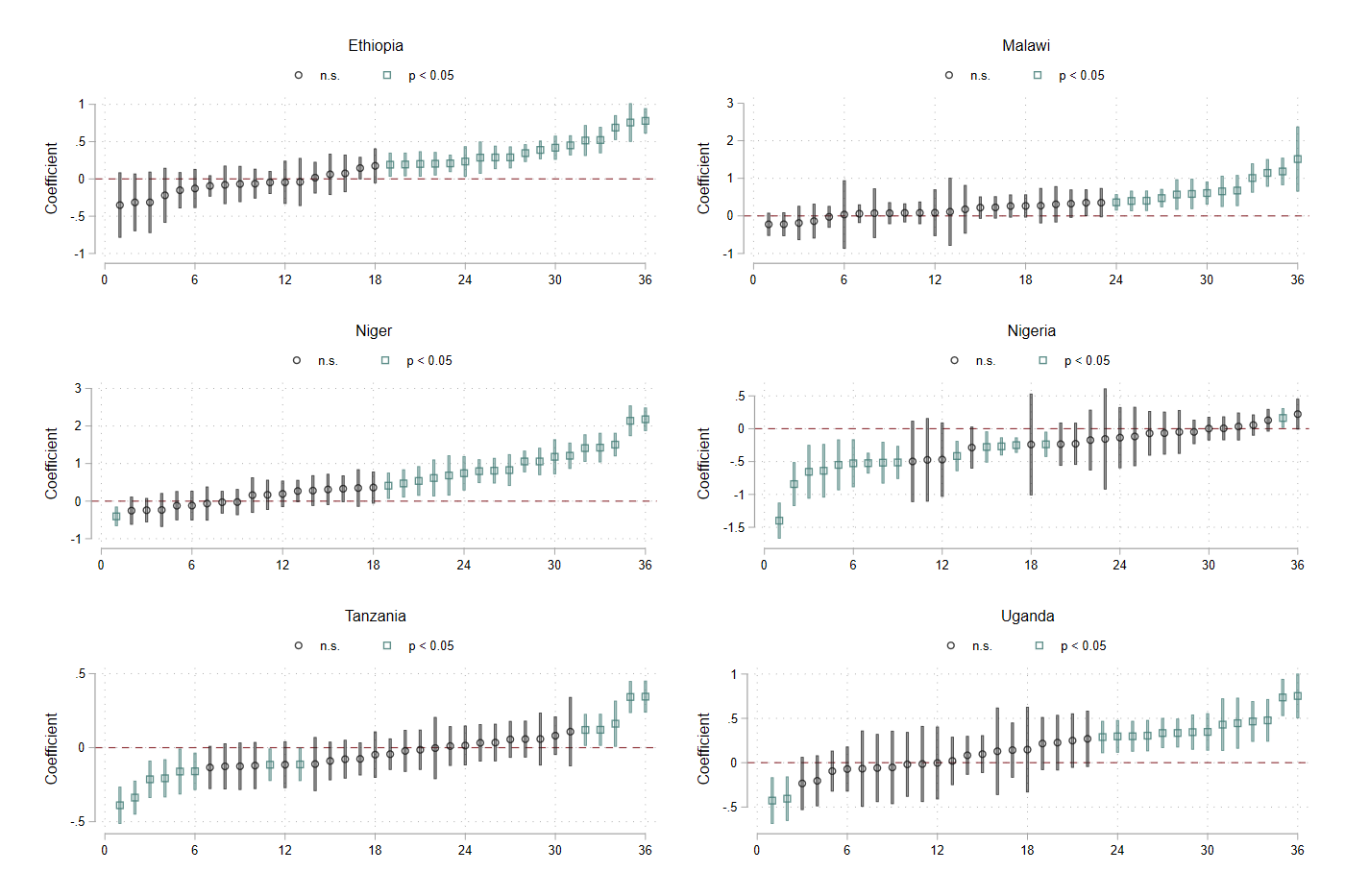}
		\end{center}
		\footnotesize  \textit{Note}: The figure presents the coefficients and confidence intervals for the skew of daily temperature, by country. Only one extraction method is included. Each panel includes 36 coefficients and confidence intervals (designated on the $x$-axis). Each column, as such, represents the findings of a single regressions, e.g. the coefficient and confidence interval itself. The significance level of these are denoted by color and shape of the identifier in the figure, as designated at the top of each panel. 
	\end{minipage}	
\end{figure}

\begin{figure}[!htbp]
	\begin{minipage}{\linewidth}		
		\caption{Coefficients and Confidence Intervals for Deviation in Growing Degree Days (GDD), by Country}
		\label{fig:v20_cty}
		\begin{center}
			\includegraphics[width=\linewidth,keepaspectratio]{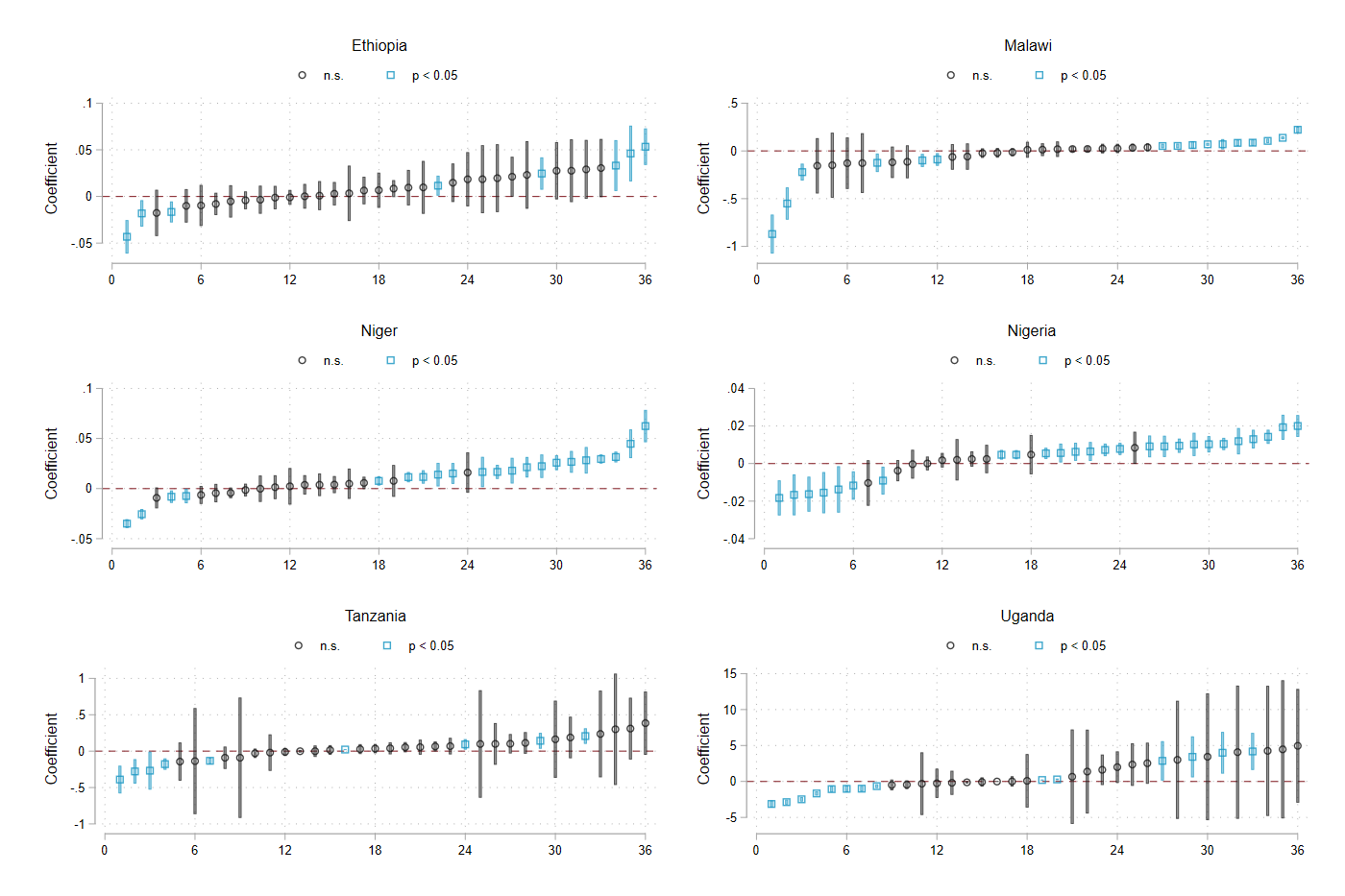}
		\end{center}
		\footnotesize  \textit{Note}: The figure presents the coefficients and confidence intervals for the deviation in growing degree days, by country. Only one extraction method is included. Each panel includes 36 coefficients and confidence intervals (designated on the $x$-axis). Each column, as such, represents the findings of a single regressions, e.g. the coefficient and confidence interval itself. The significance level of these are denoted by color and shape of the identifier in the figure, as designated at the top of each panel. 
	\end{minipage}	
\end{figure}

\begin{figure}[!htbp]
	\begin{minipage}{\linewidth}		
		\caption{Coefficients and Confidence Intervals for z-score of Growing Degree Days (GDD), by Country}
		\label{fig:v21_cty}
		\begin{center}
			\includegraphics[width=\linewidth,keepaspectratio]{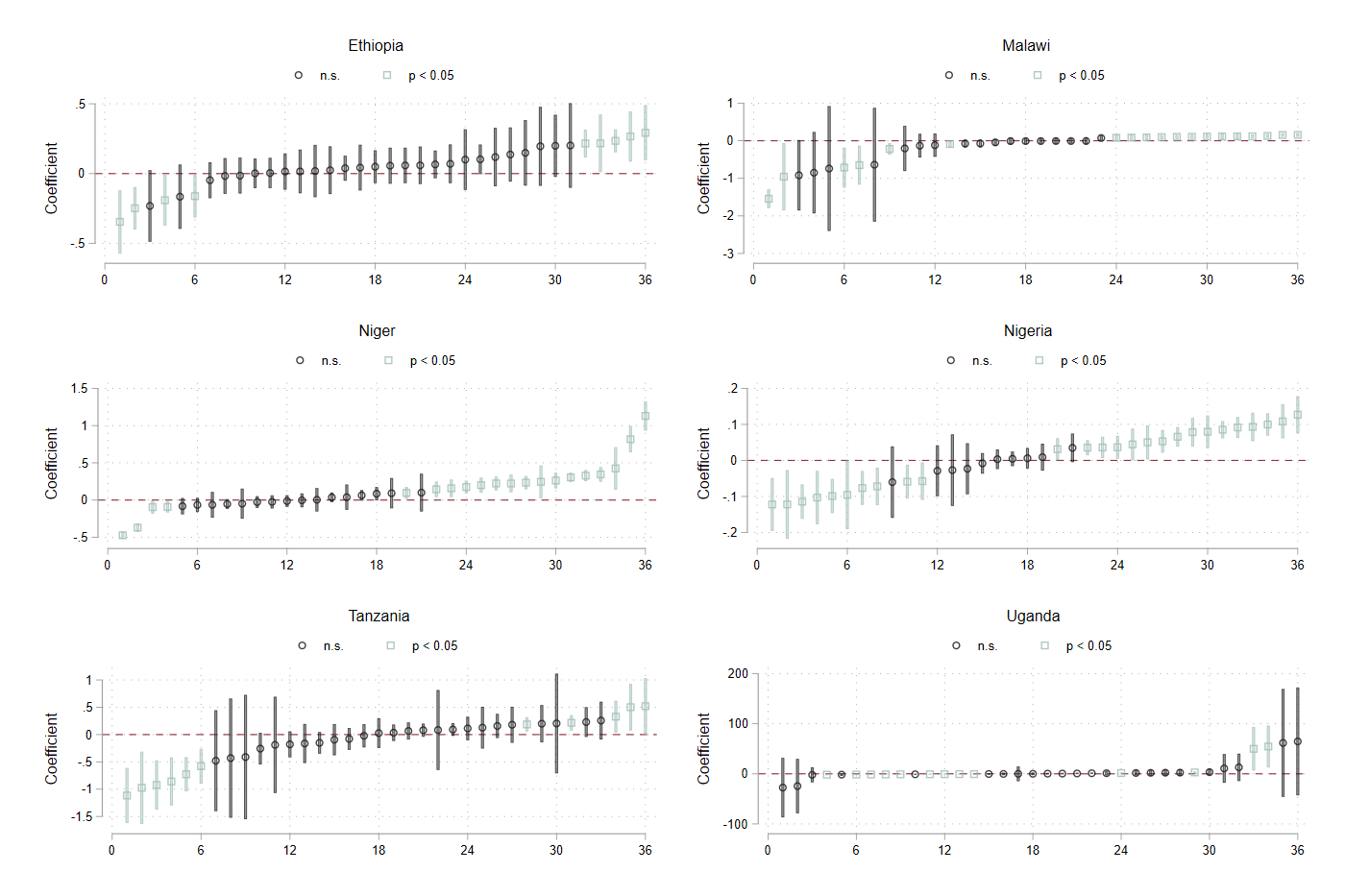}
		\end{center}
		\footnotesize  \textit{Note}: The figure presents the coefficients and confidence intervals for the z-score of growing degree days, by country. Only one extraction method is included. Each panel includes 36 coefficients and confidence intervals (designated on the $x$-axis). Each column, as such, represents the findings of a single regressions, e.g. the coefficient and confidence interval itself. The significance level of these are denoted by color and shape of the identifier in the figure, as designated at the top of each panel. 
	\end{minipage}	
\end{figure}

\begin{figure}[!htbp]
	\begin{minipage}{\linewidth}		
		\caption{Coefficients and Confidence Intervals for Maximum Daily Temperature, by Country}
		\label{fig:v22_cty}
		\begin{center}
			\includegraphics[width=\linewidth,keepaspectratio]{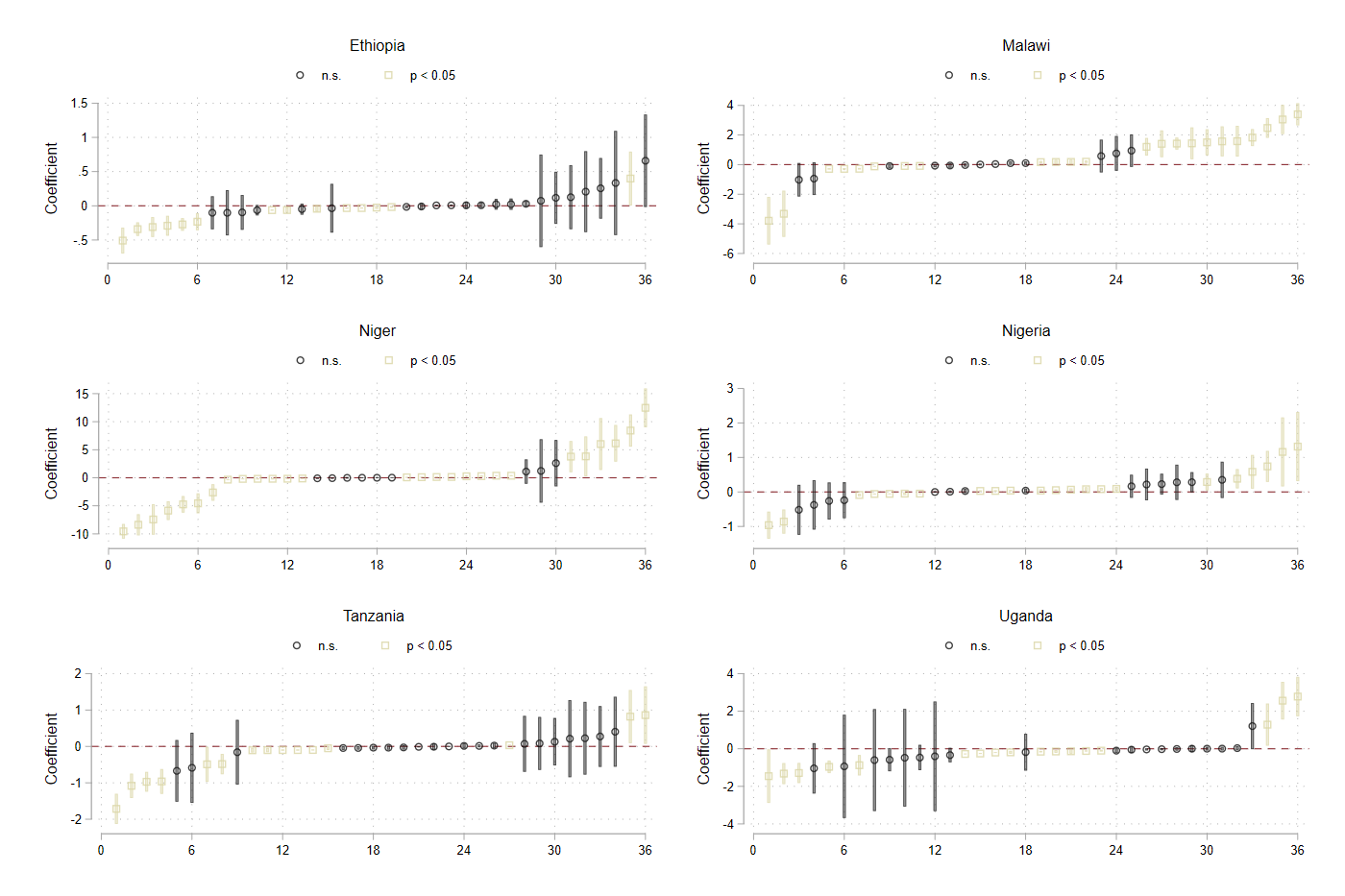}
		\end{center}
		\footnotesize  \textit{Note}: The figure presents the coefficients and confidence intervals for maximum daily temperature, by country. Only one extraction method is included. Each panel includes 36 coefficients and confidence intervals (designated on the $x$-axis). Each column, as such, represents the findings of a single regressions, e.g. the coefficient and confidence interval itself. The significance level of these are denoted by color and shape of the identifier in the figure, as designated at the top of each panel. 
	\end{minipage}	
\end{figure}


\clearpage
\newpage
\section{Remote Sensing Products}\label{sec:apsatellite}

\setcounter{table}{0}
\renewcommand{\thetable}{E\arabic{table}}
\setcounter{figure}{0}
\renewcommand{\thefigure}{E\arabic{figure}}

\begin{figure}[!htbp]
	\begin{minipage}{\linewidth}		
		\caption{Coefficients and Confidence Intervals for (1) Daily Mean, (2) Daily Median, (3) Daily Variance, and (4) Daily Skew, by precipitation product}
		\label{fig:sat_moment_rf}
		\begin{center}
			\includegraphics[width=\linewidth,keepaspectratio]{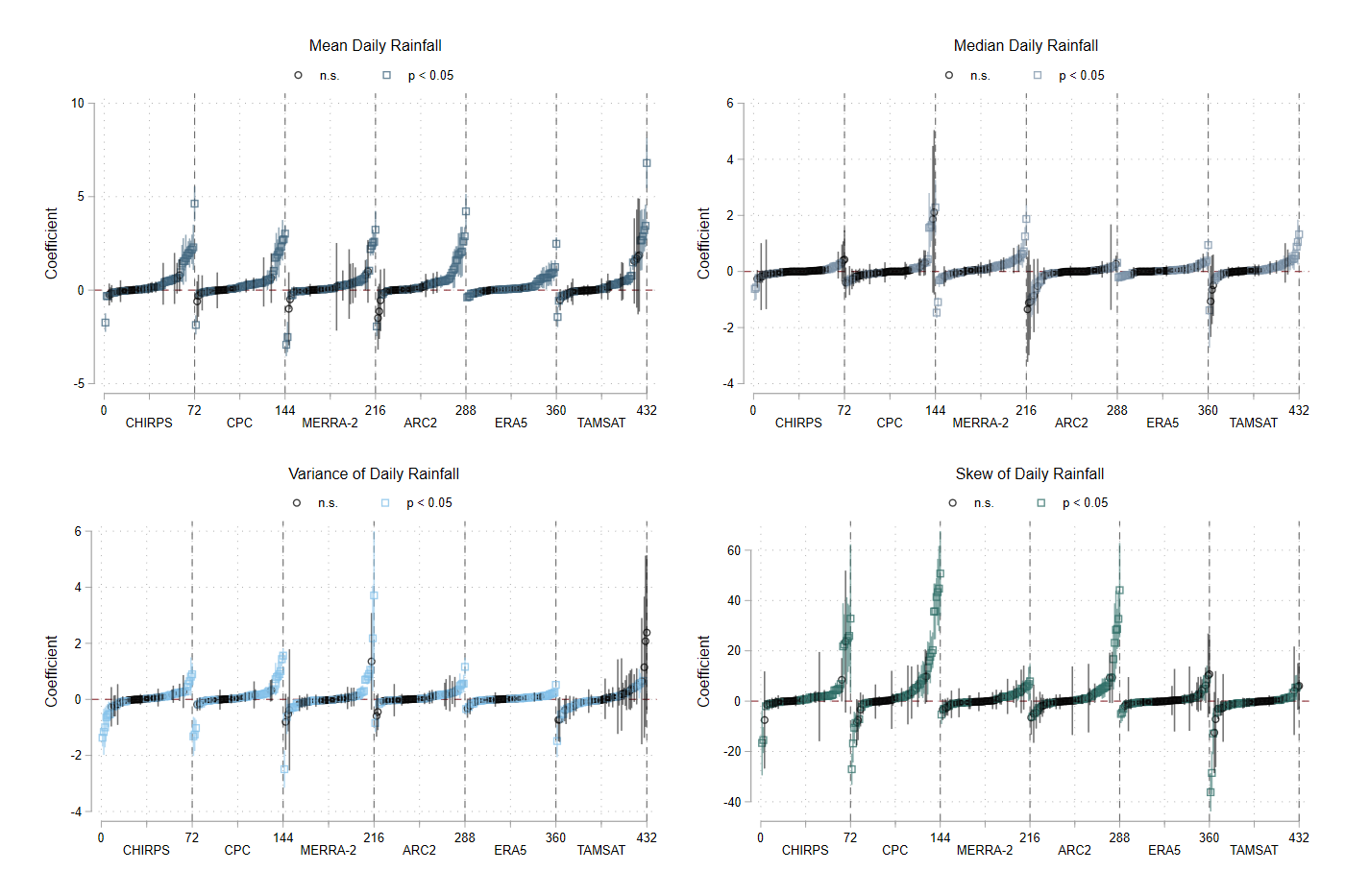}
		\end{center}
		\footnotesize  \textit{Note}: The figure presents the coefficients and confidence intervals for four rainfall metrics, by precipitation product. These are aggregated across countries, remote sensing source, outcome variable, and specification. Only one extraction method is included. Each panel includes 432 coefficients and confidence intervals (designated on the $x$-axis), with each bin (rainfall source) including 72 coefficients and confidence intervals. Each column, as such, represents the findings of a single regressions, e.g. the coefficient and confidence interval itself. The significance level of these are denoted by color and shape of the identifier in the figure, as designated at the top of each panel.  
	\end{minipage}	
\end{figure}

\begin{figure}[!htbp]
\begin{minipage}{\linewidth}		
	\caption{Coefficients and Confidence Intervals for (1) Total Season, (2) Deviation in Total Season, and (3) z-score of Total Season, by precipitation product}
	\label{fig:sat_total_rf}
	\begin{center}
		\includegraphics[width=\linewidth,keepaspectratio]{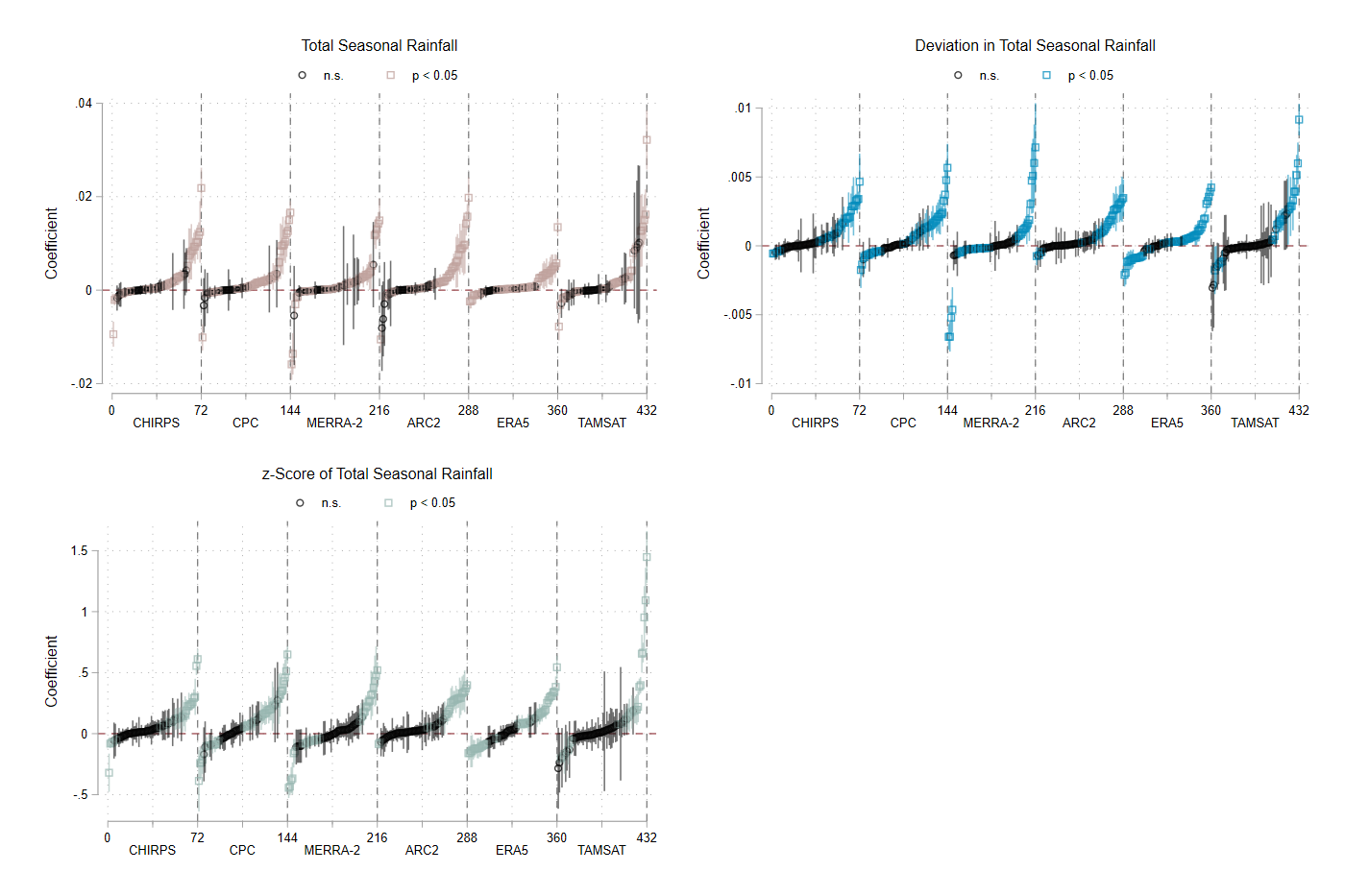}
	\end{center}
	\footnotesize  \textit{Note}: The figure presents the coefficients and confidence intervals for four rainfall metrics, by precipitation product. Only one extraction method is included. Each panel includes 432 coefficients and confidence intervals (designated on the $x$-axis), with each bin (rainfall source) including 72 coefficients and confidence intervals. Each column, as such, represents the findings of a single regressions, e.g. the coefficient and confidence interval itself. The significance level of these are denoted by color and shape of the identifier in the figure, as designated at the top of each panel.   
\end{minipage}
\end{figure}

\begin{figure}[!htbp]
\begin{minipage}{\linewidth}
	\caption{Coefficients and Confidence Intervals for (1) Number of Days with Rain, (2) Deviation in Number of Days with Rain, (3) Percentage of Days with Rain, and (4) Deviation in Percentage of Days with Rain, by precipitation product}
	\label{fig:sat_rain_rf}
	\begin{center}
		\includegraphics[width=\linewidth,keepaspectratio]{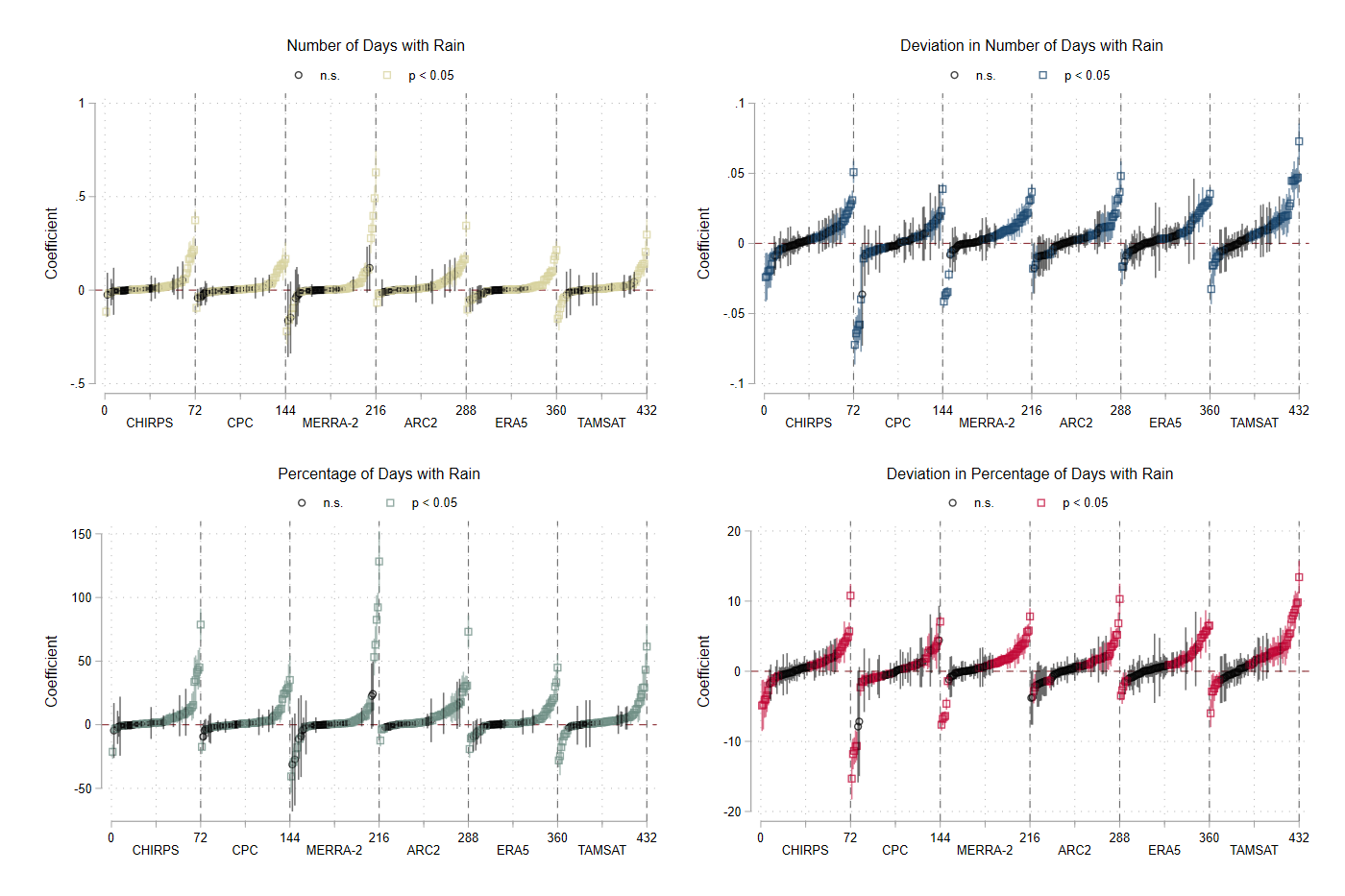}
	\end{center}
	\footnotesize  \textit{Note}: The figure presents the coefficients and confidence intervals for four rainfall metrics, by precipitation product. Only one extraction method is included. Each panel includes 432 coefficients and confidence intervals (designated on the $x$-axis), with each bin (rainfall source) including 72 coefficients and confidence intervals. Each column, as such, represents the findings of a single regressions, e.g. the coefficient and confidence interval itself. The significance level of these are denoted by color and shape of the identifier in the figure, as designated at the top of each panel. 
\end{minipage}	
\end{figure}

\begin{figure}[!htbp]
	\begin{minipage}{\linewidth}		
		\caption{Coefficients and Confidence Intervals for (1) Number of Days without Rain, (2) Deviation in Number of Days without Rain, and (3) Longest Dry Spell, by precipitation product}
		\label{fig:sat_none_rf}
		\begin{center}
			\includegraphics[width=\linewidth,keepaspectratio]{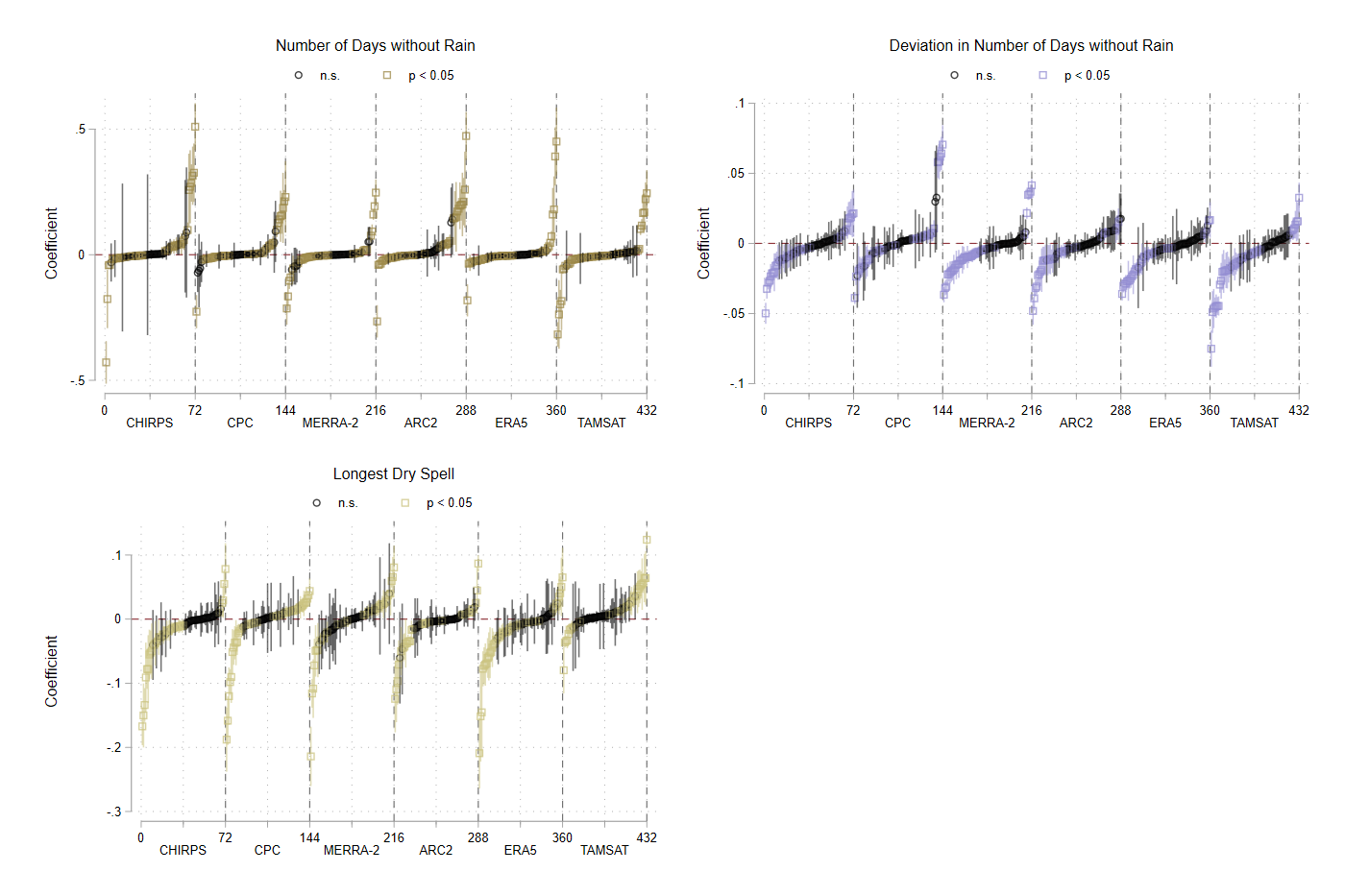}
		\end{center}
		\footnotesize  \textit{Note}: The figure presents the coefficients and confidence intervals for four rainfall metrics, by precipitation product. Only one extraction method is included. Each panel includes 432 coefficients and confidence intervals (designated on the $x$-axis), with each bin (rainfall source) including 72 coefficients and confidence intervals. Each column, as such, represents the findings of a single regressions, e.g. the coefficient and confidence interval itself. The significance level of these are denoted by color and shape of the identifier in the figure, as designated at the top of each panel. 
	\end{minipage}	
\end{figure}

\begin{figure}[!htbp]
	\begin{minipage}{\linewidth}		
		\caption{Coefficients and Confidence Intervals for (1) Daily Mean, (2) Daily Median, (3) Daily Variance, and (4) Daily Skew, by Temperature Product}
		\label{fig:sat_moment_tp}
		\begin{center}
			\includegraphics[width=\linewidth,keepaspectratio]{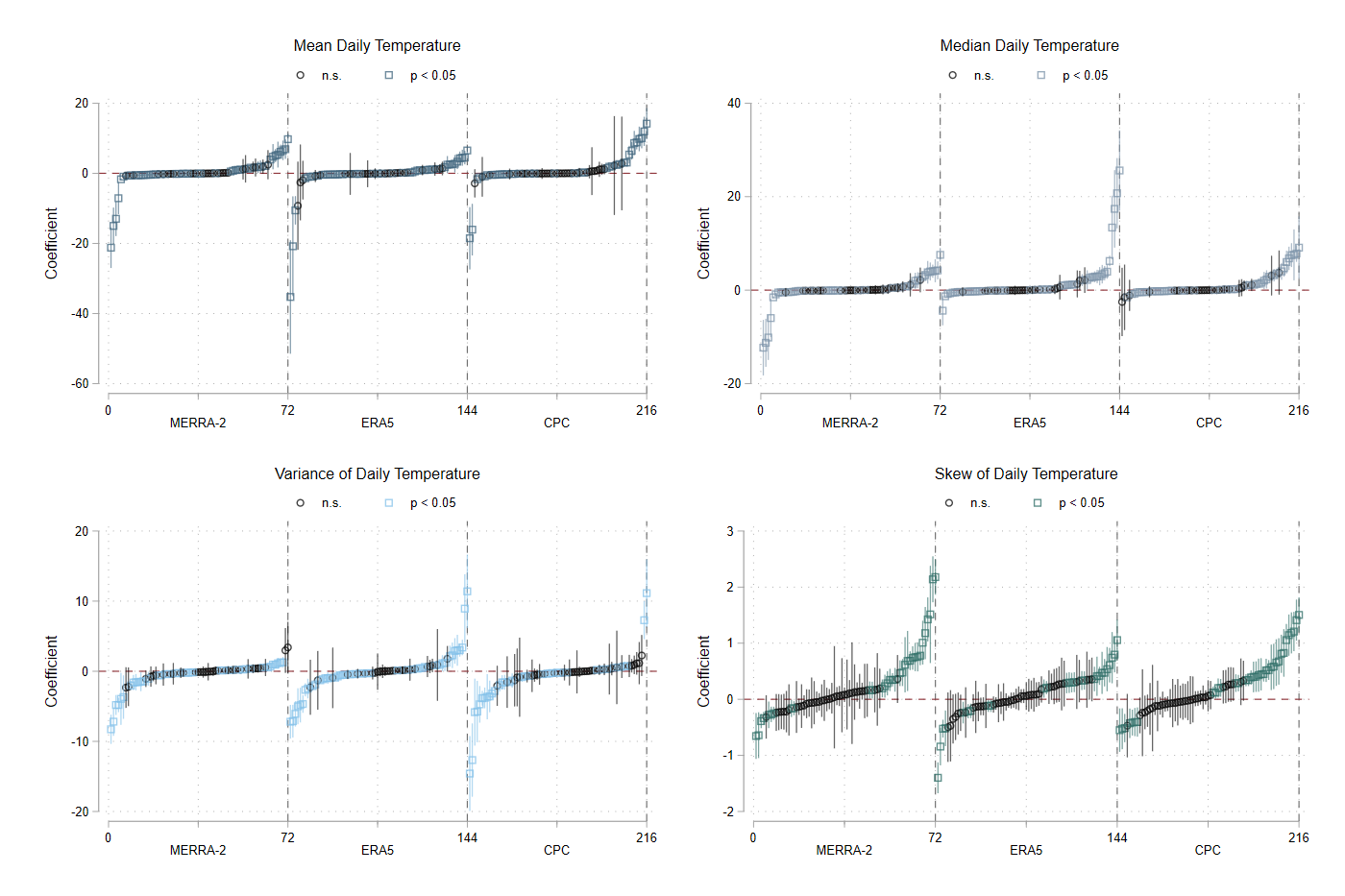}
		\end{center}
		\footnotesize  \textit{Note}: The figure presents the coefficients and confidence intervals for four temperature metrics, by temperature product. Only one extraction method is included. Each panel includes 216 coefficients and confidence intervals (designated on the $x$-axis), with each bin (temperature source) including 72 coefficients and confidence intervals. Each column, as such, represents the findings of a single regressions, e.g. the coefficient and confidence interval itself. The significance level of these are denoted by color and shape of the identifier in the figure, as designated at the top of each panel. 
	\end{minipage}	
\end{figure}

\begin{figure}[!htbp]
	\begin{minipage}{\linewidth}		
		\caption{Coefficients and Confidence Intervals for (1) Total Season, (2) Deviation in Total Season, and (3) z-score of Total Season, by Temperature Product}
		\label{fig:sat_total_tp}
		\begin{center}
			\includegraphics[width=\linewidth,keepaspectratio]{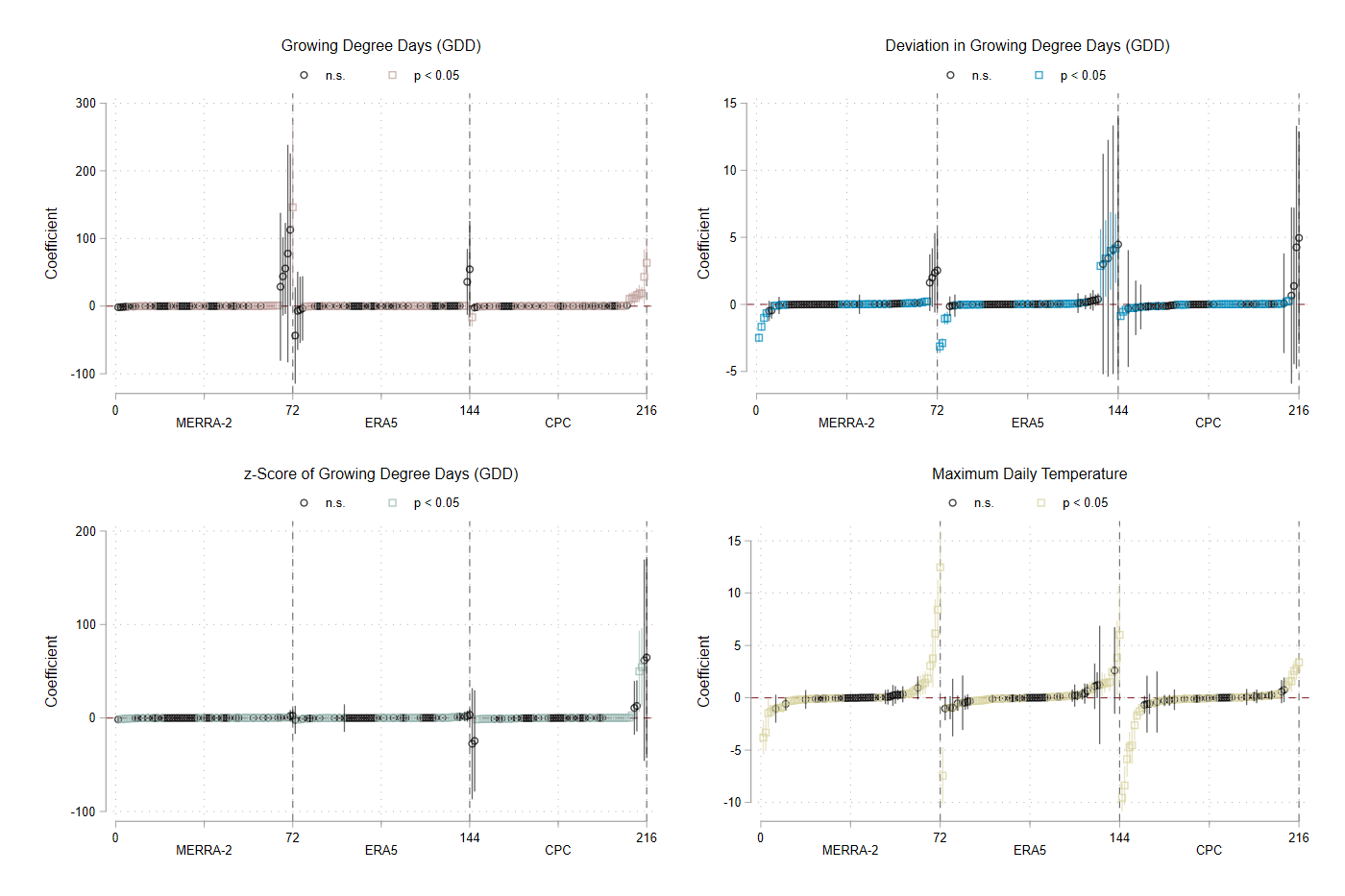}
		\end{center}
		\footnotesize  \textit{Note}: The figure presents the coefficients and confidence intervals for four temperature metrics, by temperature product. Only one extraction method is included. Each panel includes 216 coefficients and confidence intervals (designated on the $x$-axis), with each bin (temperature source) including 72 coefficients and confidence intervals. Each column, as such, represents the findings of a single regressions, e.g. the coefficient and confidence interval itself. The significance level of these are denoted by color and shape of the identifier in the figure, as designated at the top of each panel. 
	\end{minipage}
\end{figure}

To complement and extend the results presented in \ref{sec:satellitepval}, in this section, we include a series of tables and figures (Tables~\ref{tab:sig_var_eth_rf} through~\ref{tab:sig_var_uga_tp} and Figures~\ref{fig:sat_moment_rf} through \ref{fig:sat_total_tp}). First, we display Tables~\ref{tab:sig_var_eth_rf} through~\ref{tab:sig_var_uga_tp} which present the share of significant coefficients for rainfall and temperature, for each country and for each remote sensing product. Next, we display Figures~\ref{fig:sat_moment_rf} through \ref{fig:sat_total_tp}, which present the coefficients and confidence intervals for remote sensing products.


\newpage

\begin{table}\centering
\begin{threeparttable}
    \caption{Share of Significant Rainfall Coefficients in Ethiopia} \label{tab:sig_var_eth_rf}
		\estwide{tables/var_sig_eth_rf}{7}{c}
    \Fignote{\footnotesize The table presents the share of statistically significant rainfall coefficients in Ethiopia.}
\end{threeparttable}
\end{table}

\begin{table}\centering
\begin{threeparttable}
    \caption{Share of Significant Rainfall Coefficients in Malawi} \label{tab:sig_var_mwi_rf}
		\estwide{tables/var_sig_mwi_rf}{7}{c} 
    \Fignote{The table presents the share of statistically significant rainfall coefficients in Malawi.}
\end{threeparttable}
\end{table}

\begin{table}\centering
\begin{threeparttable}
    \caption{Share of Significant Rainfall Coefficients in Niger} \label{tab:sig_var_ngr_rf}
		\estwide{tables/var_sig_ngr_rf}{7}{c}
    \Fignote{The table presents the share of statistically significant rainfall coefficients in Niger.}
\end{threeparttable}
\end{table}

\begin{table}\centering
\begin{threeparttable}
    \caption{Share of Significant Rainfall Coefficients in Nigeria} \label{tab:sig_var_nga_rf}
		\estwide{tables/var_sig_nga_rf}{7}{c}
    \Fignote{The table presents the share of statistically significant rainfall coefficients in Nigeria.}
\end{threeparttable}
\end{table}

\begin{table}\centering
\begin{threeparttable}
    \caption{Share of Significant Rainfall Coefficients in Tanzania} \label{tab:sig_var_tza_rf}
		\estwide{tables/var_sig_tza_rf}{7}{c}
    \Fignote{The table presents the share of statistically significant rainfall coefficients in Tanzania.}
\end{threeparttable}
\end{table}

\begin{table}\centering
\begin{threeparttable}
    \caption{Share of Significant Rainfall Coefficients in Uganda} \label{tab:sig_var_uga_rf}
		\estwide{tables/var_sig_uga_rf}{7}{c}
    \Fignote{The table presents the share of statistically significant rainfall coefficients in Uganda.}
\end{threeparttable}
\end{table}


\newpage

\begin{table}\centering
\begin{threeparttable}
    \caption{Share of Significant Temperature Coefficients in Ethiopia} \label{tab:sig_var_eth_tp}
	    \estwide{tables/var_sig_eth_tp}{4}{c}
    \Fignote{The table presents the share of statistically significant temperature coefficients in Ethiopia.}
\end{threeparttable}
\end{table}

\begin{table}\centering
\begin{threeparttable}
    \caption{Share of Significant Temperature Coefficients in Malawi} \label{tab:sig_var_mwi_tp}
		\estwide{tables/var_sig_mwi_tp}{4}{c}
    \Fignote{The table presents the share of statistically significant temperature coefficients in Malawi.}
\end{threeparttable}
\end{table}

\begin{table}\centering
\begin{threeparttable}
    \caption{Share of Significant Temperature Coefficients in Niger} \label{tab:sig_var_ngr_tp}
		\estwide{tables/var_sig_ngr_tp}{4}{c}
    \Fignote{The table presents the share of statistically significant temperature coefficients in Niger.}
\end{threeparttable}
\end{table}

\begin{table}\centering
\begin{threeparttable}
    \caption{Share of Significant Temperature Coefficients in Nigeria} \label{tab:sig_var_nga_tp}
		\estwide{tables/var_sig_nga_tp}{4}{c}
    \Fignote{The table presents the share of statistically significant temperature coefficients in Nigeria.}
\end{threeparttable}
\end{table}

\begin{table}\centering
\begin{threeparttable}
    \caption{Share of Significant Temperature Coefficients in Tanzania} \label{tab:sig_var_tza_tp}
		\estwide{tables/var_sig_tza_tp}{4}{c}
    \Fignote{The table presents the share of statistically significant temperature coefficients in Tanzania.}
\end{threeparttable}
\end{table}

\begin{table}\centering
\begin{threeparttable}
    \caption{Share of Significant Temperature Coefficients in Uganda} \label{tab:sig_var_uga_tp}
		\estwide{tables/var_sig_uga_tp}{4}{c}
    \Fignote{The table presents the share of statistically significant temperature coefficients in Uganda.}
\end{threeparttable}
\end{table}


\clearpage
\newpage
\section{Linear Combination Results}\label{sec:aplinearcomboresults}

In addition to the primary specifications discussed in the paper, the pre-analysis plan defined a series of linear combinations of rainfall and temperature to be tested. Economists frequently use only rainfall or only temperature in analysis, particularly if weather is used as an instrument or proxy. However, there is a growing trend to use both rainfall and temperature, especially if one is looking to predict agricultural production \citep{TackEtAl12, JagnaniEtAl21}. To determine if our results hinge on the fact that we use only-rainfall or only-temperature, we explore a number of rainfall-temperature combinations. We present these results in-line with our pre-analysis plan. However, we have relegated these results to the appendix, as these do not demonstrate any clear patterns for interpretation.

We test combinations of rainfall and temperature only for the same type of obfuscation. The motivation for doing so is that any publicly available household data set used would provide the same obfuscation method for the coordinates, so no researcher would be combining rainfall and temperature with different extraction methods. For the linear combinations, we estimate each with the linear specification, but only estimate two linear combinations with the quadratic specification. For the variables using only the linear specification, this gives us 6,480 regressions per variable combination (six countries, six precipitation products, three temperature products, ten extraction methods, three specifications, two outcome variables). For the variables using the linear and quadratic specifications, this gives us 12,960 regressions per variable combination. And so, we estimate a total of 51,840 regressions for the combined variables.

\setcounter{table}{0}
\renewcommand{\thetable}{E\arabic{table}}
\setcounter{figure}{0}
\renewcommand{\thefigure}{E\arabic{figure}}


\newpage 

\begin{figure}[!htbp]
	\begin{minipage}{\linewidth}	
		\caption{Coefficients and Confidence Intervals for (1) Mean Seasonal Rainfall combined with (2) Mean Seasonal Temperature, by Country}
		\label{fig:v01_v15_cty}
		\begin{center}
			\includegraphics[width=\linewidth,keepaspectratio]{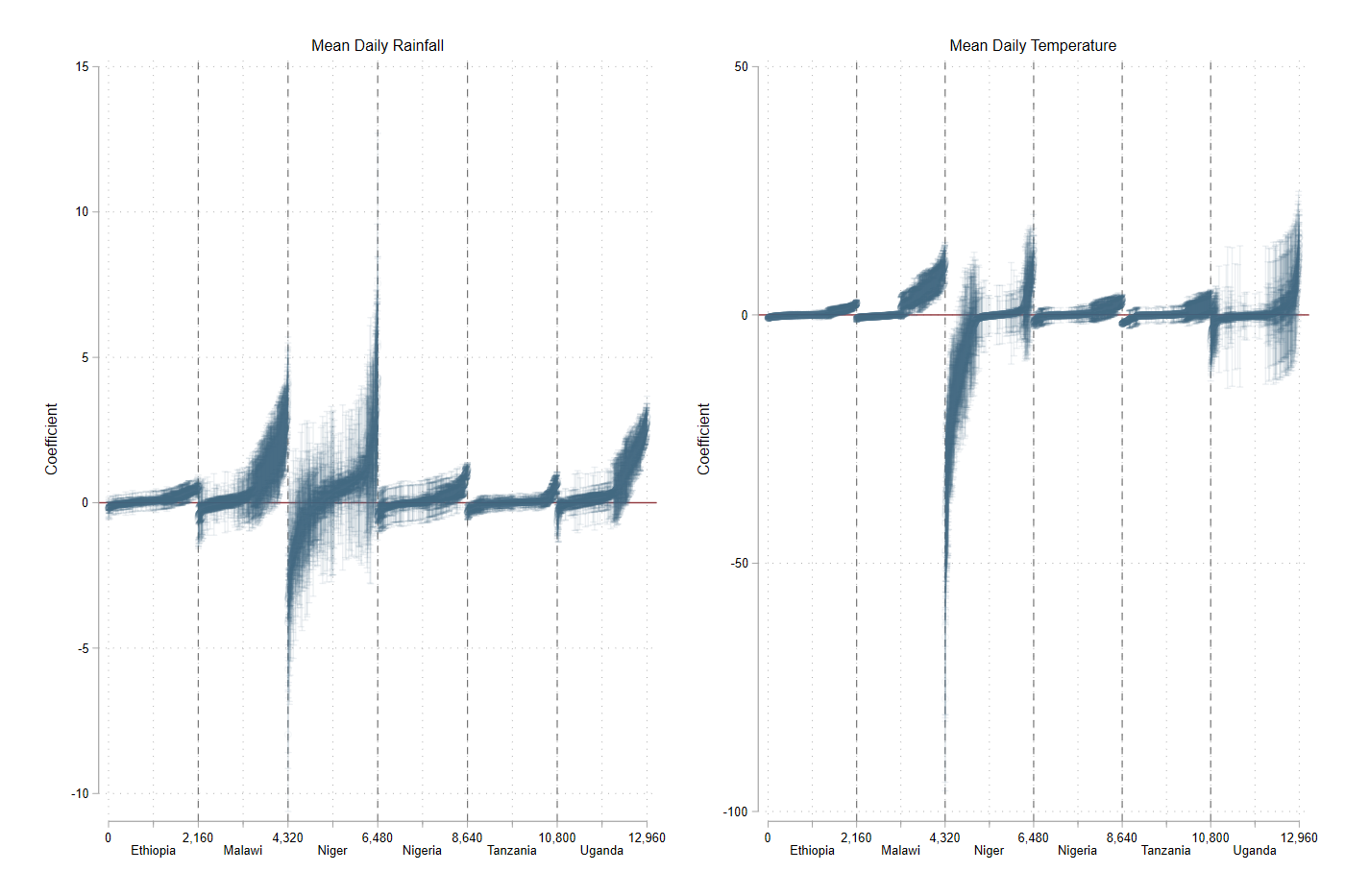}
		\end{center}
		\footnotesize  \textit{Note}: The figure presents the coefficients and confidence intervals for the regression which includes two weather metrics, by country. Each panel includes 12,960 coefficients and confidence intervals (designated on the $x$-axis), with each bin (country source) including 2,160 coefficients and confidence intervals. Each column, as such, represents the findings of a single regressions, e.g. the coefficient and confidence interval itself. 
	\end{minipage}	
\end{figure}

\begin{figure}[!htbp]
	\begin{minipage}{\linewidth}	
		\caption{Coefficients and Confidence Intervals for (1) Median Seasonal Rainfall combined with (2) Median Seasonal Temperature, by Country}
		\label{fig:v02_v16_cty}
		\begin{center}
			\includegraphics[width=\linewidth,keepaspectratio]{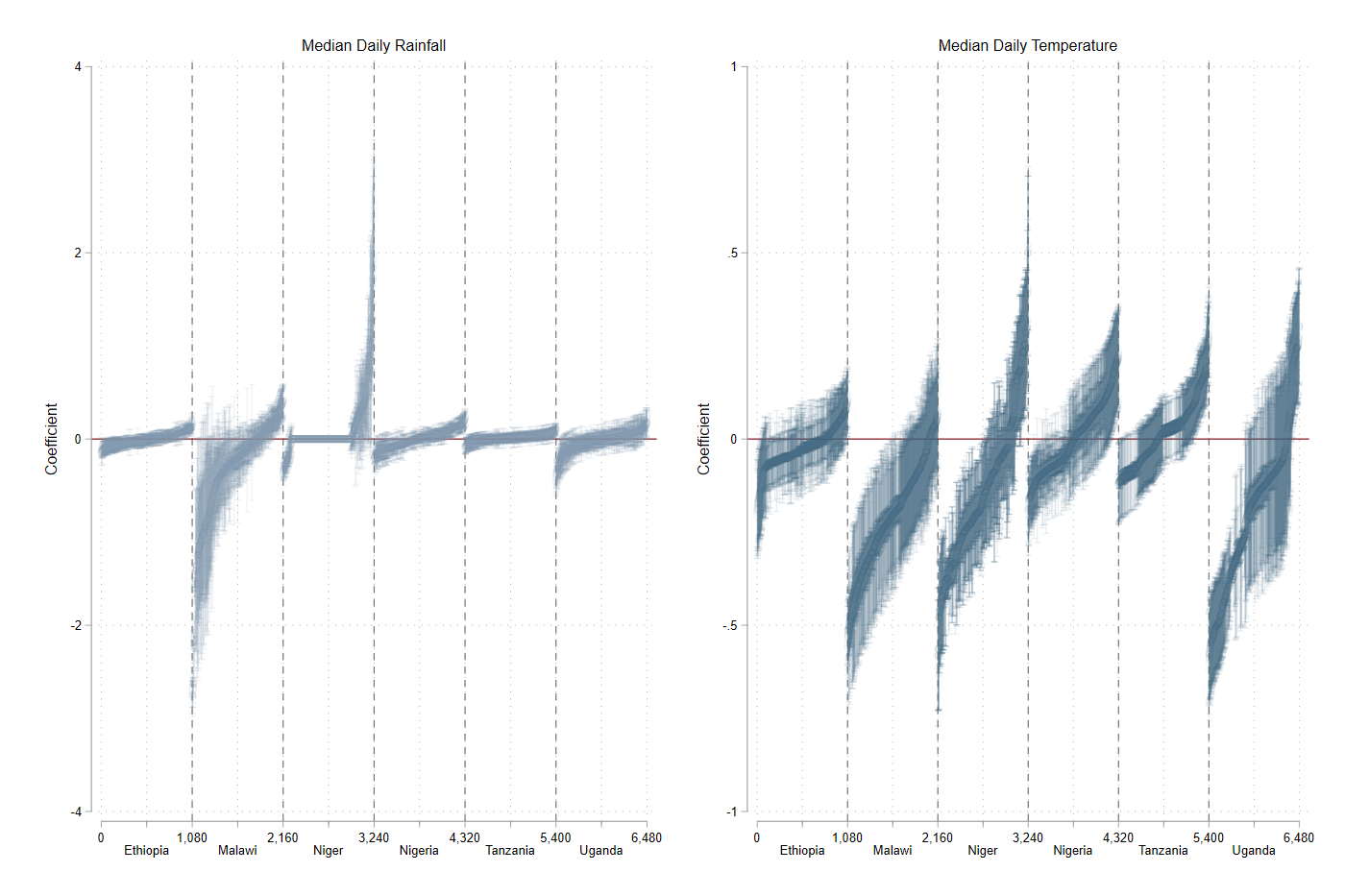}
		\end{center}
		\footnotesize  \textit{Note}: The figure presents the coefficients and confidence intervals for the regression which includes two weather metrics, by country. Each panel includes 6,480 coefficients and confidence intervals (designated on the $x$-axis), with each bin (country source) including 1,080 coefficients and confidence intervals. Each column, as such, represents the findings of a single regressions, e.g. the coefficient and confidence interval itself. 
	\end{minipage}	
\end{figure}

\begin{figure}[!htbp]
\begin{minipage}{\linewidth}	
	\caption{Coefficients and Confidence Intervals for (1) Total Seasonal Rainfall combined with (2) Growing Degree Days (GDD), by Country}
	\label{fig:v05_v19_cty}
	\begin{center}
		\includegraphics[width=\linewidth,keepaspectratio]{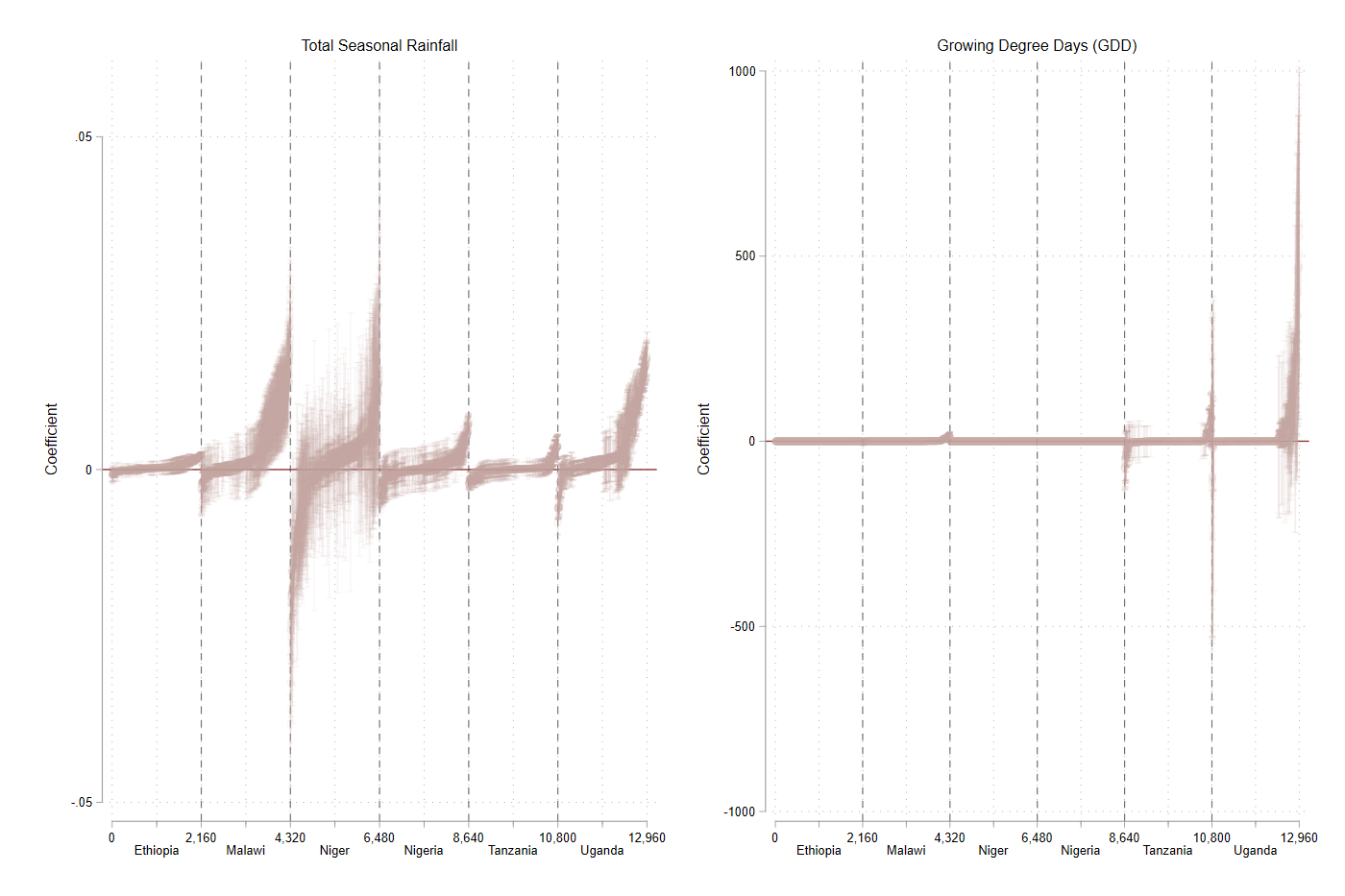}
	\end{center}
	\footnotesize  \textit{Note}: The figure presents the coefficients and confidence intervals for the regression which includes two weather metrics, by country. Each panel includes 12,960 coefficients and confidence intervals (designated on the $x$-axis), with each bin (country source) including 2,160 coefficients and confidence intervals. Each column, as such, represents the findings of a single regressions, e.g. the coefficient and confidence interval itself. 
\end{minipage}	
\end{figure}

\begin{figure}[!htbp]
\begin{minipage}{\linewidth}	
	\caption{Coefficients and Confidence Intervals for (1) z-score of Total Seasonal Rainfall combined with (2) z-score of Growing Degree Days (GDD), by Country}
	\label{fig:v07_v21_cty}
	\begin{center}
		\includegraphics[width=\linewidth,keepaspectratio]{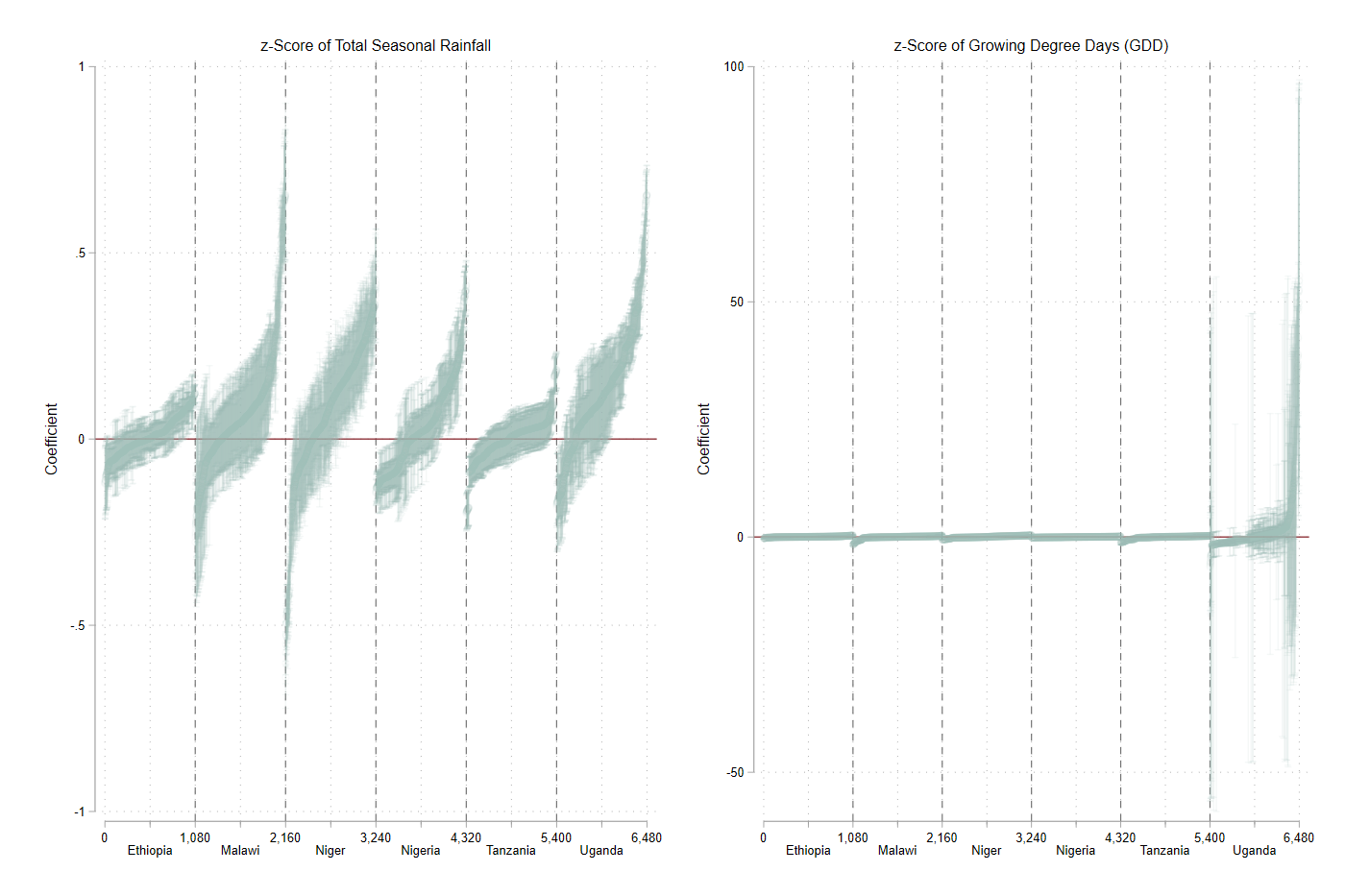}
	\end{center}
	\footnotesize  \textit{Note}: The figure presents the coefficients and confidence intervals for the regression which includes two weather metrics, by country. Each panel includes 6,480 coefficients and confidence intervals (designated on the $x$-axis), with each bin (country source) including 1,080 coefficients and confidence intervals. Each column, as such, represents the findings of a single regressions, e.g. the coefficient and confidence interval itself. 
\end{minipage}	
\end{figure}

\begin{figure}[!htbp]
\begin{minipage}{\linewidth}	
	\caption{Coefficients and Confidence Intervals for (1) First Two Moments (Mean, Variance) of Seasonal Rainfall combined with (2) First Two Moments of Seasonal Temperature, by Country}
	\label{fig:v01_v03_v15_v17_cty}
	\begin{center}
		\includegraphics[width=\linewidth,keepaspectratio]{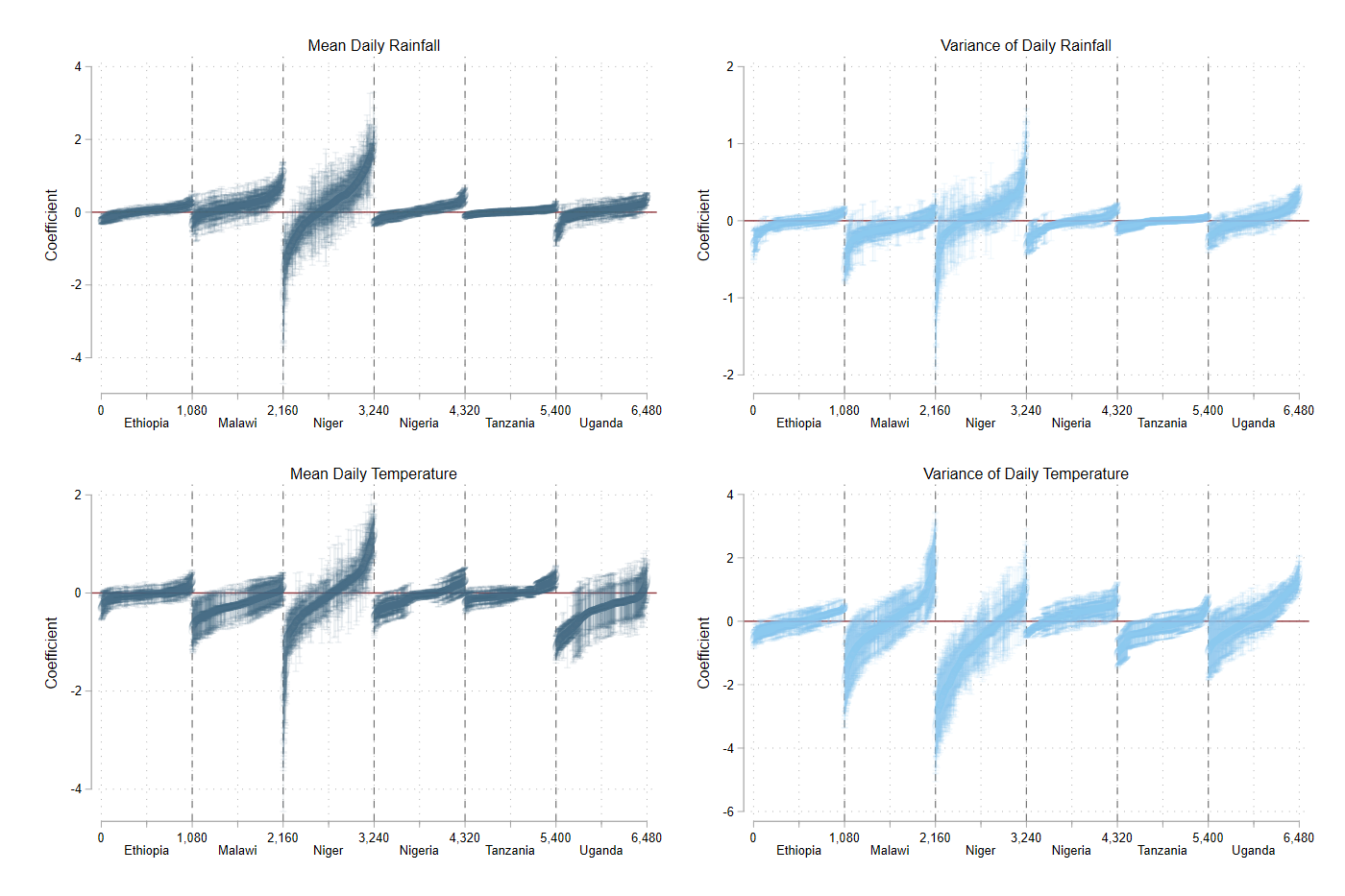}
	\end{center}
	\footnotesize  \textit{Note}:The figure presents the coefficients and confidence intervals for the regression which includes four weather metrics, by country. Each panel includes 6,480 coefficients and confidence intervals (designated on the $x$-axis), with each bin (country source) including 1,080 coefficients and confidence intervals. Each column, as such, represents the findings of a single regressions, e.g. the coefficient and confidence interval itself. 
\end{minipage}	
\end{figure}

\begin{figure}[!htbp]
\begin{minipage}{\linewidth}	
	\caption{Coefficients and Confidence Intervals for (1) First Three Moments (Mean, Variance, Skew) of Seasonal Rainfall combined with (2) First Three Moments of Seasonal Temperature, by Country}
	\label{fig:v01_v03_v04_v15_v17_v18_cty}
	\begin{center}
		\includegraphics[width=\linewidth,keepaspectratio]{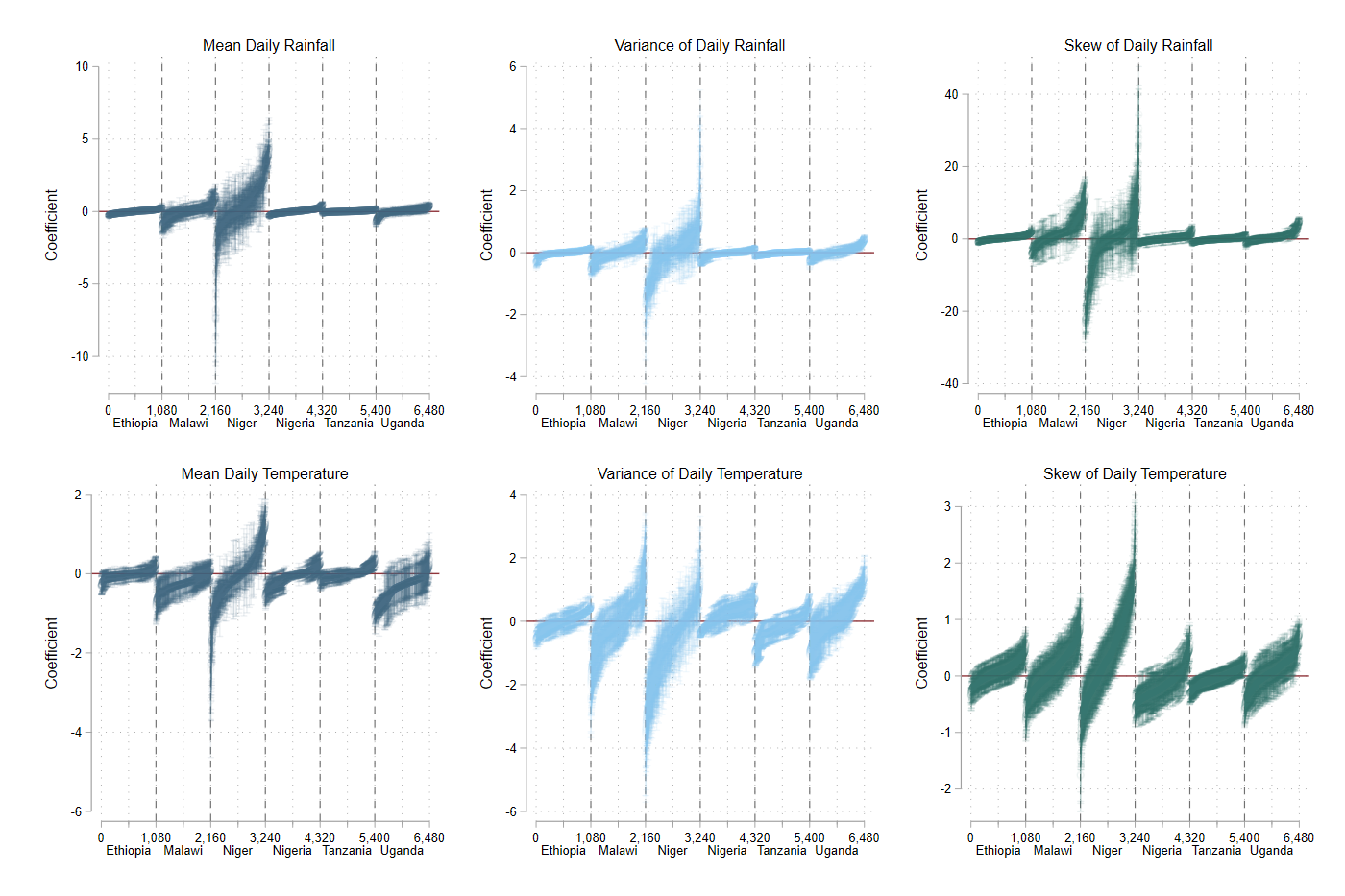}
	\end{center}
	\footnotesize  \textit{Note}: The figure presents the coefficients and confidence intervals for the regression which includes six weather metrics, by country. Each panel includes 6,480 coefficients and confidence intervals (designated on the $x$-axis), with each bin (country source) including 1,080 coefficients and confidence intervals. Each column, as such, represents the findings of a single regressions, e.g. the coefficient and confidence interval itself. 
\end{minipage}	
\end{figure}

\end{document}